\documentclass[english, msc, oneside]{layout/observatory-thesis}

\usepackage{lipsum} 
\usepackage{blindtext} 
\usepackage{pdfpages}
\usepackage{derivative}
\usepackage{physics}

\DeclareSIUnit\rad{\ensuremath{\mathrm{rad}}}
\DeclareSIUnit\elementarycharge{\ensuremath{e}}                     
\DeclareSIUnit\clight{\text {\ensuremath {c}}_{0}}                  
\DeclareSIUnit\electronmass{\text {\ensuremath {m}}_{\mathrm {e}}}  
\DeclareSIUnit\angstrom{\ensuremath{ \mathrm{{\AA}}}}
\DeclareSIUnit\parsec{\ensuremath{\mathrm{pc}}}
\DeclareSIUnit\au{\ensuremath{\mathrm{AU}}}
\DeclareSIUnit\mas{\ensuremath{\mathrm{mas}}}

\DeclareSIUnit\msun{{M\ensuremath{_\odot}}}
\DeclareSIUnit\rsun{{R\ensuremath{_\odot}}}
\DeclareSIUnit\lsun{{L\ensuremath{_\odot}}}

\DeclareSIUnit\mearth{{M\ensuremath{_\oplus}}}
\DeclareSIUnit\rearth{{R\ensuremath{_\oplus}}}
\DeclareSIUnit\learth{{L\ensuremath{_\oplus}}}

\DeclareSIUnit\mjup{{M\ensuremath{_\mathrm{J}}}}
\DeclareSIUnit\rjup{{R\ensuremath{_\mathrm{J}}}}
\DeclareSIUnit\ljup{{L\ensuremath{_\mathrm{J}}}} 

\newacronym{asteca}{ASteCA}{Automated Stellar Cluster Analysis}
\newacronym[shortplural=COM, longplural=centers of mass]{com}{COM}{center of mass}
\newacronym{gc}{GC}{globular cluster}
\newacronym{pal5}{Pal 5}{Palomar 5}
\newacronym{m5}{M5}{Messier 5}
\newacronym{m6}{M6}{Messier 6, also known as the Butterfy Cluster}
\newacronym{m31}{M31}{Messier 31, also known as the Andromeda Galaxy}
\newacronym{pca}{PCA}{principal component analysis}
\newacronym[shortplural=HR-diagrams, longplural=Hertzsprung-Russelldiagrams]{hr}{HR-diagrams}{Hertzsprung-Russelldiagram}
\newacronym{vlt}{VLT}{Very Large Telescope}
\newacronym{vlbi}{VLBI}{very-long-baseline interferometry}
\newacronym{lsst}{LSST}{Legacy Survey of Space and Time}
\newacronym{hst}{HST}{Hubble Space Telescope}
\newacronym{alma}{ALMA}{Atacama Large Millimeter Array}
\newacronym[shortplural=AGN, longplural=active galactic nuclei]{agn}{AGN}{active galactic nucleus}
\newacronym{ska}{SKA}{Square Kilometre Array}
\newacronym{2mass}{2MASS}{Two-Micron All Sky Survey}
\newacronym{cmb}{CMB}{cosmic microwave background}
\newacronym{mw}{MW}{Milky Way}

\glsxtrnewsymbol
    [type=constants,
    description={gravitational constant},
    symbol={\si{\num{6.674e-11}\unit{m^3.kg^{-1}.s^{-2}}}}]
    {c:G}
    {\ensuremath{G}}
    
\glsxtrnewsymbol
    [type=constants,
    description={speed of light},
    symbol={\si{\num{2.998e8}\unit{m.s^{-1}}}}]
    {c:c}
    {\ensuremath{c}}
    
\glsxtrnewsymbol
    [type=constants,
    description={constant of Stefan-Boltzmann},
    symbol={\si{\num{5.670e-8}\unit{W.m^{-2}.K^{-4}}}}]
    {c:sigmaSF}
    {\ensuremath{\sigma_{SB}}}
    
\glsxtrnewsymbol
    [type=constants,
    description={constant of Plack},
    symbol={\si{\num{6.626e-34}\unit{J.Hz^{-1}}}}]
    {c:plack}
    {\ensuremath{h}}

\glsxtrnewsymbol
    [type=constants,
    description={constant of Boltzmann},
    symbol={\si{\num{1.381e-23}\unit{J.K^{-1}}}}]
    {c:boltzmann}
    {\ensuremath{k}}
    
\glsxtrnewsymbol
    [type=constants,
    description={atomic mass unit},
    symbol={\si{\num{1.661e-27}\unit{kg}}}]
    {c:m0}
    {\ensuremath{m_0}}
    
\glsxtrnewsymbol
    [type=constants,
    description={rest mass of a proton},
    symbol={\si{\num{1.673e-27}\unit{kg}}}]
    {c:mp}
    {\ensuremath{m_p}}

\glsxtrnewsymbol
    [type=constants,
    description={rest mass of an electron},
    symbol={\si{\num{9.109e-31}\unit{kg}}}]
    {c:me}
    {\si\electronmass}

\glsxtrnewsymbol
    [type=constants,
    description={elementary charge},
    symbol={\si{\num{1.602e-19}\unit{C}}}]
    {c:e}
    {\si\elementarycharge}

\glsxtrnewsymbol
    [type=constants,
    description={vacuum permittivity},
    symbol={\si{\num{8.854e-12}\unit{C^2.kg^{-1}.s^2}}}]
    {c:eps0}
    {\ensuremath{\epsilon_0}}

\glsxtrnewsymbol
    [type=constants,
    description={Thomson crossection of an electron},
    symbol={\si{\num{6.652e-29}\unit{m^2}}}]
    {c:sigmaT}
    {\ensuremath{\sigma_e}}
    
\glsxtrnewsymbol
    [type=constants,
    description={gas constant},
    symbol={\si{\num{8.314}\unit{J.K^{-1}.mol}}}]
    {c:Rgas}
    {\ensuremath{R}}

\glsxtrnewsymbol
    [type=constants,
    description={constant of Avogadro},
    symbol={\si{\num{6.022e23}\unit{mol^{-1}}}}]
    {c:NA}
    {\ensuremath{N_A}}
    
\glsxtrnewsymbol
    [type=constants,
    description={absolute visual magnitude of the sun},
    symbol={\si{\num{4.83}}}]
    {c:Mvsun}
    {\ensuremath{M_{V \odot}}}

\glsxtrnewsymbol
    [type=constants,
    description={apparent visual magnitude of the sun},
    symbol={\si{\num{-26.75}}}]
    {c:mvsun}
    {\ensuremath{m_{V \odot}}}

\glsxtrnewsymbol
    [type=constants,
    description={absolute B-band magnitude of the sun},
    symbol={\si{\num{5.48}}}]
    {c:Mbsun}
    {\ensuremath{M_{B \odot}}}
    
\glsxtrnewsymbol
    [type=constants,
    description={effective temperature of the sun},
    symbol={\si{\num{5778}\unit{K}}}]
    {c:Tsun}
    {\ensuremath{T_{\odot}}}

\glsxtrnewsymbol
    [type=units,
    description={specific intensity},
    symbol={\si{\unit{J.s^{-1}.Hz^{-1}.m^{-2}.sr^{-1}}}}]
    {u:specradiancehz} 
    {\ensuremath{\mathrm{I}_\nu}} 
  
\glsxtrnewsymbol
    [type=units,
    description={specific intensity},
    symbol={\si{\unit{J.s^{-1}m^{-3}.sr^{-1}}}}]
    {u:specradiance} 
    {\ensuremath{\mathrm{I}_\lambda}} 
  
\glsxtrnewsymbol
    [type=units,
    description={intensity},
    symbol={\si{\unit{J.s^{-1}.m^{-2}.sr^{-1}}}}]
    {u:radiance} 
    {\ensuremath{\mathrm{I}}} 
    
\glsxtrnewsymbol
    [type=units,
    description={spectral flux},
    symbol={\si{\unit{J.s^{-1}.Hz^{-1}.m^{-2}}}}]
    {u:specfluxhz} 
    {\ensuremath{\mathrm{F}_\nu}} 

\glsxtrnewsymbol
    [type=units,
    description={spectral flux},
    symbol={\si{\unit{J.s^{-1}.m^{-3}}}}]
    {u:specflux} 
    {\ensuremath{\mathrm{F}_\lambda}} 

\glsxtrnewsymbol
    [type=units,
    description={flux},
    symbol={\si{\unit{J.s^{-1}.m^{-2}}}}]
    {u:flux} 
    {\ensuremath{\mathrm{F}}} 

\glsxtrnewsymbol
    [type=units,
    description={luminosity},
    symbol={\si{\unit{J.s^{-1}}}}]
    {u:lum} 
    {\ensuremath{\mathrm{L}}} 

\glsxtrnewsymbol
    [type=units,
    description={\r{A}ngstr\"{o}m},
    symbol={\si{\num{e-10}\unit{m}}}]
    {u:angstrom} 
    {\si\angstrom} 
    
\glsxtrnewsymbol
    [type=units,
    description={parsec},
    symbol={\si{\num{3.086e16}\unit{m}}}]
    {u:parsec} 
    {\si\parsec} 
    
\glsxtrnewsymbol
    [type=units,
    description={astronomical unit},
    symbol={\si{\num{1.496e11}\unit{m}}}]
    {u:au} 
    {\si\au} 

\glsxtrnewsymbol
    [type=units,
    description={opacity},
    symbol={\si{\unit{m^2.kg^{-1}}}}]
    {u:opacity} 
    {\ensuremath{\kappa}}

\glsxtrnewsymbol
    [type=units,
    description={solar mass},
    symbol={\si{\num{1.988e30}\unit{kg}}}]
    {u:msun} 
    {\si\msun}
  
\glsxtrnewsymbol
    [type=units,
    description={solar radius},
    symbol={\si{\num{6.957e8}\unit{m}}}]
    {u:rsun} 
    {\si\rsun}

\glsxtrnewsymbol
    [type=units,
    description={solar luminosity},
    symbol={\si{\num{3.828e26}\unit{J.s^{-1}}}}]
    {u:lsun} 
    {\si\lsun}
 
\glsxtrnewsymbol
    [type=units,
    description={Earth mass},
    symbol={\si{\num{5.972e24}\unit{kg}}}]
    {u:mearth} 
    {\si\mearth}
  
\glsxtrnewsymbol
    [type=units,
    description={Earth radius},
    symbol={\si{\num{6.378e6}\unit{m}}}]
    {u:rearth} 
    {\si\rearth}

\glsxtrnewsymbol
    [type=units,
    description={Jupiter mass},
    symbol={\si{\num{1.898e27}\unit{kg}}}]
    {u:mjup} 
    {\si\mjup}
  
\glsxtrnewsymbol
    [type=units,
    description={Jupiter radius},
    symbol={\si{\num{7.149e7}\unit{m}}}]
    {u:rjup} 
    {\si\rjup}  
    
\glsxtrnewsymbol
    [type=units,
    description={arcsecond},
    symbol={\si{\frac{1}{3600}\unit{\degree}}}]
    {u:arcsecond} 
    {\si\arcsecond}

\glsxtrnewsymbol
    [type=units,
    description={arcminute},
    symbol={\si{\frac{1}{60}\unit{\degree}}}]
    {u:arcminute} 
    {\si\arcminute}

\glsxtrnewsymbol
    [type=units,
    description={milliarcsecond},
    symbol={\si{\num{0.001}\unit{\arcsecond}}}]
    {u:mas} 
    {\si\mas}

\glsxtrnewsymbol
    [type=units,
    description={billion years (gigayear)},
    symbol={\si{\num{1e9}\unit{year}}}]
    {u:gyr} 
    {\ensuremath{\mathrm{Gyr}}}

\glsxtrnewsymbol
    [type=units,
    description={electronvolt},
    symbol={\si{\num{1.602e-19}\unit{\joule}}}]
    {u:ev} 
    {\si\electronvolt}

\glsxtrnewsymbol
    [type=units,
    description={density},
    symbol={\si{}\unit{kg.m^{-3}}}]
    {u:dichtheid} 
    {\ensuremath{\rho}}
\makeglossaries

\begin{document}


\title{The trans- and post-capture orbital evolution of Triton}
\author{Quirijn B. van Woerkom}
\studentid{s3366766}
\supervisor{Dr M. Rovira-Navarro}
\secondsupervisor{Dr Y. Miguel} 
\projectstart{September 2023}
\projectend{June 2024}
\affiliation{Leiden Observatory, Leiden University}
\address{P.O. Box 9513, 2300 RA Leiden, The Netherlands}
\coverimage{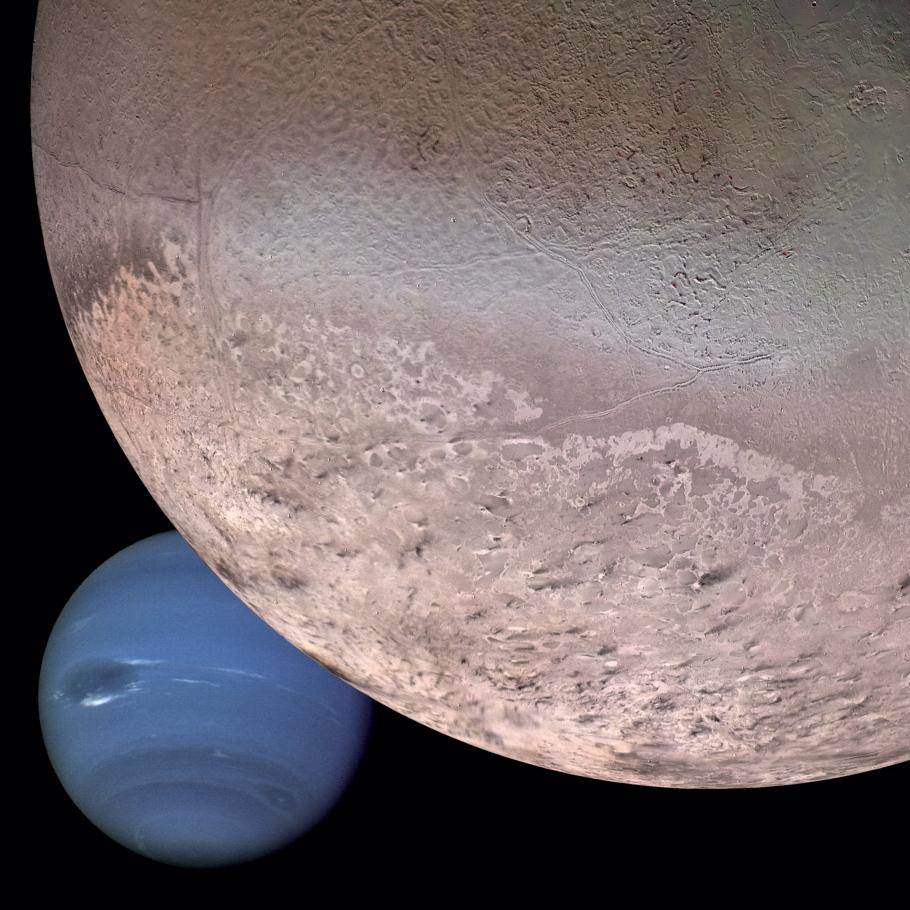}
\covercredit{NASA-PIA00340, NASA/JPL/USGS} 



\frontmatter 

\makecover 
\maketitle 

\newpage
\hspace{0pt}
\vfill
\begin{flushright}
    \textit{And [Triton] rose up from the depths in form such as he really was (\ldots); so the god, seizing hollow Argo's keel, guided her onward to the sea.}
\end{flushright}
\begin{flushright}
    Apollonius Rhodius, The Argonautica, Book IV (tr. by R. C. Seaton)
\end{flushright}
\vfill
\hspace{0pt}

\chapter{Abstract}
Triton is a unique moon in our Solar System, being the only large moon to orbit on a retrograde and highly inclined orbit. As a result, it is thought that it did not form around Neptune, but rather was captured from heliocentric orbit. The resulting capture orbit is likely to have been highly eccentric, with extreme tidal heating sufficient to melt Triton's mantle several times over as a consequence. Attempts have been made at simulating this highly dynamic epoch in the past, yet all such work has required simplifying assumptions or application (at times unwittingly) of mathematical methods outside of the domain in which they are well-behaved.

In this work, we revisit the description of this period of Triton's history, by developing methods that allow us to simulate high-eccentricity spin-orbit evolution for an arbitrary rheological model for the first time. Our aim is to provide a framework on which future work can build with more detailed planetological models, while still capturing the full intricacies of high-eccentricity tidal evolution.

This is done by starting from the Darwin-Kaula expansion in full, and foregoing the simplifying assumptions used in past work. Rather, we methodically determine the convergence properties of each infinite sum in the expansion, and truncating them appropriately. We achieve this by developing a new conservative empirical upper bound on the expansion into eccentricity functions of part of the tidal potential using analytical methods, and by introducing a novel, fast-converging power series expansion for these eccentricity functions borrowed from artificial satellite theory. Afterwards, we use this new tool to analyse the consequences of the assumptions made in previous work on Triton, and to propose a dynamically consistent history of Triton.

We find that the use of the constant time lag-rheology model, preferred in the past for the finite expressions it yields for high-eccentricity dynamical evolution, fails to match the capture into spin-orbit resonances we would realistically expect from a sufficiently viscous icy body at non-zero eccentricities. Additionally, we find that Triton can, over periods of at least millions of years, have experienced tidal heating rates orders of magnitude greater even than present-day Io. Based on these tidal heating rates, we predict that Triton likely possessed a massive Titan-like atmosphere throughout its entire tidal evolution, with a surface or thin-shell ocean. Whether this is consistent with maintenance of these tidal heating rates, or whether this would significantly extend the epoch of tidal heating will be the subject of future work.
\chapter{Acknowledgements}
What lay in front of you forms a symbolic summary of my time at Leiden University: starting with a desire to expand my background in astronomy and planetary science, and ending with what you find here. I could never have guessed how that decision has shaped my career, interests, and very likely my future.

This was made possible, fortunately, not just by my own work. No small part of this is be owed to those who guided me academically, who I therefore wish to thank explicitly: Elina, for introducing me to tidally heated (exo)moons and their community, and indulging me in my initial fascination with Triton that led to this project; Yamila, for being flexible in your capacity as internal supervisor and allowing us to take this project the way it has; and of course Marc, for guiding me on this journey, allowing me to shape it into my own while providing me with invaluable reflection and feedback during our weekly meetings. This is as much your work as it is mine. Additionally, I would like to thank all the staff at TU Delft's space department, who treated me at times more like a colleague than a student, inviting me to group meetings, discussing their research with me (and, of course, joining the space bars).

Additionally, I would like to acknowledge the help and support of my friends: from those that have been with me since childhood to those whose acquaintance is more recent, your support has been invaluable. From coffee breaks that sometimes ended up taking longer than the time-span between them to late Friday- or Wednesday-nights, memories have been wonderful, and I look forward to continuing to make them.

Finally, I am indebted to my family for supporting me through all this. To my parents: while the journey is not over yet, you have always been supportive of my decisions, even if that meant spending another three years at university studying matters whose practical applications aren't immediately clear. To my brother, I echo the sentiment you left me in your graduating work: as my mirror and my conscience, let us keep pushing each other.

\begin{flushright}
    \textit{Quirijn Benjamin van Woerkom} \\
    \textit{Delft, June 2024}
\end{flushright}
\tableofcontents


\mainmatter 

\chapter{Introduction}
\label{ch:introduction}
Triton is the only large moon of Neptune, and it orbits on a retrograde, highly inclined yet almost circular orbit. Rather than forming in a circumplanetary disc, as regular satellites of giant planets are hypothesised to do (e.g. \citealt{Canup2006APlanets, Szulagyi2018InNeptune, Batygin2020FormationSatellites}), Triton has therefore been thought to have been formed on a heliocentric orbit as part of a binary trans-Neptunian object pair \citep{Agnor2006NeptunesEncounter}, after which it was captured onto a highly eccentric ($e>0.9$) orbit around Neptune, in the process of which it will have dissipated sufficient energy to melt its ice mantle several times over, with catastrophic consequences \citep{Nogueira2011ReassessingTriton, McKinnon2014Triton}.

Additionally, several features of Triton lead us to believe it is still undergoing geological activity at present: the presence of extensive tectonic features \citep{Smith1989VoyagerResults, Croft1993TectonicsTriton}, the existence of fans and plumes on its surface \citep{Hofgartner2022HypothesesTests}, as well as a surface age less than $50$ Myr over the entire surface \citep{Schenk2007OnTriton}. Moreover, a prominent unit of surface morphology that appears to be the oldest terrain found on Triton, termed "cantaloupe terrain" for its appearance, has been hypothesised to be the result of active diapirism across the Tritonian surface \citep{Schenk1993DiapirismInstability}. The most likely culprit by which all this geological activity is forced are tides in Triton's subsurface ocean, driven by its non-zero obliquity \citep{Nimmo2015PoweringGeology}. This obliquity, in turn, is caused by the inclination of Triton's orbit \citep{Correia2009SecularTriton}, which can be traced back to its capture origin.

Hence, both the past and present conditions on Triton are intimately linked to its extra-Neptunian origin, which might have caused it to have a surface ocean overlaid with a thick Titan-like atmosphere in the past \citep{Lunine1992ATriton, McKinnon2014Triton}, but certainly allow it to maintain a large subsurface ocean and vigorous geological activity into the present \citep{Gaeman2012SustainabilityInterior, Nimmo2015PoweringGeology}. This makes Triton as a geological object, as well as its dynamical history, of major interest to planetary scientists, and so it is no surprise that it has been the subject of several dynamical (e.g. \citealt{Chyba1989TidalSystem, Agnor2006NeptunesEncounter, Correia2009SecularTriton, Nogueira2011ReassessingTriton}) and combined interior-orbital \citep{Ross1990TheTriton} studies in the past. Nonetheless, we choose to look at it again: what is our motivation?

Several observations make us aware that the work on Triton present in literature has several shortcomings and blind spots, or has as of yet failed to address new developments in the scientific community. Some (non-exhaustive) examples:
\begin{itemize}
    \item Past studies on Triton have assumed a particular interior model for Triton (the constant time lag model) not out of physical motivations, but because of the simplifications it permits mathematically (e.g. \citealt{Correia2009SecularTriton, Nogueira2011ReassessingTriton}).
    \item Commonly used tidal heating expressions (see e.g. \citealt{Ross1990TheTriton, Lunine1992ATriton}) are only valid for $e\lesssim0.1$, and underestimate tidal effects by orders of magnitude beyond that \citep{Renaud2021TidalTRAPPIST-1e} - past studies largely did not account for this, or neglected to analyse the high-eccentricity part of Triton's past as a result.
    \item Non-zero eccentricities in tidal systems with realistic rheological properties lead to capture into and rapid decay between half-integer spin-orbit resonances greater than 1:1 \citep{Walterova2020ThermalExoplanets, Renaud2021TidalTRAPPIST-1e}: this effect has not yet been analysed for early Triton.
    \item The dynamical equations originally derived by \citet{Kaula1961AnalysisSatellites, Kaula1964TidalEvolution} contain a significant error in the inclination-evolution term, that was only corrected in recent years by \citet{Boue2019TidalElements}.
    \item Recent work has shown that moon migration can lead to tilting of a host planet (e.g. \citealt{Saillenfest2022TiltingSatellite, Saillenfest2023ObliqueRadii, Wisdom2022LossRings}) - this, combined with the previous fact and the knowledge that Neptune has significant obliquity ($\sim30^{\circ}$), identifies Triton as a reasonable candidate culprit.
\end{itemize}
We therefore set out to organise a new analysis of Triton's dynamical evolution that aims to address these and other issues. In the process, we implement a novel tidal evolution model that can be applied to a satellite at arbitrary eccentricity whose tidal response is dictated by an arbitrary rheological model: we hope that this can serve as a guidebook for others with the intent to simulate the evolution of highly eccentric orbits in the future.

This report is structured as follows: we will first give a brief overview of the literature and data that exists on Triton in Ch.~\ref{ch:scientific_background}. Afterwards, we present our methods in Ch.~\ref{ch:kaula_theory}, and our results in Ch.~\ref{ch:validation}, \ref{ch:spin-orbit_chains} and \ref{ch:initial_conditions}; finally, we place our findings into the context of related literature in Ch.~\ref{ch:discussion}, and conclude the report in Ch.~\ref{ch:conclusions_recommendations}.
\chapter{Scientific Background}
\label{ch:scientific_background}
Before we dive into the descriptions of the additions this work has made to the body of knowledge on Triton and high-eccentricity tidal evolution, it is important to keep in mind the appropriate scientific background. This chapter will serve to give a brief review of the status of current knowledge on Triton, the available observations and the current hypotheses that attempt to explain those observations. To do so, we start with an overview of the observational history of Triton in Sec.~\ref{sec:observational_history}; we discuss what evidence exists for a geologically active Triton at present in Sec.~\ref{sec:geological_active_Triton}, and discuss the various explanations thereof that have been levied in Sec.~\ref{sec:explaining_interior}. Finally, we present the methods by which Triton may have been set upon the inclined, retrograde orbit on which it is currently found in Sec.~\ref{sec:producing_inclined_retrograde_triton}, and describe the elements that are to be synthesised into a consistent history for Triton in Sec.~\ref{sec:toward_consistent_history}. Finally, we present the research questions to be answered in this work, as well as the structure of the manner in which we do so, in Sec.~\ref{sec:research_questions}.

\section{The observational history of Triton}
\label{sec:observational_history}
The observational history of Triton, compared to most other large Solar System moons, is unfortunately relatively sparse. The distance of the Neptunian and Uranian systems as well as the existence of only a single large moon around Neptune make dedicated missions to the ice giants less cost-effective science-wise than visits to the closer Jovian and Saturnian systems. As a result, only \textit{Voyager 2} ever visited the Neptunian system on its grand tour of the giant planets (the trajectory of \textit{Voyager 1} had bent out of the ecliptic after visiting Titan, never to visit the ice giants), and so the history of our knowledge of Triton can roughly speaking be defined by the watershed moment of the \textit{Voyager 2} visit. We will therefore shortly discuss the observational history of Triton before (Sec.~\ref{sec:before_voyager}) and after (Sec.~\ref{sec:after_voyager}) the \textit{Voyager 2} visit.

\subsection{Triton before \textit{Voyager}}
\label{sec:before_voyager}
Guided by the predictions of Urbain Le Verrier, Neptune was first observed in 1846; fewer than three weeks later, William Lassell identified Triton orbiting it \citep{McKinnon2014Triton}. Unlike the other planets, however, which were at the time known to host multiple satellites, no other satellites were identified around Neptune in the century that followed \citep{Buratti2014PlanetarySatellites}. To add to this mystery, Triton was found to orbit retrograde to Neptune's sense of spin and orbital motion \citep{Hind1854OnNeptune}, in which it was unique at the time \citep{Buratti2014PlanetarySatellites}.

This bizarre sense of motion, and the 1930 discovery of Pluto on an orbit that crosses that of Neptune for part of the Plutonian year, led \citet{Lyttleton1936OnSystem} to hypothesise that Pluto was originally a satellite of Neptune, and that the dynamically catastrophic process in which Pluto was ejected was similarly responsible for Triton's retrograde orbit. While this is no longer the preferred theory for the origin of Triton and its orbit, this was the first in a long history of hypothesised origins for Triton that have been discussed in the past century: we will discuss these in further detail in Sec.~\ref{sec:producing_inclined_retrograde_triton}.

The distance of the Neptune-Triton system and the consequent faintness of the moon made it difficult to understand much more about the moon in the half-century that followed. It was, however, recognised that Triton had a place among the larger moons in the Solar System; in fact, brightness-based estimates in the mid-twentieth century presumed Triton to be \textit{the} largest and most massive moon in the Solar System, though these estimates turned out to overestimate Triton's mass by nearly an order of magnitude (e.g. \citealt{McCord1966DynamicalSystem}). Consequently, even the scarce information on Triton's composition that was available in the form of density estimates was unreliable (e.g. \citealt{Lewis1973CHEMISTRYSYSTEM}). The discovery of ices on Triton's surface in 1978 made clear, however, that it was in fact a smaller, brighter body than previously thought \citep{McKinnon2014Triton}. Lyttleton's hypothesis of a Plutonian origin for Triton's retrograde orbit also started to come under scrutiny at this time, as growing knowledge on Pluto required increasingly ad-hoc additions to the theory (see e.g. \citealt{Farinella1979TidalSystem}, or \citealt{Harrington1979ThePluto} and the rebuttal thereof by \citealt{Farinella1980SomePluto}). Slowly, instead, the idea of satellite \textit{capture} rather than satellite \textit{ejection} started to be discussed as the sculpting mechanism for the Neptunian system (e.g. \citealt{McCord1966DynamicalSystem, Pollack1979GasCapture, Farinella1980SomePluto}), until \citet{McKinnon1984OnPluto} published a convincing treatise on the necessity of Triton being captured, which he recognised would have spectacular thermal consequences. Later, \citet{McKinnon1988PlutosNebula} showed that Pluto's makeup is chemically more consistent with a heliocentric rather than circumplanetary origin, strengthening this alternative history for Triton.

In recognition of this spectacular past, and in anticipation of the \textit{Voyager 2} visit, several groups then considered the implications of a tidally active past of Triton in the months leading up to the \textit{Voyager 2} encounter (e.g. \citealt{Goldreich1989NeptunesStory, Chyba1989TidalSystem}). Notably, to the author's knowledge, \citet{Jankowski1989OnTriton} made the first mention of obliquity tides as a potential contributor to tidal activity on Triton. More detailed consideration of these matters would have to wait until after the \textit{Voyager 2} visit to Triton, however, which would finally yield resolved imagery of Triton.

\subsection{Triton after \textit{Voyager}}
\label{sec:after_voyager}
And \textit{Voyager 2} delivered as promised; the data it sent back finally gave a face to the moon that William Lassell had discovered almost 146 years earlier. Triton turned out to be more reflective and therefore even smaller than had been thought prior, still; finally, \textit{Voyager 2} fixed the mass, size and consequently the average density of Triton, and its bulk composition was found to be nearly identical to that of Pluto \citep{Smith1989VoyagerResults, McKinnon2014Triton}, comprising roughly a mixture of 70\% rock, 30\% ice; Triton was richer in rock than the icy satellites of the other planets, seemingly indicative of formation in the Kuiper Belt. An overview of some defining properties for Triton is given in Tab.~\ref{tab:bulk_parameters}.

\begin{table}[]
\caption{An overview of the bulk parameters, orbital properties and some internal structure estimates as given by \citet{McKinnon2014Triton}.}
\label{tab:bulk_parameters}
\begin{tabular}{@{}llll@{}}
\toprule
\textbf{Quantity}         & \textbf{Value}      & \textbf{Unit(s)} & \textbf{Remarks}           \\ \midrule
Radius                    & $1352$                & km               & -                          \\
Hydrosphere thickness     & $\sim400$           & km               & Estimate                   \\
Silicate mantle thickness & $\sim350$           & km               & Estimate                   \\
Metallic core thickness   & $\sim600$           & km               & Estimate                   \\ \midrule
Mass                      & $2.140\cdot10^{22}$ & kg               & -                          \\
Mean density              & $2060$                & kg/m$^3$         & $\sim60-70\%$ rock + metal \\
Surface temperature       & $\sim40$            & K                & Season-dependent                          \\ \midrule
Semi-major axis           & $354.8$               & Mm            & 14.33 Neptune radii                          \\
Orbital period            & $5.877$               & d                & -                          \\
Inclination               & $156.8$               & $^{\circ}$       & -                          \\ \bottomrule
\end{tabular}
\end{table}

After a long wait, sufficient information was now known about Triton to start considering its internal makeup and its structural organisation, and as a result modellers could at last get to work on Triton. \citet{Ross1990TheTriton} used the interior models proposed by \citet{Smith1989VoyagerResults} on the basis of \textit{Voyager} observations to give a first estimate of the coupled thermal and orbital evolution of Triton after capture, which up to then could not be considered. Even though their models only hold up to first order in eccentricity (as they already recognised at the time), \citet{Ross1990TheTriton} showed that the spectacular thermal event promised by \citet{McKinnon1984OnPluto} should indeed have taken place if Triton were captured and circularised by tidal dissipation alone. Importantly, their results also showed the severe effect interior parameters have on orbital evolution in this case, showcasing the shortcomings of work undertaken before the availability of \textit{Voyager} data.

Aside from bulk parameters, \textit{Voyager 2}, for the first time, gave us a window into Tritonian geology and surface morphology. What it found was unprecedented: while by now a geologically (hyper-)active Triton was no longer wholly unexpected to have existed in the distant past, \textit{Voyager 2} imagery showed a Triton that was in fact geologically active \textit{now}, comparable in scale only to the inner satellites of Jupiter, Io and Europa \citep{Smith1989VoyagerResults}. The implications of this geological activity were far-reaching: either Triton's putative capture happened so recently that the associated heat was still driving resurfacing on Triton, or some other endogenic process was at play. This therefore motivates us to consider in further detail the geological evidence for a presently active Triton.

\section{Geological evidence for a presently active Triton}
\label{sec:geological_active_Triton}
Before we continue to interpret this geological activity and its consequences, it is instructive that we discuss the various pieces of evidence that make up the conclusion that Triton is currently geologically active, as these may also give us insight into the underlying mechanisms.

\subsection{Surface morphology and age}
Initial examinations of the Tritonian surface and its deficiency in impact craters concluded that its surface seemed to be young, like that of Io and Europa; various morphological units showed evidence of eruptive activities and endogenic processes on a global scale \citep{Smith1989VoyagerResults}. This was surprising: whereas the eccentricities forced by the Laplace resonance in the Galilean satellites were by this time known to allow this activity in Io and Europa, Triton had long been known to have a very nearly circular orbit. While the tidal melting of Io had in fact been predicted even before the \textit{Voyager 1} visit to the Jovian system \citep{Peale1979MeltingDissipation}, no such predictions had therefore been made for current Triton.

A crater count comparison with the Uranian moon Miranda turned out indicative of similar impactor source populations between the two bodies, in which case the relative dearth of impact craters by an order of magnitude on Triton suggested an extremely young surface \citep{Smith1989VoyagerResults, Strom1990TheTriton}; later revisiting of \textit{Voyager 2} results by \citet{Stern2000TRITONSACTIVITY} showed that the resurfacing rate implied by crater counts was indeed only secondary to Io and Europa, and moreover that the resulting surface ages are in fact no greater than 0.1-0.3 Gyr old. Later still, \citet{Schenk2007OnTriton} put an upper limit of 50 Myr on the surface age everywhere, with a plausible surface age of less than 10 Myr if the majority of cratering is a result of planetocentric impactors. Clearly, endogenic activity was still widespread on Triton on recent timescales that cannot be explained by a late-time capture, unless we currently observe Triton at a unique time shortly post-capture. \citet{Schenk2007OnTriton} argue on the basis of cratering rate estimates that the probability of a capture within the last 4 billion years is $\sim1\%$, however, so that Triton be captured and circularised within the past couple million years seems exceedingly unlikely.

\subsection{Tectonic activity and cryovolcanism}
Tectonic features were soon mapped on the \textit{Voyager} data, and found to be indicative of global tectonic activity in the Tritonian shell \citep{Croft1993TectonicsTriton}; though \citet{Croft1993TectonicsTriton} interpreted these tectonic features to be a result of tidal despin and circularisation of Triton's orbit after capture (which we now know is incompatible with its surface age), it is these global 
tectonic yielding features that would later lead \citet{Nimmo2015PoweringGeology} to argue that convection in the lower ice shell is responsible for Triton's recent resurfacing, which we shall discuss in further detail in Sec.~\ref{subsec:tidal_heating}.

But this tectonic network was not the only indication of endogenic activity on Triton. The dimpled terrain that covers much of the equatorial regions of Triton, termed `cantaloupe terrain', had not been seen on any other icy satellite \citep{Schenk1993DiapirismInstability, Smith1989VoyagerResults}. \citet{Smith1989VoyagerResults} initially proposed that these may be the result of collapse of surface landforms by sublimation; \citet{Schenk1993DiapirismInstability} consequently drew a parallel between the diapirs observed in the Great Kavir on Earth and this cantaloupe terrain. Under this explanation, the cantaloupe terrain results from a density inversion in the upper layer of the Tritonian ice shell; \citet{Schenk1993DiapirismInstability} explain this density inversion as the result of a dense ammonia-water mixture layer deposited by cryovolcanism, the existence of which had been interpreted from the \textit{Voyager 2} images through geological arguments \citep{Croft1990PhysicalTriton, Schenk1992VolcanismTriton}. In this case the fact that this diapirism is so widespread would therefore constitute direct evidence of global-scale cryovolcanic activity such as is required to explain the extremely young surface age everywhere. It should be noted, however, that while diapirism is considered the leading hypothesis for the cantaloupe terrain \citep{McKinnon2014Triton}, alternative hypotheses have also been raised (e.g. \citealt{Boyce1993ATriton}); more recently, \citet{Hammond2018CompactionTriton} proposed catastrophic resurfacing by upward migration of a low-density, ammonia-enriched ocean. Notably, however, all leading explanations seem to be endogenic in nature \citep{Boyce1993ATriton, McKinnon2014Triton, Hammond2018CompactionTriton}.

Consequently, the surface features identified on Triton seem to be consistent with what surface morphology and crater density-based age estimates would also have us think: Triton is currently still experiencing extensive resurfacing and geological activity.

\subsection{Plumes and fans}
\label{subsec:plumes_and_fans}
A tertiary piece of evidence for endogenic activity is formed by two active plumes that were observed by \textit{Voyager 2}, and the dark streaks (termed fans in some other sources) that have been attributed to past or lower-altitude plumes \citep{Smith1989VoyagerResults}. Based on the position of these plumes near the subsolar point at the time of the \textit{Voyager 2}, it was argued at the time that these plumes are likely solar-driven (e.g. \citealt{Smith1989VoyagerResults, Hansen1990SurfaceTriton, Soderblom1990TritonsCharacterization, Brown1990EnergyPlumes, Kirk1990SubsurfaceTriton}). Recently, however, this hypothesis has come under scrutiny, and \citet{Hofgartner2022HypothesesTests} argue that an endogenic origin for the plumes cannot be excluded.

Whatever the case, as argued by \citet{Stern2000TRITONSACTIVITY}, the resurfacing rate required to explain Triton's crater density cannot be supplied by these plumes, nor by escape-loss erosion or aeolian processes. We are therefore forced to conclude that the observed resurfacing must thus necessarily be geological in nature \citep{Stern2000TRITONSACTIVITY, Schenk2007OnTriton}. This then begs the question; what can cause the Tritonian interior to have remained hot and active into the present?


\section{Explaining the active Tritonian interior}
\label{sec:explaining_interior}
Of course, we are not the first to ask this question. There are a variety of energy sources that can contribute to this interior activity, a number of which hold for icy satellites in general, and some for Triton in particular. 

While one may naively want to exclude this first class of heating mechanisms, given the fact that the geological activity we are trying to explain is unique to Triton, we shall treat them nonetheless; discussing why these mechanisms are unlikely to allow for a geologically active Triton is instructive, and in some cases not entirely trivial. Of course, even though these mechanisms clearly do not force present geological activity on other icy satellites, they may still play an auxiliary role in Triton's geology. These mechanisms constitute solar heating, primordial heat and radiogenic heating, and they will be discussed in Sec.~\ref{subsec:solar_heating}, Sec.~\ref{subsec:primordial_heat} and Sec.~\ref{subsec:radiogenic_heating}, respectively.

A second class of heating mechanisms is more particular to the circumstances of Triton; one would therefore think it likely that this is where the mechanism behind Triton's geology can be found. These mechanisms are provided by heat remaining from capture around Neptune, whatever the mechanism, and tidal heating. These will be discussed in Sec.~\ref{subsec:capture_heat} and Sec.~\ref{subsec:tidal_heating}, respectively.

\subsection{Solar heating}
\label{subsec:solar_heating}
Given the small amount of insolation Triton receives, with its correspondingly low surface temperature, it may seem bizarre to even posit that solar heating might be responsible for the internal activity of Triton. However, solar heating has not yet been excluded as the driving mechanism behind, for example, the plumes observed on Triton (e.g. \citealt{Hofgartner2022HypothesesTests}; see Sec.~\ref{subsec:plumes_and_fans} for a discussion). Hence, in the interest of a holistic understanding of the geology of Triton, we must consider solar heating. Comparative planetology fortunately allows a clean way out; Triton is the furthest large satellite from the Sun, and clearly no other satellite closer to the Sun experiences the type of interior activity that Triton does without another clearly identifiable cause. Even still, Tritonian seasons are extreme: however, fortunately, Triton's twin body Pluto experiences similarly extreme seasons at a similar orbital distance from the Sun, yet it has regions of age comparable to the Solar System (though Sputnik Planitia is notably of comparable age to the Tritonian surface; \citealt{Stern2018TheHorizons}). We may therefore assume that solar heating plays no significant role in Triton's current active geology.

\subsection{Radiogenic heating}
\label{subsec:radiogenic_heating}
Aside from solar heating, another currently active source of energy is radiogenic heating; given Triton's large rock-to-ice ratio compared to other icy satellites (e.g. \citealt{Hussmann2010ImplicationsSatellites}), it is not unreasonably to be expected that radiogenic heating might play a major role in the behaviour of the ices overlaying the rocky layer. However, as with solar heating, comparative planetology provides a way out: Pluto, with a similar (though slightly lower) rocky-to-ice ratio and similar proportions (e.g. \citealt{McKinnon2017OriginFlyby, Bierson2018ImplicationsContrast}), does not have nearly as young a surface globally (e.g. \citealt{Moore2016TheHorizons, Stern2018TheHorizons}). While locally comparable geological activity can be found on Pluto \citep{Stern2018TheHorizons}, global geological activity seems to require a greater source of heating still. Numerical modelling supports this idea: \citet{Nimmo2015PoweringGeology} show using numerical models that the observed surface yielding cannot be achieved by radiogenic heating alone.

\subsection{Primordial heat}
\label{subsec:primordial_heat}
Aside from current energy fluxes, Triton will have become heated during accretion; a part of this primordial heat may still remain if Triton has been sufficiently insulated to retain it\footnote{\citet{Nimmo2015PoweringGeology} include capture heat in their notion of ``primordial" heat. As this (1) would implicitly suggest an early capture for Triton and (2) the mechanism behind capture heat is distinctly defining for and unique to Triton, we will treat it as separate from accretional heat, which is common to heliocentric and planetocentric-born objects.}. Though the precise amount of potential energy released during this accretion is difficult to pin down precisely, especially as the degree of differentiation of Triton before capture cannot be known, \citet{Hussmann2010ImplicationsSatellites} arrive at a total accretion heat of $\sim 1.5\cdot10^{28}$ J assuming a two-layer rock-ice model, which seems plausible for pre-capture Triton (which was possibly further differentiated post-capture). As the amount of heat released during capture is equivalent to roughly $2\cdot 10^{29}$ J \citep{McKinnon2014Triton, Nimmo2015PoweringGeology}, whatever primordial heat remained at capture was surely overshadowed by the thermal consequences of capture. We will therefore defer this discussion to Sec.~\ref{subsec:capture_heat}, noting that primordial heat may only be of (minor) interest to current geological activity if capture heat is.

\subsection{Capture heat}
\label{subsec:capture_heat}
As mentioned, based on orbital specific energy arguments, one can show that the heat released during capture of Triton into planetocentric orbit will have released $\sim2\cdot 10^{29}$ J of energy, or a mean temperature change of $\sim 10^4$ K \citep{McKinnon2014Triton, Nimmo2015PoweringGeology}: sufficient to melt Triton's entire ice mantle several times over. Additionally, the consequent synchronisation of Triton's spin rate to its orbit will have released a further $\sim 2\cdot 10^{26}$ J \citep{Hussmann2010ImplicationsSatellites}; while this latter figure is seemingly negligible in the greater picture, it may have provided a rapid influx of heat depending on how fast and when in the capture process this synchronisation took place. If this energy is dissipated at a constant rate over the age of the Solar System (though it surely was not), that produces $\sim1.4\cdot 10^{12}$ W at the present, which would be only a factor 2 more than required to produce surface yielding as per \citet{Nimmo2015PoweringGeology}; given that this dissipation is temperature-dependent through thermal radiation and thus self-damping as a result, it is certainly much lower at present. Indeed, \citet{Nimmo2015PoweringGeology} show that after $\sim3$ Gyr, an initially hot and cold conductive Triton, even including radiogenic heating, are thermally indistinguishable. Unless an early atmosphere like that proposed by \citet{Lunine1992ATriton} persisted over Gyr timescales or longer, it seems implausible that primordial heat can therefore today be responsible for a resurfacing Triton.

It therefore seems that unless Triton was captured at a geologically recent time or insulated over Gyr-timescales, capture heat cannot be the driving force behind Triton's present geological activity, and consequently, neither can primordial heat. We have exhausted all but one source of geological activity: tidal heating.

\subsection{Tidal heating}
\label{subsec:tidal_heating}
We have concluded that, unless Triton was captured recently, which is unlikely both for dynamical reasons as well as for the simple fact that the KBO-population was far greater in the early Solar System (see e.g. \citealt{Vokrouhlicky2008IrregularReactions}), we require it to be experiencing tidal heating at present. Two main mechanisms exist for this to occur; tidal heating due to eccentricity, and tidal heating due to obliquity (e.g. \citealt{Bagheri2022TidalOverview}). Here, we discuss why obliquity tides must act at present; a discussion of tidal heating in general (i.e. also for early Triton) will be deferred to later, and will include both eccentricity and obliquity tides.

\subsubsection{Eccentricity tides}
The most well-known form of tidal heating, eccentricity tides, are what forces volcanic activity on Io (e.g. \citealt{Peale1979MeltingDissipation}) and geological phenomena on Enceladus (e.g. \citealt{Nimmo2016OceanSystem}), the youngest known surfaces in the Solar System. However, these are self-damping unless sustained by an external forcing factor such as the Laplace resonance \citep{Nimmo2016OceanSystem}, which cannot exist for the lone Neptunian moon. As the currently observed eccentricity of Triton is very low \citep{McKinnon2014Triton}, eccentricity tides therefore cannot explain the observed resurfacing unless we observe Triton at a special time (i.e. just after capture), which we have previously discussed to be unlikely.

While we should therefore not expect eccentricity tides to be of any major influence at present, it must be noted that this is a very important contribution to the heating of early post-capture Triton, however. In fact, it was the only tidal-heating term accounted for in nearly all early considerations of post-capture Triton (e.g. \citealt{Goldreich1989NeptunesStory, Chyba1989TidalSystem, Ross1990TheTriton, Lunine1992ATriton}). Yet, for the currently observed resurfacing, we must therefore defer to obliquity tides as a final explanation.

\subsubsection{Obliquity tides}
While Triton's obliquity has never been measured, if it occupies a Cassini state it would possess a significant obliquity, both with and without a subsurface ocean \citep{Nimmo2015PoweringGeology}; importantly, the difference between these two scenarios can be up to roughly a factor two \citep{Chen2014TidalOceans}. Hence, the existence or production of a subsurface ocean may well drive or have driven Triton toward more vigorous obliquity tides.

\citet{Nimmo2015PoweringGeology} have convincingly shown that Triton's young surface requires that obliquity tides are currently operating in its ocean. As these obliquity tides are (in a Cassini state) the result of the inclination of Triton's orbit, which is exceptional among the large Solar System satellites, this directly links the presently active Tritonian geology to its exceptional origin, whatever this may be. Notably, standard moon formation through accretion in a circumplanetary disc (e.g. \citealt{Canup2002FORMATIONACCRETION, Canup2006APlanets, Szulagyi2018InNeptune, Batygin2020FormationSatellites}) naturally leads to systems of prograde satellites on equatorial orbits, and so explanation of Triton requires an additional event or set of events. This therefore motivates us to discuss what mechanism might have deposited Triton in the orbit it is currently found in.

\section{Producing an inclined, retrograde Triton}
\label{sec:producing_inclined_retrograde_triton}
Given that it appears that the retrograde, inclined orbit of Triton is ultimately responsible for its current geological activity, it is natural that we discuss what might have caused it to orbit Neptune in this manner. We will discuss two (exhaustive) possibilities: planetocentric formation followed by a catastrophic dynamic event (Sec.~\ref{subsec:dynamic_interactions}) and formation in heliocentric orbit followed by capture into planetocentric orbit (Sec.~\ref{subsec:heliocentric_capture}). Let us treat these in turn.

\subsection{Dynamic interactions between a circumplanetary-born Triton and third bodies}
\label{subsec:dynamic_interactions}
With the discovery of Pluto on a Neptune-crossing orbit in 1930, the notion that the orbits of the planets were stable and unchanging seemed untenable. Though the orbits do not intersect at present due to Pluto's inclination, one could well argue that Neptune and Pluto may well have suffered one or more encounters in the past as their orbits precessed. On this basis, \citet{Lyttleton1936OnSystem} proposed that in fact Pluto had been a satellite of Neptune, ejected in the same event that caused Triton's orbit to turn retrograde, an idea upon which later authors would elaborate (e.g. \citealt{Kuiper1957FURTHERPLUTO, Dormand1977InteractionsSystem, Dormand1980ThePluto}). However, more accurate mass measurements of Pluto made possible by the discovery of the existence of Charon, its moon, showed that Pluto was not massive enough to have effected a reversal of Triton's orbit \citep{Farinella1979TidalSystem}; while later observations of Triton showed that it too was less massive than previously thought, this was still not by enough to allow for the reversal that \citet{Lyttleton1936OnSystem} had envisioned. The death blow for a circumneptunian origin for Pluto came in a landmark paper by \citet{McKinnon1984OnPluto}: McKinnon reinforced the conclusion that no orbital interactions exist that can reverse Triton's orbit for any plausible mass ratio of Triton and Pluto with more detailed calculation, but moreover showed that the angular momentum state of the Pluto-Charon binary system is inconsistent with a circumplanetary origin. Later work on Pluto showed that its composition was not compatible with circumplanetary formation either \citep{McKinnon1988PlutosNebula, Stern2014Pluto}, which would seem to lay this hypothesis to rest; it seems that Pluto and Triton did not share a common circumneptunian heritage.

However, this would not yet put to rest the idea of Triton alone originating as a Neptunian satellite, with its sense of orbit the result of some catastrophic event. \citet{Harrington1979ThePluto} discussed the possibility that this catastrophic event may have been an encounter between Neptune and a third as of yet unseen body. This body would have had a mass of 2-5 Earth masses, corresponding what we would now call a super-Earth, which was consequently either ejected from the Solar System or launched into a highly elliptical long-period orbit. However, their research was motivated by a misinterpretation of work by \citet{McCord1966DynamicalSystem}\footnote{\citet{Harrington1979ThePluto} cited \citet{McCord1966DynamicalSystem} as showing that the greatest orbital distance achieved by Triton is $400$ Mm on the condition that Neptune's tidal $Q$ (a measure of the efficiency of tidal dissipation) is less than $10^3$, but it was in fact Triton's $Q$ that is bounded by $10^3$, a far more reasonable value.}, as discussed by \citet{Farinella1980SomePluto}, who moreover note that the ad-hoc nature of the argument by \citet{Harrington1979ThePluto} makes it probabilistically untenable. More recently, similar work has been done on the basis of the five-planet Nice model; in this altered version of the Nice model (see \citealt{Gomes2005OriginPlanets, Morbidelli2005ChaoticSystem, Tsiganis2005OriginSystem}), a third ice giant inhabited the early Solar System and was later ejected or set upon a far-out, long-period orbit (e.g. \citealt{Nesvorny2011YoungPlanet, Roig2015THEMODEL}). It was then proposed that Triton originated as a moon of Neptune disturbed by an interaction with the third ice giant \citep{Li2020TheEncounter} or as a moon of the additional planet that was captured by Neptune in such an interaction \citep{Li2020CaptureNereid}. Neither scenario, however, solves the criticisms that \citet{Farinella1980SomePluto} voice on the work by \citet{Harrington1979ThePluto}; moreover, no circumplanetary-origin scenario can yet explain the similarities in the compositions of Triton, Pluto and other Kuiper belt objects, which seem to suggest a heliocentric rather than planetocentric origin \citep{McKinnon2014Triton, Stern2014Pluto}.

\subsection{Capture from heliocentric orbit}
\label{subsec:heliocentric_capture}
It therefore seems that no mechanism exists by which Triton might have formed in Neptunian orbit to have consequently been deposited onto its current orbit with its current composition. The logical alternative is that Triton did not form around Neptune, and was only captured into its current orbit later. Indeed, this was the conclusion also drawn by \citet{McKinnon1984OnPluto}. Nonetheless, the mechanism by which this effected was not yet clear; while \citet{McCord1966DynamicalSystem} had already shown that tidal dissipation might allow for circularisation of an orbit from near-parabolic initial conditions, capturing Triton entirely through this process would require fine-tuned initial conditions that do not lead to Triton ending up as it is currently observed (e.g. \citealt{McKinnon1995GasTriton}). Another mechanism must therefore be responsible at least for the initial capture of Triton\footnote{It is of interest to note that the ability of purely tidal dissipation to effectively capture larger (ice giant-size or greater) objects, however, has been demonstrated by \citet{Hamers2018I} and \citet{Ochiai2014ExtrasolarScattering}.} - further dissipation can then proceed through tides as described by \citet{McCord1966DynamicalSystem}, potentially aided or influenced by third-body perturbations \citep{Benner1995OrbitalOrbit, Nogueira2011ReassessingTriton}. Let us therefore discuss the four mechanisms that have been proposed to allow for this: disc drag, collision with a primordial satellite, pull-down capture and binary dissociation.

\subsubsection{Disc drag}
The first mechanism, which was already put forth by \citet{McKinnon1984OnPluto}, was capture by gas drag in a circumneptunian disc or extended atmosphere. This idea had been in vogue at the time, with similar origins being proposed for the various groups of irregular moons of Jupiter \citep{Pollack1979GasCapture}, our own Moon \citep{Nakazawa1983ORIGINATMOSPHERE} and for individual retrograde satellites of Jupiter \citep{Huang1983THESATELLITES} around this period; it is no surprise that this is what McKinnon came up with. 

However, gas drag as main mechanism for capture and consequent circularisation of Triton is difficult to interpret consistently with the existence of the other satellites of Neptune closer-in than Triton (e.g. \citealt{Goldreich1989NeptunesStory}). Moreover, the absence of a large Uranian-like satellite system at present cannot be reconciled with the capture of Triton during the formation of such a system, as recent moon formation results indicate that satellite formation is relatively fast - occurring on timescales of a couple $100$ kyr \citep{Cilibrasi2018SatellitesMoons, Szulagyi2018InNeptune, Batygin2020FormationSatellites}. Unless the capture of Triton happened as the disc was waning, or itself set on this waning, one should therefore expect a satellite system of some sort to have formed. The first is probabilistically unlikely given the timescales involved: circumplanetary discs live up to several Myr (e.g. \citealt{Szulagyi2018InNeptune, Cilibrasi2018SatellitesMoons, Batygin2020FormationSatellites}), as opposed to the aforementioned $\sim100$ kyr formation timescale of satellites. The second is improbable given that in-situ growth and migration of satellites occurs several times throughout the lifetime of a satellite-forming circumplanetary disc \citep{Canup2006APlanets, Cilibrasi2018SatellitesMoons, Batygin2020FormationSatellites}. As Neptune likely never had an extended atmosphere \citep{Jewitt2007IrregularSystem}, capture by gas drag in an extended atmosphere must also be ruled out.

When the possibility of collision with a primordial satellite was later discussed, it was proposed by \citet{Cuk2005CONSTRAINTSTRITON} that the debris of perturbed protosatellites may instead induce the required drag, playing the role that might in the other scenarios have been played by the circumplanetary disc or extended atmosphere. However, this has been disputed by \citet{Rufu2017TritonsSystem}, who show that the associated reaccretion timescale of any perturbed protosatellites is too short to allow for significant evolution, which thus rules out this mechanism. This does leave open the opportunity that the initial collision may have allowed a capture into Neptunian orbit, however: we will review this mechanism separately.

We are forced to conclude that gas or dust drag is not a viable mechanism by which Triton might have been captured, though we note that none of the arguments we have levied expressly forbid this mechanism from capturing moons in general. It may therefore be an interesting mechanism to study in the context of captured (exo-)moons in a more general sense in the future.

\subsubsection{Pull-down capture}
Another mechanism that has been considered is pull-down capture; originally proposed by \citet{Heppenheimer1977NewCapture}, pull-down capture involves capture of a satellite from a loosely bound initial orbit into a permanent orbit by growth of the Hill sphere as a result of mass growth of the parent body. While this has been shown an effective mechanism by which super-Jupiters in exoplanetary systems may capture co-orbital protocores of would-be giant planets \citep{Hansen2019FormationCapture}, the runaway mass growth speed necessary to allow for this to plausibly happen was likely never reached by the ice giants \citep{Jewitt2007IrregularSystem}, and so we are forced to conclude that pull-down capture cannot be responsible for Triton's orbit.

\subsubsection{Collision with a primordial satellite}
\citet{Goldreich1989NeptunesStory} argue that collision with an inner regular protosatellite of Neptune instead provides a mechanism by which a satellite might be captured, on the basis of an order-of-magnitude calculation showing that the probability of such a collision occuring at some point in Neptune's history is on the order of several percent. However, later work on planet formation shows that the assumptions made by \citet{Goldreich1989NeptunesStory} on the size distribution of Kuiper belt objects do not hold (e.g. \citealt{Morbidelli2009ConsiderationsTrojans}); additionally, Triton's current inclination outside the Laplace plane excludes disruptive impacts (e.g. \citealt{McKinnon1995GasTriton, Rufu2017TritonsSystem}), putting restrictive bounds on the size of the putative colliding protosatellite, which drastically lower the probability of a Triton capture in this manner. While this capture mechanism is thus not expressly forbidden, it is exceedingly unlikely. In the absence of another explanation, one might be inclined to accept this improbability; the Solar System is bound to be inhabited by some statistical unlikelihoods. However, this conclusion cannot be drawn until it is clear that no other plausible explanations exist.

\subsubsection{Dissociation of a trans-Neptunian binary}
And just such an explanation was put forward by \citet{Agnor2006NeptunesEncounter}; if a set of binary objects passes sufficiently close to a planet, it may be disrupted, with a non-marginal probability that half of the pair is captured. \citet{Agnor2006NeptunesEncounter} showed that the properties of such a binary were reasonable, that the capture orbits upon which Triton would be set are compatible with that on which it is currently observed, and that the predicted capture rate is consistent with the number of such binary pairs observed in the outer Solar System. Their explanation requires no ad-hoc objects or dynamical events and no restrictive timing of the capture event. More remarkably still, \citet{Agnor2006NeptunesEncounter} showed using a back-of-the-envelope estimate that probabilistically their mechanism proved far more favourable than other contenders.

The importance of this result was immediately recognised at the time (see e.g. \citealt{Morbidelli2006InterplanetaryKidnap}), and consequent examinations of this mechanism soon followed \citep{Vokrouhlicky2008IrregularReactions, Nogueira2011ReassessingTriton}, confirming the plausibility of this scenario as a precursor to modern Triton. Interest in this mechanism spread beyond just explanation of Triton's orbit; it was also noted that this may be a mechanism by which extrasolar moons of detectable and habitable sizes may be produced \citep{Williams2013CapturePlanets}. Binary dissociation is nowadays considered the canonical method of capture for Triton (e.g. \citealt{McKinnon2014Triton}), though this has not stopped authors from examining other capture mechanisms or hybrid scenarios still (e.g. \citealt{Rufu2017TritonsSystem, Li2020TheEncounter, Li2020CaptureNereid}).

There are several issues with this form of capture, however. It appears that such capture is most probable if Triton was the smaller of a pair with a more massive companion \citep{Agnor2006NeptunesEncounter, Vokrouhlicky2008IrregularReactions}, yet none such objects are currently observed; \citet{Nogueira2011ReassessingTriton} note, however, that such objects may well be expected to have existed in the early Solar System based on estimates by \citet{Morbidelli2009ConsiderationsTrojans}. Aside from the probabilistic argument, there is no particular planetological reason to prefer binary dissociation over collisional capture: as non-disrupting collisional capture and binary dissociation will both be followed by rapid and catastrophic tidal heating during initial circularisation, there is likely not much that can distinguish the two events, and so for our purposes it will be safe to assume that binary dissociation was the capture process for Triton.

\section{Toward a consistent history of the Tritonian orbit and interior}
\label{sec:toward_consistent_history}
From the preceding, we conclude that it thus seems that (1) the capture origin of Triton can be linked both to the continued existence of a subsurface ocean through obliquity tides, and to the plausible existence of a (near-)surface ocean and large atmosphere in the distant past, following capture, and (2) that plausible mechanisms to deposit Triton onto a captured orbit which can then be circularised by tidal dissipation exist. While a reliable linked interior-orbital evolution model is perhaps out of reach without the use of several constraining assumptions, in particular as whatever evidence may have existed of the past Tritonian surface is wiped clean by geological activity, it does seem as though we have a sufficient part of the whole story to start piecing together what might become a consistent history of Triton's orbit and interior.

It is thus worthwhile to examine what existing models and research tell us about Triton's past following this circularisation, if anything: this can then inform the questions we set out to answer in this work. We will summarise the current state of knowledge on Triton's interior in Sec.~\ref{subsec:tritonian_interior}, and that of knowledge on its orbital evolution in Secs.~\ref{subsec:orbital_evolution_background}; finally, we will discuss some recent developments in the tidal theory of (exo)planetary systems, and how they might affect our understanding of early Triton, in Sec.~\ref{subsec:developments_tidal_theory}.

\subsection{Description and evolution of the Tritonian interior}
\label{subsec:tritonian_interior}
Little is known of the Tritonian interior, particularly in its early history; this can largely be attributed to the fact that the vast majority of geological information available for Triton is limited to the results obtained by \textit{Voyager 2}. Indeed, the study of Triton's interior largely kicks off after \textit{Voyager 2}'s visit to the Neptune-Triton system, with the main results summarised in Sec.~\ref{sec:geological_active_Triton}. What models for Triton's interior evolution do exist have largely been based on this data: indeed, the first interior models were produced in anticipation of the \textit{Voyager 2} visit \citep{McKinnon1989ThePrediction}, and \citet{Smith1989VoyagerResults} yield the first properly observation-constrained density- and pressure-profile predictions. These are then followed up on with qualitative or toy-model evaluations of Triton's thermal-interior history by \citet{McKinnon1990TritonsHistory} and \citet{Ross1990TheTriton}; quantitative evaluations of the interior evolution with time would have to wait at least until the work by \citet{Gaeman2012SustainabilityInterior} and \citet{Nimmo2015PoweringGeology}.

Though \citet{Nimmo2015PoweringGeology} show at least superficially that its past interior state is likely of little direct influence on the present interior, it is still of interest to consider this evolution given the effect the interior has on the orbital evolution of Triton. Conversely, constraining the past orbital evolution of Triton may allow us to say something about the past evolution of the interior, any evidence of which is erased by geological activity at present. That being said, the existence of an ocean in the past is a near certainty, given the heat released during capture, and its continued existence into the present is likely \citep{Gaeman2012SustainabilityInterior, Nimmo2015PoweringGeology, Hammond2018CompactionTriton, Soderlund2024TheMoons}, though not confirmed observationally.

Previous interior models considered only a static orbit over the circularised phase (e.g. \citealt{Gaeman2012SustainabilityInterior, Nimmo2015PoweringGeology, Hammond2018CompactionTriton}), with the notable exception of \citet{Ross1990TheTriton}; their use of lower-order approximations to the eccentricity functions at higher eccentricities seemingly invalidates their results, however. It is likely that the initial combination of high eccentricity and high obliquity in Triton heated the Tritonian interior to large temperatures, and so the orbital and interior evolution cannot be analysed separately at this epoch. Consequently, no attempts besides that of \citet{Ross1990TheTriton} have been made at an evaluation of Triton's interior evolution throughout this initial epoch, aided in part by the fact that the expressions required for high-eccentricity orbital evolution for non-trivial rheological models have only been developed in recent years (we will discuss this in further detail in Sec.~\ref{subsec:orbital_evolution_background}). In particular, previous expressions that remain valid up to high eccentricities (e.g. \citealt{Correia2009SecularTriton, Nogueira2011ReassessingTriton}) required the use of a constant time-lag rheology, intrinsically limiting the validity of these results. This is particularly problematic, as it has been shown that the constant time-lag rheology is not accurate for icy and rocky bodies \citep{Bagheri2022TidalOverview}. Additional rheological problems arise when considering the coupled spin-orbit evolution: we will discuss this in Sec.~\ref{subsec:developments_tidal_theory}. Before doing so, we will treat the general orbital evolution of Triton, however.

\subsection{Orbital evolution of Triton}
\label{subsec:orbital_evolution_background}
Given Triton's dynamically interesting history, it is no surprise that a variety of simulations of the orbital evolution of the Neptunian moon exist. Unfortunately, as was the case for the treatment of the orbital evolution in examinations of the interior evolution, the majority of treatments of the orbital evolution do not account for variation in Triton's interior, largely due to a lack of constraints on the latter, especially at early times. In fact, a significant number of them forego realistic interior models altogether, preferring those that allow for simplification of the dynamical equations in question.

\subsubsection{Examinations of Triton's orbital evolution before and following \textit{Voyager 2}}
Examinations of Triton's orbital evolution first arise with the work of \citet{Lyttleton1936OnSystem}; while his theory for Pluto's origin as an ejected satellite of Neptune is now largely discounted, this is the first time that the possibility of a time-altered architecture of the Neptunian satellite system is considered. With the advent of the landmark work on tidal dissipation by \citet{Kaula1961AnalysisSatellites, Kaula1964TidalEvolution}, \citet{McCord1966DynamicalSystem} seems to have been the first to consider capture followed by evolution onto a circular orbit by tidal dissipation and to numerically compute it, though with a better understanding of Triton's mass and rheological properties his calculations have become outdated.

\citet{Farinella1979TidalSystem} further considered the capture hypothesis, as well as the plausibility of Lyttleton's hypothesis for Pluto's circum-Neptunian origin, but now with an initially heliocentric Triton that is captured onto an orbit around Neptune: in this scenario, it is the subsequent shrinking of Triton's orbit that eventually ejects Pluto, though they do not base this on any orbital history beyond a qualitative description. After this, the first numerical evaluations of Triton's orbital evolution commensurate with our current understanding start arising: \citet{McKinnon1984OnPluto} predicts that Triton was captured around Neptune without any Plutonian involvement, and gives the first estimates of the duration of this circularisation (on the order of $\sim$Myrs).

In anticipation of \textit{Voyager 2}
s imminent Triton flyby, \citet{Jankowski1989OnTriton}, \citet{Chyba1989TidalSystem} and \citet{Goldreich1989NeptunesStory} publish computation-based predictions on Triton's orbital evolution in rapid succession. \citet{Jankowski1989OnTriton} expect that Triton may be caught in the Cassini 2 state, which would excite significant obliquity tides; \citet{Chyba1989TidalSystem} publish a time-evolution prediction based on the constant phase-lag model (see Sec.~\ref{sec:CTL_CPL}), as do \citet{Goldreich1989NeptunesStory}. A notable difference is the fact that \citet{Goldreich1989NeptunesStory} consider the evolution directly post-capture, driven by tides in Triton, while \citet{Chyba1989TidalSystem} consider the long-term evolution, driven by tides in both bodies.

With the arrival of \textit{Voyager 2} data on Triton, these predictions are put to the test: Triton's locking into Cassini state 2 is not consistent with observations, and so those scenarios as described by \citet{Jankowski1989OnTriton} and \citet{Chyba1989TidalSystem} are ruled out. The observations do not yield any definitive ruling on the validity of the other orbit simulations, however. In the time that followed, orbit evolution examinations were mostly driven by the uncertainty in the mechanism of Triton's capture (see Sec.~\ref{subsec:heliocentric_capture}), and attempts to make various capture mechanisms work (e.g. \citealt{McKinnon1995GasTriton, Cuk2005CONSTRAINTSTRITON}). A notable exception is the work by \citet{Ross1990TheTriton}, who produce a first model coupling a thermally evolving interior to the orbital evolution. Their use of a constant time-lag-like model, based on the work by \citet{MacDonald1964TidalFriction} and \citet{Goldreich1966QSystem}, as well as their use of tidal evolution equations well outside their region of validity, however, mean that it is difficult to trust their results beyond the purpose of showing the added value of including a temperature-variable tidal response (which was, admittedly, their aim).

\subsubsection{Recent results on Triton's high-eccentricity evolution}
With the capture mechanism of Triton seemingly resolved by \citet{Agnor2006NeptunesEncounter}, most studies start to assume binary dissociation as the canonical capture mechanism, and tidal dissipation as the consequent mechanism of circularisation (see e.g. \citealt{Vokrouhlicky2008IrregularReactions, Gaeman2012SustainabilityInterior, Nimmo2015PoweringGeology}). Notable exceptions are given by \citet{Rufu2017TritonsSystem}, \citet{Li2020TheEncounter}, and \citet{Li2020CaptureNereid}, who explore other scenarios. We remark that \citet{Rufu2017TritonsSystem} neglect tidal dissipation, while \citet{Li2020TheEncounter} and \citet{Li2020CaptureNereid} use the tidal model given by \citet{Correia2009SecularTriton}. \citet{Correia2009SecularTriton} uses a constant time-lag model (for which analytical, finite tidal evolution expressions exist at arbitrarily high eccentricities) to produce the first fully mathematically consistent simulation of Triton's circularisation from a high-eccentricity capture orbit. Finally, \citet{Nogueira2011ReassessingTriton} added Kozai cycles to Correia's model, showing that these have a significant disturbing effect at high separations.

\subsection{Developments in the tidal theory of highly-eccentric objects}
\label{subsec:developments_tidal_theory}
This is then where the current examination of Triton remains: this does not mean, however, that this is where the development of high-eccentricity tidal evolution expressions ends. In particular, we want to emphasise developments in the general tidal theory, as well

\subsubsection{General developments in tidal theory}
\citet{Boue2019TidalElements} present a full, consistent re-derivation of the expressions initially derived by \citet{Kaula1961AnalysisSatellites, Kaula1964TidalEvolution}: in doing so, they correct a significant error in Kaula's derivation of the inclination-evolution term, which had until that point seemingly not been caught; as recent results seem to indicate that moon migration can tilt the obliquity of host planets significantly (e.g. \citealt{Wisdom2022LossRings, Saillenfest2022TiltingSatellite, Saillenfest2021TheTitan}), this thus might warrant a re-investigation of the inclination-evolution of Triton's orbit (and the associated evolution of Neptune's obliquity).

The manner in which \citet{Boue2019TidalElements} present the development of the Kaula expansion allows insertion of arbitrary rheological models, as opposed to the formalism derived by \citet{Hut1981TidalSystems} as used by \citet{Correia2009SecularTriton} (which mandates a constant time-lag model). As the understanding of the orbit-related implications of various interior (rheological) models has vastly surpassed the constant time-lag and constant phase-lag models (see e.g. \citet{Bagheri2022TidalOverview} for an overview relevant to Triton-like bodies), this is a major development; the use of more realistic rheological models has led to the exploration of higher-order spin-orbit resonances as stable system states for eccentric systems, as is the case for Mercury (e.g. \citealt{Noyelles2014Spin-orbitRevisited}). As Triton will have gone through a significant high-eccentricity phase, it is worth considering what the ramifications may be.

\subsubsection{The coupled spin-orbit evolution of Triton}
Orbital evolution as has been considered in the past does not yet give the full dynamical picture, however. The stable equilibrium spin state for eccentric tidal systems with realistic viscoelastic behaviour may be a half-integer spin-orbit resonance greater than the 1:1 resonance (e.g. \citealt{Goldreich1966Spin-OrbitSystem, Makarov2012ConditionsResonances, Noyelles2014Spin-orbitRevisited}); additionally, recent work has shown that the transition between such equilibria may be rapid compared to other tidal evolution \citep{Walterova2020ThermalExoplanets, Renaud2021TidalTRAPPIST-1e}. While the spin energy dissipated in Triton is of the order of 1/1000-th that dissipated in orbital energy during capture (see Sec.~\ref{subsec:capture_heat}), meaning that the dissipation of spin energy is not likely a large effect energy-wise, the rapidity with which it is dissipated might make it a significant effect in the process of spin-orbit transitions. The constant time lag model (which we note has been used in most high-eccentricity evaluations of Triton's orbital evolution), notably, does not predict such resonances to occur altogether (for an analytical formula for the equilibrium rotation rate for constant time lag, see Eq.~\ref{eq:CTL_eq_rotrate}). As this phenomenon is unique to eccentric systems, it might well be an explaining factor in some of the geological features unique to Triton, and so we deem it worth exploring.

\section{Conclusions}
\label{sec:research_questions}
With this, we conclude our overview of the current status of knowledge on Triton: we have discussed the observational history of Triton, and the corresponding evidence that we have gathered to support the idea that Triton is currently still geologically active; we have consequently treated the evidence that leads us to think that obliquity tides in Triton's oceans are responsible for this current activity, as well as the fact that this is likely linked to Triton's inclined orbit, which is in turn a consequence of its capture origin. Finally, we have considered the available evidence for binary dissociation-capture followed by tidal circularisation, and have given a brief overview of the current status of knowledge of Triton's interior and orbit throughout its history, as well as developments that have been ongoing in the field of tidal dynamics. With this background in mind, we can start considering which gaps in this knowledge we would like to fill. The primary question we will strive to answer will then be:
\begin{quote}
    \textbf{What constraints can be put on the tidally forced trans- and post-capture dynamical and thermal evolution of Triton?}
\end{quote}
To answer this question as completely as possible, we formulate an additional set of questions that, when answered, can be combined to give a satisfactory answer to the question above. Where useful, we additionally define another subset of questions.
\begin{itemize}
    \item How can we effectively and accurately model the high-eccentricity orbital evolution of Triton over astronomical timescales?
    \item How do we appropriately model the spin-orbit evolution of Triton?
    \begin{itemize}
        \item What interior model do we need to properly model spin-orbit evolution?
        \item Is the release of spin-energy in spin-orbit resonance transitions a relevant thermal-evolution effect?
    \end{itemize}
    \item What bounds can we put on the Neptune-Triton system's past and future dynamical evolution?
    \begin{itemize}
        \item Did Triton pass through spin-orbit resonances during post-capture circularisation?
        \item Did Triton tilt Neptune to its present-day obliquity?
        \item How long did the circularisation phase for Triton last?
        \item What does the dynamical evolution during Triton's circularised phase look like?
    \end{itemize}
\end{itemize}
By answering the first question, we set up the mathematical (and programmatic) framework necessary to answer the other two questions: the necessary work will be presented in Ch.~\ref{ch:kaula_theory} and \ref{ch:validation}. Through answering the second question (and its subset of questions), we can constructively challenge existing results on the spin-orbit evolution of Triton where necessary, and know when to trust existing results generated using different methods than those presented here: the associated results are presented in Ch.~\ref{ch:spin-orbit_chains}. Finally, the third question will allow us to propose a consistent dynamical history for Triton throughout its life: we will discuss the relevant simulation work in Ch.~\ref{ch:initial_conditions}.

\chapter{Methodology}
\label{ch:kaula_theory}
To propagate Triton's orbit over time, we will require a mathematical model of the interactions governing Triton, Neptune and their combined interior and orbital evolution. We give an overview of the elements required of such a model in general and motivate the chosen approach for this analysis in Sec.~\ref{sec:modelling_Triton_Neptune}. Consequently, we will introduce the chosen dynamical-evolution formalism in Sec.~\ref{sec:darwin-kaula_expansion} and give a brief overview of the associated interior models in Sec.~\ref{sec:quality_function_homogeneous}.

\section{Modelling coupled dynamical-interior evolution}
\label{sec:modelling_Triton_Neptune}
Modelling the dynamical-interior\footnote{While we will often use the term ``interior" to refer to the planetological evolution of bodies, it should be noted that this may, in some cases, extend to atmospheric evolution too.} evolution of planetary bodies is in general a complicated, multidisciplinary problem, and no one-size-fits-all approach exists. However, we can identify several elements that are (explicitly or implicitly) common to all approaches: it is in filling in these elements that the approaches used in various studies differ. We will therefore briefly emphasise the generalities that arise in such studies in Sec.~\ref{subsec:modelling_dynamical_interior_interactions}; afterwards, we will justify our choice of approach in Sec.~\ref{subsec:formalism_selection_neptune_triton}.

\subsection{Modelling coupled dynamical-interior interactions in general}
\label{subsec:modelling_dynamical_interior_interactions}
With the variety in astrophysical bodies undergoing dynamical-interior interactions, an equal variety of approaches to modelling these interactions has arisen. On a dynamical level, examples of such interactions modelled using various formalisms are binary star-star interactions (e.g. \citealt{Hut1981TidalSystems}), (supermassive) black hole-star interactions (e.g. \citealt{Frank1976EffectsSystems, Young1977THEQSOs}), planet-star tides (e.g. \citealt{Luna2020TheSystem, Walterova2020ThermalExoplanets, Wu2024TidalDissipation}), planet-moon interactions (e.g. \citealt{Kaula1964TidalEvolution, Mignard1979THEI, Mignard1980THEII, NeronDeSurgy1997OnEarth, Musotto2002NumericalSatellites, Fuller2016ResonanceSystems}), planet-star-moon interactions (e.g. \citealt{Williams1997HabitablePlanets, Barnes2002STABILITYPLANETS, Scharf2006THEEXOPLANETS, Grishin2017GeneralizedInclinations, Makarov2023PathwaysExoplanets}), star-planet-asteroid interactions (e.g. \citealt{Kozai1962SecularEccentricity}), planet-moon-moon interactions (e.g. \citealt{Hussmann2004Thermal-orbitalEuropa}) and even star-planet-atmosphere interactions (e.g. \citealt{Gold1969AtmosphericVenus, Valente2023SpinApproach}). In some cases, such dynamical models have been augmented with coupled or evolving interior models for (one of) the bodies in question (e.g. \citealt{Ross1990TheTriton, Hussmann2004Thermal-orbitalEuropa, Bland2009TheGanymede, Meyer2010CoupledMoon, Walterova2020ThermalExoplanets}).

\begin{figure}
    \centering
    \includegraphics[width=1\linewidth]{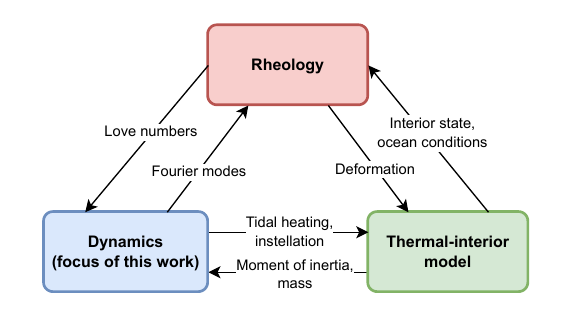}
    \caption{Schematic overview of the three main components comprising a dynamical-interior evolution model, with examples of interactions between the models along the edges.}
    \label{fig:schematic_components}
\end{figure}

While these different bodies and their interactions can be governed by various physical processes (and an arbitrary number of bodies may be present), we identify in general (though sometimes only implicitly, or present in a very simplified fashion) three elements that make up these models: a dynamical model that describes the (perturbed) evolution of the orbits and spin states of the bodies in question, a thermal-interior model that models the evolution of the bodies' planetological state (e.g. composition, pressure and density profiles, atmospheric interactions whenever applicable, etc.), and a rheological model that translates a given thermal-interior state into the parameters required by the dynamical model. Their relations and some examples of interactions between the three elements are drawn in Fig.~\ref{fig:schematic_components}. Of course, the boundaries between these three elements may blur: for example, the rheological model and thermal-interior model tend to become integrated when considering three-dimensional tidal models (e.g. \citealt{Roberts2008TidalEnceladus}), and tend to be combined into simple, singular expressions whenever admitted by certain assumptions (e.g. in the case of a homogeneous body with a Maxwell rheology in Darwin-Kaula theory; see e.g. \citealt{Bagheri2022TidalOverview}). In fact, as we will see in Sec.~\ref{sec:quality_function_general}, the form of the rheological model sometimes imposes a certain form of the thermal-interior model, and choosing a convenient rheological model may at times allow us to simplify the dynamical model significantly. Nonetheless, on a conceptual level it is useful to make the distinction between these modules.

In this work, the main emphasis will be on extending existing dynamical models to work at high eccentricities such as those that exist for early Triton. Consequently, we will not place as much emphasis on the thermal-interior and rheological models: these will be the subject of future work enabled by the development of a generally applicable dynamical model. Keeping this interaction between the three models in mind, however, we will aim to keep the requirements on the other two modules as loose as possible. With this framework in mind, we can start giving shape to each of these three modules. Before we do so, however, we need to consider the "language" in which these modules will be defined: in this process, we must consider the particular needs and properties of the Neptune-Triton system.

\subsection{Choosing a suitable formalism for the evolution of the Neptune-Triton system}
\label{subsec:formalism_selection_neptune_triton}
The motion of objects on a perturbed Keplerian orbit is in general described by the Lagrange planetary equations (see e.g. \citealt{Boue2019TidalElements} for a modern derivation and overview). These equations describe the evolution of dynamical quantities (orbital elements and spin state) of a system as a result of some arbitrary perturbing potential with time: additionally, the tidal heating rate follows directly from the spin-orbit energy lost due to the tidal potential of a given body (though finding the spatial distribution of that energy is in general a more involved problem; see e.g. \citealt{Roberts2008TidalEnceladus, Roberts2008Near-surfaceAnomaly}).

In principle, one could directly integrate these equations from here, and this is the approach taken for integration of spacecraft orbits or for estimation of ephemerides from or for use in space missions (e.g. \citealt{Fayolle2022DecoupledMission, Fayolle2023CombiningMoons}): the trouble with doing this over astronomical timescales, however, is the fact that the short-period oscillating motions (the mean motion and nodal and apsidal precession) dominate the error terms in integrator step size selection, while this variation is not the variation we are after. Consequently, long-term integration of these equations directly is computationally far too expensive. Instead, we would like to predict solely the long-term evolution: we are not concerned with estimating the exact position of any of the objects in the system at a given point in time, but rather with estimating the behaviour of their orbits and interior (and the energy transferred between them) over longer timescales.

This is, fortunately, not a new problem. In fact, it was precisely this realisation that led \citet{Kaula1964TidalEvolution} to take an approach he had taken previously to decompose the perturbing potential for satellite orbits into a mixture of periodic and secular (long-term changing) terms \citep{Kaula1961AnalysisSatellites}, and discard the periodic terms: what part of the expansion that remains is, naturally, the secular change in orbital elements, and will result in changes over astronomical timescales. As a partial sum of this series had previously been derived by \citet{Darwin1879OnEarth, Darwin1880OnPlanet}, this result has been termed the Darwin-Kaula expansion, and it has become the de facto standard for analysis of the evolution of celestial bodies evolving due to a perturbing potential over astronomical timescales. While this perturbing potential is commonly taken to be tidal in nature, this approach works equally well for any perturbing potential that is well-described using a decomposition in spherical harmonics (e.g. spherical oblateness or triaxiality; \citealt{Luna2020TheSystem}).

With this framework in mind, we can start giving shape to our implementation of the evolution of Triton. We will start by expanding on the dynamical model in Sec.~\ref{sec:darwin-kaula_expansion}, before moving on to our particular choices for the rheological and thermal-interior models in Sec.~\ref{sec:quality_function_homogeneous}.

\section{The Darwin-Kaula expansion of the tidal potential}
\label{sec:darwin-kaula_expansion}
To start with, one must consider the disturbing function to be inserted into the planetary equations (see e.g. \citealt{Boue2019TidalElements} for a more complete explanation and derivation, though here we adopt a variation on the notation used by \citealt{Luna2020TheSystem}):
\begin{equation}
    \mathcal{R} = -\frac{M_N + M_T}{M_N M_T} \left[M_N \Tilde{U}_T(\boldsymbol{r}_{T\to N}, \boldsymbol{r}_{T\to N}^*) + M_T \Tilde{U}_N(-\boldsymbol{r}_{N\to T}, -\boldsymbol{r}_{N\to T}^{\star} )\right]
\end{equation}
where $M$ is the mass of a body, $\Tilde{U}$ is the additional tidal potential generated by tidal disturbances and the subscripts $N$ and $T$ refer to Neptune and Triton, respectively. $\boldsymbol{r}_{T\to N}$ is the position vector pointing from Triton to Neptune (and $\boldsymbol{r}_{T\to N}$, naturally, that from Neptune to Triton), and the asterisk and star superscripts denote that these variables are those of the perturber, which can be set to be equal to the variables without the superscript after the required derivatives of $\mathcal{R}$ have been taken.

The main idea underpinning Darwin-Kaula tidal theory is to decompose the tidal response of the gravitational potential of a body to a perturbing second body of mass $M$ into spherical harmonics components. For a perturber in some point $\boldsymbol{r}^*$, the resulting additional tidal potential in another point $\boldsymbol{r}$ becomes (e.g. \citealt{Boue2019TidalElements, Luna2020TheSystem}):
\begin{align}
    \Tilde{U}(\boldsymbol{r}, \boldsymbol{r}^*) = - \frac{\mathcal{G}M}{a^*}\sum_{l\geq2}\sum_{m=0}^{l} \sum_{p=0}^{l} \sum_{q=-\infty}^{\infty}\sum_{h=0}^{l}\sum_{j=-\infty}^{\infty}U_{lmpqhj}
\end{align}
with the response in each such mode being characterised by a dynamic Love number $k_l(\omega_{lmpq})$ and a phase lag $\epsilon_l(\omega_{lmpq})$, with $\omega_{lmpq}$ the tidal Fourier mode:
\begin{align}
    U_{lmpqhj} = &\left(\frac{R}{a}\right)^{l+1}\left(\frac{R}{a^*}\right)^l \frac{(l-m)!}{(l+m)!}(2-\delta_{0m}) \\ \nonumber
    &\times\; F_{lmp}(i^*) G_{lpq}(e^*) F_{lmh}(i)G_{lhj}(e) \\ \nonumber
    &\times\; k_l(\omega_{lmpq}) \cos\left(\left(v_{lmpq}^* - m \theta^*\right) - \left(v_{lmhj} - m \theta\right) - \epsilon_l(\omega_{lmpq})\right) \\
    v_{lmpq}^* \equiv& (l-2p)\omega^* + (l-2p+q)\mathcal{M}^* + m\Omega^* \\
    v_{lmhj} \equiv& (l-2h)\omega + (l-2h+j)\mathcal{M} + m\Omega.
\end{align}
Here $\mathcal{G}$ is the gravitational constant, $R$ is the radius of the disturbed body, $(a, e, i, \omega, \Omega, \mathcal{M})$ denote the usual Keplerian elements expressed in the coprecessing frame of the primary (see Section~2.1 in \citealt{Boue2019TidalElements} for a comprehensive treatment), $\theta$ is the rotation angle of the disturbed body, $F_{lmp}$ and $G_{lpq}$ are the inclination and eccentricity functions, respectively (e.g. \citealt{Kaula1964TidalEvolution}; our method for computation of the inclination functions is given in Sec.~\ref{app:inclination_functions}), and all quantities marked with an asterisk $(*)$ correspond to the perturber.

There are several ways of computing the eccentricity functions: though we relay a larger discussion thereof to the appendix, in Secs.~\ref{sec:eccentricity_function_integral} and \ref{sec:power_series_appendix}, we will briefly discuss some here. Our approach will consist of using a novel power series given by \citet{Proulx1988SeriesCoefficients} over part of the eccentricity domain, which converges faster than the conventionally used power series for the eccentricity functions (e.g. \citealt{Izsak1964ConstructionComputer, Hughes1981THECOEFFICIENTS}), and using a modified numerical integral expression comparable to that used by \citet{Gooding1989ExplicitOrbits} over the part of the domain where the other approach fails. The main advances in these two approaches that we employ compared to those used in the past can be summarised by their elimination of the pole at $e=1$, which greatly enhances efficiency both of the power series and numerical integral approach.

In the equations of motion, $U(\boldsymbol{r}, \boldsymbol{r}^*)$ will appear in a derivative, and in the scenario under consideration (where the only two tidally involved bodies are Neptune and Triton) the perturbing body is in any case precisely the body whose orbital evolution is altered by the perturbed potential, meaning that the variables marked with an asterisk can be set equal to those without, though only after the necessary derivatives have been taken. After these are set equal, we have for the tidal Fourier modes $\omega_{lmpq}$ that
\begin{align}
\label{eq:omega_lmpq_def}
    \omega_{lmpq} &= (l-2p)\dot{\omega} + (l-2p+q)n + m(\dot{\Omega} - \dot{\theta}) \\
    &\approx (l-2p+q)n - m\dot{\theta} \label{eq:omega_lmpq_def_approx}
\end{align}
where in the second equality we assume that the precession of the argument of periapse and of the ascending node are negligible compared to the Keplerian mean motion and sidereal rotation of the body; this approximation is useful as it greatly diminishes the number of potential derivatives to be evaluated if we furthermore approximate the mean motion by its Keplerian (non-perturbed) value. Do note (see e.g. App.~B in \citealt{Efroimsky2014TidalMethods}) that this approximation requires that we also assume that the perturbed (anomalistic) mean motion $\derivative{\mathcal{M}}{t}$ is close to the Keplerian mean motion $n$. For the difference in the cosine we have that
\begin{equation}
    \left(v_{lmpq}^* - m \theta^*\right) - \left(v_{lmpq} - m \theta\right) = 2(h-p)\omega + (2h-2p+q-j)\mathcal{M}.
\end{equation}
Given that we are interested in the secular behaviour of Triton's orbit, this latter expression will become useful once we average over the fast angles (the mean anomaly, argument of pericentre and the longitude of the node), as these prescribe which components of the sum are secular and which are oscillatory in nature. This means that we can dispose of all terms except for those that have $h=p$, $q=j$: hence, we obtain a simpler sum in the form
\begin{equation}
    \Tilde{U}(\boldsymbol{r}, \boldsymbol{r}^*) = - \frac{\mathcal{G}M}{a^*}\sum_{l\geq2}\sum_{m=0}^{l} \sum_{p=0}^{l} \sum_{q=-\infty}^{\infty}U_{lmpq}
\end{equation}
where
\begin{align}
    U_{lmpq} = &\left(\frac{R}{a}\right)^{l+1}\left(\frac{R}{a^*}\right)^l \frac{(l-m)!}{(l+m)!}(2-\delta_{0m}) \\ \nonumber
    &\times\; F_{lmp}(i^*) G_{lpq}(e^*) F_{lmp}(i)G_{lpq}(e) \\ \nonumber
    &\times\; k_l(\omega_{lmpq}) \cos\left(\epsilon_l(\omega_{lmpq})\right)
\end{align}
We then note that, under assumption of  Eq.~\ref{eq:omega_lmpq_def_approx} with additionally the mean motion approximated by its Keplerian value, the derivatives of $\mathcal{R}$ required to evaluate the secular (i.e. long-term averaged) tidal rates of the semi-major axis, eccentricity and inclination of a body are in general solely those in the mean anomaly, the arguments of pericentre of the primary and secondary and the longitudes of the nodes (see e.g. \citealt{Boue2019TidalElements, Renaud2021TidalTRAPPIST-1e} or \citealt{Luna2020TheSystem}), appearing in the following expressions:
\begin{align}
    \left<\derivative{a}{t}\right> &= 2na \left(\frac{M_N}{M_T}\left<\pdv{U_T}{\mathcal{M}}\right> + \frac{M_T}{M_N}\left<\pdv{U_N}{\mathcal{M}}\right>\right) \\
    \left<\derivative{e}{t}\right> &= n\frac{\sqrt{1-e^2}}{e} \left[\sqrt{1-e^2}\left(\frac{M_N}{M_T}\left<\pdv{U_T}{\mathcal{M}}\right> + \frac{M_T}{M_N}\left<\pdv{U_N}{\mathcal{M}}\right>\right) \right. \nonumber \\
    &\quad - \left.\left(\frac{M_T}{M_N}\left<\pdv{U_N}{\varpi_N}\right> + \frac{M_N}{M_T}\left<\pdv{U_T}{\varpi_T}\right>\right)\right] \\
    \sin{i_j}\left<\derivative{i_j}{t}\right> &= \frac{M_k}{M_j}\left[\frac{\mathcal{G} M_j M_k}{aC_j \dot{\theta}_j}\left(\left<\pdv{U_j}{\varpi_j}\right> - \cos{i_j}\left<\pdv{U_j}{\Omega_j}\right>\right) \right. \nonumber \\
    &\quad \left. - \frac{n}{\sqrt{1-e^2}}\left(\left<\pdv{U_j}{\Omega_j}\right> - \cos{i_j}\left<\pdv{U_j}{\varpi_j}\right>\right)\right] \\
    \left<\derivative{\dot{\theta}_j}{t}\right> &= -\frac{\mathcal{G} M_k^2}{a C_j}\left<\pdv{U_j}{\Omega_j}\right>
\end{align}
where we have followed the subscript convention as used by \citet{Renaud2021TidalTRAPPIST-1e} that $j$ is any of the two objects and $k$ its partner, and moreover introduce the dimensionless, positive potential $U_j$:
\begin{equation}
    U_j = -\frac{a}{\mathcal{G}M_k}\Tilde{U}_j
\end{equation}
so as to be consistent with the notation used by \citet{Luna2020TheSystem}. Note that in defining $i_N$ and $i_T$ in particular we follow the unfortunate convention that is usual in literature on Kaula's expansion (though it is often a source of confusion); the subscript denotes the object of which the equator is in that case the reference point of the inclination of the orbit. In particular, this means that $i_N$ is the inclination of the mutual orbit with respect to the equator of Neptune (the inclination $i$ that is common), and $i_T$ is the inclination of the mutual orbit with respect to the equator of Triton. They are therefore better thought of as obliquities (with respect to the orbit) rather than inclinations, so as to avoid confusing $i_T$ for the conventional inclination of Triton's orbit (which is equivalent to $i_N$ instead).

The angled brackets $\left<\cdot\right>$ denote, as in \citet{Luna2020TheSystem}, an averaging over the fast angles (the mean anomaly, arguments of pericentre and the longitudes of the ascending node) on the left-hand sides of these equations, while on the right-hand side they denote the consideration solely of the secular terms in these derivatives. Consequently, the derivatives in $\mathcal{M}$, $\varpi_T$, $\varpi_N$, $\Omega_T$ and $\Omega_N$ are the only derivatives of $U_N$ and $U_T$ that we need, and we can express them (and the tidal heating rate) as (e.g. \citealt{Renaud2021TidalTRAPPIST-1e}):
\begin{align}
\label{eq:partial_potential}
    \begin{bmatrix}
    \partial{U_{j}}/\partial{\mathcal{M}} \\
    \partial{U_{j}}/\partial{\varpi_j} \\
    \partial{U_{j}}/\partial{\Omega_j} \\
    \dot{E}_j
    \end{bmatrix} =& -\sum_{l\geq2}\left(\frac{R_j}{a}\right)^{2l+1}\sum_{m=0}^{l}  \frac{(l-m)!}{(l+m)!}(2-\delta_{0m}) \sum_{p=0}^{l} F_{lmp}^2(i_j) \nonumber \\
    &\times\; \sum_{q=-\infty}^{\infty}G_{lpq}^2(e) K_{l,j}(\omega_{j,lmpq})
    \begin{bmatrix}
        l-2p+q \\
        l-2p \\
        m \\
        -\omega_{j,lmpq}n^2a^2\beta M_k/M_j
    \end{bmatrix}
\end{align}
with the "quality function" $K_{l,j}(\omega_{j,lmpq})$ defined as (see e.g. \citealt{Bagheri2022TidalOverview})
\begin{equation}
    K_{l,j}(\omega_{j,lmpq}) = k_{l,j}(\omega_{j,lmpq})\sin{\left(\epsilon_{l,j}(\omega_{j,lmpq})\right)}.
\end{equation}
where we note that it is customary to drop the subscripts $j$ and $lmpq$ in these quantities whenever this promises no confusion (or when the distinction is meaningless); additionally, we have introduced the reduced mass $\beta=\frac{M_T M_N}{M_N+M_T}$. It is these preceding potential derivatives that we will have to approximate by a truncated version of the sums to perform our simulations; we must thus examine the convergence properties of these sums. To do so for each of these sums as well as for all possible configurations of Triton would be a laborious task, and so we shall approach the matter in a more systematic and general fashion, which is discussed in Sec.~\ref{sec:truncating_kaula_general}.

To forego any computation stability issues due to numerical (though not actual) singularities owing to the appearance of $e$ and $\sin i$ in several denominators, we will employ the substitutions used by \citet{Luna2020TheSystem}; that is, we will replace all occurrences of $i_j$ and $e$ by
\begin{align}
    x_j &= \cos{i_j} \\
    \xi &= \sqrt{1-e^2}
\end{align}
such that
\begin{align}
    -\sin{i_j}\derivative{i_j}{t} &= \derivative{x_j}{t} \\
    -\frac{e}{\sqrt{1-e^2}}\derivative{e}{t} &= \derivative{\xi}{t}.
\end{align}
The resulting equations of motion are then
\begin{align}
    \left<\derivative{a}{t}\right> &= 2na \left(\frac{M_T}{M_N}\left<\pdv{U_N}{\mathcal{M}}\right> + \frac{M_N}{M_T}\left<\pdv{U_T}{\mathcal{M}}\right>\right) \label{eq:semimajor_axis} \\
    \left<\derivative{\xi}{t}\right> &= n\left[\left(\frac{M_T}{M_N}\left<\pdv{U_N}{\varpi_N}\right> + \frac{M_N}{M_T}\left<\pdv{U_T}{\varpi_T}\right>\right) - \xi\left(\frac{M_T}{M_N}\left<\pdv{U_N}{\mathcal{M}}\right> + \frac{M_N}{M_T}\left<\pdv{U_T}{\mathcal{M}}\right>\right)\right] \label{eq:eom_eccentricity} \\
    \left<\derivative{x_j}{t}\right> &= \frac{M_k}{M_j}\left[\frac{n}{\xi}\left(\left<\pdv{U_j}{\Omega_j}\right> - x_j\left<\pdv{U_j}{\varpi_j}\right>\right) - \frac{\mathcal{G} M_j M_k}{aC_j \dot{\theta}_j}\left(\left<\pdv{U_j}{\varpi_j}\right> - x_j \left<\pdv{U_j}{\Omega_j}\right>\right)\right] \label{eq:eom_obliquity} \\
    \left<\derivative{\dot{\theta}_j}{t}\right> &= -\frac{\mathcal{G} M_k^2}{a C_j}\left<\pdv{U_j}{\Omega_j}\right>. \label{eq:eom_rotation_rate}
\end{align}
Note that the terms to do with the bodies' oblateness are contained in the terms we neglected in the approximation made in Eq.~\ref{eq:omega_lmpq_def_approx} (see e.g. \citealt{Luna2020TheSystem}). The terms describing the bodies' triaxiality would appear in the expressions we use, yet modern Triton is an oblate ellipsoid without triaxial component to within measurement error \citep{Thomas2000TheProfiles}; as our assumptions on early Triton's triaxiality would have to be speculative, we will neglect this term. The principal moments of inertia for Neptune are generally not determined for the non-polar components, and studies tend to assume an axially and hemispherically symmetric mass distribution (e.g. \citealt{Helled2010UranusRotation, Podolak2012WhatNeptune, Neuenschwander2022EmpiricalNeptune}), so we will assume that the resulting triaxility effects can be ignored.

To avoid suffering from the short-period nature of the spin rate-evolution (which is precisely the type of phenomenon we were trying to avoid by resorting to Kaula's expansion) while not losing our ability to track its progression, we will resort to the method used by \citet{Walterova2020ThermalExoplanets}: initially, we will propagate the rotation rate explicitly, until it equilibriates. Afterwards, we will assume it remains in equilibrium and compute the rotation rate that way. This assumption does (implicitly) neglect the spin energy dissipated between spin-orbit transitions in the tidal heating rate given by Eq.~\ref{eq:partial_potential}, and so we must manually correct for it a posteriori (see Sec.~\ref{sec:equilibrium_dissipation_transition} for details).

There are some assumptions in this formulation that we wish to make explicit: the inner moons of Neptune (which comprise the only moons of a mass $>10^{-4}M_T$, with the exception of Nereid; see e.g. \citealt{Holman2004DiscoveryNeptune, Brozovic2020OrbitsNeptune}) would effectively increase the value of $J_2$ by a marginal amount, and so their effects are assumed negligible implicitly by the approximation in Eq.~\ref{eq:omega_lmpq_def_approx}, which neglects $J_2$. The outer moons are not massive enough to have any major effect, with the possible exception of Nereid. As this latter moon is thought to be a remnant of Neptune's proto-satellite system (from before Tritonian capture; see e.g. \citealt{Brozovic2020OrbitsNeptune}), including it in the simulation would be of scientific interest. Nonetheless, the inclusion of outer perturbations would require the computation of a whole new set of terms in Kaula's expansion of the perturbing potential, and given the small size (compared to Triton) of Nereid and the loose constraints on its initial orbit, we opt to neglect it.

To propagate these equations of motion, we need to choose an appropriate rheological model to represent the tidal response of both Triton and Neptune: that is, we need to choose an expression for the tidal quality functions $K_{l,T}(\omega)$ and $K_{l,N}(\omega)$.

\section{The tidal quality function}
\label{sec:quality_function_general}
With the dynamical block in Fig.~\ref{fig:schematic_components} now defined, we need only find suitable expressions for the rheological and interior components of our model. As the primary purposes of this work is to showcase the possibility of going to extreme eccentricities even with an arbitrary rheological model, we will refrain from choosing too convoluted an interior model: keeping it fixed will be sufficient to do so, without sacrificing too much of the explaining power of this representation. With that set, we need only choose a rheology for Triton, parametrised by the tidal quality function $K_l(\omega)$.

In recent years, a number of new expressions for the tidal quality function have arisen: these have generally come into use to compensate for the deficiencies in modelling power that older models have turned out to possess. However, we will find that it is still convenient or equally valid to use simplified tidal quality functions in certain scenarios; additionally, such simplified tidal quality functions are what has been used in a large part of classical literature, and they are all that has been used to describe Triton thus far. Hence, we will want to compare the performance of more advanced models against the simpler models. As such, we introduce two simplified expressions for the tidal quality function in Sec.~\ref{sec:CTL_CPL}, followed by an explanation of more advanced models in Sec.~\ref{sec:quality_function_homogeneous}.

\subsection{Simplified expressions for the quality function}
\label{sec:CTL_CPL}
Before we explore more advanced rheological models, it is useful to give a brief overview of the two simple rheological models that have been in use in much of the discourse concerning tidal dissipation in general, and Triton in particular.

The first such model is the constant phase lag model (sometimes also called the constant geometric lag model: commonly abbreviated to CPL); in his original, it seems that \citet{Kaula1961AnalysisSatellites, Kaula1964TidalEvolution} implicitly assumed that the Love numbers $k_l$ and their associated phase lags $\epsilon_l$ were constant with frequency. While they have turned out not to be \citep{Boue2019TidalElements, Bagheri2022TidalOverview}, a large number of studies in the second half of the last century have implicitly or explicitly followed this approach. In this case, the quality function is conventionally written
\begin{equation}
    K_l(\omega) = \textrm{sign}(\omega)\frac{k_l}{Q_l},
\end{equation}
where $k_l$ is the dynamic Love number (assumed independent of frequency), and $Q_l$ is the tidal quality factor (also assumed independent of frequency). The sign out front determines whether the tidal mode in question is lagging the primary-secondary line or leading it, and therefore whether it will drag an orbit down or raise it. For issues with this model, see the work by \citet{Efroimsky2013TidalTorque}. Note additionally that there are various models which can be and have been termed the constant phase lag model; generally, they are derived by inserting the above expression into Kaula's expansion, and neglecting small terms where applicable.

The second simple model is the constant time lag (CTL) model. It originates from the study of binary stars undergoing tidal interaction, and arose there by assuming that the tidal friction is comparatively mellow \citep{Hut1981TidalSystems}. Consequently, it is sometimes called the weak friction model; in general, it can be thought of as a viscously damped harmonic oscillator \citep{Efroimsky2013TidalTorque}, and it holds relatively well for bodies without (visco)elastic components, such as stars or gas giants \citep{Renaud2018IncreasedExoplanets}. Whether it is applicable to ice giants still remains an open matter, and depends on the dominance of the frozen core, if any exists \citep{Renaud2021TidalTRAPPIST-1e}. Rather than keeping the phase lag constant, it is now assumed that the tidal response of the body is delayed by an equal time lag for every tidal mode. In this case, the tidal quality function is instead given by:
\begin{equation}
    K_l(\omega) = k_l\sin(\omega \Delta t_l) \approx k_l\omega \Delta t_l
\end{equation}
where the small-angle approximation is in practice oftentimes the expression that is preferred, as it allows a further set of simplifications (see e.g. \citealt{Correia2022TidalCoefficients}). \citet{Makarov2013NoMoons} give an in-depth treatment of the issues with this model, and the regions where and the objects for which it can apply.


\subsection{The quality function for a homogeneous body}
\label{sec:quality_function_homogeneous}
To supplement the simple quality functions, we also discuss briefly an additional set of quality functions that model more complex material behaviour. Applications of these using analytical expressions requires that we assume a homogeneous body, however: computation of the tidal response for layered bodies (though more realistic) requires a significantly more advanced and computationally expensive rheological module, such as the propagator matrix technique \citep{Sabadini2016GlobalEdition}. As the aim of this work is to showcase the feasibility of high-eccentricity tidal dynamics, we will stick with homogeneous bodies, which can at the very least provide qualitative results (e.g. \citealt{Efroimsky2012TidalSuper-earths, Luna2020TheSystem, Renaud2021TidalTRAPPIST-1e}), though we do note that \citet{Walterova2020ThermalExoplanets} have recently made a foray into the territory of layered bodies.

The quality function of a homogeneous body can in general be given as (e.g. \citealt{Bagheri2022TidalOverview}):
\begin{equation}
    K_l(\omega) = - \frac{3}{2(l-1)}\frac{\Tilde{\mu}\mathcal{B}_l\:\textrm{Im}(\Tilde{J})}{\left(\textrm{Re}(\Tilde{J})+\Tilde{\mu}\mathcal{B}_l\right)^2 + \left(\textrm{Im}(\Tilde{J})\right)^2}
\end{equation}
where
\begin{align}
    \Tilde{J} &= \Bar{J} \mu, \\
    \mathcal{B}_l &= \frac{1}{l}(2l^2+4l + 3), \\
    \Tilde{\mu} &= \frac{\mu}{\rho g R} = \frac{4\pi\mu R^4}{3\mathcal{G} M^2}
\end{align}
with $\Bar{J}$ the complex compliance, $\mu$ the rigidity, $\rho$ the density of the body in question, $g$ its surface gravitational acceleration and $R$ its radius. We will refer to $\Tilde{\mu}$ as the effective rigidity, explicitly deviating from the nomenclature in e.g. \citet{Renaud2018IncreasedExoplanets}: losing the factor $19/2$ out front removes the implicit assumption of the $l=2$-case. By introducing the effective rigidity we do not require strong assumptions on $\mu$, $\rho$ and $g$, but only on their ratio. We then only require a rheological law $\Tilde{J}(|\omega|)$ (noting that in general the rheological law is even in the argument; henceforth we shall use $\chi=|\omega|$ to denote this). The most commonly used realistic rheological models are the Maxwell, Burgers, Andrade and Sundberg-Cooper rheologies $\Tilde{J}_M, \Tilde{J}_B$, $\Tilde{J}_A$ and $\Tilde{J}_{SB}$ \citep{Bagheri2022TidalOverview}:
\begin{align}
    \Tilde{J}_M(\chi) &= 1 - \frac{j}{\chi \tau_M} \\
    \Tilde{J}_B(\chi) &= \left(1+\frac{\Delta}{1+\chi^2\tau^2}\right) + \left(\frac{\Delta \chi\tau}{1+\chi^2\tau^2}\right)j - \frac{j}{\chi\tau_M}\\
    \Tilde{J}_{A}(\chi) &= 1 + \Gamma(1+\alpha)(\chi \tau_A)^{-\alpha}e^{\alpha j\pi/2} - \frac{j}{\chi\tau_M} \\
    \Tilde{J}_{SB}(\chi) &= 1 + \Gamma(1+\alpha)(\chi \tau_A)^{-\alpha}e^{\alpha j\pi/2} + \frac{\Delta}{1+\chi^2\tau^2} + \left(\frac{\Delta \chi\tau}{1+\chi^2\tau^2} - \frac{1}{\chi\tau_M}\right)j
\end{align}
where $\tau_M=\frac{\eta}{\mu}$ is the Maxwell time, $j$ is the complex number (as $i$ is reserved for the inclination), $\Delta$ is the anelastic relaxation strength, $\tau$ is the characteristic time for the anelastic response, $\Gamma(x)$ is the Gamma function, $\alpha$ the Andrade parameter, and $\tau_A$ the Andrade time. As the Burgers, Andrade, and Sundberg-Cooper models reduce to the Maxwell model for appropriate values of the corresponding parameters (and their behaviour to first order is dictated by the Maxwell-terms), we will employ the Maxwell rheology as a model that is qualitatively representative of these more advanced rheological models, such that we can contrast them against the commonly employed simplified models presented in Sec.~\ref{sec:CTL_CPL}.

As a final point of comparison: it is possible to obtain the CTL model from the Maxwell model when $\mathcal{B}_l\Tilde{\mu}\gg1$ \citep{Bagheri2022TidalOverview}: in this case, over the region $|\omega|<\omega_{\textrm{peak}}=|\frac{1}{\mathcal{B}_l\tau_M\Tilde{\mu}}|$ we obtain
\begin{equation}
\label{eq:Maxwell_CTL_approx}
    K_l(\omega)\approx \frac{3\mathcal{B}_l\tau_M\Tilde{\mu}}{2(l-1)}\omega \implies \Delta t_l k_l = \frac{3\mathcal{B}_l\tau_M\Tilde{\mu}}{2(l-1)}
\end{equation}
This is particularly useful, as it allows us to generate a CTL model that directly corresponds to a given Maxwell model, so as to provide the most representative comparison. In this case, we note that the corresponding value of $\Delta t_l k_l$ for realistic values of an icy Maxwell body becomes extremely high: while this steep slope is not problematic for the Maxwell body (it will not last forever, after all), this same slope for a Maxwell body gives rise to extreme tidal potentials for high eccentricities, as the active tidal modes will then not all fall in the regime corresponding to $|\omega|<\omega_{\textrm{peak}}$. Consequently, we cannot blindly apply Eq.~\ref{eq:Maxwell_CTL_approx} to find CTL approximations for any body at any eccentricity: this will become important in Ch.~\ref{ch:spin-orbit_chains}.

An important final note is that each of these advanced rheologies result in a quality function that grows sub-linearly everywhere (for some appropriate scaled linear function): this will allow us to later set a conservative upper bound on the number of terms we must account for to properly represent the innermost infinite sums in Eq.~\ref{eq:partial_potential}.

\section{Conclusions}
While there many ways in which dynamical-interior evolution of planetary bodies can be conducted, no single approach need have any preference for physical reasons. From the approaches that exist we derive a framework that seemingly describes the components that are, explicitly or implicitly, common to each of the approaches taken in literature: a dynamical, interior and rheological model. On the basis of this framework and the notion that our model must be practical even when simulating over astronomical timescales, we find that the Darwin-Kaula formalism is most appropriate to model Triton, which agrees with the approach taken in literature, and give a mathematical description of the methods and assumptions to be applied in this work. From here on, we can thus start moving towards an implementation of a novel model for Triton.

\chapter{Truncating Kaula's expansion}
\label{ch:validation}
To account for the dynamical and thermal consequences of the tides raised in Triton by Neptune and vice-versa, we can thus employ a development of the tidal potential into spherical harmonic terms as pioneered by \citet{Darwin1879OnEarth, Darwin1880OnPlanet} and expanded upon in great detail by \citet{Kaula1961AnalysisSatellites, Kaula1964TidalEvolution}. While this allows, in principle, for a full accounting for the tidal-dynamical effects caused by mutual deformations between two orbiting bodies, the necessary transformation from the spherical harmonics expression to the familiar orbital elements introduces three infinite series (see Eq.~\ref{eq:partial_potential}): the first, because it requires evaluation of the eccentricity functions, which require an expansion into an infinite series in eccentricity $e$, the second because those eccentricity functions themselves appear in an infinite sum, and the third from the infinite sum over degrees $l$ due to the spherical harmonics expansion that one started with.

The chief objective of this chapter shall therefore be to motivate the point up to which these expansions must be computed to be able to properly represent early Triton. To this end, we will first derive some more general results in this respect in Sec.~\ref{sec:truncating_kaula_general}, before validating these results and deriving empirical support for our truncation decisions in Secs.~\ref{sec:truncating_degree} and \ref{sec:validation_truncation_prescriptions}. This is done using a Python 3.11.7-based code that implements the formalism presented in Ch.~\ref{ch:kaula_theory}: a verification of this code is given in App.~\ref{ch:verification}.

\section{Truncating the Kaula expansion in a general setting}
\label{sec:truncating_kaula_general}
We wish to truncate the infinite expressions in Eq.~\ref{eq:partial_potential}: it is clear that the outermost sum must be truncated to some degree $l_{\max}$, and that the sum over $q$ must also be truncated. Another truncation choice that has to be made, though it is not immediately clear from Eq.~\ref{eq:partial_potential}, is the truncation of the eccentricity functions $G_{lpq}(e)$, as those are customarily expressed in power series (though numerical integral representations also exist). A treatment of a numerical integral expression is given in Sec.~\ref{sec:eccentricity_function_integral}, while the power series expressions, their properties and its implementation in our code is discussed in Sec.~\ref{sec:power_series_appendix}. While other authors often tend to truncate both the inner sum in $q$ and the eccentricity functions $G_{lpq}$ simultaneously by truncating their combined sum to a certain power of $e$ (see e.g. the work by \citealt{Renaud2021TidalTRAPPIST-1e}), the coefficients involved in the power series expansion of $G_{lpq}$ grow to considerable magnitude for larger values of $q$, and so an order-of-magnitude estimate based solely on the power of $e$ present in a term is prone to error. Hence, we will prefer to treat the convergence of the power series for the $G_{lpq}$ separately from the convergence of the sum in $q$.

Due to the low present-day eccentricity of most Solar System objects of note, for most evaluations of tidal heating an evaluation up to only the $e^2$-terms can be used, or at least has been so historically (see e.g. \citealt{Peale1979MeltingDissipation, Segatz1988TidalIo, Makarov2014TidalB, Rovira-Navarro2021, Kleisioti2023TidallyCharacterization}, or for examples relating to Triton in particular see e.g. \citet{Chyba1989TidalSystem, Goldreich1989NeptunesStory, Ross1990TheTriton, Nimmo2015PoweringGeology}. However, Triton's eccentricity is certain to have been large in the past (see e.g. \citealt{Ross1990TheTriton}), and recent work on the topic has shown that expansions up to $e^{20}$ are necessary for the Pluto-Charon or Mars-Phobos-Deimos systems \citep{Renaud2021TidalTRAPPIST-1e, Bagheri2021DynamicalProgenitor, Bagheri2022TheSystem}, all of which are likely to have possessed similar to if most likely even lower eccentricities than early Triton (which probably had $e>0.9$; see e.g. \citealt{Ross1990TheTriton, McKinnon2014Triton} for evolution-based arguments, or e.g. \citealt{Nogueira2011ReassessingTriton} for capture-based arguments supporting such high values). Disappointingly, none of these analyses show a robust convergence analysis, however.

To be confident that our finitely truncated implementation accurately reflects the reality of Eq.~\ref{eq:partial_potential}, we will therefore discuss separately the approximation of the eccentricity functions $G_{lpq}$ in Sec.~\ref{subsec:truncating_G_lpq} and that of the full expansion in $q$ in Sec.~\ref{subsec:truncating_q}, both in a general sense depending only on the eccentricities encountered in the scenario in question. In Sec.~\ref{sec:validation_truncation_prescriptions} we will then evaluate the truncation of these two quantities in a practical setting in the particular scenario of early Triton orbiting Neptune by numerical experiments. The truncation in the degree $l$ is, unfortunately, less receptive to a general consideration: in fact, most authors assume that the terms $l\geq 3$ can be neglected as a result of the factor $(R/a)^{2l+1}$, though numerical experimental work has previously shown that this reasoning does not hold for non-negligible eccentricities \citep{Renaud2021TidalTRAPPIST-1e, Bagheri2021DynamicalProgenitor, Bagheri2022TheSystem}. For lack of an acceptable analytical bound, we will also explore this the truncation in the degree empirically in Sec.~\ref{sec:truncating_degree}.

\subsection{Truncating the expansion in $|q|$}
\label{subsec:truncating_q}
Let us first concern ourselves with truncating the expansion in $q$. It can be shown (see Sec.~\ref{app:limiting_behaviour_G_lpq_large_q}) that the eccentricity functions will decay and become arbitrarily small beyond some $|q|$, as hypothesised by \citet{Szeto1982ONORBITS}: our approach here is therefore to bound the convergence rate of the general inner sum in Eq.~\ref{eq:partial_potential} (i.e. that in $q$) by a set of infinite sums for which finite expressions exist. The convergence criterion for an approximation to the infinite sums to be sufficiently accurate can then be evaluated by comparing the analytical result to the truncated sum for various truncation levels.

It can be shown that the convergence rate of the sum over $q$ in Eq.~\ref{eq:partial_potential} is always bounded by a multiple of a sum of the following three terms, for which finite analytical expressions exist (see Sec.~\ref{sec:q_series_convergence} for a more detailed treatment):
\begin{align}
    \sum_{q=-\infty}^{\infty}G_{lpq}^2 &= X^{-2(l+1),0}_0 \label{eq:q^0_analytic} \\
    \sum_{q=-\infty}^{\infty}(l-2p+q)G_{lpq}^2 &= (l-2p)\sqrt{1-e^2}X^{-2l-4,0}_0 \label{eq:q^1_analytic} \\
    \sum_{q=-\infty}^{\infty}(l-2p+q)^2G_{lpq}^2 &= (l+1)^2\left(2X^{-2l-5,0}_0 - X^{-2l-4,0}_0\right) \nonumber \\
    &+ \left[(l-2p)^2-(l+1)^2\right](1-e^2)X^{-2l-6,0}_0. \label{eq:q^2_analytic}
\end{align}
where the $X^{n,k}_s$ are the Hansen coefficients (see Sec.~\ref{subsec:hansen_coeff_intro}). So as to be able to perform this analysis robustly and independently of the analysis in Sec.~\ref{subsec:truncating_G_lpq}, we will compute the eccentricity functions and/or Hansen coefficients using their numerical integral representations as discussed in Sec.~\ref{sec:eccentricity_function_integral}, which can be computed to any desirable accuracy (though at significant computational cost). Though it is possible there is some asymmetry in the values of $q$ to take into account for a given level of accuracy between the negative and positive values of $q$ (which seems to be the case e.g. from the analysis presented by \citealt{Szeto1982ONORBITS}), we opt to examine only a symmetric maximum considered value of $|q_{\max}|$, to simplify the analysis to be one-dimensional. 

\begin{figure}
    \centering
    \includegraphics[width=1\linewidth]{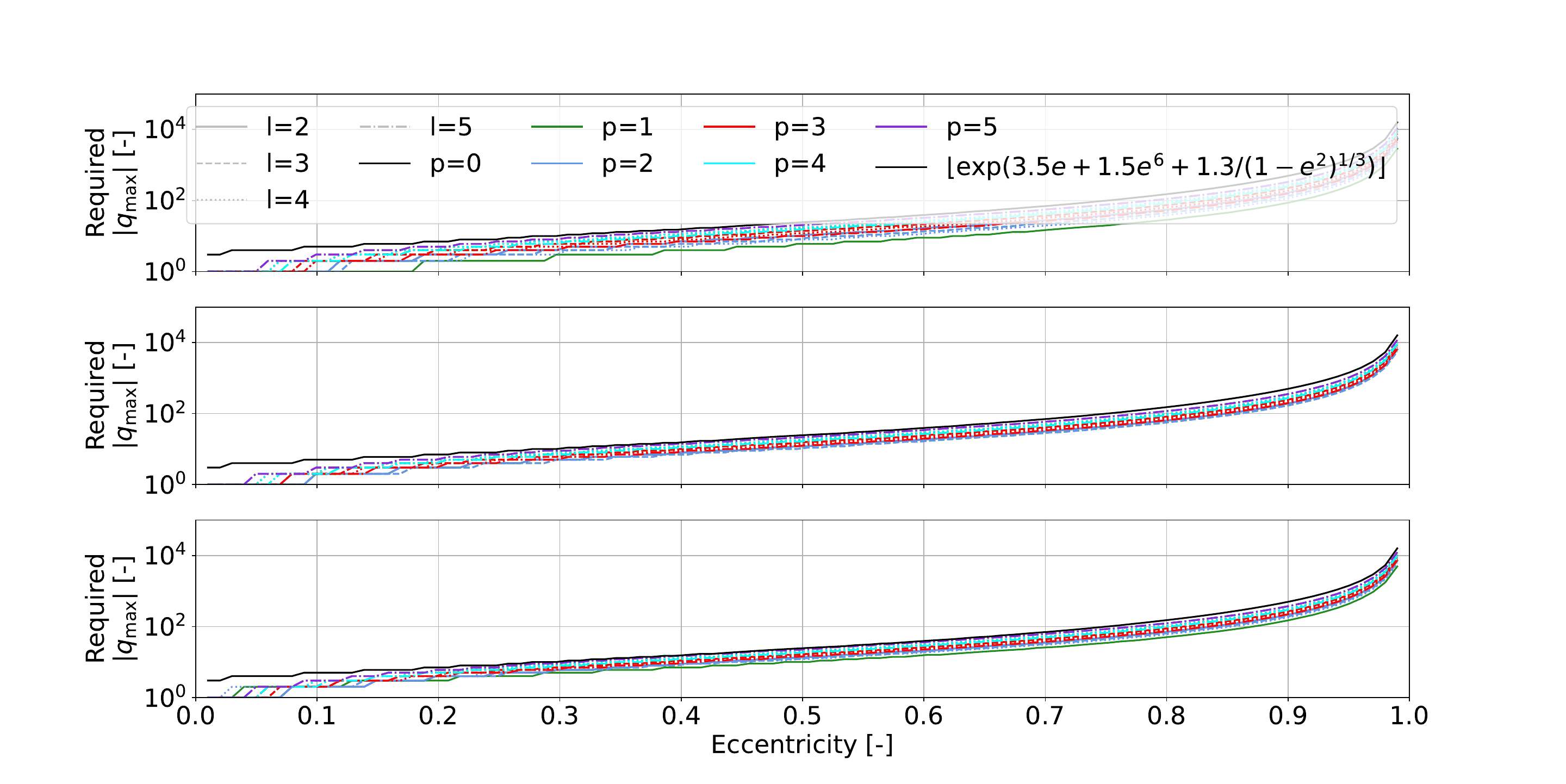}
    \caption{Required $|q_{\max}|$ to compute the left-hand side of Eqs.~\ref{eq:q^0_analytic}-\ref{eq:q^2_analytic} (from top to bottom) to within 1\% of the right-hand side as a function of eccentricity.}
    \label{fig:1percent_q}
\end{figure}

\begin{figure}
    \centering
    \includegraphics[width=1\linewidth]{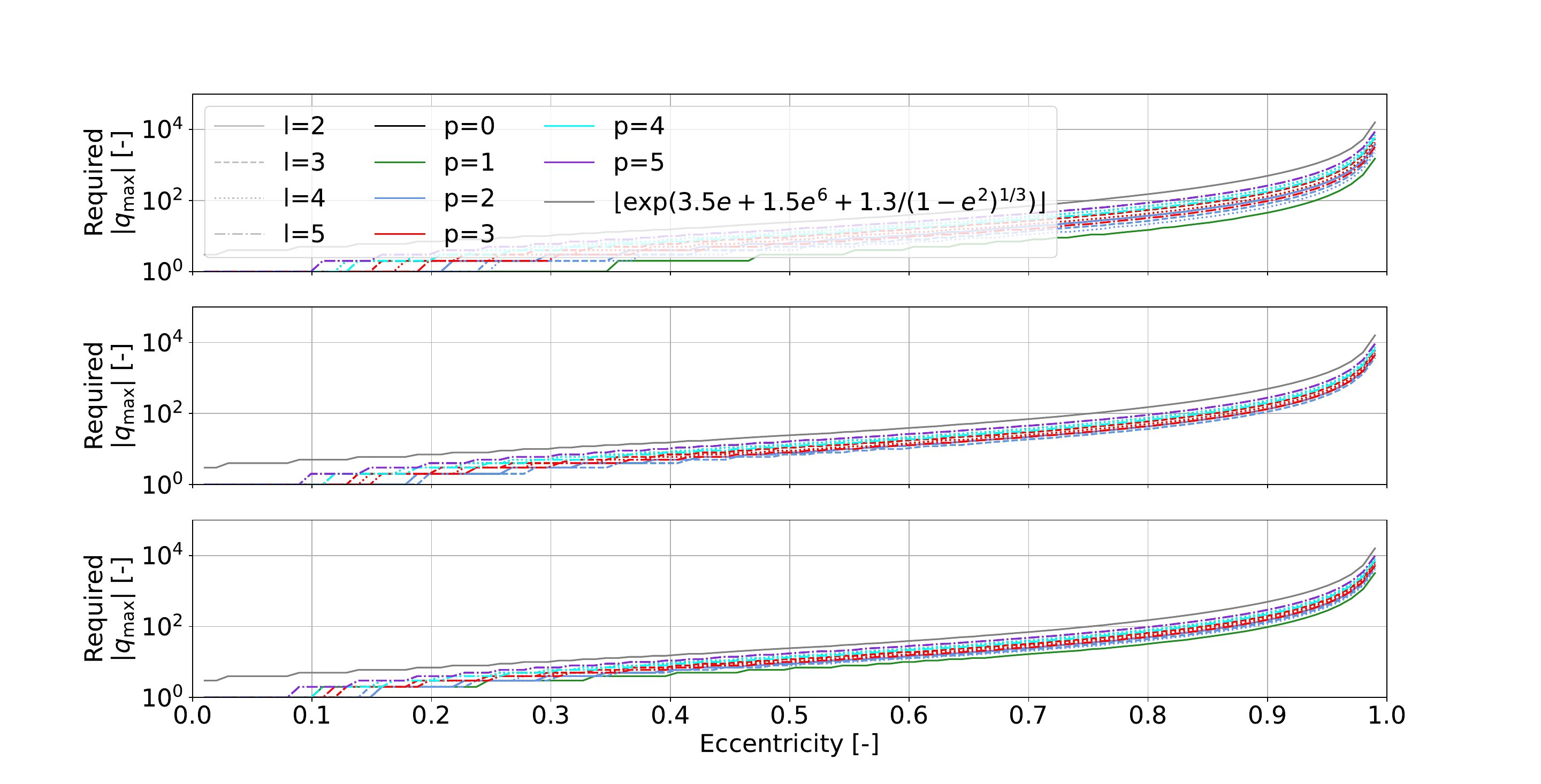}
    \caption{Required $|q_{\max}|$ to compute the left-hand side of Eqs.~\ref{eq:q^0_analytic}-\ref{eq:q^2_analytic} (from top to bottom) to within 10\% of the right-hand side as a function of eccentricity.}
    \label{fig:10percent_q}
\end{figure}

For a given eccentricity $e$, one can then compute out to which absolute value $|q_{\max}|$ the index $q$ must be expanded to yield a given accuracy. We compute the required value for $1\%$ and $10\%$ (to show that the convergence is well-behaved) accuracy for all values of $l$, and $p$ up to degree $l=5$, and display them in Figs.~\ref{fig:1percent_q} and \ref{fig:10percent_q} respectively. The values for individual combinations of $l$ and $p$ are not particularly important; rather, it becomes clear that there is no significant functional dependence on degree $l$ more than as a scaling factor, while some stratification appears to happen with $p$, where larger values of $p$ (which come with larger values of $l$) require a slightly greater number of terms to capture fully, though this effect is minor compared to the effect of increasing $e$. Additionally, an empirical functional form for the upper bound of the required $|q_{\max}|$ to be accurate to better than $1\%$ can then be found to be as follows, holding out to at least $e=0.99$ and at least $l \leq 5$ (though note that it will likely have to be loosened for $l>5$):
\begin{equation}
\label{eq:q_max_req}
    |q_{\max}|(e)\leq \lfloor\exp\left(3.5e+1.5e^6+1.3(1-e^2)^{-1/3}\right)\rfloor.
\end{equation}
Note that this upper bound then requires no assumptions on the particular system under consideration: all that is required is that the magnitude of the quality function of both bodies grows sub-linearly with $q$ through its argument $\omega_{lmpq}$, as is the case for the rheologies discussed in Sec.~\ref{sec:quality_function_homogeneous}.

\subsection{Truncating the expansion of the eccentricity functions}
\label{subsec:truncating_G_lpq}
Now that we can bound the required values of $q$ to take into account, we must also determine how many terms to take into account in the expansion of the eccentricity functions $G_{lpq}$. While previous work has largely made use of a simple power series in $e^2$ to approximate the $G_{lpq}$, doing so already requires terms up to $e^{20}$ or greater for moderate eccentricities observed in some exoplanetary systems \citep{Renaud2021TidalTRAPPIST-1e}, with no guarantee of convergence even then. However, \citet{Proulx1988SeriesCoefficients} showed that factoring out a pole of the form $(1-e^2)^{n+3/2}$ before expanding the Hansen coefficients into a power series of the form $e^2$ can greatly enhance its rate of convergence (see Sec.~\ref{sec:power_series_appendix} for an overview of our implementation of their recursive method by which to compute the power series coefficients). Unfortunately, their manuscript appears to have contained a typographical error which we correct in Eq.~\ref{eq:recurrence_Y}: we were able to reproduce their printed tables of convergence rates, and so we presume this error was only typographical. Nonetheless, no other studies have (to the knowledge of the authors) used this improved method, possibly as a result. As the eccentricity functions are a subset of the Hansen coefficients (see Sec.~\ref{subsec:hansen_coeff_intro} for their relation), this approach can, however, greatly reduce the computational cost of evaluation of the eccentricity functions, by requiring fewer terms than the simple power series in $e^2$ when going out to eccentricities closer to $e=1$.

\begin{figure}
    \centering
    \includegraphics[width=1\linewidth]{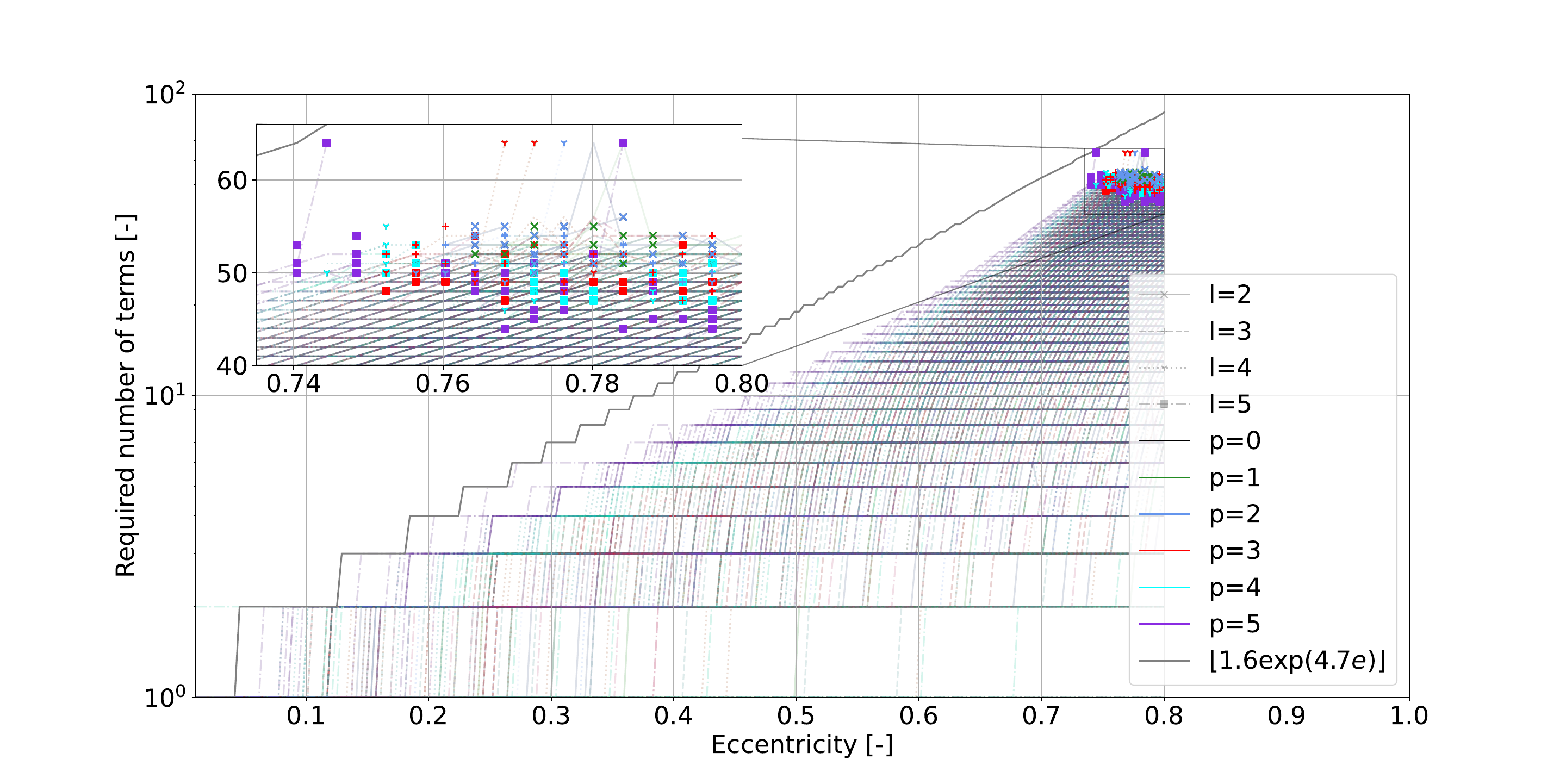}
    \caption{Number of power series terms $r_{\max}$ required to compute the individual eccentricity functions $G_{lpq}$ to within 10\% as a function of $l$, $p$ and eccentricity ($q$ is also varied over the range dictated by \ref{sec:q_series_convergence}, but its value is not marked on the graph as no clear effects due to $q$ occur). Markers denote the eccentricities beyond which a given eccentricity function $G_{lpq}$ expressed as power series no longer converges to within $10\%$ due to round-off error.}
    \label{fig:power_series_convergence}
\end{figure}

As our approach in the Sec.~\ref{subsec:truncating_q} dictates which eccentricity functions must be included for a given level of accuracy, we can now guarantee convergence of our summation terms for a given value of $l$ to within a specified accuracy level if we can show the eccentricity functions $G_{lpq}$ to be approximated well enough by a power series of a given number of coefficients. As we have a numerical integral representation of these eccentricity functions, which can be computed to arbitrary precision, we can in fact compute precisely how far out we need to go to do this! The result of this analysis is shown in Fig.~\ref{fig:power_series_convergence}, along with (1) markers denoting the point beyond which round-off error prevents convergence to within $10\%$ and (2) an empirical, conservative upper bound on the required number of power series terms for convergence to within $10\%$:
\begin{equation}
    r_{\max}(e) \leq \lfloor1.6\exp(4.7e)\rfloor.
\end{equation}
Similar to for the index $q$, a stratification appears with the terms due to greater $p$ requiring more terms to converge; for larger $e$, however, this stratification seems to break down, and by the point round-off error starts to play a role it has mostly disappeared.
Once we exceed eccentricities of $e\approx 0.74$, the power series coefficients start to grow sufficiently large that round-off error becomes a problem, and power series start to fail to converge to the specified $10\%$; initially this only affects the power series for terms of $l\geq 3$, but at $e\approx 0.76$ this also happens for the terms due to $l=2$. Additionally, as the inset in Fig.~\ref{fig:power_series_convergence} highlights, there is no clear pattern in which $G_{lpq}$ fail when as a function of $l$ or $p$. We may therefore expect that the power series representation by \citet{Proulx1988SeriesCoefficients} fails for all $G_{lpq}$ for $e\gtrsim 0.75$ when implemented in a standard 64-bit precision setting.

Consequently, beyond this point the power series become unreliable and cannot be expected to represent the true value of the individual $G_{lpq}$ well, and we will either require a different power series representation, higher-precision arithmetic (at correspondingly greater computational cost) or we must rely on the numerical integral representation: in the remainder of this work, we will do the latter.

\begin{figure}
    \centering
    \includegraphics[width=1\linewidth]{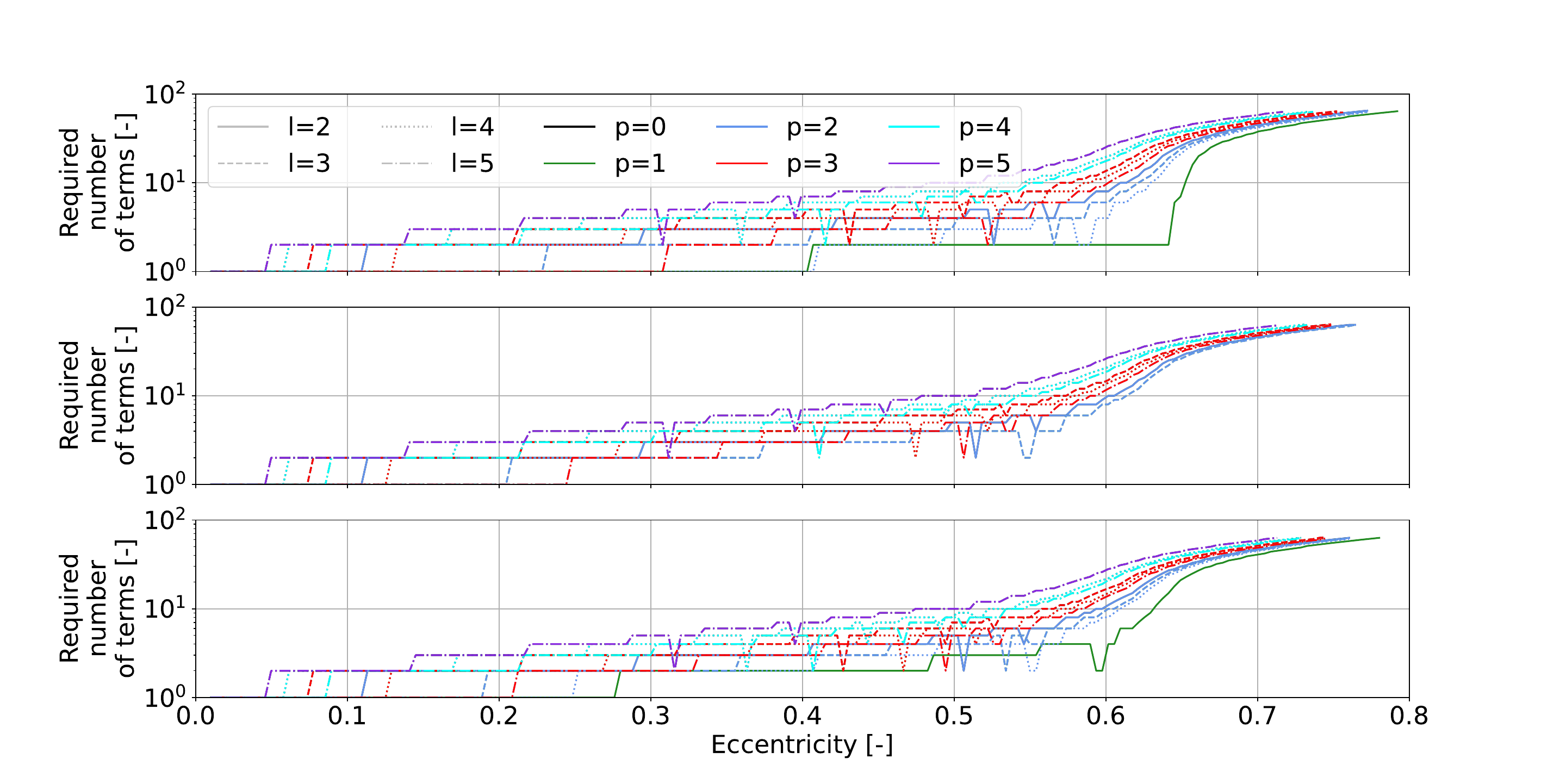}
    \caption{Number of power series terms required for the $G_{lpq}$ to compute the left-hand side of Eqs.~\ref{eq:q^0_analytic}-\ref{eq:q^2_analytic} (ordered from top to bottom) to within $10\%$ of the right-hand side while truncating $q$ according to Eq.~\ref{eq:q_max_req}.}
    \label{fig:power_series_sum}
\end{figure}

To see whether this analysis is robust when applied in a situation similar to Eq.~\ref{eq:partial_potential}, we compute the number of terms required (when truncating all $G_{lpq}$ to the same number of terms) to compute the left-hand sides of Eqs.~\ref{eq:q^0_analytic}-\ref{eq:q^2_analytic} to within of the right-hand side $10\%$ when approximating the left-hand side with the power series while truncating at $|q_{\max}|$ prescribed by Eq.~\ref{eq:q_max_req} in Fig.~\ref{fig:power_series_sum}. Indeed, here too we see a roughly constant exponential increase in the number of terms required to accurately represent the sums up to $e\approx0.5$, from whence onward a sharp increase starts to occur of which we can now most likely appoint round-off error to be the culprit.

This therefore allows us to set, in general, a bound on the number of terms (1) in the inner expansion in $q$ and (2) in the Proulx-McClain power series expression for $G_{lpq}$ to include in the evaluation of the equations of motion for a high-eccentricity body as a function only of its eccentricity. All that then remains is to evaluate the degree $l_{\max}$ out to which to expand Eq.~\ref{eq:partial_potential}; as this does not appear to admit any similar isolated evaluation (that is, without requiring further assumptions on the body in question), we will evaluate this empirically using the fully implemented equations of motion to propagate a representative scenario for the Triton-Neptune system in Sec.~\ref{sec:truncating_degree}. We will also use the same scenario in Sec.~\ref{sec:validation_truncation_prescriptions} to validate that our (presumably general) results hold at least when applied to the Triton-Neptune system.

\section{Empirically truncating the expansion in the degree $l$}
\label{sec:truncating_degree}
While we have been able to provide truncation prescriptions for the index $q$ and for the power series expressions for the eccentricity functions in a more general setting using analytical expressions, this does not appear feasible for the truncation of the degree $l$ of the expansion. To truncate this, we will therefore examine the evolution of a representative Triton-Neptune-like system for various truncation levels $l_{\max}\leq 5$, and see by which point the truncation of the degree ceases to play a significant role.

The scenario will start at the maximum allowable eccentricity of 0.74 for which we previously found the eccentricity functions to still converge\footnote{While we can in principle exceed this eccentricity by using the numerical integral representation of the eccentricity functions, the associated computational cost is prohibitive for the repeated computations required in this analysis.} for all $l\leq5$ in Sec.~\ref{subsec:truncating_G_lpq}, and the associated semi-major axis is chosen by assuming that Triton's orbit conserved its angular momentum over the course of its eccentricity damping; a posteriori we will see that this is a reasonable assumption. By choosing a zero-eccentricity (``final") semi-major axis of $a_{f}=16$ $R_N$, this scenario will reach Triton's present-day orbit with zero eccentricity after roughly the age of the Solar System ($\sim4$ Gyr). The corresponding initial semi-major axis is then $a_0=a_f/(1-e^2)\approx 35.4$ $R_N$. Triton is started on an initially inclined orbit with its present-day inclination of $156.8^{\circ}$ and at zero obliquity; any obliquity that it starts with is quickly damped out in our simulations. Finally, Neptune is started with its present-day spin rate of $\sim 540^{\circ}$/d as given by \citet{Jacobson2009TheNeptune}, while Triton is started with a spin period of $8$ h that is typical of trans-Neptunian objects (see e.g. \citealt{Perna2009RotationsObjects, Thirouin2014RotationalBelt}).

Finally, to simulate the interior properties of the two bodies, Neptune is endowed with a constant time-lag rheology with the constant $k_2\approx0.407$ and time lag of 1.02 s used by \citet{Correia2009SecularTriton} (and Neptune's tidal potential is thus always truncated at $l=2$); Triton is given a Maxwell rheology as presented in Sec.~\ref{sec:quality_function_homogeneous}, with a Maxwell time of $\tau_{M}\approx 5.79$ h and an effective rigidity of $\Tilde{\mu}\approx 2.21$, corresponding to a viscosity of $10^{14}$ Pa s and a rigidity of $4.8$ GPa (as used by \citealt{Bagheri2022TheSystem} for the Pluto-Charon system).

To evaluate the suitability of a certain truncation level for accurate assessment of the evolution of each of the orbital elements, we will then briefly discuss the error behaviour for the evolution of each of the relevant orbital elements. To start with, there are some overall comments to be made, however.

\subsection{Qualitative error behaviour}
\label{subsec:qualitative_error_behaviour}
\begin{figure}
    \centering
    \includegraphics[width=1\linewidth]{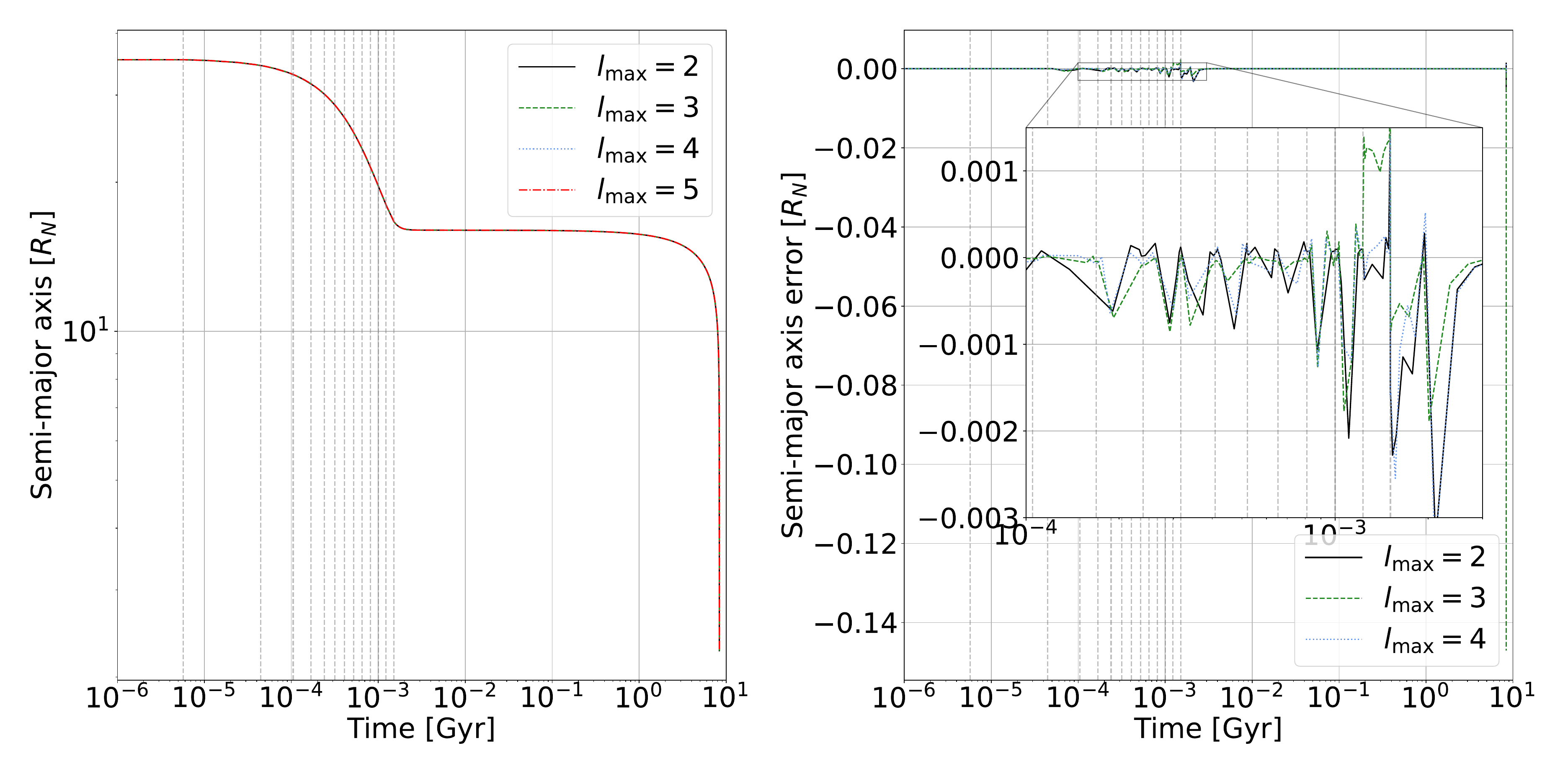}
    \caption{Evolution of the semi-major axis for the reference scenario described in Sec.~\ref{sec:truncating_degree} for various values of $l_{\max}$, as well as the error with respect to the $l\leq 5$ case. The dashed grey lines indicate the epochs at which Triton's rotational rate drops between resonances.}
    \label{fig:semi_major_axis_l5}
\end{figure}

\begin{figure}
    \centering
    \includegraphics[width=1\linewidth]{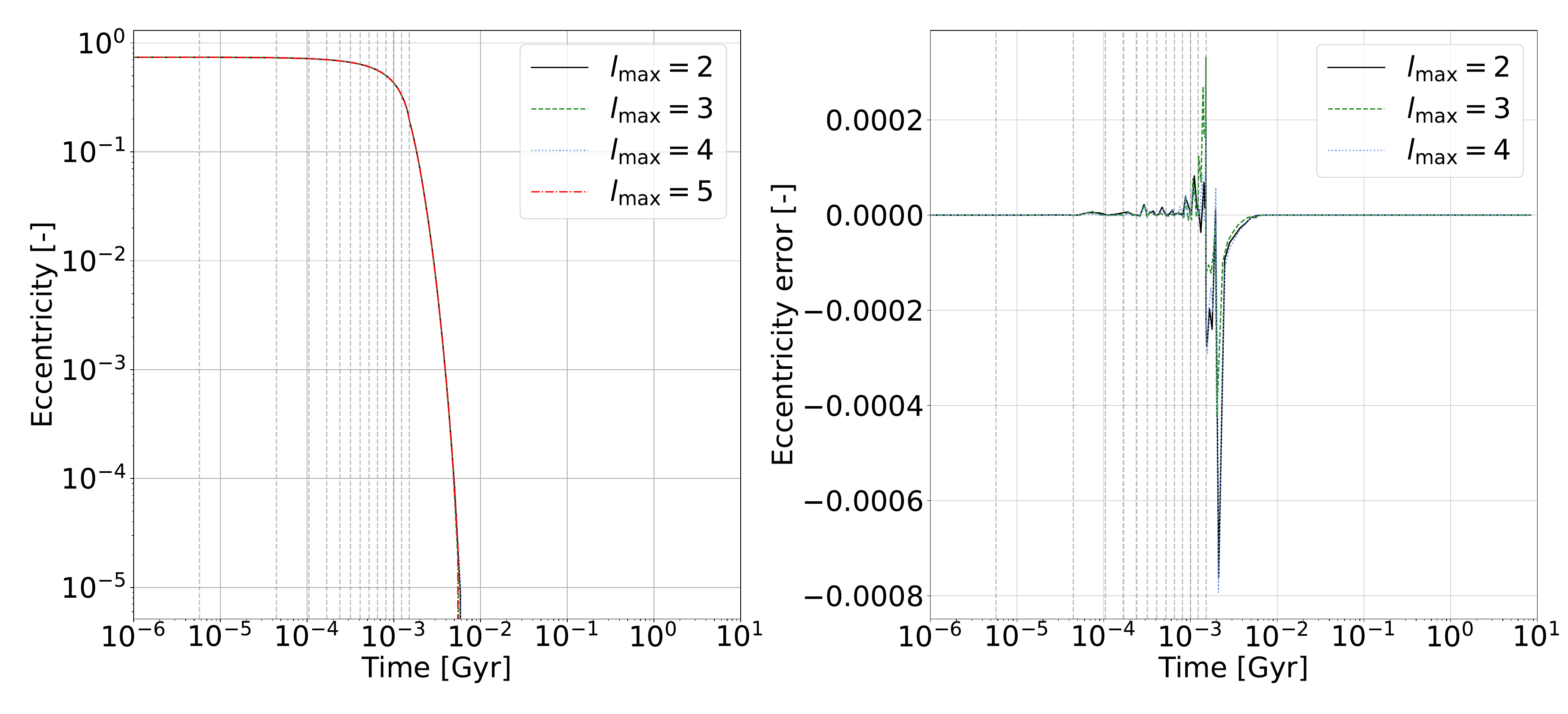}
    \caption{Evolution of the eccentricity for the reference scenario described in Sec.~\ref{sec:truncating_degree} for various values of $l_{\max}$, as well as the error with respect to the $l\leq 5$ case. The dashed grey lines indicate the epochs at which Triton's rotational rate drops between resonances.}
    \label{fig:eccentricity_l5}
\end{figure}

\begin{figure}
    \centering
    \includegraphics[width=1\linewidth]{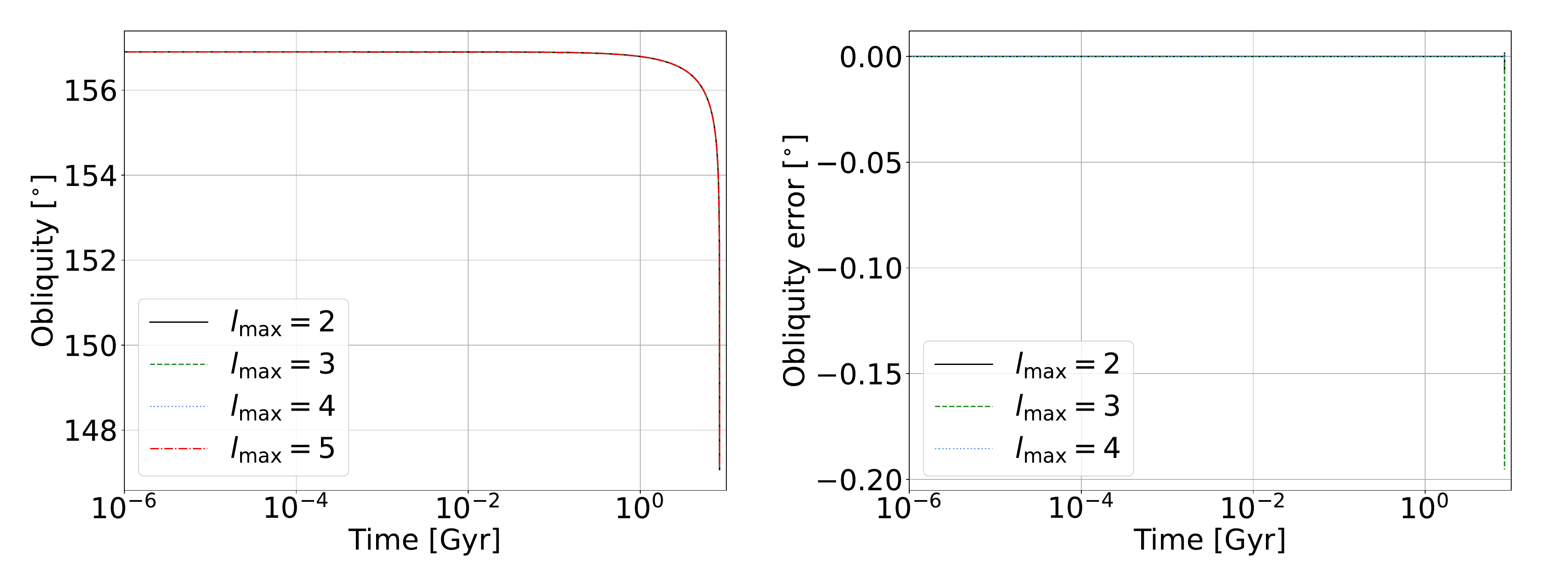}
    \caption{Evolution of the obliquity of Neptune with respect to the orbit (equivalent to the inclination of Triton's orbit) for the reference scenario described in Sec.~\ref{sec:truncating_degree} for various values of $l_{\max}$, as well as the error with respect to the $l\leq 5$ case. The lack of fine structure over the first $\sim8.3$ Gyr is not an artefact of the scale of the plot, but a true reflection of the fact that the difference between the various runs is negligible.}
    \label{fig:obliquity_l5}
\end{figure}

\begin{figure}
    \centering
    \includegraphics[width=1\linewidth]{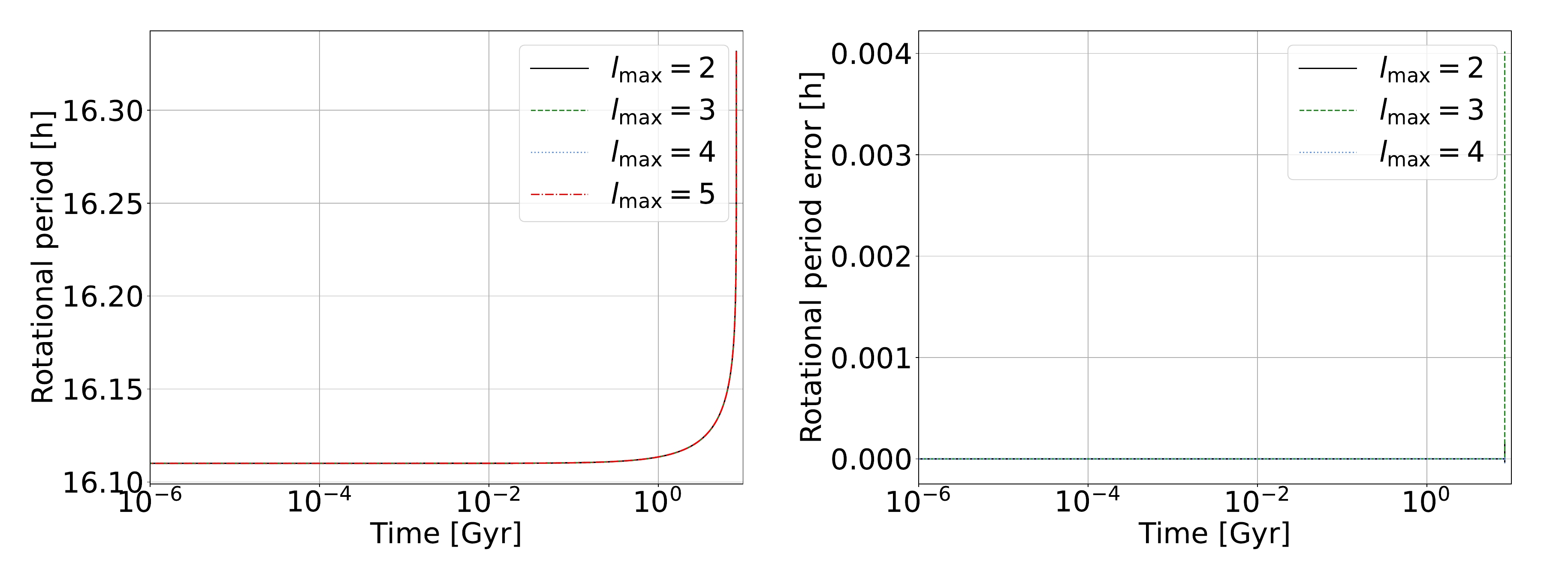}
    \caption{Evolution of the period of revolution of Neptune for the reference scenario described in Sec.~\ref{sec:truncating_degree} for various values of $l_{\max}$, as well as the error with respect to the $l\leq 5$ case. The lack of fine structure over the first $\sim8.3$ Gyr is not an artefact of the scale of the plot, but a true reflection of the fact that the difference between the various runs is negligible.}
    \label{fig:rot_period_l5}
\end{figure}

The evolution and errors with respect to the $l=5$-case for the semi-major axis and eccentricity are shown in Figs.~\ref{fig:semi_major_axis_l5} and \ref{fig:eccentricity_l5}; those for the obliquity and rotational period of Neptune are given in Figs.~\ref{fig:obliquity_l5} and \ref{fig:rot_period_l5}, respectively, and finally the evolution of the rotational rate of Triton and its error are shown in Fig.~\ref{fig:rot_rate_l5}. The error is produced by smoothly interpolating the data for the highest value of $l_{\max}$ using \textit{SciPy}'s interpolate.PchipInterpolator function, and interpolating it onto the timestamps for the other integration datasets.

Overall, we can see that the evolution is qualitatively identical for each of the truncation levels; in all cases, Triton drops between spin-orbit resonances at comparable times, damps out its eccentricity at a similar rate and drops into Neptune after $\sim8.4$ Gyrs. Wherever major error occurs, it is then due to a (small) mismatch in timing between rapid-evolution events (i.e. spin-orbit drops, or the eventual decay into Neptune), not due to qualitatively different behaviour; outcomes, it appears, are not changed by inclusion of higher degrees.

Though our results in Secs.~\ref{subsec:truncating_q} and \ref{subsec:truncating_G_lpq} do not extend beyond $l=5$, we performed another set of exploratory runs going out to $l=7$, the results of which are available to the interested reader in Sec.~\ref{sec:appendix_truncation_l7}; these support the ideas presented here, but additionally show that there is no apparent variation anymore between the $l_{\max}=6$- and $l_{\max}=7$-cases.

\subsection{Repercussions for the semi-major axis and eccentricity}
The error behaviour for the semi-major axis and eccentricity, shown in Figs.~\ref{fig:semi_major_axis_l5} and \ref{fig:eccentricity_l5}, appears to be dominated by the error incurred due to small differences in the times at which the various runs decay to lower spin-orbit resonances or, at the end of Triton's life, slight differences in the times of descent into Neptune. This is consistent with the overall qualitative behaviour we discussed in Sec.~\ref{subsec:qualitative_error_behaviour}.

One other significant point of error (though still minor in a relative sense: note that the eccentricity is of order $e\approx0.1$ during this time) appears for both the semi-major axis and eccentricity at $\sim 2$ Myr, well after Triton has decayed into the 1:1 spin-orbit resonance; it can therefore not be attributed to differences in timing of this resonance decay. Given (1) the relatively quick onset of these errors and (2) the lack of eccentricity errors of this magnitude earlier and later on, it seems likely that these errors are real consequences of cutting the expansion at different degrees, and not integration artefacts. In this case, this may well be one of the finer modes that \citet{Renaud2021TidalTRAPPIST-1e} showed to exist for higher degrees in the Pluto-Charon system in their Fig.~12; in any case, these do not appear to result in any appreciable nor observable change in the spin-orbit state of Triton at any point in its history.

\subsection{Repercussions for the obliquity and period of revolution of Neptune}
The evolution of the obliquity and period of revolution of Neptune is shown in Figs.~\ref{fig:obliquity_l5} and \ref{fig:rot_period_l5}, respectively. There is no significant difference between the various degrees, except for some error at the end due to a negligible difference in the timing of onset of the decay into Neptune; noticeably, these quantities are, in contrast to the other quantities, not perturbed by the spin-orbit resonance progression, and so they do not experience any error due to discrepancy in timing of the various resonance drops. This should not be surprising, as the equations of motion for the obliquity and rotation rate of Neptune (Eqs.~\ref{eq:eom_obliquity} and \ref{eq:eom_obliquity}, respectively) do not include any terms due to Triton's tidal potential. Consequently, we can reasonably assume that the major influence of higher-degree terms will lay in the spin-orbit resonance progression of Triton, and any associated damping of the eccentricity and semi-major axis. The obliquity and period of revolution of Neptune are not affected.

\begin{figure}
    \centering
    \includegraphics[width=1\linewidth]{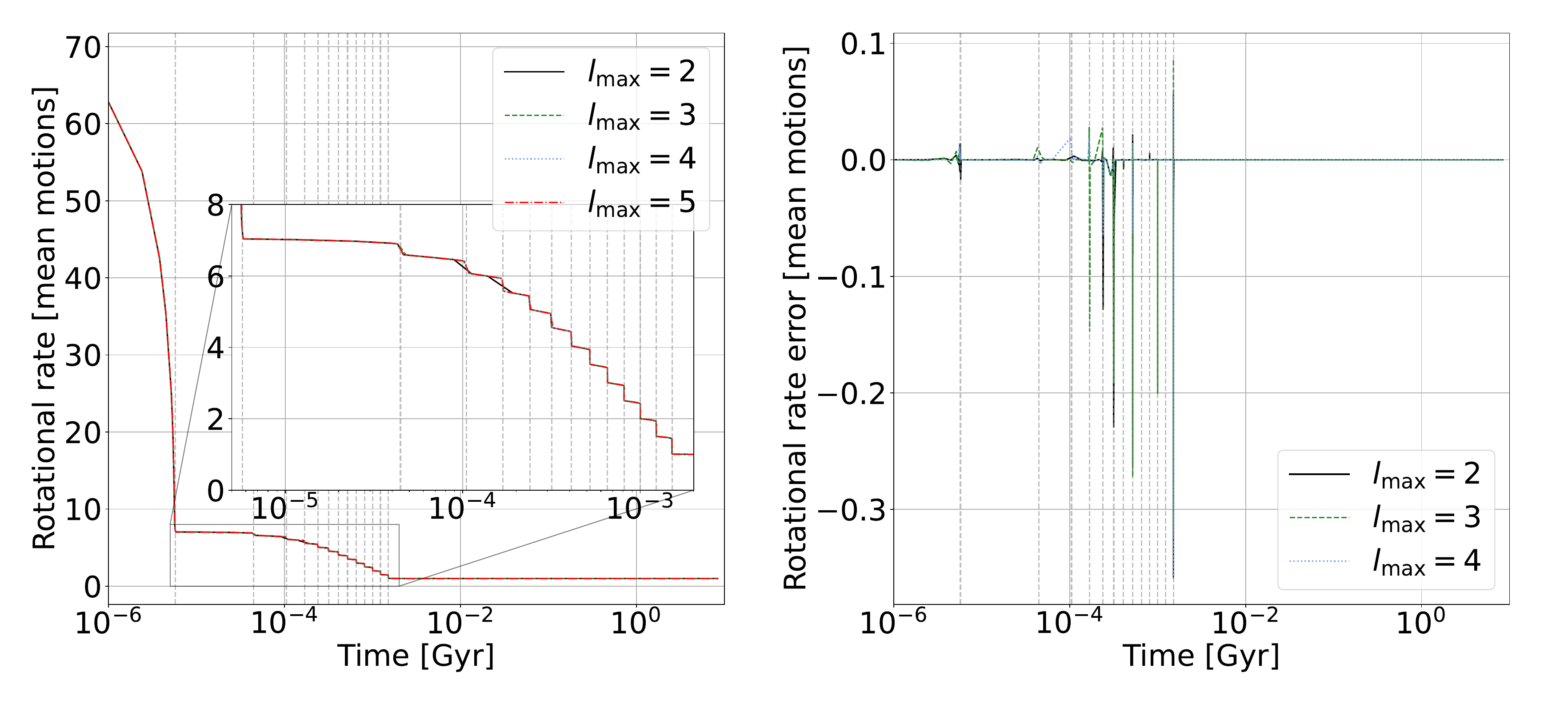}
    \caption{Evolution of the rotational rate of Triton for the reference scenario described in Sec.~\ref{sec:truncating_degree} for various values of $l_{\max}$, as well as the error with respect to the $l\leq 5$ case. The dashed grey lines mark the epochs at which Triton's rotational rate drops between resonances. Note that the discrepant feature that occurs from $\sim 10^{-6}-10^{-5}$ Gyr is a sampling artefact, and not a true discrepancy between the various degrees.}
    \label{fig:rot_rate_l5}
\end{figure}

\subsection{Repercussions for the rotational rate of Triton}
The evolution of the rotational rate of Triton for the reference scenario is shown in Fig.~\ref{fig:rot_rate_l5}. Note that the feature that appears from $10^{-5}-10^{-6}$ Gyr seems to be a sampling artefact, and when the data is smoothly interpolated it disappears; hence its absence (or rather, vastly reduced presence) on the right panel.

Aside from this, no unexpected features are present. On this graph, it becomes exceedingly clear that for the most part, slight mismatches in the time of spin-orbit resonance decay are responsible for the larger discrepancies between the various values of $l_{\max}$ wherever those appear. There does not appear to be any particular bias in whether the decay happens earlier or later for higher values of $l_{\max}$, fortunately, and so accounting only for the first couple of degrees is not likely to result in any significant inaccuracies on a quantitative, not just qualitative level.

\subsection{Conclusions on the truncation of the degree}
In conclusion, it thus appears that we can truncate Triton's tidal potential to $l=2$ in this case without losing any significant detail or events in the resulting tidal evolution, and without introducing any offset in terms of the timeline of Triton's evolution. That is not to say that the higher-degree terms do not carry any weight in high-eccentricity orbit evolution in general: in Triton's case, the high-eccentricity evolution happens sufficiently far from Neptune that the $(R_T/a)^{2l+1}$-term in Eq.~\ref{eq:partial_potential} damps out any large values of the eccentricity functions. In scenarios where the high eccentricity coincides with a closer approach to the host body, this may not be the case; similarly, the tidal potential for Neptune, if described with a model including higher-degree terms (which the constant time-lag model we use does not include), is damped for higher eccentricities only as $(R_N/a)^{2l+1}$. Such a model might therefore also require truncation at a higher degree.

\section{Validation and loosening of the truncation prescriptions}
\label{sec:validation_truncation_prescriptions}
In Sec.~\ref{sec:truncating_kaula_general}, we provided a set of conservative upper bounds on the value of $|q|$ at which to truncate Kaula's expansion of the tidal potential for a given eccentricity using analytical expressions available for the constant time-lag model. When using a more realistic tidal model, however, the linear behaviour of the quality function only holds out roughly until a peak frequency, beyond which its value decreases again\footnote{The exact behaviour beyond this point depends on the tidal model that is used: in any case, the resulting values beyond this peak frequency are sublinear (see e.g. \citealt{Bagheri2022TidalOverview} for an overview of more advanced models currently in use in literature), and so the constant time-lag model will overestimate the quality function in this regime.}. The constant time-lag model therefore (significantly) overestimates the contributions of high-frequency terms, which correspond to those with large values of $|q|$; hence, we expect that the truncation prescriptions derived from our results in Sec.~\ref{sec:truncating_kaula_general} can be loosened significantly when dealing with realistic rheological models. The aim of this section is to examine whether such a loosened truncation prescription can be constructed in a manner that is motivated by the properties of the used rheological model. The reference scenario here will be the same as that used in Sec.~\ref{sec:truncating_degree}.

\subsection{Proposed alternative truncation prescriptions}
We will examine several alternative truncation prescriptions: we will discuss each of these first, along with a short justification for why these truncation prescriptions seem reasonable. The hope is that well-performing truncation laws can be motivated physically, which might aid in their application to other scenarios.

The first proposed truncation rule is derived from the rotation-rate results shown in Sec.~\ref{sec:truncating_degree}: we see that, in this scenario, Triton is only captured into spin-orbit resonances $\dot{\theta}_T/n < 10$. As such resonances can only occur near zero-crossings of one of the (odd) quality functions $K_{2,T}(\omega_{T,2mpq})$, this seems to suggest that the magnitudes of the terms due to the $\omega_{T,2mpq}$ corresponding to higher resonances are not significant enough to induce a change of sign of the spin rate: this motivates us to try neglecting these, as higher resonances are in part induced by $\omega_{T,2mpq}$ with higher values of $q$. To see to which resonance a certain $\omega_{T,2mpq}$ corresponds, we observe that:
\begin{equation}
    \omega_{T,2mpq}\approx (2-2p+q)n - m\dot{\theta}_T = 0 \implies \frac{\dot{\theta}_T}{n} = \frac{2-2p+q}{m}
\end{equation}
whence it can be seen that the largest-q term contributing to the 10:1 resonance is that for which $2-2p+q=20$, $m=2$, with $p=2$ such that $q=22$. Taking this conservatively, truncating at $|q|\leq 25$ therefore seems reasonable.

A second truncation rule follows from a similar analysis, but instead by noting that we may wish to neglect the tidal response as described by the quality function $K_{2,T}(\omega_{T,2mpq})$ beyond a certain frequency. For some threshold frequency $\omega_t$ beyond which we will neglect terms, we can then see that
\begin{equation}
\label{eq:threshold_frequency}
    \omega_{T,2mpq} \leq \omega_t\; \forall m,p \leq 2 \implies |q|\leq \frac{\omega_t}{n} + 2\left(\frac{\dot{\theta}_T}{n}+1\right).
\end{equation}
Taking as threshold frequency $\omega_t$ 25 times the peak frequency of the Maxwell rheology (which is, admittedly, somewhat arbitrary, but beyond this point the quality factor is less than roughly $10\%$ of its peak value) used in Sec.~\ref{sec:truncating_degree}, setting $n$ equal to the initial mean motion and setting $\dot{\theta}_T/n=10$ (as the rotation rate is quickly damped into resonance from the initial rotation rate of 8 hours), only terms with $|q|\leq 40$ yield a frequency within this cutoff threshold $\omega_t$.

A third truncation rule follows by fitting an empirical bound to the expansion required to approximate the sums examined in Sec.~\ref{sec:truncating_kaula_general} to within $10\%$ solely for the terms that correspond to $l=2$. One such bound, which is thus less restrictive than the bound given in Eq.~\ref{eq:q_max_req}, is given by
\begin{equation}
    |q_{\max}|(e) \leq \lfloor2\exp(3.5e+1.5e^6+1.3/(1-e^2)^{1/3})/5\rfloor.
\end{equation}

Finally, two more pragmatic bounds are taken so as to compare to those often found in literature: \citet{Luna2020TheSystem} include all terms $|q|\leq 10$ to study the evolution of a system with $e<0.1$, so we take a fourth truncation rule to include all terms with $|q|\leq 10$. A fifth and final truncation rule is set to include only $q=-1, 0, 1$, to evaluate the effects of truncating the power series expansion of the Hansen coefficients at terms of order $e^2$.

\begin{figure}
    \centering
    \includegraphics[width=1\linewidth]{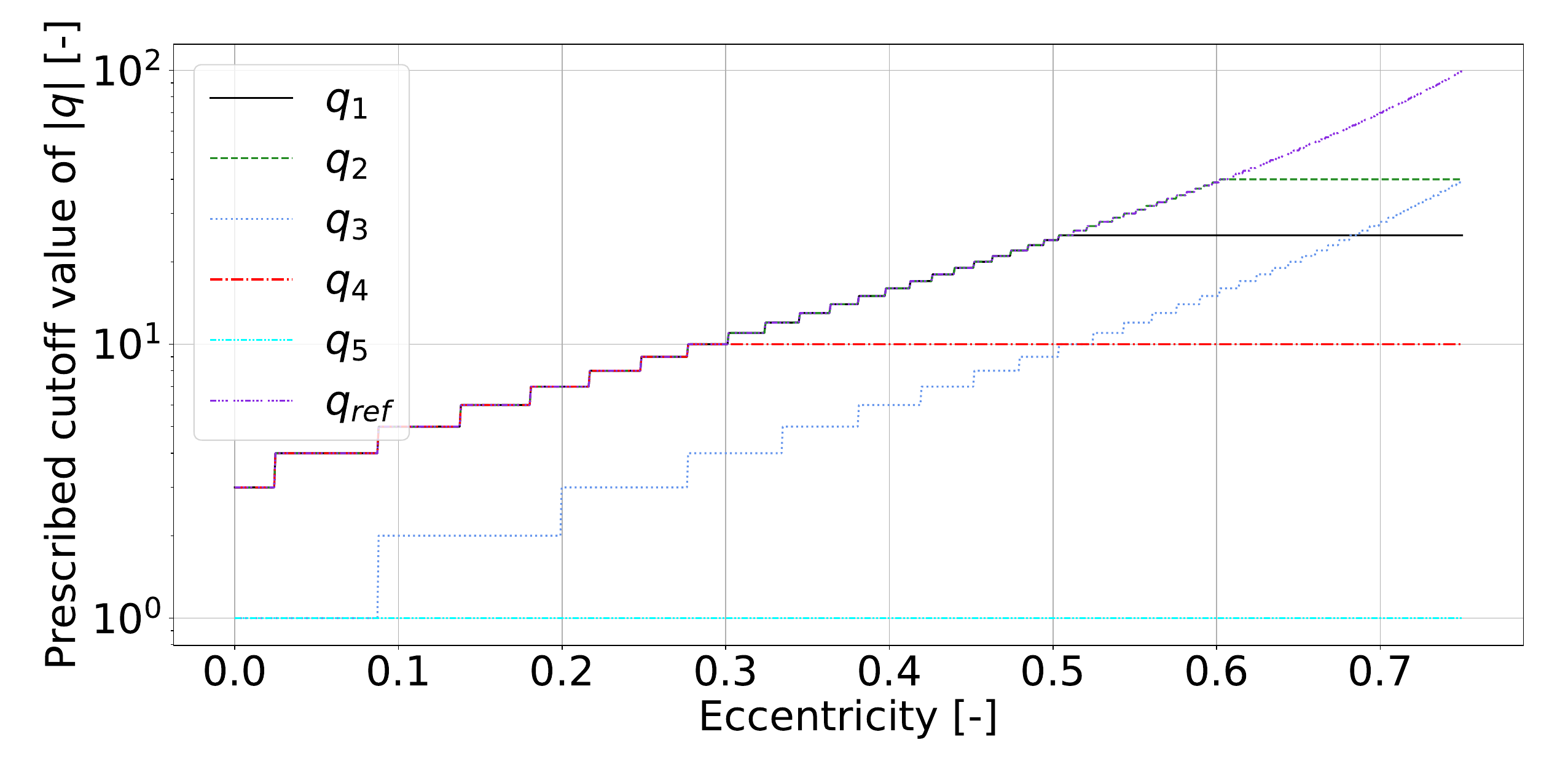}
    \caption{The cutoff values for $|q|$ as prescribed by the various expressions given in Eq.~\ref{eq:cutoff_functions}.}
    \label{fig:q_cutoff_functions}
\end{figure}

As the truncation prescription derived empirically in Sec.~\ref{sec:truncating_kaula_general}  (which we will call $q_{\max,\textrm{ref}}(e)$) is an upper bound, we can then take the minima between these truncation levels and this bound. Hence, we obtain the following truncation laws:
\begin{align}
\label{eq:cutoff_functions}
    q_1(e) &= \min(25, q_{\max,\textrm{ref}}(e)) \\
    q_2(e) &= \min(40, q_{\max,\textrm{ref}}(e)) \\
    q_3(e) &= \lfloor2\exp(3.5e+1.5e^6+1.3/(1-e^2)^{1/3})/5\rfloor \\
    q_4(e) &= \min(10, q_{\max,\textrm{ref}}(e)) \\
    q_5(e) &= 1 \\
    q_{\max,\textrm{ref}}(e) &= \lfloor\exp\left(3.5e+1.5e^6+1.3(1-e^2)^{-1/3}\right)\rfloor.
\end{align}
These values prescribed by these functions over the interval $[0, 0.75]$ are shown in Fig.~\ref{fig:q_cutoff_functions}.
As for the truncation in degree, let us then examine what the consequences are of truncating the terms in Kaula's expansion according to the various cutoff laws for $|q|$. As we have shown that the degree 2-terms sufficiently capture the evolution of Triton even at high eccentricities in Sec.~\ref{sec:truncating_degree}, we will limit the expansion to just $l=2$.

\begin{figure}
    \centering
    \includegraphics[width=1\linewidth]{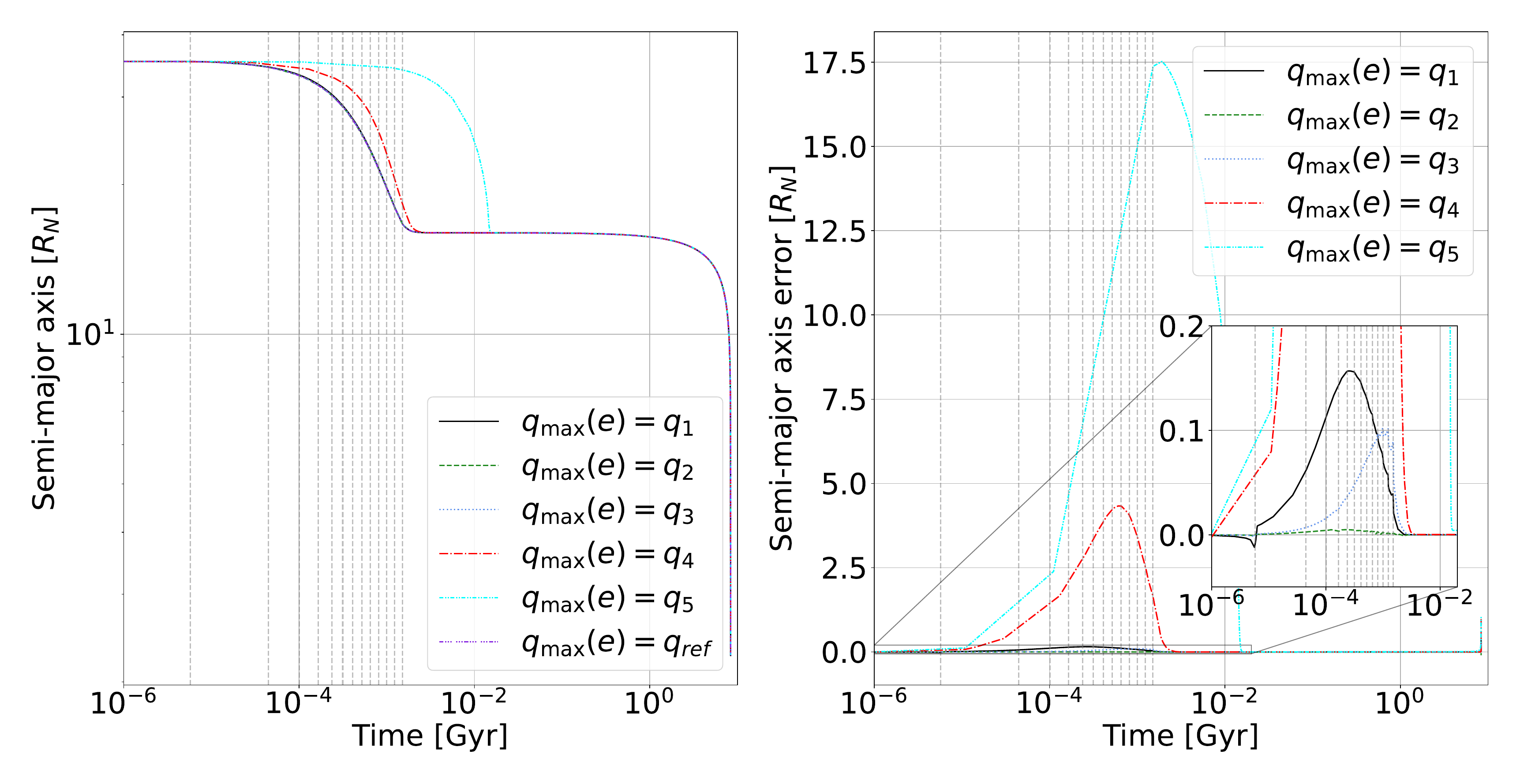}
    \caption{Evolution of the semi-major axis for the reference scenario described in Sec.~\ref{sec:truncating_degree} with the various truncation laws, as well as the error with respect to the nominal case. The dashed grey lines indicate the epochs at which Triton's rotational rate drops between resonances.}
    \label{fig:trunc_q_a}
\end{figure}

\begin{figure}
    \centering
    \includegraphics[width=1\linewidth]{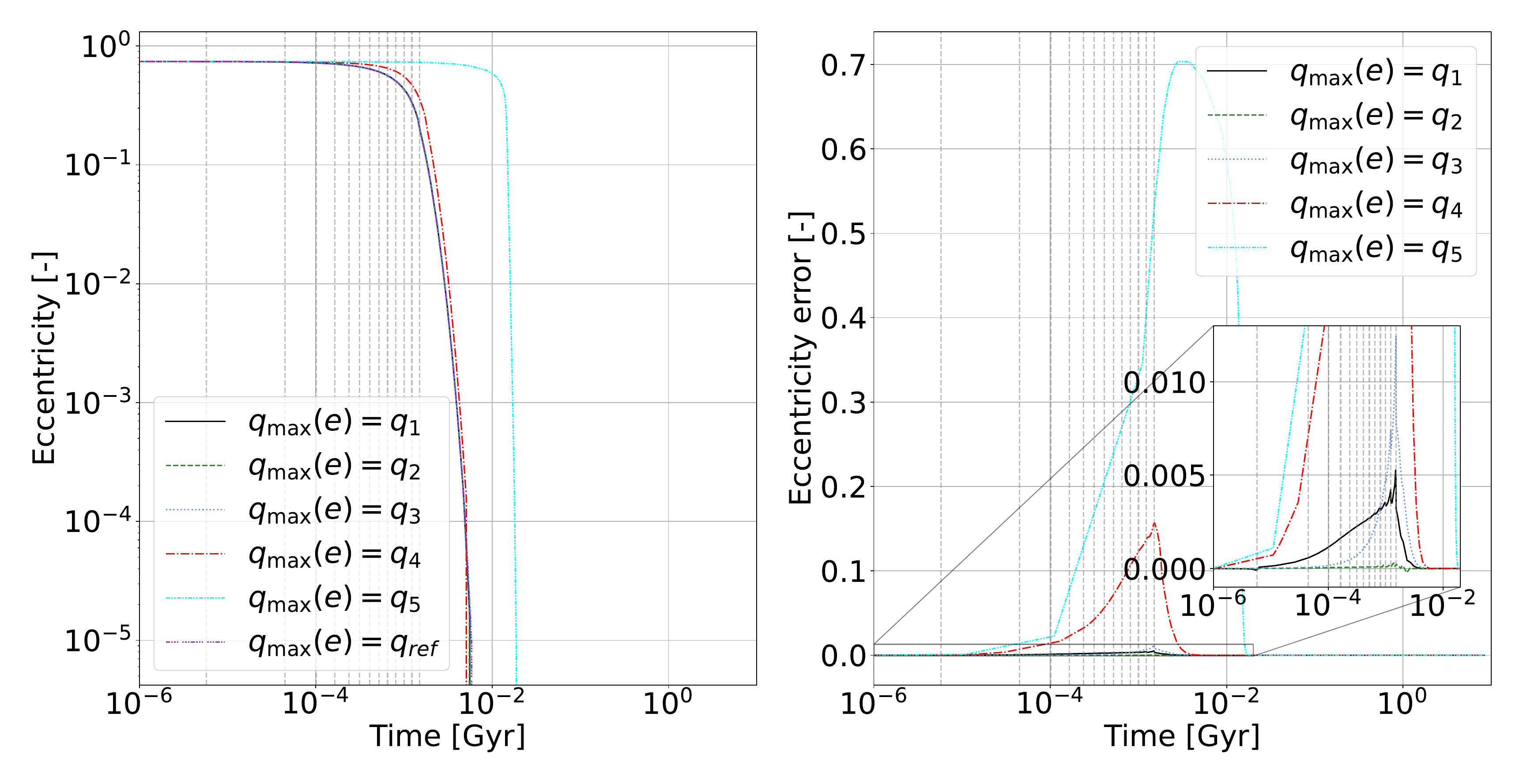}
    \caption{Evolution of the eccentricity for the reference scenario described in Sec.~\ref{sec:truncating_degree} with the various truncation laws, as well as the error with respect to the nominal case. The dashed grey lines indicate the epochs at which Triton's rotational rate drops between resonances.}
    \label{fig:trunc_q_e}
\end{figure}

\subsection{Repercussions for the semi-major axis and eccentricity}
The evolution of the semi-major axis and eccentricity in the scenarios simulated with the various truncation laws are shown in Figs.~\ref{fig:trunc_q_a} and \ref{fig:trunc_q_e}, respectively. The behaviour for truncation in $|q|$ is more predictable than that for truncation in the degree: overall, the errors are significantly larger, and the looser truncation laws tend to underestimate tidal effects, thus leading to an overestimation of the eccentricity and semi-major axis at a given point in time. Additionally, the resulting error is not so much due to variation in the time at which events occur, but rather a true error due to the aforementioned underestimation of tidal effects.

Importantly, however, the outcomes are not altered significantly, and even the loosest truncation law (accounting only for $q=-1,0,1$) predicts the time at which Triton plunges into Neptune without major error; while the behaviour during the initial couples million years is significantly different, the history beyond that coincides again for all cases. This can be attributed to the fact that the duration of this high-eccentricity history is very short compared to the low-eccentricity period that follows. Even if the error incurred during the high-eccentricity period is catastrophic, it is a small comfort knowing that a misjudgement in the truncation in $|q|$ will not result in a dramatically different outcome of the low-eccentricity phase or lack of conservation of angular momentum, however.

\begin{figure}
    \centering
    \includegraphics[width=1\linewidth]{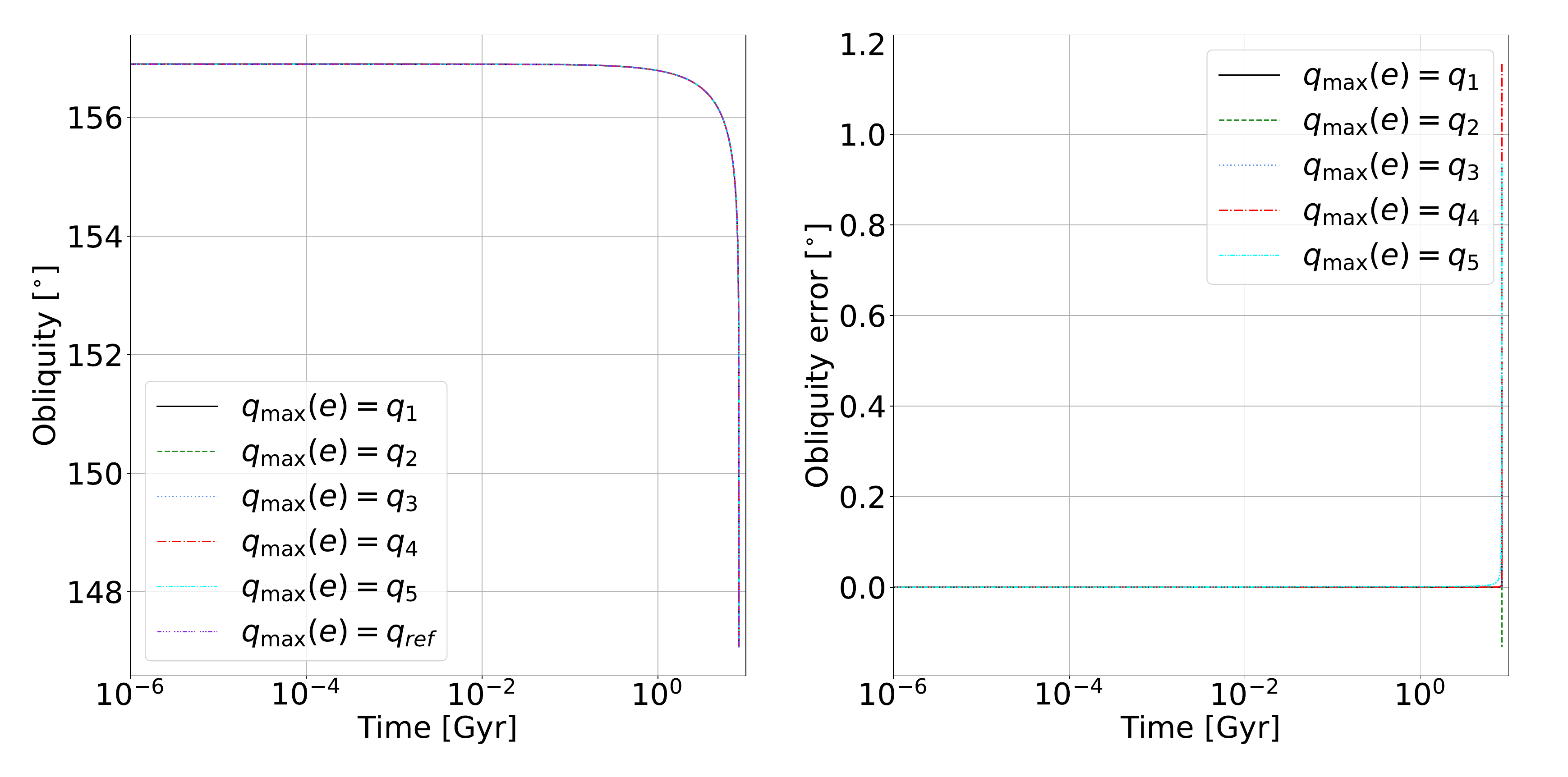}
    \caption{Evolution of the obliquity of Neptune for the reference scenario described in Sec.~\ref{sec:truncating_degree} with the various truncation laws, as well as the error with respect to the nominal case.}
    \label{fig:trunc_q_i}
\end{figure}

\begin{figure}
    \centering
    \includegraphics[width=1\linewidth]{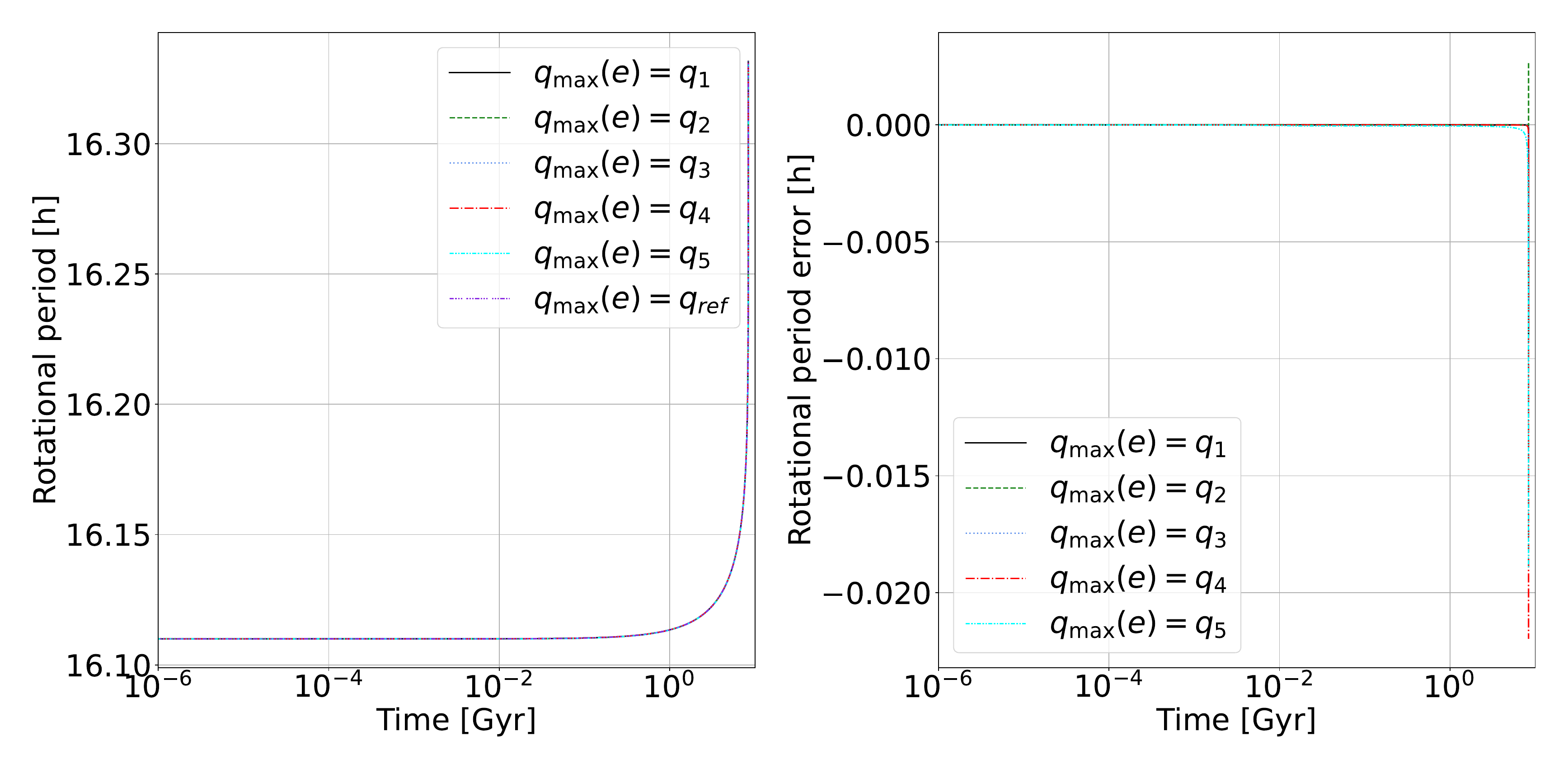}
    \caption{Evolution of the period of revolution of Neptune for the reference scenario described in Sec.~\ref{sec:truncating_degree} with the various truncation laws, as well as the error with respect to the nominal case.}
    \label{fig:trunc_q_P}
\end{figure}

\subsection{Repercussions for the obliquity and period of revolution of Neptune}
The evolution of the obliquity and period of revolution of Neptune are shown in Figs.~\ref{fig:trunc_q_i} and \ref{fig:trunc_q_P}, respectively; as was the case with the truncation in degree, there is hardly any difference between the various cases, with the sole exception of error incurred due to small variations in the time at which Triton eventually decays into Neptune. The earlier truncation levels in general therefore seem to underestimate the tidal effects, as is the case for the semi-major axis and eccentricity, but in this case this full effect can be attributed to the slower decay of the orbit due to underestimation of the damping of the semi-major axis.

\begin{figure}
    \centering
    \includegraphics[width=1\linewidth]{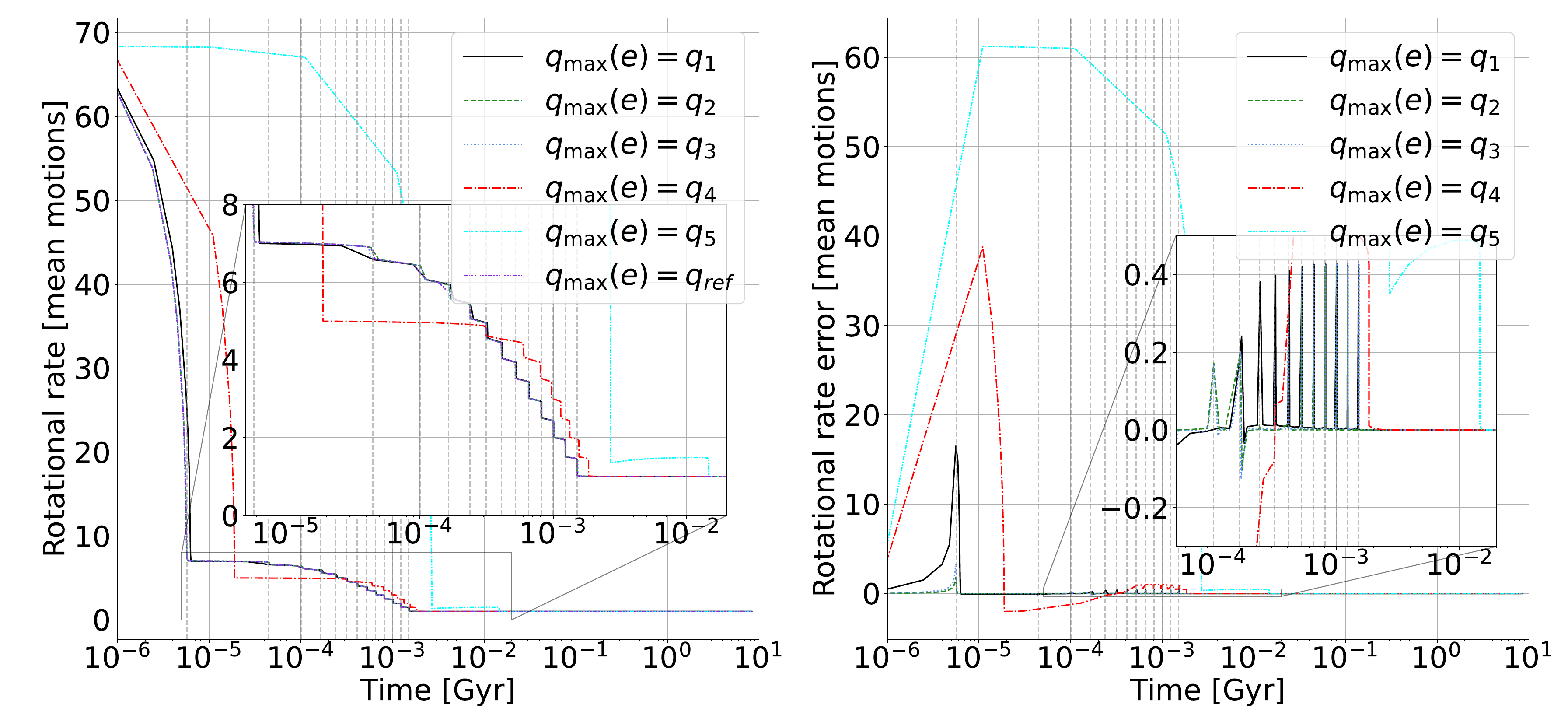}
    \caption{Evolution of the rotation rate of Triton for the reference scenario described in Sec.~\ref{sec:truncating_degree} with the various truncation laws, as well as the error with respect to the nominal case. The dashed grey lines indicate the epochs at which Triton's rotational rate drops between resonances.}
    \label{fig:trunc_q_rotrate}
\end{figure}

\subsection{Repercussions for the rotational rate of Triton}
Finally, the rotational rate; its evolution with the various truncation prescriptions is shown in Fig.~\ref{fig:trunc_q_rotrate}. In contrast with the different truncation prescriptions for $l_{\max}$, the different truncation schemes yield qualitatively different results, and not just shifts in the time at which Triton decays into consequent spin-orbit resonances. The $q_4$- and $q_5$-schemes miss out on some resonances altogether, while the other schemes, though qualitatively similar, vary slightly in the rate at which they decay into the first stable resonance. Once they have, however, their behaviour is nearly identical. Most interesting is the discrepancy between the scheme $q_1$, which was based on including all relevant resonances, and schemes $q_2$, $q_3$ and the reference scheme $q_{\max, \textrm{ref}}$; while the former does indeed (as it was designed to do) fully describe the capture of Triton into the first resonance, it underestimates the rate at which it does compared to the latter schemes.

\subsection{Conclusions on the truncation in $|q|$}
Based on these results, we can derive some conclusions on the appropriate truncation schemes to use, and how they may be derived. $q_4$ and $q_5$ are positively worthless; fortunately, they do showcase the fact that a misjudgement or inaccuracy of the truncation law for $|q|$ will not, on a dynamical level, affect the eventual outcome of the scenario\footnote{Do note that the underestimation of tidal heating and the associated artificial lengthening of the high-eccentricity phase may well lead to dramatically different outcomes if the thermal and interior evolution of the body is taken into account, however.}. The underestimation of the tidal effects described by \citet{Renaud2021TidalTRAPPIST-1e} once truncating at too low a power of $e^2$ can therefore seemingly be attributed to the implicit neglect of terms in $q$ that this corresponds to; it may be interesting to investigate in future work whether the false sign reversal that they observe for $e>0.8$ at poor truncation levels is a result of this same effect, or of the truncation of the eccentricity functions that this is also associated with.

Moreover, the fact that $q_1$ misjudges the pace at which Triton drops into the first resonance suggests that terms cannot be excluded solely because the spin-orbit resonance they correspond to is not significant; the method employed to derive the scheme $q_2$, on the contrary, does seem an appropriate approach by which to determine at which point to truncate. One might perhaps consider truncating in a manner similar to $q_1$ if the simulation already starts in resonance, however.

If, rather than a constant truncation bound, one wants to set a truncation bound that varies with $e$, it seems that the methods employed in Sec.~\ref{sec:truncating_kaula_general} are most robust; the fact that the results for $q_3$ coincide so well with those for $q_{\max,\textrm{ref}}$ seem to indicate that by the point that the expressions in Sec.~\ref{sec:truncating_kaula_general} have converged to within $10\%$ of their true value, the potential terms in practice will likely have converged to within a tighter margin already. Nonetheless, $q_3$ does still perform considerably worse than $q_2$; a suitable compromise may be to continuously evaluate Eq.~\ref{eq:threshold_frequency} at each timestep, or it may be possible to devise some hybrid approach that can account both for the decay of the eccentricity functions with increasing $q$ and that of the quality function for large enough frequency. We leave the evaluation of such an approach to future work, however, and in the following will simply use the analytically derived empirical bounds from Sec.~\ref{sec:truncating_kaula_general}.

\section{Conclusions}
We have thus determined how to go from the infinite sum in Eq.~\ref{eq:partial_potential} to finite sums that can be implemented programmatically. All in all, it appears that the neglect of terms with $l>2$ will have no significant effect other than perhaps some minor alterations to the times at which the moon drops between resonances; neglecting significant terms in $|q|$, however, will generally yield an underestimation of tidal effects, but the eventual fate of Triton nor its low-eccentricity phase is not significantly altered from a dynamical standpoint. We can thus now state with relative confidence that a proper treatment of Triton's evolution does not require the incorporation of terms corresponding to $l>2$. Secondly, we can say that the truncation laws derived in Sec.~\ref{sec:truncating_kaula_general} seem to work well in the case of Triton, and the corresponding truncation levels for $q$ as a function of $e$ need no revision upward or downward.

Aside from these statements on the scenario for Triton in particular, there are some more general conclusions or suggestions that arise: first of all, it seems that the truncation prescriptions set for $q$ in Sec.~\ref{sec:truncating_kaula_general} are a promising approach, and while the resulting truncation laws are conservative, they do not seem to be conservative to the point of inefficiency. Secondly, it seems that truncating the sum in $q$ prematurely will, over the eccentricity range we explored, not lead to a significant change in outcomes, only in timescales. Small inaccuracies in the truncation law used for $q$ are therefore only likely to qualitatively alter results in highly time-sensitive scenarios.

\chapter{Spin-orbit evolution}
\label{ch:spin-orbit_chains}
With the mathematical framework to describe Triton's evolution set up in Ch.~\ref{ch:kaula_theory} and manipulated into a numerically viable implementation in Ch.~\ref{ch:validation}, we can now move on to analysing the physics of the Neptune-Triton scenario: to start with, we will look at the coupled spin-orbit evolution that occurs while Triton has appreciable eccentricity (and consequently, before it decays into the 1:1 spin-orbit resonance), and the thermal consequences that accompany the resulting tidal dissipation. In particular, we will first examine whether the rheologies that have been used in the past can be trusted to reproduce the more realistic behaviour exhibited by more advanced rheological models in Sec.~\ref{sec:simplified_models}. Afterwards, we will examine the consequences that might arise from severe variation in rheological parameters due to thermal evolution of Triton under the force of the extreme eccentricity tides it initially experiences, as well as any effects due to spin-orbit resonance transitions, in Sec.~\ref{sec:rheological_parameters}.

\section{Consequences of simplified rheologies}
\label{sec:simplified_models}
Previous work on Triton has mostly relied, explicitly or implicitly, on the use of simplified rheologies or equations derived using such simplifications, either as no advanced rheological models were in widespread use in literature at the time (e.g. \citealt{Chyba1989TidalSystem, Goldreich1989NeptunesStory}) or as such approaches allow for elimination of the infinite sum in Kaula's expansion of the potential (e.g. \citealt{Correia2009SecularTriton, Nogueira2011ReassessingTriton}). Now that we have a set of simulation tools that can assess the tidal evolution of Triton for an arbitrary tidal model, we can examine what the consequences of such assumptions are for the resulting evolution.

\begin{figure}
    \centering
    \includegraphics[width=1\linewidth]{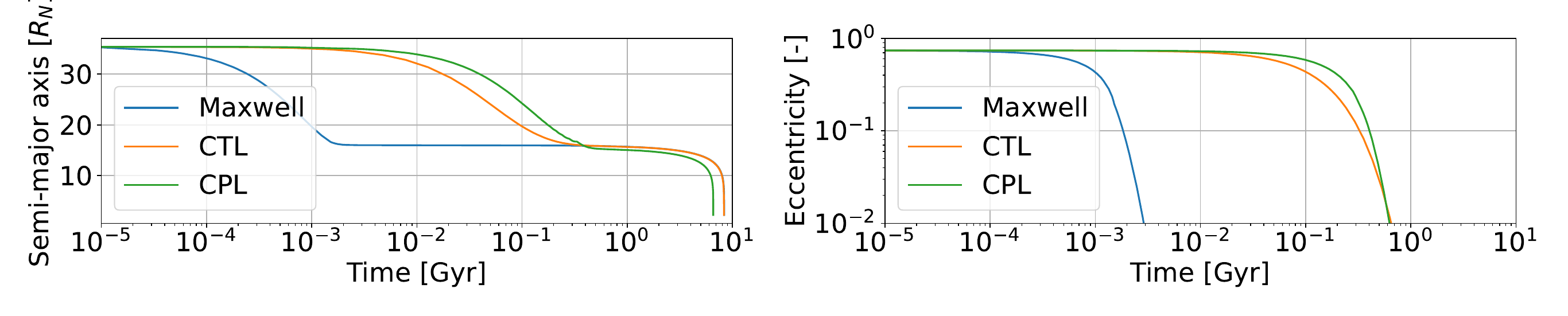}
    \caption{The time histories of the semi-major axis and eccentricity of Triton assuming CPL, CTL and Maxwell rheologies.}
    \label{fig:e_a_history_CTL_CPL}
\end{figure}

\begin{figure}
    \centering
    \includegraphics[width=1\linewidth]{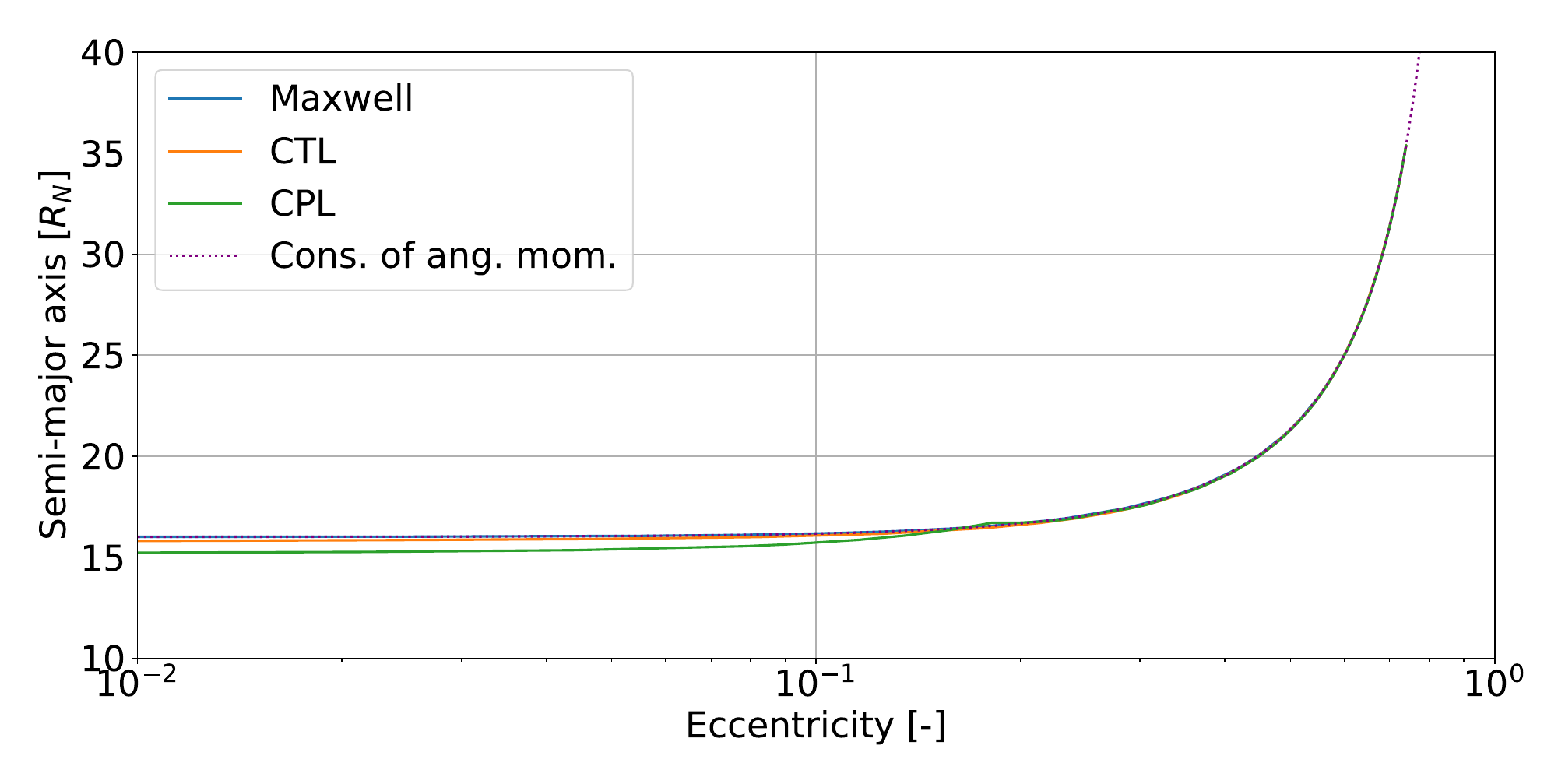}
    \caption{The phase history of Triton with a CPL, CTL and Maxwell rheology; additionally, the line corresponding to conservation of orbital angular momentum is shown.}
    \label{fig:e_a_phase_history}
\end{figure}

To assess this, we propagate three different rheologies: one version of Triton endowed with a homogeneous Maxwell rheology with the same parameters used in Sec.~\ref{sec:truncating_degree} (which, we remember, assumes no more than a homogeneous body with icy composition), a second endowed with a constant time lag (CTL) rheology with a time lag of $808$ s and a Love number $k_{2,T}=0.1$ as in \citet{Correia2009SecularTriton}\footnote{While it might be tempting to use Eq.~\ref{eq:Maxwell_CTL_approx} to produce a CTL-Triton that corresponds exactly to the Maxwell Triton that we use, it is important to remember that this is not possible at higher eccentricities (see the explanation accompanying Eq.~\ref{eq:Maxwell_CTL_approx} for details).}, and a third endowed with a constant phase lag (CPL) rheology with the same Love number, but a quality factor of $Q_T=100$ as used by \citet{Nogueira2011ReassessingTriton} and \citet{Correia2009SecularTriton} to derive the time lag in their constant time lag-model.

We propagate these models starting from a reference starting scenario with eccentricity $e_0=0.74$, a post-damping semi-major axis of $a_f=16R_N$ such that the initial semi-major axis is $a_0=a_f/(1-e_0^2)$ and an initial obliquity of Neptune with respect to Triton's orbit equal to its present-day value of $i_0=156.9^{\circ}$. Note that this eccentricity is not representative of the eccentricity that Triton likely started with, which while poorly constrained is likely to be far in excess of 0.74 (see e.g. \citealt{Nogueira2011ReassessingTriton}); as in Ch.~\ref{ch:validation}, the associated computational cost of computing the eccentricity functions using numerical integration is unfortunately too restrictive, and so we stay with $e<0.74$. We will see that even over this domain significant differences start to arise, and so there is no need to go any further.

The resulting, seemingly drastically different histories for the semi-major axis and eccentricity are shown in Fig.~\ref{fig:e_a_history_CTL_CPL}; however, when comparing the three in phase-space in Fig.~\ref{fig:e_a_phase_history}, it is clear that the major difference occurs in time. Indeed, this is a result of the fact that Triton's small mass compared to Neptune means that its orbital angular momentum is conserved well over the relatively short period during which its eccentricity is damped; this assumption is poorest for CPL Triton, which damps out its eccentricity slowest such that its eccentricity tides operate more efficiently closer to Triton, and as a result it drops into Neptune earlier than the other two models. This, in turn, can be attributed to the CPL model's lack of inclusion of the sensitivity of the tidal response to the forcing frequency: it will, in this case and with the given parameters, underestimate the tidal response at greater orbital distances, as a result wrongfully assuming that Triton approaches Neptune with a significant eccentricity which will have it drop into Neptune earlier than expected.

One might therefore be inclined to explain the differences in predicted history by the effective magnitude of the tidal response elicited by each of the three tidal models, while assuming that the state evolution between the three is roughly equal; as the CTL model underestimates the tidal response less severely at greater orbital distances, one might be inclined to think that its results can be trusted to represent Triton's evolution well on a qualitative level.

\begin{figure}
    \centering
    \includegraphics[width=1\linewidth]{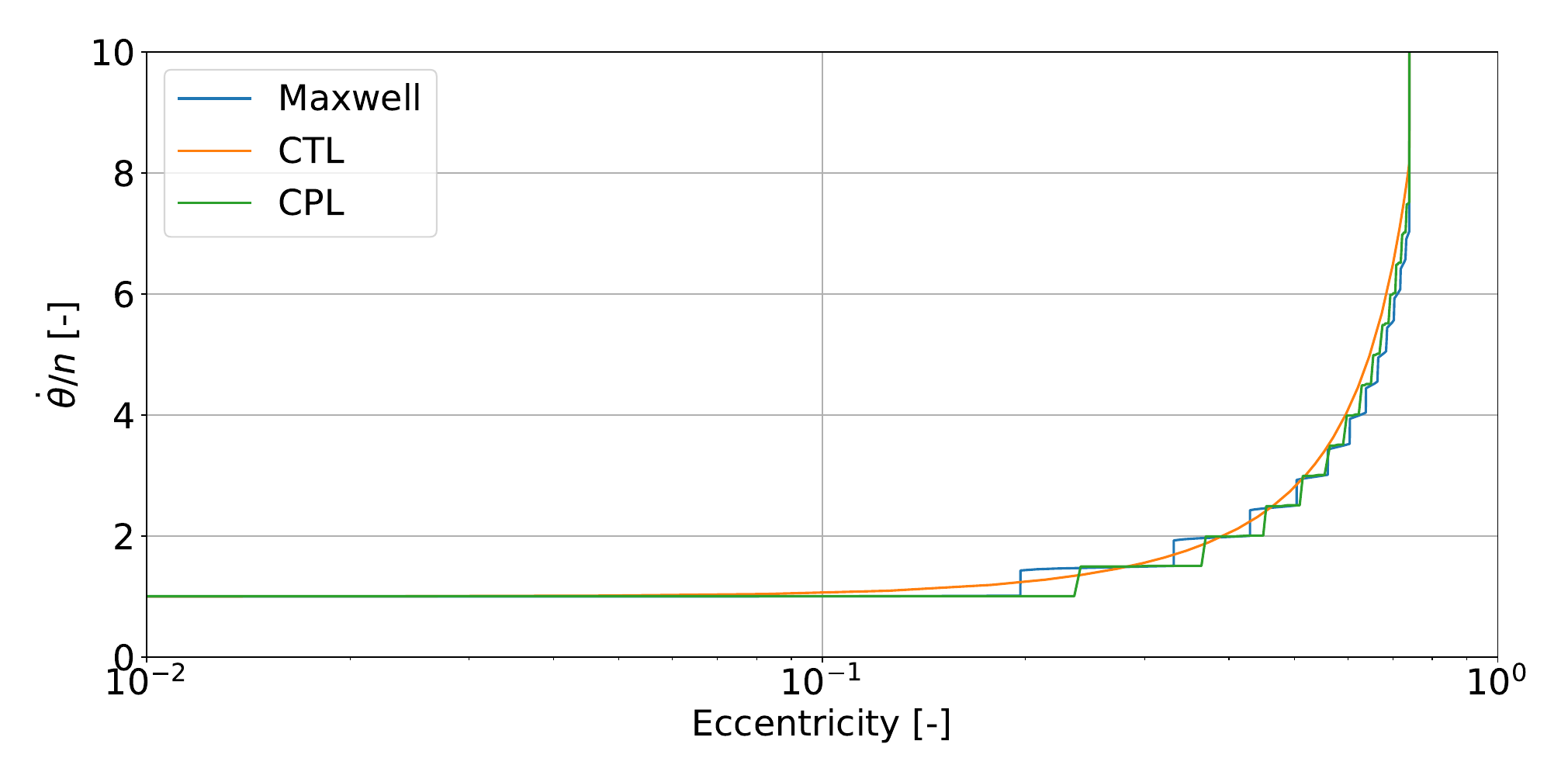}
    \caption{History of the rotational rate of Triton with a CPL, CTL and Maxwell rheology.}
    \label{fig:e_rotrate_comparison}
\end{figure}

\begin{figure}
    \centering
    \includegraphics[width=1\linewidth]{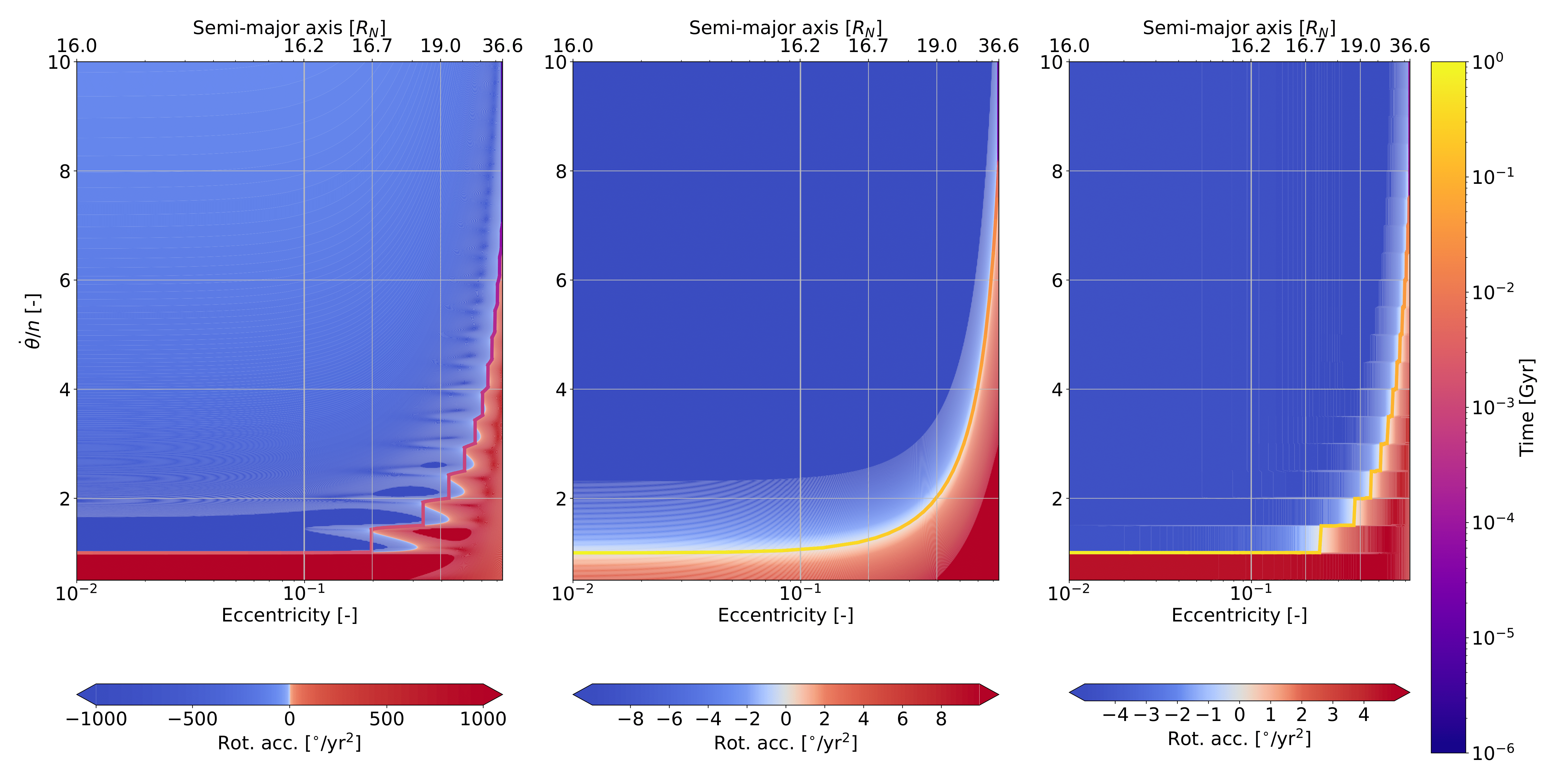}
    \caption{Evaluation of the rotational acceleration throughout phase space for the Maxwell (left), constant time lag (middle) and constant phase lag (right) models, with additionally the state evolution propagated through the equations of motion. Note that the colour-scale differs for each panel. The semi-major axis is coupled to the eccentricity through conservation of angular momentum.}
    \label{fig:phase_all}
\end{figure}

A different story is told by the rotational rate of Triton, however, as shown in Fig.~\ref{fig:e_rotrate_comparison}, which shows the results of the numerical integration of the equations of motion expressed in rotational rate-eccentricity space, and in Fig.~\ref{fig:phase_all}, which shows an exploration of the rotational acceleration throughout rotational rate-eccentricity phase space (mind the different colour-scales): while the predicted rotational rate at any given eccentricity is quantitatively similar between the three models, the qualitative behaviour is significantly different between the three. In particular, we note that the CTL Triton does not display the spin-orbit resonance progression that the CPL Triton and Maxwell Triton do exhibit: instead, it follows a smooth curve down to the 1:1 resonance. Where CTL Triton therefore seemingly approximates the orbital evolution of Triton well enough (though at the wrong timescales), the resonance progression of its combined spin-orbit behaviour is lost entirely. Indeed, the evolution described by \citet{Correia2009SecularTriton}, who uses a constant time lag model, misses out on the spin-orbit resonances that Triton passes through. A partial saving grace can be awarded to the constant time-lag model for the fact that its equilibrium rotation rate has an analytical, finite expression, which we give without proof as Eq.~\ref{eq:CTL_eq_rotrate} in App.~\ref{app:mathematical_addenda}; while the constant time lag model does not appropriately model spin-orbit resonances, its equilibrium rotation rate (rounded to the nearest half-integer) can seemingly at least be used as a first estimate for the true equilibrium rotation rate at a given eccentricity for icy bodies. As we will see in Sec.~\ref{sec:rheological_parameters}, this is a property that is unique to icy (low-viscosity) bodies, however.

The Maxwell model, finally, seems at least capable of reproducing qualitatively the behaviour that we expect spin-orbit wise: the single peak its quality function possesses is the fundamental cause for the stable and relatively aggressive equilibriation of the spin-orbit state of the moon. It additionally allows for a computation of the quality function from fundamental material properties, which the CTL and CPL models do not allow. More advanced rheological models, such as the Andrade, Burgers or Sundberg-Cooper rheologies incorporate additional material behaviours, but do not qualitatively alter the shape of the quality function (see Sec.~\ref{sec:quality_function_homogeneous}); more advanced planetological models, incorporating a layered body of different materials and phases, will give rise to a more complicated shape of the quality function (as different layers will have different peak frequencies and viscoelastic behaviour), but require a full thermal-interior model of the body. The Maxwell model therefore strikes a good balance between qualitatively reproducing the tidal response we expect in general and requiring the fewest additional assumptions on Triton's composition, interior structure and thermal history. Indeed, our spin-orbit evolution for the Maxwell model seems qualitatively similar to the homogeneous-body Sundberg-Cooper-based results obtained by \citet{Renaud2021TidalTRAPPIST-1e}, the layered-body Sundberg-Cooper-based results obtained by \citet{Bagheri2022TheSystem} and the layered-body Andrade-based results obtained by \citet{Walterova2020ThermalExoplanets}. Notably, it appears that Maxwell Triton inhabits a transitionary regime between the sharp and smoothed spin-orbit progression exhibited by the rocky and icy TRAPPIST-1e in Figs.~6 and 8 of \citet{Renaud2021TidalTRAPPIST-1e}, respectively.

Compared to the Mawell model, the constant time lag and constant phase lag models are thus, despite the simplifications they admit for high eccentricities, not appropriately equipped to give any meaningful and complete results on the Tritonian evolution through the high-eccentricity phase of its history; the shortcomings of the CPL model mean that it is not equipped to deal with the wide range of forcing frequencies that Triton encountered, while those of the CTL model mean that it will not adequately describe the spin evolution that accompanied Triton's early evolution. Additionally, the CTL model cannot accurately reproduce even the linear-regime behaviour of the Maxwell model due to the presence of tidal modes that exceed this linear regime at higher eccentricities. It is therefore suitable neither at low eccentricities (where, in reality, spin-orbit resonances might occur) nor at higher eccentricities (even though spin-orbit resonances might smooth out again here). This underlines the results obtained in a more general context by \citet{Renaud2021TidalTRAPPIST-1e}. Consequently, we need to progress to more advanced rheological models, where a full treatment of Kaula's expansion is required to assess convergence of the infinite sums.

\section{Consequences of uncertainty in rheological parameters}
\label{sec:rheological_parameters}
We can (and will, from now on) thus assume that the solid-body tidal response of Triton is qualitatively well-described by a homogeneous Maxwell body. Determining the evolution of the rheological parameters of this Maxwell Triton, that is, the Maxwell time and the effective rigidity, requires a thermal-interior model that is outside the scope of this study. Nonetheless, as the viscosity is severely temperature-dependent and can vary over several orders of magnitude, it is instructive to examine what effects such viscosity variations might induce. To assess this, we vary the viscosity from a nominal value of $10^{14}$ Pa s (as in \citealt{Deschamps2021ScalingEuropa, Bagheri2022TheSystem}; see Sec.~3.2.2 in the latter for a discussion as to why the tidal response for a Pluto-like body is dominated by the ice layer), to a lower value of $10^{12}$ Pa s, which might correspond to a warmer Triton, and a higher value of $10^{16}$ Pa s, which might correspond to a particularly cold Triton or one whose tidal response is muted by its silicate core; this corresponds roughly to the ranges explored in \citet{Deschamps2021ScalingEuropa, Bagheri2022TheSystem} for Europa and Pluto-Charon, respectively.

As before, we propagate starting from the reference starting scenario with eccentricity $e_0=0.74$, a post-damping semi-major axis of $a_f=16R_N$ such that the initial semi-major axis is $a_0=a_f/(1-e_0^2)$ and an initial obliquity of Neptune with respect to Triton's orbit equal to its present-day value of $i_0=156.9^{\circ}$.

\subsection{Differences in state history due to viscosity variation}
\label{subsec:viscosity_state_history}
\begin{figure}
    \centering
    \includegraphics[width=1\linewidth]{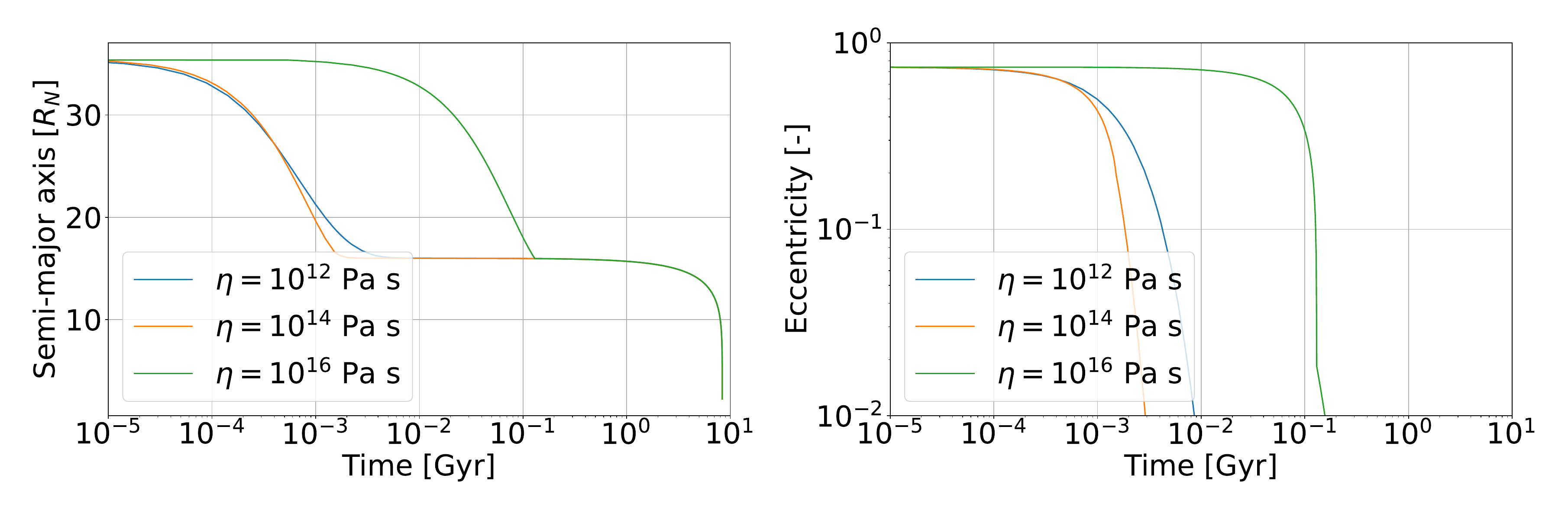}
    \caption{The time histories of the semi-major axis and eccentricity of Triton assuming various viscosities.}
    \label{fig:e_a_hist_viscosity}
\end{figure}

\begin{figure}
    \centering
    \includegraphics[width=1\linewidth]{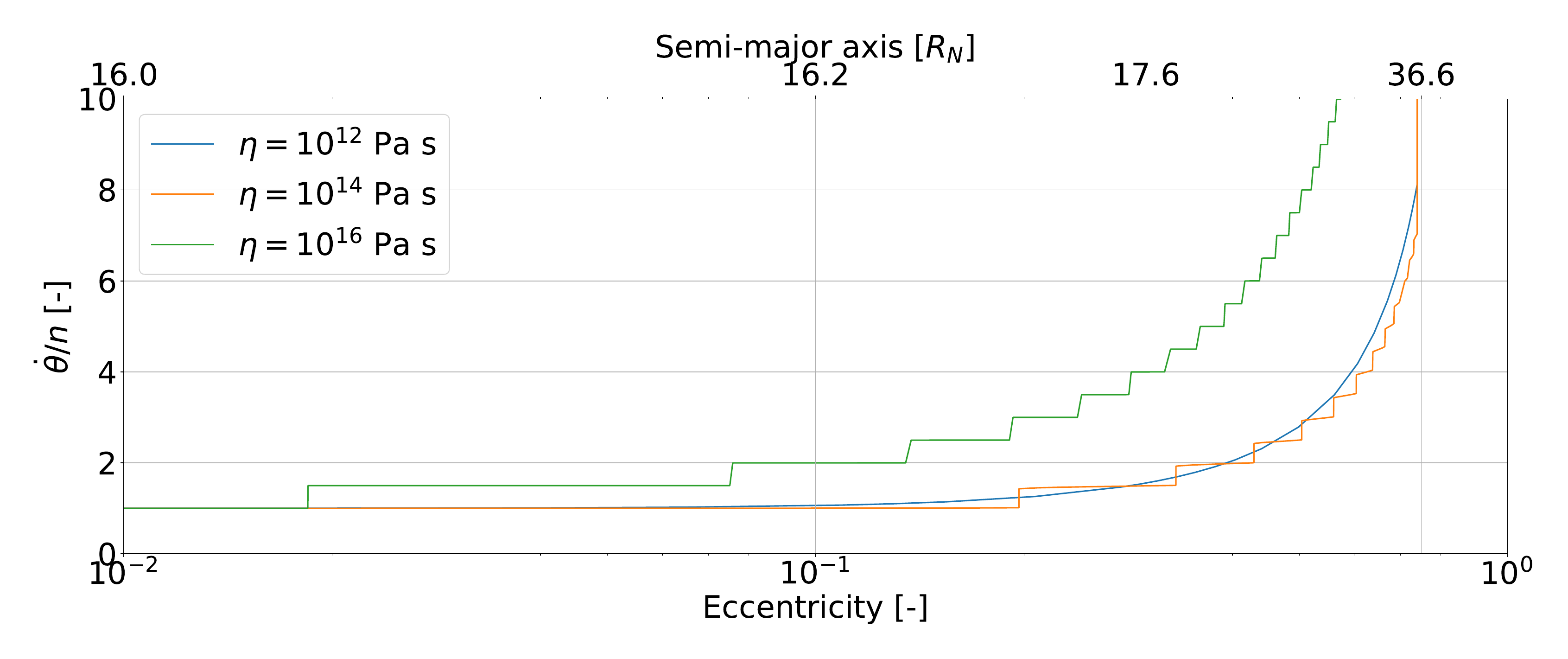}
    \caption{History of the rotational rate of Triton as a function of eccentricity for various viscosities}
    \label{fig:e_rotrate_viscosity}
\end{figure}

\begin{figure}
    \centering
    \includegraphics[width=1\linewidth]{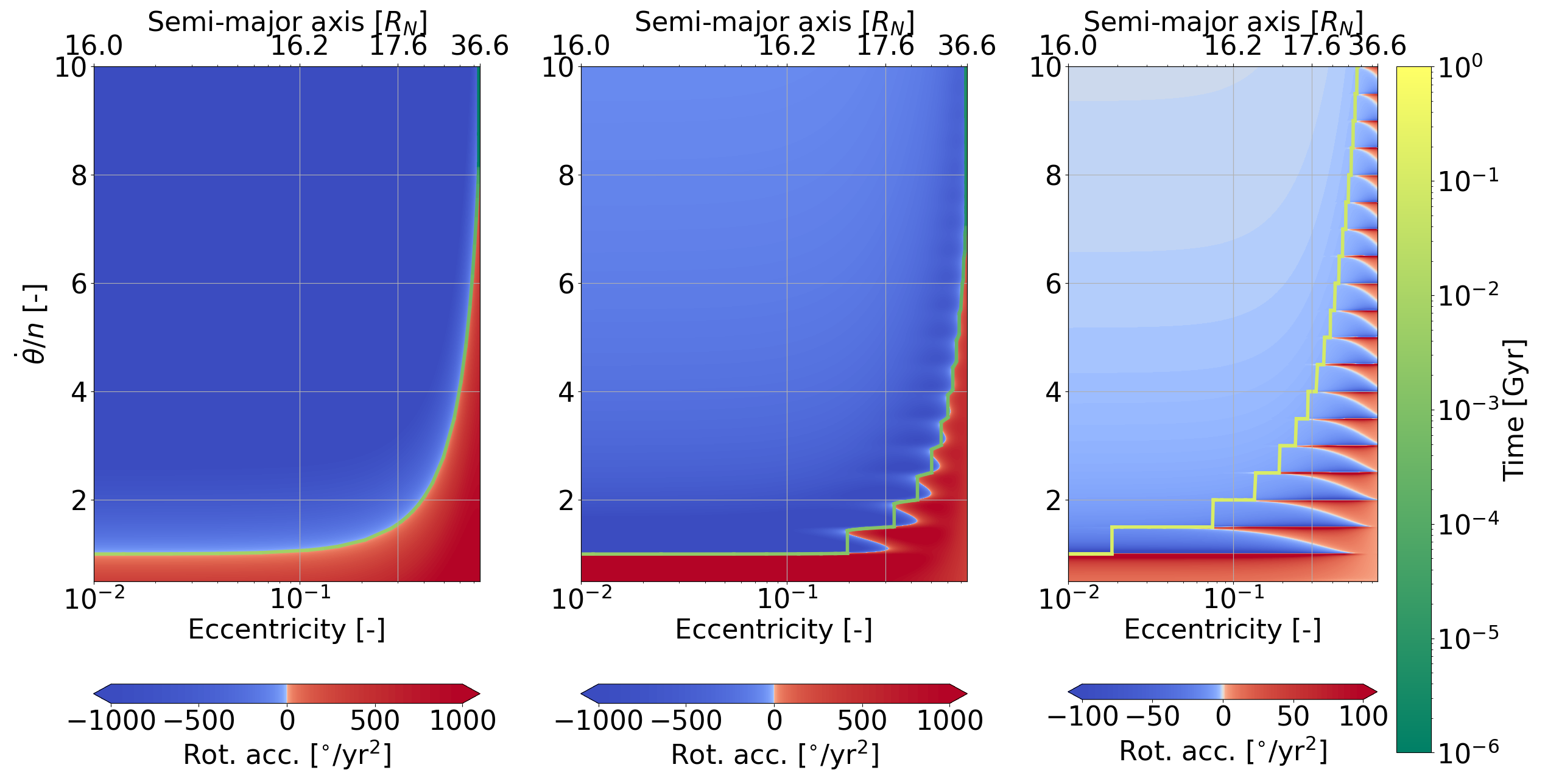}
    \caption{Evaluation of the rotational acceleration throughout phase space for viscosities $\eta=10^{12}$ Pa s (left), $\eta=10^{14}$ Pa s (middle), and $\eta=10^{16}$ Pa s (right), with overlaid the state evolution as propagated through the equations of motion. Note the different colourbar scale on the rightmost panel.}
    \label{fig:phase_space_viscosity}
\end{figure}

The resulting time history of the semi-major axis and eccentricity is shown in Fig.~\ref{fig:e_a_hist_viscosity}; as before, the likeness of the models is obfuscated by timescale differences, and so we also plot the phase space-evolution for the combination of eccentricity and rotational rate in Fig.~\ref{fig:e_rotrate_viscosity}. In all cases, the angular momentum of the orbit is conserved excellently, and so we do not show separately the phase history through the $a-e$ plane; instead, the semi-major axis corresponding to a given eccentricity assuming conservation of angular momentum is shown in Fig.~\ref{fig:e_rotrate_viscosity}. These phase histories are overlaid on a phase cross-section of the eccentricity (and accompanying semi-major axis) and rotation rate in Fig.~\ref{fig:phase_space_viscosity}.

A major difference that occurs for higher viscosities is the fact that spin-orbit resonances can be maintained at far lower eccentricities; these resonances, while stable, are fragile (as is evident from the rightmost panel in Fig.~\ref{fig:phase_space_viscosity}), and it seems unlikely that an icy body could maintain the required high viscosity at significant eccentricity for very long (as a result of the associated heating). One should therefore expect a transition to one of the lower-viscosity regimes and the accompanying lower stable spin-orbit resonance once heating reaches some threshold. At a lower viscosity, the spin-orbit resonances are lost altogether, and the graph resembles that for the constant time-lag model; the peak frequency of the Maxwell rheology has, for this viscosity, moved out sufficiently far that all active tidal modes are situated on the linear part of the Maxwell quality function. A notable difference with the CTL scenario explored in Sec.~\ref{sec:simplified_models} is the magnitude of the equilibriating acceleration of the rotational rate, which is two orders of magnitude greater than the CTL case as examined by \citet{Correia2009SecularTriton}; in assuming an Io-like value of the quality function, it seems they have underestimated the tidal response of a viscous Triton, and correspondingly we find that a realistic CTL-like Triton will have circularised about a factor of 100 faster than they predict. Wherever more Io-like conditions hold (noting that the viscosity for a rocky body is far greater still at $\sim10^{22}$ Pa s, see e.g. \citealt{Renaud2018IncreasedExoplanets}), spin-orbit resonances appear and so the CTL model is no longer an appropriate approximation.

As a final remark, it must be noted that at the higher viscosity the spin-orbit resonance is exact (which agrees with the results shown by \citealt{Renaud2021TidalTRAPPIST-1e, Walterova2020ThermalExoplanets}), while for viscosities that are realistic for ice the resonances occupy a broader band, and there is a small but significant spin-slowing between the initial capture into a resonance and the moment just before transition into a lower one. This can be attributed to the broader peak that is present in the lower-viscosity Maxwell quality functions; as a result, the terms belonging to resonances are still relevant while occupying another spin-orbit resonance. Whenever higher-degree terms are relevant (though we have shown that this is not the case for Triton in Ch.~\ref{ch:validation}), one might expect that similar behaviour occurs even for more viscous bodies. In fact, this might well explain the discrepancies observed with higher degrees in Sec.~\ref{sec:truncating_degree}.

\subsection{Differences in thermal evolution due to viscosity variation}
\begin{figure}
    \centering
    \includegraphics[width=1\linewidth]{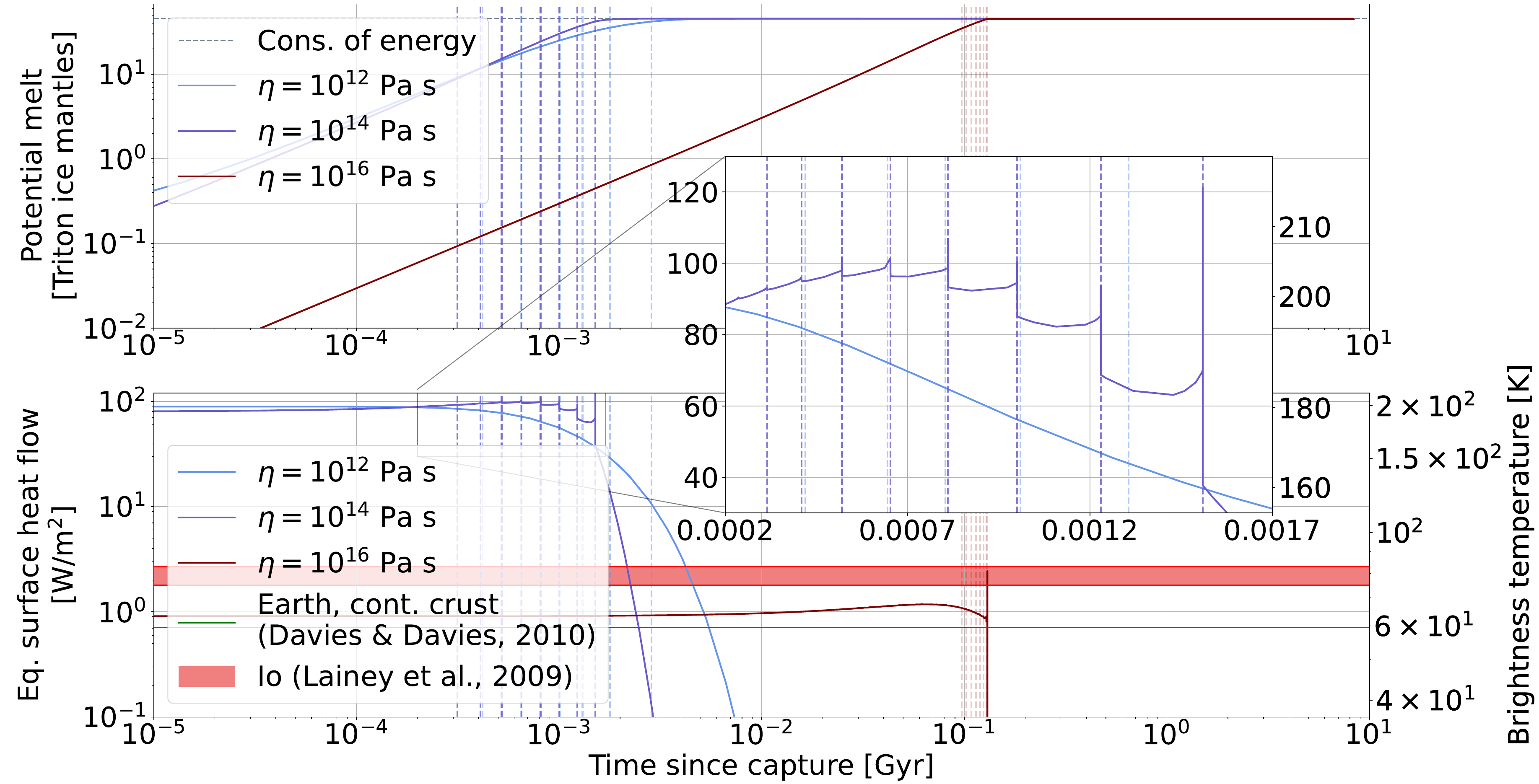}
    \caption{The total dissipated tidal energy and the rate of dissipation as a function of time for the three viscosities. The vertical dashed lines denote the transitions between the spin-orbit resonances starting from the drop down from the 5:1 resonance (computed as the times at which Triton's spin-rate crosses some multiple $m+1/4$ of its mean motion, with $m$ a half-integer), and the horizontal dashed line indicates the dissipated energy predicted by conservation of orbital energy between the initial and final states, equivalent to $\sim45$ Triton mantles. Note that the inset has linear, not logarithmic axes. Earth (continental crust) and Io heat flows taken from \citet{Davies2010EarthsFlux} and \citet{Lainey2009StrongObservations}.}
    \label{fig:dissipated_energy}
\end{figure}

In addition to the state evolution, we compute the associated tidal dissipation and its rate, which are shown in Fig.~\ref{fig:dissipated_energy}. The dissipated energy is expressed in terms of the energy required to raise a mass of ice equivalent to Triton's mantle (at most $\sim35\%$ of its total mass, \citealt{McKinnon2014Triton}) from an ambient temperature of 30 K to 270 K. In reality, the melting temperature may be significantly lowered by impurities in the ocean, commonly modelled as ammonia (e.g. \citealt{Bagheri2022TheSystem}), which can lower it as far as 190 K \citep{Leliwa-Kopystynski2002TheSatellites}; also, the starting temperature may be higher if Triton still contained primordial heat. This figure is therefore conservative.

The dissipated spin energy is added a posteriori over the rotational equilibrium-regime (see Sec.~\ref{sec:equilibrium_dissipation_transition} for a justification), and the associated dissipation rate is computed by adding the mean dissipation rate over the time interval in which the spin-orbit drop happens. As the integration timesteps are in general larger than the timescales over which this spin-orbit resonance transition happens, the associated dissipation shown is therefore an underestimate.

Nonetheless, it is clear that the spin-orbit transitions significantly enhance the (already significant) "continuum" dissipation. Interestingly, these transitions do not occur for the low-viscosity model, and they are relatively muted for the high-viscosity model (with the exception of the drop into the 1:1 resonance). Determining which of these models is most realistic would require a detailed thermal modelling of the Tritonian interior and its (multi-layer) tidal response, and possibly a modelling of ocean dynamics. Whether Triton progressed through the spin-orbit resonances seen in the nominal scenario therefore heavily depends on whether it maintained an ice shell despite its severe heating or not, and if it did, on the viscosity in this shell, and would require detailed thermal-interior modelling.

\begin{figure}
    \centering
    \includegraphics[width=1\linewidth]{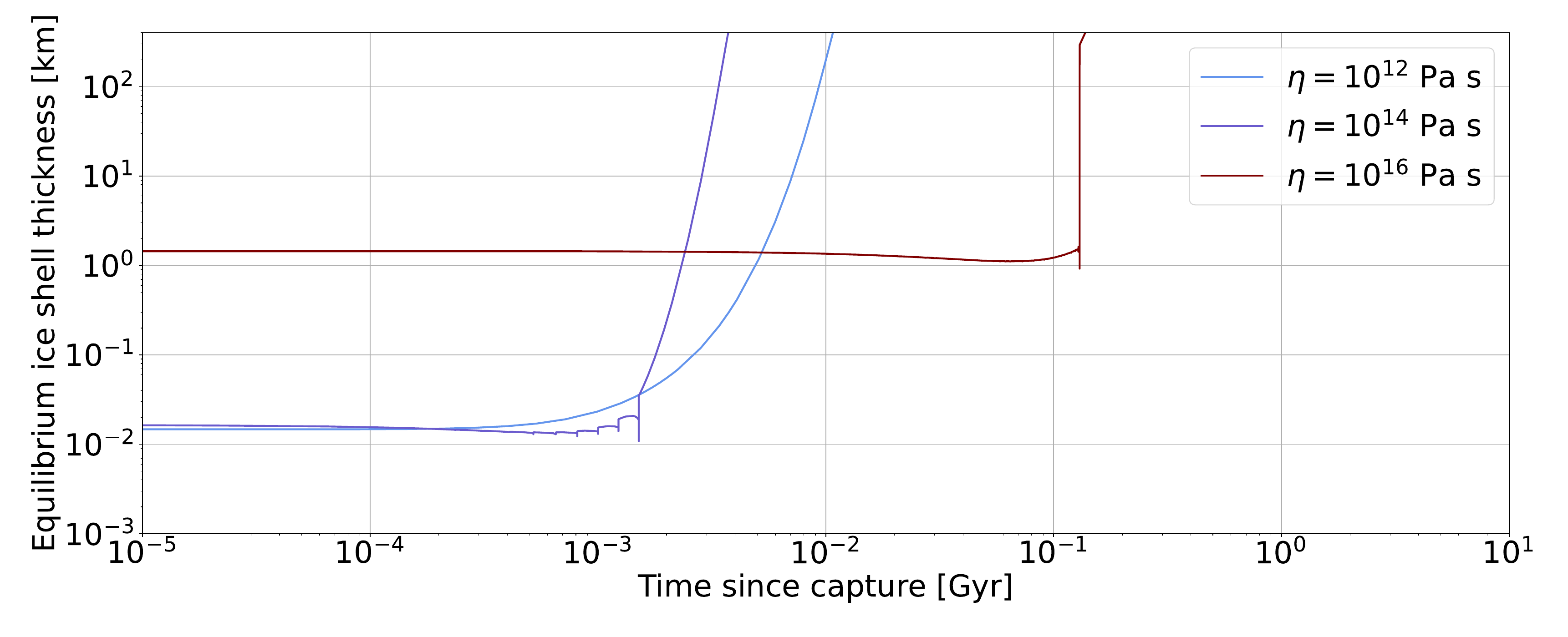}
    \caption{Equilibrium ice thickness calculated using Eq.~7 in \citet{Quick2015ConstrainingCooling} for the three different viscosities, assuming a conductive ice shell with a (conservative) surface temperature of 30 K. Note that the total hydrosphere for Triton spans $\sim 400$ km.}
    \label{fig:shell_thickness_viscosity}
\end{figure}

Whatever the case, it is clear that solid-body tides in the core and ice shell are able to produce heating at least on the order of the $\sim2$ W/m$^2$ surface heat flux that is seen on present-day Io (see e.g. \citealt[Fig.~2]{Lainey2009StrongObservations} for an overview), and possibly even orders of magnitudes greater, and maintain those rates over timescales of 1-100 Myrs. Even starting at an eccentricity of only $e_0=0.74$, as we do in Fig.~\ref{fig:dissipated_energy}, for all viscosity-values the dissipated energy is sufficient to melt Triton's ice mantle several times over within a timescale of $0.1-10$ Myrs; if Triton started at an even greater eccentricity of $e=0.95-0.995$ (as happened in $\sim70\%$ of cases explored by \citealt{Nogueira2011ReassessingTriton}), the total dissipated energy is several times greater still than that displayed here. Consequently, it seems inevitable that Triton's mantle will have melted for a large part, if not entirely. To give a first-order estimate of the resulting component of Triton's shell that will have molten, we compute the equilibrium ice shell thickness using Eq.~7 as given by \citet{Quick2015ConstrainingCooling}, assuming the present-day surface temperature of $\sim 30$ K and a conductive ice shell: the result is shown in Fig.~\ref{fig:shell_thickness_viscosity}, and we deduce that the associated heat flow will likely have the ice shell dwindle to be only several tens of meters thick for the two lower-viscosity scenarios unless tidal dissipation severely drops.

If a sufficiently viscous shell remains (though from Fig.~\ref{fig:shell_thickness_viscosity} it becomes clear that we require more advanced interior models to evaluate whether this is the case), the dissipated spin energy is largely dumped into Triton in short bursts, visible as the peaks in Fig.~\ref{fig:dissipated_energy}, as opposed to the orbital energy lost due to eccentricity-damping of the semi-major axis, which is lost in a more continuous fashion. The consequences of such burst-like dissipation will be further discussed in Sec.~\ref{sec:discussion_discrete_spin_orbit_evolution}.

A final note must be made by remarking that it seems that, for all scenarios in terms of viscosity, the evolution due to the tides raised in Triton by Neptune and that due to the tides raised in Neptune by Triton can largely be separated in time; the evolution over the initial $\sim 1-100$ Myr is governed by the eccentricity-driven damping of the eccentricity, and thus affected by the chosen model for Triton's interior. Afterwards, the remaining evolution is driven by the tidal bulge raised on Neptune dragging down retrograde-orbiting Triton, which is independent of the assumed interior for Triton and largely proceeds after the orbit has been damped to have zero eccentricity. While the details of the initial evolution are therefore heavily dependent on the thermal evolution of Triton and difficult to constrain, it seems likely that the initial conditions of the circularised phase \textit{can} be determined within reason, while needing only an assumed interior model for Neptune. Hence, this is precisely what we will concern ourselves with in the following chapter.

\section{Conclusions}
We have examined the consequences of using a simplified rheological model over a more advanced, physically motivated model for a viscoelastic body. From doing so, we conclude that high-eccentricity evolution is generally not modelled well by the constant phase lag and constant time lag models. To add to the epistemological objections levied by \citet{Efroimsky2013TidalTorque, Makarov2013NoMoons}, we have shown that neither model is in practice capable of reliably reproducing, even on a qualitative level, the behaviour of more involved models for a Triton-like body over the wide range of forcing frequencies it encounters. The CTL-model, moreover, fails on two additional fronts: (1) it fails to predict the existence of half-integer spin-orbit resonances for non-zero eccentricities, and (2) there is no CTL model that can accurately match the behaviour of a Maxwell(-like) body once the activated tidal modes include terms beyond the Maxwell body's peak frequency; it is therefore, despite its seeming approximation of the Maxwell rheology's properties, ill-suited to represent a solid body's rheology at any non-zero eccentricity.

To follow up on this, we have studied the variety of spin-orbit scenarios a body might undergo on a early Triton-like orbit, based on its viscosity. This has shown that even solely the solid-body tides experiences by early Triton will have been significantly dependent on the thermal conditions it experiences, though a first estimate might be given by assuming a nominal value of the viscosity of $10^{14}$ Pa s, as done e.g. by \citet{Deschamps2021ScalingEuropa} and \citet{Bagheri2022TheSystem}, and the ice shell will likely have thinned to at most km-scales even in the most viscous scenario. Additionally, we have observed that, if a sufficiently viscous shell remained throughout Triton's mid-to-high-eccentricity evolution ($e\sim0.5$), it can have undergone rapid transitions between spin-orbit resonances; moreover, that these spin-orbit transitions might have temporarily enhanced the tidal heating rates significantly.

\chapter{Triton's history and future}
\label{ch:initial_conditions}
We have now established that the simplified models that oftentimes found use in previous work cannot be trusted to produce a reliable history for bodies that have experienced at least mildly eccentric epochs. Additionally, we have produced results that allow us to understand the context within which to place any results obtained using fixed-interior models whenever we expect that the interior will realistically have undergone significant variation. We will now apply our findings to produce realistic initial conditions for Triton conditional on an interior for Neptune in Sec.~\ref{sec:inclination_a_circularised_phase}, and propagate Triton forward from these initial conditions with an appropriate internal model for Triton in Sec.~\ref{sec:complete_history}. While it is unlikely that the resulting history for Triton is fully true to reality, it will allow us to make qualitative predictions as a starting point for further studies and pave the way for a complete and self-consistent history for Triton.

\section{Evolution of the semi-major axis and inclination through the circularised phase}
\label{sec:inclination_a_circularised_phase}
Throughout the circularised phase, the semi-major axis and inclination vary relatively slowly and predictably, and as the rotation of Triton has synchronised and its eccentricity and obliquity are negligible by this point, the tidal evolution is dominated by Neptune's interior model, and our model for Triton's interior becomes irrelevant (at least for the dynamical evolution). We can therefore make relatively reliable predictions on the semi-major axis and inclination into which Triton must have settled right after circularisation by only invoking assumptions on Neptune's interior. Gas giants are in general well-described by a constant time-lag model, but it is still debated whether this holds for ice giants, too \citep{Renaud2021TidalTRAPPIST-1e}; for lack of a better model, we will stick with the constant time lag assumption. \citet{Zhang2008OrbitalDespina} find a quality factor for Neptune that varies between $9\cdot10^3$ and $3.6\cdot10^4$; \citet{Correia2009SecularTriton} used the lower end of these values, which is what we have been using so far. More recently, \citet{James2024ThermalNeptune} produced an evolving thermal model for Neptune and found a tidal Q that was far greater during the initial $3$ Gyr of Neptune's evolution, after which it dropped to a value of $\sim5\cdot10^{3}$ with the appearance of a frozen core, to grow to a value of $\sim3.5\cdot10^{4}$ at present. We will not examine a time-evolving model for Neptune, but from this we can gather that $9\cdot10^3$ is probably a reasonably representative lower bound for the mean tidal Q, while $3.6\cdot10^4$ is likely a reasonable upper bound. The time evolution for these two values should then bound the feasible past states that can be occupied by a Triton evolving under influence of a time-varying Neptune that will still lead to the presently observed state. We will not attempt to bound the evolution of Neptune's rotational period, as the uncertainty on current measurements of Neptune's bulk rotation are still uncertain on the order of hours (see \citealt{Neuenschwander2022EmpiricalNeptune} for a discussion); in all cases, we will simply use its present-day rotational period as derived by \citet{Jacobson2009TheNeptune} as a starting value. As the evolution of Neptune's period is in all cases on the order of minutes at most, this should not impact our results.

\begin{figure}
    \centering
    \includegraphics[width=1\linewidth]{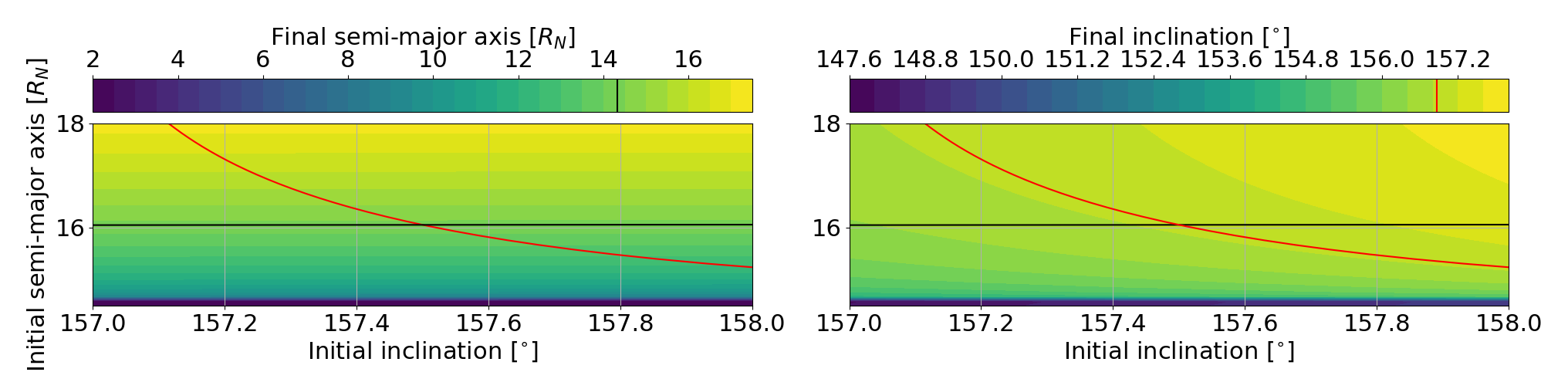}
    \caption{Initial semi-major axis and inclination and the resulting semi-major axis (left) and inclination (right) after 4.5 Gyr for $Q=9\cdot10^3$. The red line corresponds to the constraint on the present inclination, and the black line corresponds to the constraint on the present semi-major axis.}
    \label{fig:q_9000}
\end{figure}

\begin{figure}
    \centering
    \includegraphics[width=1\linewidth]{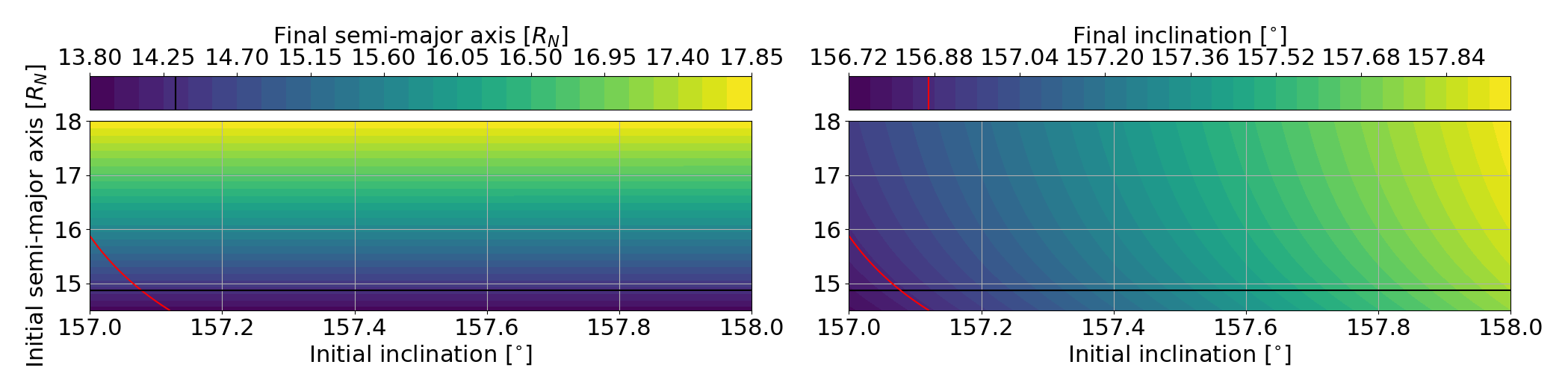}
    \caption{Initial semi-major axis and inclination and the resulting semi-major axis (left) and inclination (right) after 4.5 Gyr for $Q=3.6\cdot10^4$. The red line corresponds to the constraint on the present inclination, and the black line corresponds to the constraint on the present semi-major axis.}
    \label{fig:q_36000}
\end{figure}

To find what these constraints on the semi-major axis and inclination are, we perform a grid search for the initial semi-major axis and inclination-combinations that will lead to observing modern Triton in its present-day orbit. The results are shown in Figs.~\ref{fig:q_9000} and \ref{fig:q_36000}, and they yield an initial semi-major axis and inclination of $16.05$ $R_N$ and $157.50^{\circ}$ or $14.87$ $R_N$ and $157.08^{\circ}$ for the $Q=9\cdot10^3$ and $Q=3.6\cdot10^4$ cases, respectively.


\begin{figure}
    \centering
    \includegraphics[width=1\linewidth]{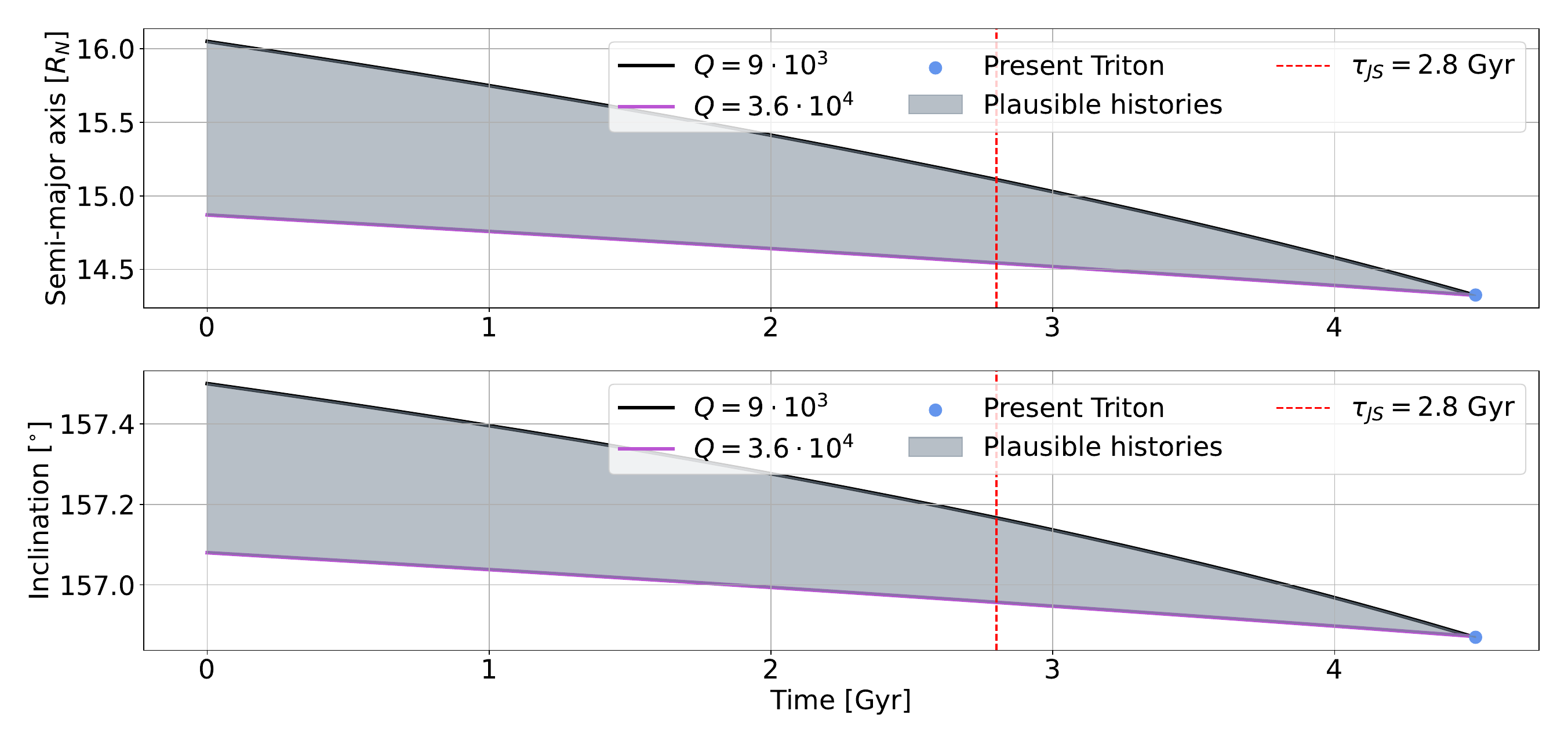}
    \caption{Constraints on the history of Triton's orbit throughout the circularised part of its history. The dashed red line corresponds to the time at which Neptune's core freezes in the thermal-interior model by \citet{James2024ThermalNeptune}; they argue that little to no tidal evolution would have happened before that time.}
    \label{fig:circularisation_past}
\end{figure}

These two cases, providing an upper and lower bound on the tidal response of Neptune, will thus bound the region of plausible state histories for the Tritonian orbit with time for the zero-eccentricity (circularised) phase of its history. This region is displayed in Fig.~\ref{fig:circularisation_past}; we additionally mark the time at which Neptune's core freezes in the thermal-interior model by \citet{James2024ThermalNeptune}. They predict that Neptune's tidal quality factor is sufficiently large before that that no significant tidal evolution would have taken place as a result of the tides raised in Neptune; in that case, the plausible initial conditions for the circularised phase can be found along this red line. In the scenario where tidal evolution did also take place before that, the plausible initial conditions are thus a semi-major axis of $\sim14.9-16$ $R_N$, which is comparable to the value of $15.6$ $R_N$ derived analytically by \citet{Nogueira2011ReassessingTriton}, and an inclination of $\sim157.1-157.5^{\circ}$: in either case, the lower bound corresponds to $Q_N=3.6\cdot{10^4}$, while the upper bound corresponds to $Q_N=9\cdot10^3$. Broadly speaking, we can thus bound Triton's semi-major axis and inclination at the start of the circularised phase to have roughly $a/R_N\in[14.3, 16]$ and $i\in[156.9^{\circ}, 157.5^{\circ}]$. 

Before the circularised phase, as shown in Ch.~\ref{ch:spin-orbit_chains}, the dependence of the evolution on the used interior model is too severe to place any meaningful constraints on the spin-orbit evolution of Triton without placing some assumptions on the interior model. An exception can be made for the quantities related to the spin state of Neptune, which vary relatively slowly; in particular, this concerns the rotation rate of Neptune and the inclination of Triton's orbit with respect to Neptune. The equations governing these quantities only rely upon the perturbed potential of Triton in an indirect manner (see Eqs.~\ref{eq:eom_obliquity} and \ref{eq:eom_rotation_rate}), and so we can discuss how they evolve at least on a qualitative level.

The evolution of the rotational period of Neptune, though not shown, was also computed for all scenarios explored in Ch.~\ref{ch:spin-orbit_chains}, yet the resulting slowing of its spin period remains at most on the order of minutes over the full timespan, and was fully negligible over the progression through spin-orbit resonances during circularisation. Similarly, the inclination change (though sometimes significant) occurs over the circularised phase, not the initial circularisation phase. However, at higher eccentricities the growth of the magnitude of the eccentricity functions with $\sim(1-e^2)^{-3/2}$ may well overpower the factors of $n/\xi$ (which, for constant angular momentum, is proportional to $1/a$) and $M_T/M_N$ which serve to suppress the evolution of the inclination and Neptune's rotational rate (see Eqs.~\ref{eq:eom_obliquity} and \ref{eq:eom_rotation_rate}). Our best guess for a starting point for Triton's evolution is thus to assume that the inclination of Triton's orbit and Neptune's rotational period did not significantly change over the initial circularisation phase: we can then check these assumptions a posteriori.

\section{A complete history and future for Triton's evolution}
\label{sec:complete_history}
Hence, we have estimates for the initial inclination of Triton's orbit, initial period of revolution of Neptune, and the semimajor axis for a given initial eccentricity (the latter by conservation of angular momentum from the zero-eccentricity semimajor axis into which Triton settled at the start of the circularisation phase; the former, we will see, do not vary over the initial phase). For the starting eccentricity we take $e=0.97$, which is the median starting eccentricity obtained by \citet{Nogueira2011ReassessingTriton}, such that the initial semi-major axis is $\sim251.6$ $R_N$ in the $Q_N=3.6\cdot10^4$ case, and $\sim271.6$ $R_N$ in the $Q_N=9\cdot10^3$ case; the associated initial inclinations are $157.08^{\circ}$ and $157.5^{\circ}$, respectively. All that is left is to set the initial rotation rate and obliquity of Triton, and we will have a complete set of initial conditions to propagate from. Trial runs indicate that in the regime of $e\sim1$, contrary to the assertion made by \citet{Correia2009SecularTriton}, the obliquity is damped before the rotational rate is. In fact, the obliquity seems to in general be the first thing to be damped out, and so we assume that it is zero to start with; as we do not model the possible capture of Triton into Cassini states, any non-zero results would likely not be trustworthy regardless. This then leaves the rotational rate: as we do not know whether the rotational rate locks into a spin-orbit resonance right away, we will start with a value roughly representative of trans-Neptunian binaries at a rotational period of 10 h (see e.g. \citealt{Perna2009RotationsObjects, Thirouin2014RotationalBelt}).

To then assess the past and future evolution of Triton, we propagate two initial scenarios, corresponding to the different values of $Q$ explored in Sec.~\ref{sec:inclination_a_circularised_phase} over a timespan of 10 Gyr. For eccentricities greater than $0.74$, the associated values of the eccentricity functions are computed not using the power series expressions from Sec.~\ref{sec:power_series_appendix}, but using the (far more computationally expensive) numerical integral representation given in Sec.~\ref{sec:eccentricity_function_integral}. We will present the results for the dynamical and thermal behaviour over the initial, eccentric phase in Sec.~\ref{subsec:high_eccentricity_evolution}, and discuss the long-lasting, circularised phase that follows in Sec.~\ref{subsec:long_term_evolution}.

\subsection{High-eccentricity evolution}
\label{subsec:high_eccentricity_evolution}
\begin{figure}
    \centering
    \includegraphics[width=1\linewidth]{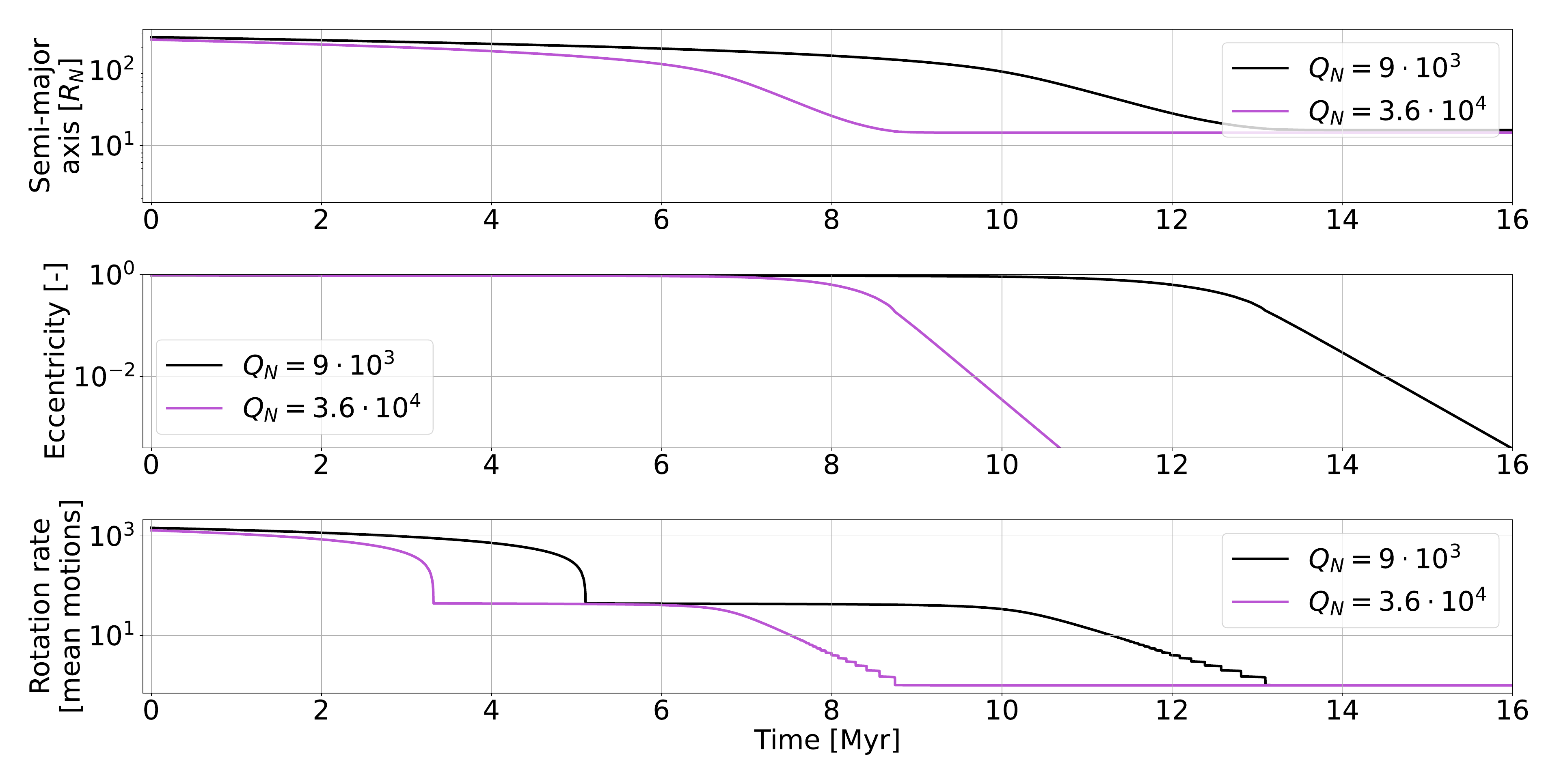}
    \caption{The spin-orbit elements over the circularisation phase for the two bounding values for $Q_N$, starting from $e=0.97$ in either case. Note that the initial conditions between the two are different: see text for details.}
    \label{fig:spin_orbit_elements_circularisation}
\end{figure}

\begin{figure}
    \centering
    \includegraphics[width=1\linewidth]{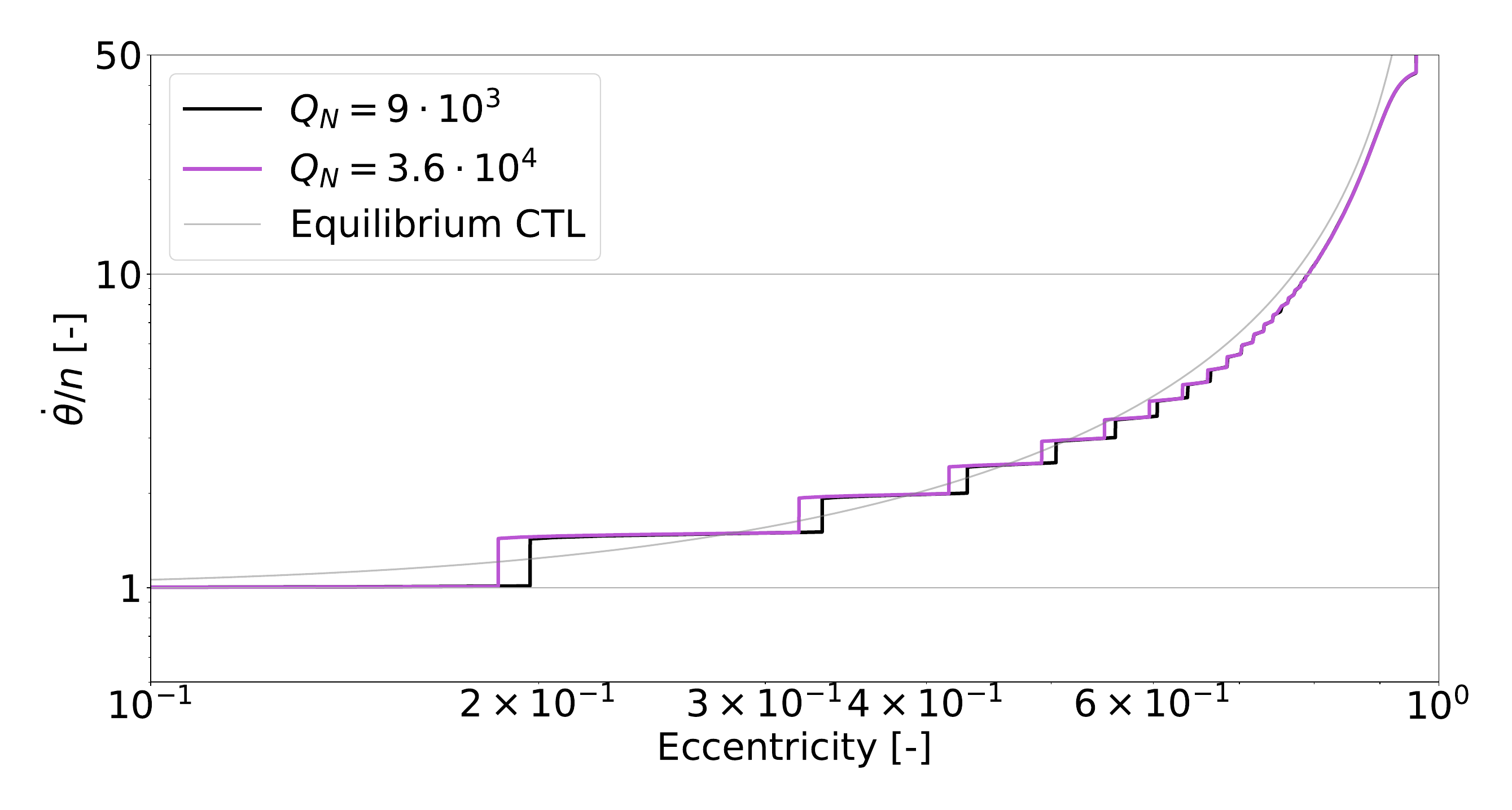}
    \caption{The phase-space evolution of the eccentricity and rotation rate; also drawn is the analytical equilibrium evolution for a constant time-lag rheology. The disappearance of the step-like behaviour at higher spin-orbit multiples is not an artefact of the logarithmic scale; the spin-orbit evolution is smooth, not step-like, for $\dot{\theta}/n\gtrsim10$.}
    \label{fig:e_rotrate_phase_full}
\end{figure}

Let us first explore the results for the dynamically active initial circularisation phase: Figs.~\ref{fig:spin_orbit_elements_circularisation} and \ref{fig:e_rotrate_phase_full} show the evolution of the spin-orbit elements over the initial $16$ Myr and the phase-space evolution of the spin rate in mean motions, respectively: the inclination and Neptune's rotation rate remain unperturbed, and so are not shown. Qualitatively, the two scenarios seem relatively similar, only somewhat stretched out in time. This can seemingly be attributed to the evolution of the $Q=9\cdot10^3$-scenario occurring out at higher semi-major axes, such that correspondingly the factor of $n$ in the equations of motion decreases. 

Notably, the evolution remains relatively mellow for high eccentricities, only speeding up once the eccentricity reaches the regime explored in Ch.~\ref{ch:spin-orbit_chains}; the spin rate takes several Myr to equilibriate, yet reaches similar equilibrium values between the two scenarios, as is evident from Fig.~\ref{fig:e_rotrate_phase_full}. Additionally, the initial spin evolution is smooth, and seemingly does not contain step-like behaviour in the spin-orbit transitions (or the resolution of our integrations is too rough to observe them). While this is reminiscent of the equilibrium behaviour of the constant time-lag rheology, we can clearly see in Fig.~\ref{fig:e_rotrate_phase_full} that it is still not a perfect approximation even over this regime. We note that the eccentricity and non-synchronous rotation rate are damped out after $\sim 16$ Myrs at the latest: what remains is the long-term evolution at zero eccentricity and with synchronous rotation. This phase will be presented in Sec.~\ref{subsec:long_term_evolution}.

\begin{figure}
    \centering
    \includegraphics[width=1\linewidth]{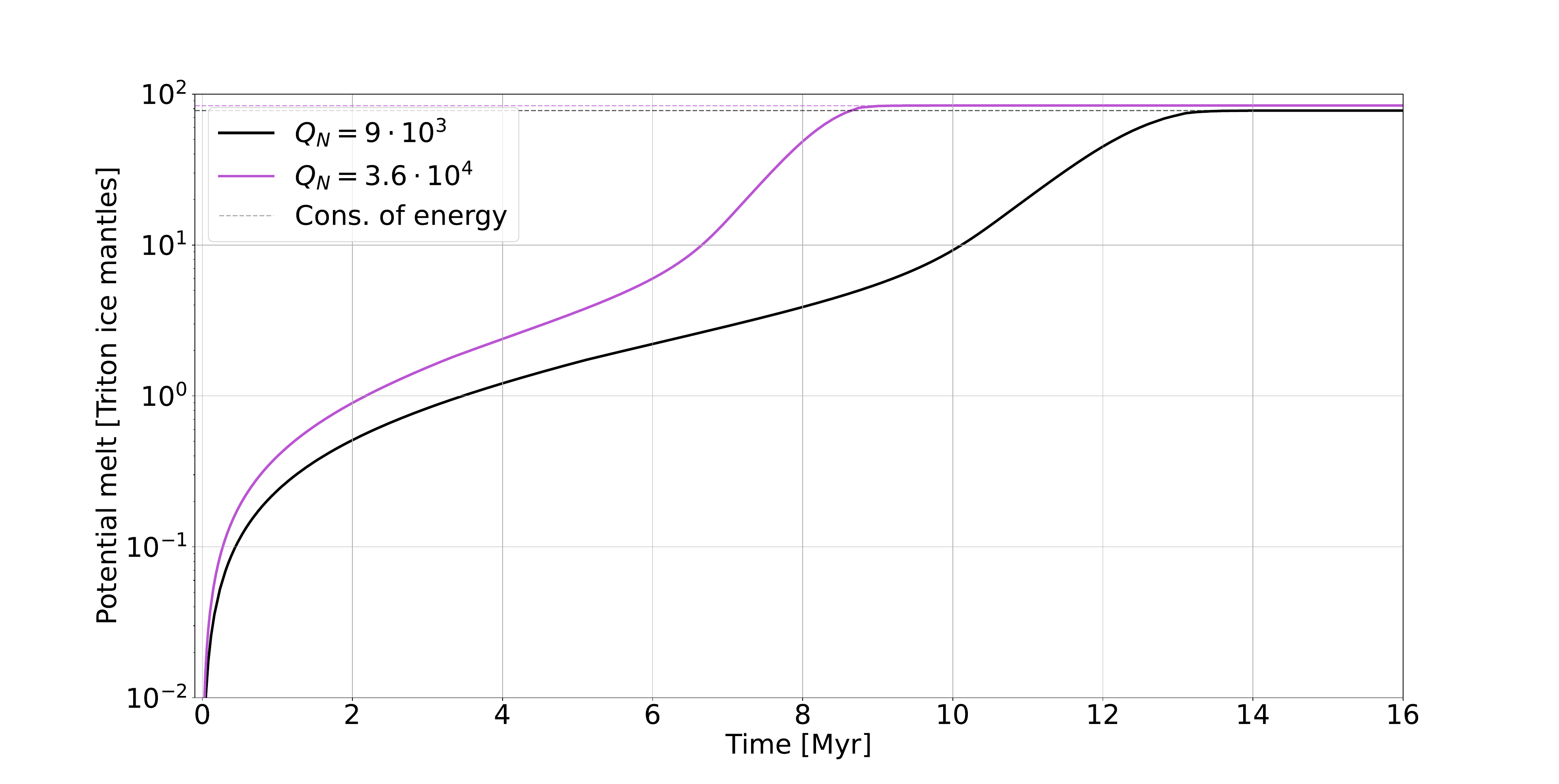}
    \caption{The tidal energy dissipated in Triton over the initial, highly eccentric epoch. The dashed lines mark the energy difference between the initial, eccentric orbit and the circularised orbit (assuming conservation of angular momentum in Triton's orbit): this difference is fully deposited into Triton. }
    \label{fig:pot_melt_energy}
\end{figure}

\begin{figure}
    \centering
    \includegraphics[width=1\linewidth]{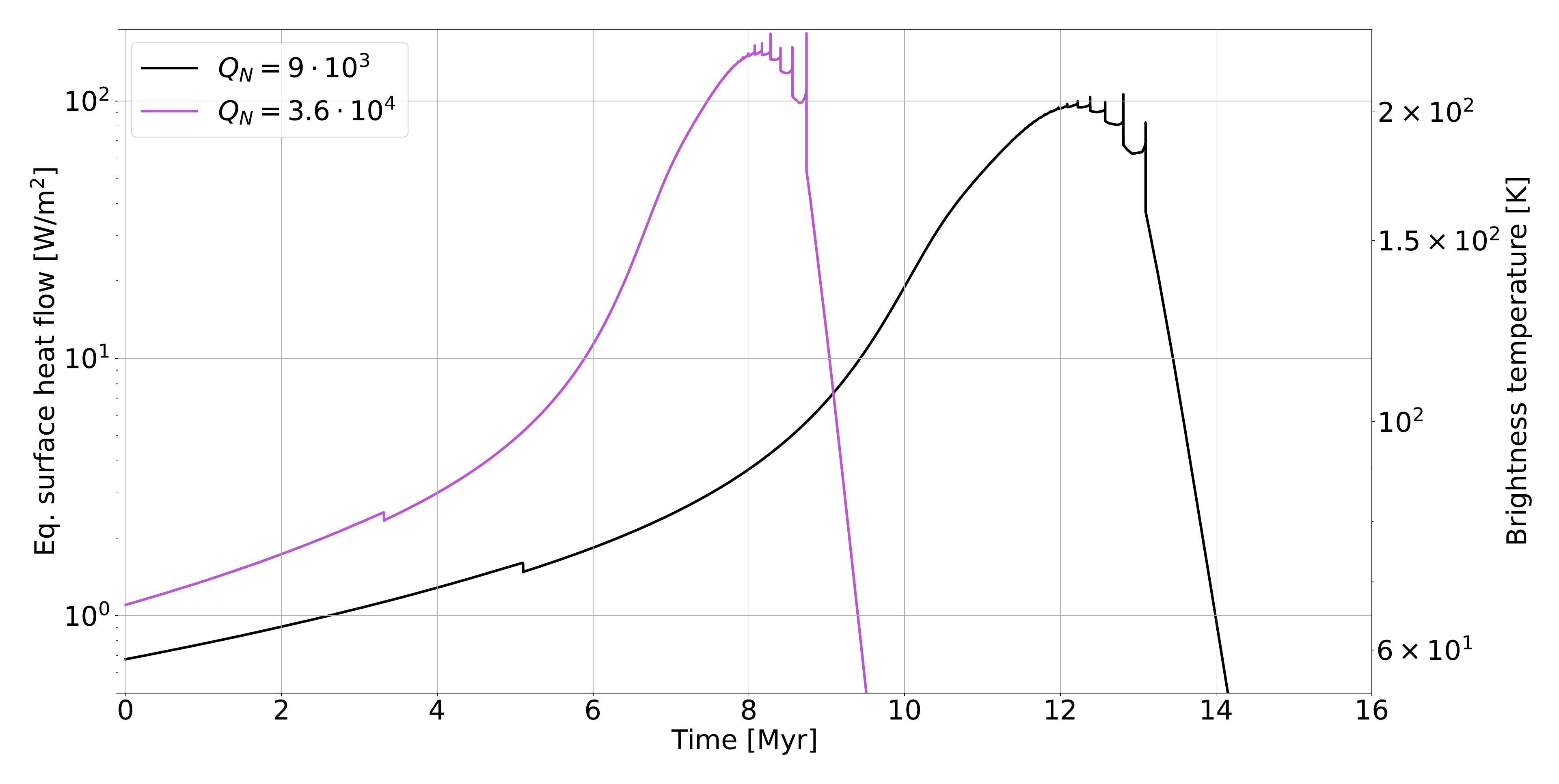}
    \caption{The energy dissipation rate in Triton over the initial eccentric epoch, expressed as equivalent surface heat flow and the corresponding brightness temperature. The dips at $\sim3$ and $\sim5$ Myr correspond to equilibriation of the rotation rate, and comprise a difference of roughly 10\%. The spikes from $8-9$ and $12-13$ Myr correspond to the spin-orbit resonance transitions, and are not integration artefacts.}
    \label{fig:dissipation_rate_full}
\end{figure}

\begin{figure}
    \centering
    \includegraphics[width=1\linewidth]{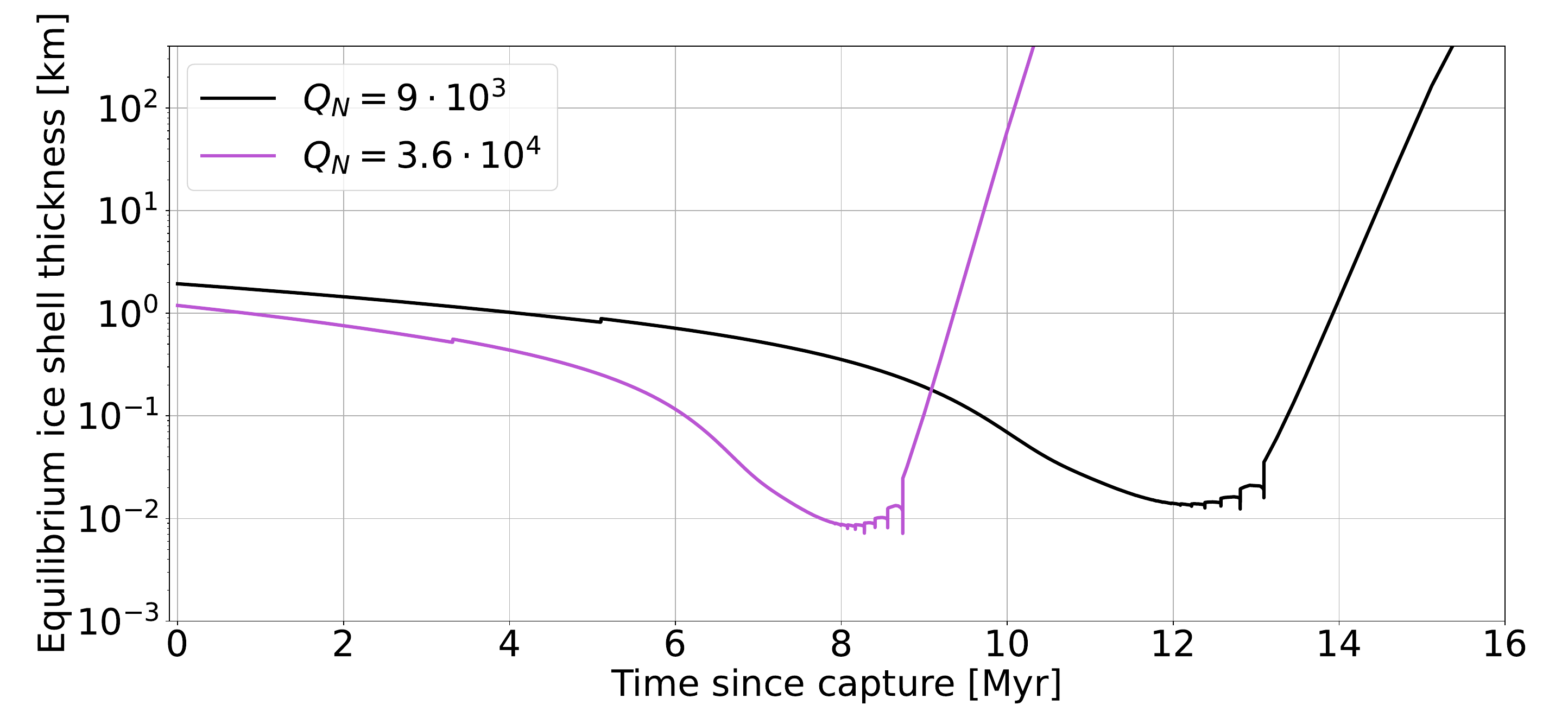}
    \caption{The equilibrium ice shell thickness corresponding to the energy dissipation rates given in Fig.~\ref{fig:dissipation_rate_full} according to Eq.~7 given by \citet{Quick2015ConstrainingCooling}, with a conductive shell and a (conservative) surface temperature of $30$ K. The spikes from $8-9$ and $12-13$ Myr correspond to the spin-orbit resonance transitions, and are not integration artefacts.}
    \label{fig:ice_shell_thickness_full}
\end{figure}

Even though the dynamic evolution is relatively mellow initially, it is the thermal evolution that will most affect the planetological evolution of Triton. This is especially important, as this will affect whether Triton is likely to enter into the region where we expect rapid evolution and discrete spin-orbit transitions with a viscous, thick ice shell, with an ocean overlaid by a thin ice shell or without any remaining ice shell altogether. We therefore also examine the tidal energy dissipated into Triton: the total dissipated energy is displayed in Fig.~\ref{fig:pot_melt_energy}, its rate in Fig.~\ref{fig:dissipation_rate_full} and the resulting equilibrium ice shell thickness in Fig.~\ref{fig:ice_shell_thickness_full}. 

Examining Fig.~\ref{fig:pot_melt_energy}, we see that the total energy dissipated in Triton varies relatively little between the two scenarios, and corresponds in either case excellently with the orbital energy difference between the eccentric and circularised orbits. Qualitatively, we see that the behaviour is relatively similar: the majority of the energy is dissipated in Triton over a timespan on the order of 1-2 Myr in length, and the major difference between the two scenario is a stretching out of the dissipation over a timespan $\sim1.5\times$ as long. In either case, the energy is well in excess of the energy required to melt Triton's mantle, and so the question becomes not if, but when the mantle will start to thin. We can answer this question by examining the energy dissipation rates at this time, and what those would mean for the thickness of any ice shell.

The dissipation rate (converted to an equivalent surface flux) and the associated equilibrium ice shell thickness are shown in Figs.~\ref{fig:dissipation_rate_full} and \ref{fig:ice_shell_thickness_full}, respectively. We note that, again, these look qualitatively very similar for the two different sets of initial conditions: it seems that, to first order, the circularisation phase follows a qualitatively similar trajectory, with only the timescales changing (and the rates changing accordingly). More interestingly, as sufficient energy to melt Triton's entire mantle is deposited by at most 2.2-3.5 Myr, it seems likely that the ice shell will reach the equilibrium ice shell thickness before peak tidal heating starts. As this ice shell thickness is already on the order of kilometres, it seems plausible that Triton enters the epoch of extreme tidal heating with a large subsurface ocean overlaid by a thin, low-viscosity ice shell. It is then probable that our rheological model does not accurately represent the tidal response of such a body, which should have a particularly large value for the tidal quality factor, and so it is likely that the timescale of tidal evolution is extended accordingly. Our results should therefore be interpreted as a lower bound.

Interestingly, the equivalent surface heat fluxes ($\gtrsim1$ W/m$^2$) would already be sufficient to raise a thick atmosphere almost immediately, according to the results by \citet{Lunine1992ATriton}. The consequences thereof are further discussed in Sec.~\ref{subsec:early_atmosphere}, but we can thus be certain that a coupled interior-tidal model with a thermal model for the early atmosphere is necessary to provide any more trustworthy predictions.

Finally, we can remark that the assumptions used to yield the used initial conditions (that is, a negligible variation in Neptune's rotation rate and Triton's inclination over the high-eccentricity phase) are now indeed validated a posteriori, meaning that our history is dynamically consistent with the present-day orbit of Triton. There are, however, still some remarks to be made on the long-term evolution.

\subsection{Long-term evolution}
\label{subsec:long_term_evolution}
\begin{figure}
    \centering
    \includegraphics[width=1\linewidth]{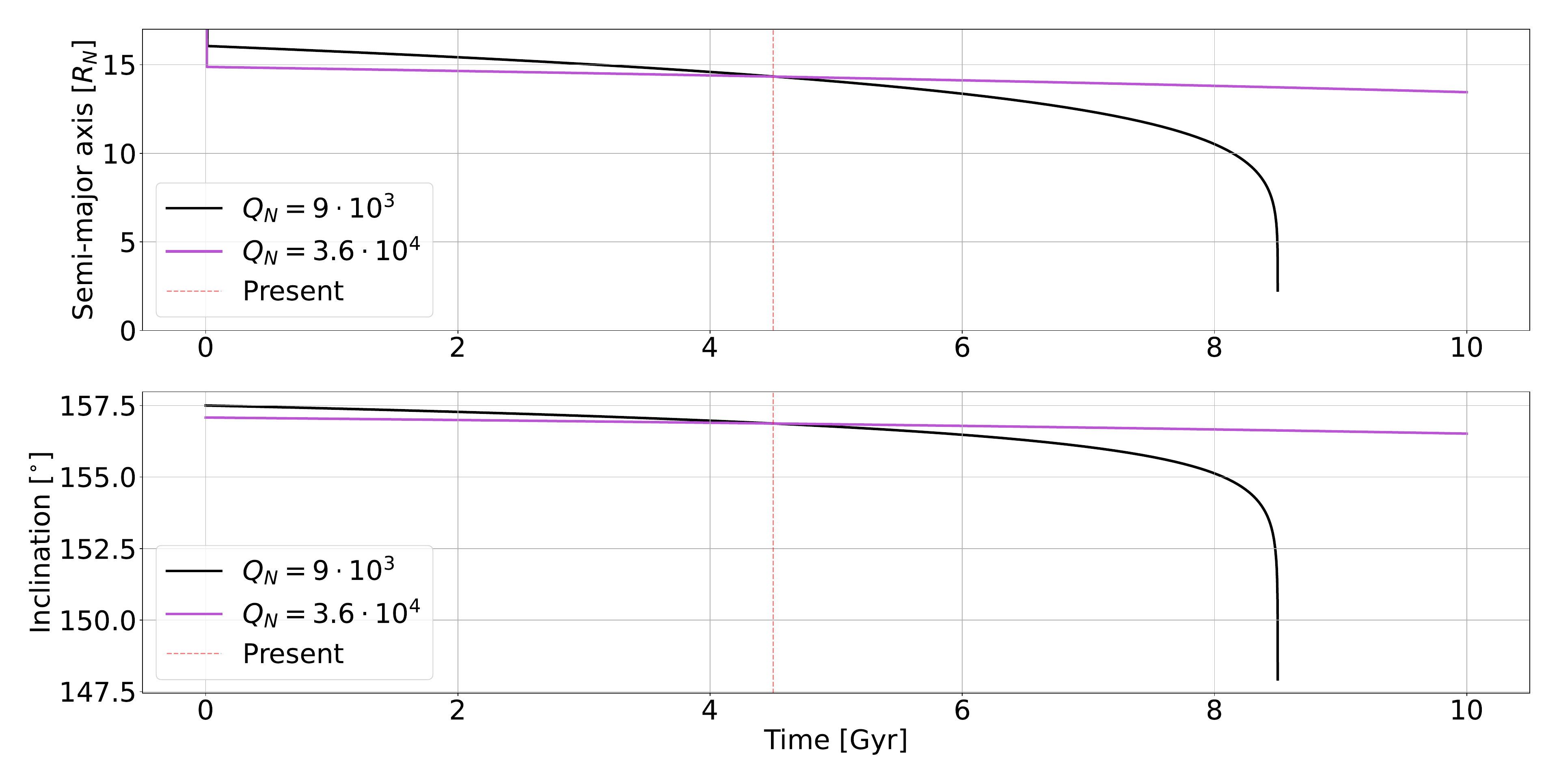}
    \caption{The long-term evolution of the semi-major axis and inclination of Triton for the two Neptune-models. At the 4.5 Gyr-point, we see that both models place Triton on its present-day orbit (14.3 $R_N$) with its present-day inclination ($156.9^{\circ}$).}
    \label{fig:a_i_full_history}
\end{figure}

After this initial high-eccentricity phase, the eccentricity is damped out, and the rotation rate has synchronised. The only remaining evolution will then occur in the semi-major axis and the inclination (not counting short period-precessing quantities that we do not model, the argument of pericentre and the longitude of the ascending node) over Gyr-timescales, shown in Fig.~\ref{fig:a_i_full_history}. Here, we will discuss in particular two things: the decay of Triton into Neptune, and the potential for Neptune-tilting to occur as it does so.

While the semi-major axis and inclination have apparently not varied by much over the 4.5 Gyr following the circularisation phase, the slow encroachment of Triton on Neptune will speed up this process with time. Whether this leads to decay into Neptune within the next $\sim5$ Gyr or not depends on the interior model of Neptune, and cannot be determined with any certainty yet: within the bounds on Neptune's tidal response set by \citet{Zhang2008OrbitalDespina}, either possibility is still allowed, while with the evolving Neptune-model produced by \citet{James2024ThermalNeptune} it seems more likely that Triton will survive.

As the inclination change over the initial 4.5 Gyr (including the circularisation phase) is negligible, it is not to be expected that Neptune's obliquity with respect to the ecliptic (which is of order $\sim30^{\circ}$) can be caused by Triton's migration. Indeed, a quick back-of-the-envelope calculation shows that the ratio of the orbital angular momentum contained in Triton and the spin angular momentum in Neptune is $\leq1$ over the full circularised phase, and so a change in the inclination of Triton's orbit will be accompanied by a smaller change in the obliquity of Neptune's spin axis. Consequently, we may expect to see an obliquity change by this point in time: it can be computed that, if Neptune reorients to absorb all angular momentum present in Triton\footnote{This is a plausible assumption if all material present in Triton ends up either in Neptune or as part of a prograde, equatorial set of rings around it, while not imparting significant angular momentum change to any of the other satellites of Neptune. This is not implausible, but would realistically require verification through dynamical simulations of such a disc.}, the associated obliquity change would be $\sim25^{\circ}$. Interestingly, Triton seems close to a critical point in this behaviour: if Triton were twice as massive but captured on the same orbit, the associated obliquity change would be sufficient to tilt Neptune to Uranus-like obliquities $>90^{\circ}$. We will discuss the consequences thereof in Sec.~\ref{subsec:oblique_exorings}.

\section{Conclusions}
We have made plausible that Triton remained on an orbit comparable to that it finds itself on in the present over the 4.5 Gyr following circularisation of its orbit, provided that this circularisation happened sufficiently fast that the tides Triton raises in Neptune are of negligible magnitude. From this, we hypothesise that Triton was captured on an inclination very close to its present-day value, undergoing little inclination-evolution up to now, and with a semi-major axis that is commensurate to its eccentricity by conservation of angular momentum from this near-modern orbit.

Starting from these initial conditions, we then perform a full simulation of Triton's evolution assuming a fixed interior: here, we see that Triton will, for the range of viscosities that its mantle may have had if it remained (partially) solid, have experienced surface heat flows orders of magnitude greater than those experienced by present-day Io immediately following capture, and therefore will likely have entered into the phase of most extreme tidal heating already covered by an ocean and atmosphere. The circularisation phase that follows will have lasted $\gtrsim10$ Myrs, but this estimate may well be extended by inclusion of a rheological-interior model that accurately captures the realities of a partially-molten Triton. If Triton maintained sufficient viscosity, it will have passed through spin-orbit resonances during this circularisation, until eventually it reaches a synchronously-rotating, circular orbit close to its present one. Following this circularisation, Triton migrates slowly toward its present-day position, and it will continue to slowly decay towards Neptune, dropping into the planet in $\gtrsim3$ Gyr if Neptune's tidal quality factor is on the low end of plausible values, producing an inclined set of rings.

\chapter{Discussion}
\label{ch:discussion}
While the results presented in the preceding chapters will in principle allow us to answer all the questions we set out to answer, it is important that the findings be placed in the correct scientific context. Consequently, we will present additional discussion of the results and how they relate to existing literature in the following. In particular, we will highlight some shortcomings of the applied model in Sec.~\ref{sec:shortcomings}. Nonetheless, we can infer that the history we presented for Triton will have implications for Triton in a planetological context, which we will treat in Sec.~\ref{sec:planetological_implications}; we can also place these results in a broader context, and so we will dedicate Sec.~\ref{sec:implications_neptune_system} to a discussion of the implications they have on our state of knowledge of the Neptunian system as a whole. Finally, Sec.~\ref{sec:consequences_exoplanetary_systems} concerns itself with the repercussions the existence of a Triton according to the history we propose has on the expectations we should have for exoplanetary systems.

\section{Model shortcomings}
\label{sec:shortcomings}
As any other, our model contains a set of shortcomings or potential weaknesses: some of these have already found exploration in literature, while others have not found any such appreciation as of yet. Nonetheless, we will briefly go over some of these that we deem important to consider in future work, for one reason or another. These concern, in no particular order, our neglect of the asphericity of Triton (Sec.~\ref{subsec:asphericity}), our use of a fixed interior model (Sec.~\ref{subsec:evolving_interior}), the restriction of tidal effects to solid-body tides (Sec.~\ref{subsec:ocean_tides}), the absence of third-body perturbations in our simulations (Sec.~\ref{subsec:kozai}) and possible inaccuracies in Neptune's quality function (Sec.~\ref{subsec:resonance_spikes}).

\subsection{Oblateness, Cassini states, and triaxiality}
\label{subsec:asphericity}
To start with the first: over the entire circularised period, the tidal evolution, as presented in Sec.~\ref{subsec:long_term_evolution}, is dominated by the tides Triton raises in Neptune. This can be explained by by noting that the rotation rate of Triton is equilibriated, and so $\pdv{U_T}{\Omega_T}=0$ (see Eq.~\ref{eq:partial_potential}). However, for $\cos{i_T}=1$, only the $m=l-2p$ in Kaula's expansion are non-zero, and so $\pdv{U_T}{\varpi_T}=0$, too; but for $e=0$, only the terms with $q=0$ in the Kaula expansion are non-zero, and so $\pdv{U_T}{\mathcal{M}}=0$. Consequently, our model predicts that Triton no longer experiences tidal dissipation altogether.

This is not entirely true to reality, however: had our model included the dynamical effects of an asymmetry in the moments of inertia of Triton (e.g. due to spherical oblateness), Triton would have been captured in a Cassini state, with a non-zero obliquity \citep{Correia2009SecularTriton}. The resulting obliquity tides in Triton's ocean have been presumed responsible for the present-day geological activity seen on the satellite (see Sec.~\ref{sec:explaining_interior}): even at a dissipation rate of 3 GW (in excess of the greatest value predicted by \citealt{Nimmo2015PoweringGeology}), the associated change in semi-major axis measures only in the hundreds of metres per Myr, and so it is likely that our neglect of any such Cassini state does not invalidate the results we found. It would be valuable to include this term in future work, however, if only because it could then be used to accurately model Triton's geological activity throughout the circularised epoch, and potentially support any more specific planetological predictions that could inform future missions (see e.g. Sec.~\ref{sec:geological_active_Triton}).

Additionally, if Triton originally solidified while under the tidal influence of its binary companion, it may well have entered into orbit around Neptune possessing some permanent triaxiality. While it is unlikely that this will have survived the melting of its ice shell (and indeed, present-day Triton is at least geometrically no longer triaxial; \citealt{Thomas2000TheProfiles}), it may thus have possessed some triaxiality to start with. In that case, this will have introduced libration \citep{Frouard2017TidesTheory}, and possibly have expedited Triton's capture into a spin-orbit equilibrium state (which, itself, may have been altered from the purely tidal spin-orbit resonances as a result of the triaxiality).
Capturing the intricacies of any oblateness and triaxiality requires that the mean motion, as well as the nodal and apsidal rates also be computed at each timestep, as it is in these terms that the oblateness and triaxiality enter the Lagrange planetary equations: consequently, we would have to do away with the approximation for the tidal Fourier modes given by Eq.~\ref{eq:omega_lmpq_def_approx}.

\subsection{Evolving interior models}
\label{subsec:evolving_interior}
From our history produced in Sec.~\ref{subsec:high_eccentricity_evolution}, it has become clear that, paradoxically, it is unlikely that a viscous ice shell like the one we have assumed in our fixed-interior model can be maintained. \citet{Ross1990TheTriton} already showed that a temperature-dependent rheology affected by a steep heat influx such as experienced by Triton can result in changes in orbital evolution, and our results indicate that it is likely that the required significant temperature- and phase-changes have occurred for early Triton. As our method can in principle work for an arbitrary rheological (and consequently, interior) model which need not be fixed in time, it therefore seems as though an evolving interior model is a logical next step, such that the dynamical and thermal conditions of modelled Triton are self-consistent.

From the work by \citet{Nimmo2015PoweringGeology} on Triton and that by \citet{Bagheri2022TheSystem} on the roughly comparable Pluto-Charon system we can deduce that such a model will at least have to account for radiogenic heating, convection in the lower ice shell and potentially even evolution of ocean composition. Additionally, the work by \citet{Nolan1988SomeTriton} and \citet{Lunine1992ATriton} suggests that a greenhouse atmosphere is likely, and so would also have to be accounted for: we will discuss some consequences of such an atmosphere in Sec.~\ref{subsec:early_atmosphere}.

As a byproduct of such a thermal-interior model, we would additionally be able to constrain the size of Triton's ocean, and consequently make an estimate of the tides therein: we will discuss this in further detail in Sec.~\ref{subsec:ocean_tides}. The consequences of a large ocean on solid-body tides can also already be severe: \citet{Bagheri2022TheSystem} find that Pluto's tidal Love numbers are more than an order of magnitude greater than Charon's, in part explained by the persistence of a large ocean on Pluto, while Charon's has frozen over, though it is likely that the quality factor will rise in excess of this increased Love number, given its significant dependence on viscosity (e.g. \citealt{Bagheri2022TheSystem}). We should therefore expect that the tidal response of a partially- or fully-molten Triton is muted compared to that for a Triton possessing a fully frozen ice shell, and so the high-eccentricity tidal evolution may be dampened and stretched out in time. How much cannot be said with any certainty until this phase is modelled appropriately, but if the eccentricity damping timescale is raised sufficiently the action of the tides raised in Neptune by Triton might allow Triton to approach Neptune with some residual eccentricity. As a result, the eccentricity tides in Triton will operate more efficiently than they would further away, and we can expect Triton to potentially undergo a delayed but even more severe tidal heating spike than those found in Sec.~\ref{subsec:high_eccentricity_evolution}. The analytical expressions given by \citet{Matsuyama2018OceanShells} might provide a good starting point to explore such thin shell- or surface ocean-driven tidal evolution without having to resort to a fully coupled evolving thermal-interior model straight away.


Aside from an evolving model for Triton, a time-evolving model for Neptune may be a worthwhile addition. Though the temperatures attained in Neptune make the energy dissipated in it by Triton's tides pale in comparison (e.g. \citealt{James2024ThermalNeptune}), such that we do not expect any significant feedback between Tritonian capture and migration and Neptune's interior evolution, Neptune does apparently cool and evolve significantly over time. Results by \citet{James2024ThermalNeptune} show that a frozen core in Neptune may, potentially, only arise after $\sim3$ Gyr, meaning that the tidal response of Neptune before that is almost negligible. Afterwards, the growing size of this frozen core increases the tidal quality factor by several-fold over time, while the cooling of Neptune as a whole lowers the tidal Love number $k_2$. We can thus realistically expect that Neptune's tidal response will have varied significantly with time, possibly transitioning from a Jovian-like weak-friction tidal response to a viscoelastically-dominated tidal response once a frozen core arises.



\subsection{Ocean tides}
\label{subsec:ocean_tides}
With the presence of a large subsurface ocean almost certain and the existence of a past surface ocean on Triton seemingly likely, we may also want to account for ocean tides in the future. While the obliquity tides raised in present-day Triton's oceans contribute relatively little compared to the past eccentricity tides from an energy-perspective \citep{Nimmo2015PoweringGeology}, the simultaneous muting of the solid-body response due to a melting ice shell and enhancing of the ocean tidal response by the slow lifting of the damping effect of this ice shell might mean that there is a turning point beyond which ocean tides dominate Triton's tidal response, such as is the case on Earth \citep{Matsuyama2018OceanShells}. Uniquely, the existence of surface ocean tides on Triton would be an exciting parallel to the modern Earth, though it would likely violate the thin-ocean assumption that commonly underlies ocean tidal theory. With such a thick ocean, stratification seems likely, and so we should expect internal gravity waves to accompany the familiar surface gravity waves also seen for unstratified oceans (e.g. \citealt{Rovira-Navarro2023Thin-shellWorlds}). The resonances possibly encountered by such ocean tides can therefore be expected to be complex, and probably worthy of study on their own.

\subsection{Absence of third-body perturbations}
\label{subsec:kozai}
We also did not account for third-body perturbations: in this case, these might be solar perturbations, but also due to other planets or Neptunian (proto-)satellites. \citet{Cuk2005CONSTRAINTSTRITON}, \citet{Nogueira2011ReassessingTriton} and \citet{Rufu2017TritonsSystem} accounted for the Kozai mechanism by solar perturbations, and found it to be significant when $>70$ $R_N$ from Neptune; we should therefore expect behaviour qualitatively similar to that showcased by \citet{Nogueira2011ReassessingTriton} for very high eccentricities, where the tidal evolution is otherwise relatively mellow. In that case, we expect that the Kozai mechanism might rapidly push Triton into the region where tidal evolution is more rapid, after which the Kozai mechanism damps out and Triton "freezes" into an angular momentum-conserving orbital evolution.

Additionally, it has been shown that inward or outward satellite migration in combination with spin-orbit resonances of the host planet can cause major tilting of the planet \citep{Saillenfest2021FutureTilting, Saillenfest2022TiltingSatellite, Wisdom2022LossRings}. As Neptune's obliquity with respect to the ecliptic is significant at $\sim30^{\circ}$, it is tempting to assume that this mechanism operated on the Neptune-Triton system: however, it can equally well be explained by tilting due to likely spin-orbit resonances with a massive circumplanetary disc \citep{Rogoszinski2020TiltingResonance}, and so we need not invoke a chance configuration of the early Solar System to explain this obliquity. There is therefore no strict necessity for this mechanism to have acted to explain the currently observed system (and our results are consistent with a primordial obliquity for Neptune). Yet, future work might do well to examine the possibility.

Finally, we did not account for any proto-Neptunian satellite system. This is mainly out of necessity, as we have little to no constraints on the existence thereof, with the exception that Nereid is thought to be a primordial satellite of Neptune \citep{Nogueira2011ReassessingTriton}. If the proto-Neptunian system was Uranus-like (as made plausible by e.g. \citealt{Rufu2017TritonsSystem}), it is likely that the eccentric and wide orbit of freshly-captured Triton will disrupt its original satellites \citep{Cuk2005CONSTRAINTSTRITON}, with Triton likely experiencing some, but no disruptive impacts \citep{Rufu2017TritonsSystem}. Given the low mass of the present-day non-Tritonian satellites of Neptune, it is unlikely that the mass of any proto-satellite system was sufficient to cause major orbital perturbations other than through such collisions, making the disruptions by captured Triton mostly a one-way street.

\subsection{Resonance spikes in Neptune's tidal quality function}
\label{subsec:resonance_spikes}
The fast outward migration of Io \citep{Lainey2009StrongObservations} and Saturn's satellites \citep{Lainey2012StrongAstrometry, Lainey2017NewData, Lainey2020ResonanceTitan} as determined from astrometric observations indicates that the tidal evolution of satellites in the Solar System is happening faster than previously thought. This has led \citet{Fuller2016ResonanceSystems} to put forward the idea of resonance locking, a mechanism originally proposed for binary stars by \citet{Witte1999TidalLocking}, as a plausible explanation of this fast migration; results by \citet{Lainey2020ResonanceTitan} seem to agree with this mechanism.

In the paradigm of resonance locking, the internal structure of a planet induces narrow resonance spikes in its quality function; as the planetary interior cools and evolves with time, these resonance spikes move to lower frequencies. In the process these resonance spikes sweep up satellites, which correspondingly start migrating outward at an accelerated rate.

While this mechanism has been demonstrated to be a plausible explanation for the outward migration rates of the satellites of the Solar System gas giants (e.g. \citealt{Lainey2020ResonanceTitan}) that is consistent with the expected peaks in their tidal quality functions (e.g. \citealt{Ogilvie2004TIDALPLANETS}), it is not yet clear whether this same mechanism is relevant for ice giants, for which it appears the tidal response is dominated by viscoelastic layers (e.g. \citealt{Storch2014ViscoelasticMigration}). By assuming a simple CTL quality function for Neptune, we have implicitly assumed that resonance locking is not important: as resonance locking can only occur for outward-moving moons when a planet is spinning down \citep{Fuller2016ResonanceSystems}, we do not expect that this mechanism is likely to affect Triton's orbital evolution (unless Triton was captured as Neptune was still accreting gas and correspondingly spinning up). Consequently, these resonance spikes may only drive short-lived period of accelerated inward migration, which are unlikely to affect the qualitative evolution of Triton, and we expect that the neglect of any such resonance spikes in Neptune's quality function is justified.

\section{Planetological implications for Triton}
\label{sec:planetological_implications}
Our results have several implications for the planetological understanding we have of Triton's past. There are two in particular that we wish to highlight: the consequences of any possible discrete spin-orbit evolution (in Sec.~\ref{sec:discussion_discrete_spin_orbit_evolution}), and the repercussions of the potential existence of an early atmosphere and hot ocean on Triton (in Sec.~\ref{subsec:early_atmosphere}).

\subsection{The consequences of discrete spin-orbit evolution}
\label{sec:discussion_discrete_spin_orbit_evolution}
We have shown in Ch.~\ref{ch:spin-orbit_chains} that, if Triton maintained a sufficiently viscous tidal response, it may well have passed through a number of spin-orbit transitions during its mid-eccentricity ($e\sim0-0.5$) phase. While the energy dissipated by the synchronisation of Triton's spin state is negligible compared to that dissipated by the decay of its semi-major axis, the discrete nature of these spin-orbit transitions (which take on the order of kyrs) means that Triton may have experienced epochs of tidal heating even more severe than that caused solely by its orbital migration. While these epochs are short-lived, their consequences may thus be significant, and we propose that these be explored as potential originators of the crustal layering required to explain its cantaloupe terrain by diapirism (e.g. \citealt{Schenk1993DiapirismInstability}).

\subsection{An early atmosphere and hot ocean on Triton}
\label{subsec:early_atmosphere}
While the orbital evolutions at early high eccentricities is relatively moderate (see Sec.~\ref{subsec:high_eccentricity_evolution}), the associated tidal heating is already comparable to that of modern-day Io. By the results of \citet{Nolan1988SomeTriton} and \citet{Lunine1992ATriton} an optically thick, massive atmosphere can be raised by such temperatures, and then maintained over $\sim100$ Myr or longer even in the absence of tidal heating. If the material necessary to raise this atmosphere can be outgassed in the initial several Myrs preceding the extreme tidal heating-phase, it is likely that Triton had an insulating greenhouse atmosphere while undergoing it. The effects of tidal heating over this period may thus be significantly amplified. Notably, \citet{Nimmo2015PoweringGeology} argue on the basis of Triton's diffusion timescale that Triton's primordial heat conditions are of little relevance today: this argument is potentially invalidated by the presence of an early atmosphere, and so primordial heat may still play a role today, in which case it could potentially yet be used to constrain Triton's time of capture.

An additional consequence of the plausible existence of an atmosphere and hot (surface) ocean on early Triton is in the possible chemical phenomena it might enable. \citet{Shock1993HydrothermalTriton} already proposed that hydrothermal processing might explain some compositional peculiarities of Triton's CO abundance, which does not match what is expected from a body that formed from the protosolar nebula \citep{Stevenson1990PuzzlesTriton}. Other species might similarly be reprocessed in a hot ocean or atmosphere or through ocean-core interaction, by analogy with processes ongoing at Enceladus \citep{Hsu2015OngoingEnceladus}, though those are at present still not fully understood \citep{Choblet2017HeatMoons}. These ocean-core, but also atmosphere-ocean and atmosphere-space environment interactions may cause significant alterations in composition that could potentially remain detectable today, while geological evidence for this hot-Triton epoch is now long obscured by ongoing geological processes. In an extreme example, \citet{Barnes2013TidalHeating} propose that a runaway greenhouse effect induces by tides might rid otherwise habitable terrestrial exoplanets orbiting M-dwarfs of their water, so compositional change due to tidal heating is not unprecedented. Modelling the compositional consequences of hot capture would thus be a valuable avenue through which we might be able to constrain Triton's past observationally, despite the lack of geological signatures more ancient than 50 Myr ago.

\section{Implications for the Neptunian system}
\label{sec:implications_neptune_system}
Our results also have a bearing on the Neptunian system in a broader sense, beyond just Triton. We wish to highlight three notions in particular: we will discuss the implications of our results on the plausible capture mechanism for Triton in Sec.~\ref{subsec:capture_mechanism_triton}, the dynamical architecture we expect to be plausible for the evolving Neptunian system in Sec.~\ref{subsec:dynamical_architecture}, and finally how Triton's hot, early phase might have contaminated the early Neptunian system in Sec.~\ref{subsec:contamination}.

\subsection{The capture mechanism for Triton}
\label{subsec:capture_mechanism_triton}
We have shown that Triton, with a physically plausible rheology, can have been captured at the eccentricities expected from binary dissociation \citep{Nogueira2011ReassessingTriton}, followed by circularisation over timescales of $\sim10$ Myr. Whatever non-zero eccentricity Triton has at present can be attributed to cometary impacts \citep{Chyba1989TidalSystem}. Even if the capture process is extended by orders of magnitude due to structural changes in Triton due to tidal heating, it can thus still be deposited on its present orbit at the present day without issue.

While \citet{Nogueira2011ReassessingTriton} foresaw issues for the production and stability of Nereid as a captured object on its current orbit in this scenario, Nereid's surface composition appears to be more consistent with it being an original regular satellite of Neptune perturbed by Triton \citep{Brown1998DETECTIONNEREID}, as proposed by \citet{Goldreich1989NeptunesStory}. It is therefore not necessary to invoke any additional mechanisms, such as collisions (e.g. \citealt{Goldreich1989NeptunesStory, Rufu2017TritonsSystem}), or a gas or dust disc (e.g. \citet{McKinnon1995GasTriton, Cuk2005CONSTRAINTSTRITON}), to explain Triton's present-day orbit. In principle, our results are also consistent with capture by non-disruptive collisions followed by tidal circularisation (as it is, with our current observations of Triton, functionally indistinguishable from binary dissociation): from a probabilistic viewpoint the binary dissociation scenario seems more likely, however, and should be preferred.

\subsection{Dynamical architecture of the early and current Neptunian system}
\label{subsec:dynamical_architecture}
The binary dissociation capture scenario followed by tidal circularisation over $>$Myr timescales gives credence to the history of Triton's inner satellites given by \citet{Banfield1992ASatellites}, who suggest that Neptune's inner satellites are not primordial, but rather formed from the remnants of the proto-Neptunian satellite system that was rapidly disrupted by Triton's capture. This agrees with the results by \citet{Rufu2017TritonsSystem}, and means that not only do we not need a debris disc to capture Triton, but additionally that it is not feasible for a debris disc to have persisted long enough to meaningfully interact with captured Triton. While \citet{Rufu2017TritonsSystem} echo the concerns raised by \citet{Nogueira2011ReassessingTriton} with respect to Nereid's stability in this scenario, we have discussed in Sec.~\ref{subsec:capture_mechanism_triton} that this is not problematic if Nereid is a regular satellite of Neptune that was scattered by Triton, as is consistent with its surface composition.

Additionally, our results indicate that Neptune's obliquity is primordial, not excited by Triton's migration, unless this migration was paired with some fortuitous spin-orbit resonance between the planets (e.g. as in \citealt{Saillenfest2021TheTitan}). The fact that this is the case for the Neptune-Triton system does not mean that we can exclude such tilting due to captured moons as a general mechanism, however: we will discuss this in further detail in Sec.~\ref{subsec:oblique_exorings}.

\subsection{Contamination of the early Neptunian system}
\label{subsec:contamination}
The prospect of a thin ice shell or even a surface ocean overlaid by a thick atmosphere on early Triton indicates that we may reasonably expect the Neptunian environment to be contaminated by ejecta in plumes or gas escaping from the atmosphere. Indeed, the mass loss rate due to atmospheric escape predicted by \citet{Lunine1992ATriton} is well in excess of the rates seen in Enceladan plumes \citep{Hansen2011ThePlume}, which result in a water torus surrounding Saturn \citep{Hartogh2011DirectHerschel}. The detectability of such a putative torus for early Triton is discussed in Sec.~\ref{subsec:detecting_exo_tritons}.

As the material from the Enceladan plumes is responsible for water vapour in Saturn's upper atmosphere \citep{Hartogh2011DirectHerschel} and the coating of other Saturnian satellites in reflective material which enhances their albedo \citep{Verbiscer2007Enceladus:Act}, we might reasonably expect early Triton to have contaminated Neptune's atmosphere as well as the surfaces of the other Neptunian satellites. Particularly the latter, if these surfaces have then not been eroded by other processes since, might mean that some clues to Triton's past have not yet been obscured by geological processes on other satellites like they have on Triton, and so are still accessible by any future mission. Evaluating the plausibility of this requires accurate estimates on the rate of precipitation of this material onto the other satellites.

\section{Implications for exoplanetary systems}
\label{sec:consequences_exoplanetary_systems}
Aside from the implications of these results on the Neptune-Triton system, we must also consider how these findings might generalise. Indeed, the results by \citet{Agnor2006NeptunesEncounter}, \citet{Vokrouhlicky2008IrregularReactions} and \citet{Nogueira2011ReassessingTriton} seem to indicate that Triton's existence as a large, captured satellite is not a mere happenstance: such captures are not unlikely to occur whenever planets meet TNO-like binary objects. We will consider what the prospects are for detection of Triton-like objects in exoplanetary systems (exo-Tritons) in Sec.~\ref{subsec:detecting_exo_tritons}, and propose Triton-like objects as possible progenitors for oblique exorings observed in exoplanetary systems in Sec.~\ref{subsec:oblique_exorings}.

\subsection{Detectability of exo-Tritons}
\label{subsec:detecting_exo_tritons}
As we expect capture of exo-Tritons to be reasonably common, it would not be surprising if we encounter them when we eventually start confidently observing exomoons. While proposed exomoon detection methods are in general not positively or negatively impacted by the sense of orbital motion of a moon (e.g. \citealt{Kipping2009TransitExomoon, Heller2014DetectingEffect, Agol2015THEEXOMOONS, Oza2019SodiumExoplanets, Limbach2021OnObjects, Lazzoni2022DetectabilityDwarfs, Kleisioti2023TidallyCharacterization, vanWoerkom2024TheExomoons}), some methods allow for determination of the direction of an exosatellite's orbit (e.g. \citealt{Kipping2009TransitII, Heller2014HowMotion}), and so we can envision a future survey allowing for determination of the prevalence of exo-Tritons.

What is perhaps more interesting, however, is the fact that tidal heating of exo-Tritons, according to our results, can support brightness temperatures in excess of $200$ K. This approaches the temperatures deemed detectable around $\epsilon$ Eridani b by \citet{Kleisioti2023TidallyCharacterization} (though their results concern rocky bodies), while exceeding the required brightness temperatures presented by \citet{vanWoerkom2024TheExomoons}. Spectroastrometry is particularly promising, as it works especially well for warm, eccentric satellites on wide orbits around faint planets. It will not escape the reader that these are precisely the qualities describing early Triton.

A second method by which the presence of exo-Tritons might be inferred is indirectly: the clouds hypothesised to have been present around early Neptune due to material ejected or escaping from Triton in Sec.~\ref{subsec:contamination} may well allow for detection in a manner analogous to the method proposed by \citet{Oza2019SodiumExoplanets}. Additionally, if sufficient material is deposited onto Neptune's upper atmosphere, this may also be detectable in exoplanet spectra.

\subsection{Production of oblique exorings}
\label{subsec:oblique_exorings}
A number of exoplanets have been observed with sub-Neptune masses but surprisingly large radii (see the introduction of \citealt{Saillenfest2023ObliqueRadii} for an overview); while some of these can be explained by thermally inflated atmospheres, some such planets are too far from their host star for this to be a viable explanation. \citet{Akinsanmi2020Can} propose that these examples can be explained by a ring system that is inclined to the orbit of the planet; this only defers the problem instead, however, as we are then left to explain how such an oblique ring came to exist in the first place. \citet{Saillenfest2023ObliqueRadii} propose that such rings might be explained by (1) the tilting of a planet due to migration of a moon and (2) the consequent destruction of that moon as it approaches the planet to within its Roche limit; their mechanism requires that this moon migration be accompanied by certain orbit precession resonances, however.

\begin{figure}
    \centering
    \includegraphics[width=1\linewidth]{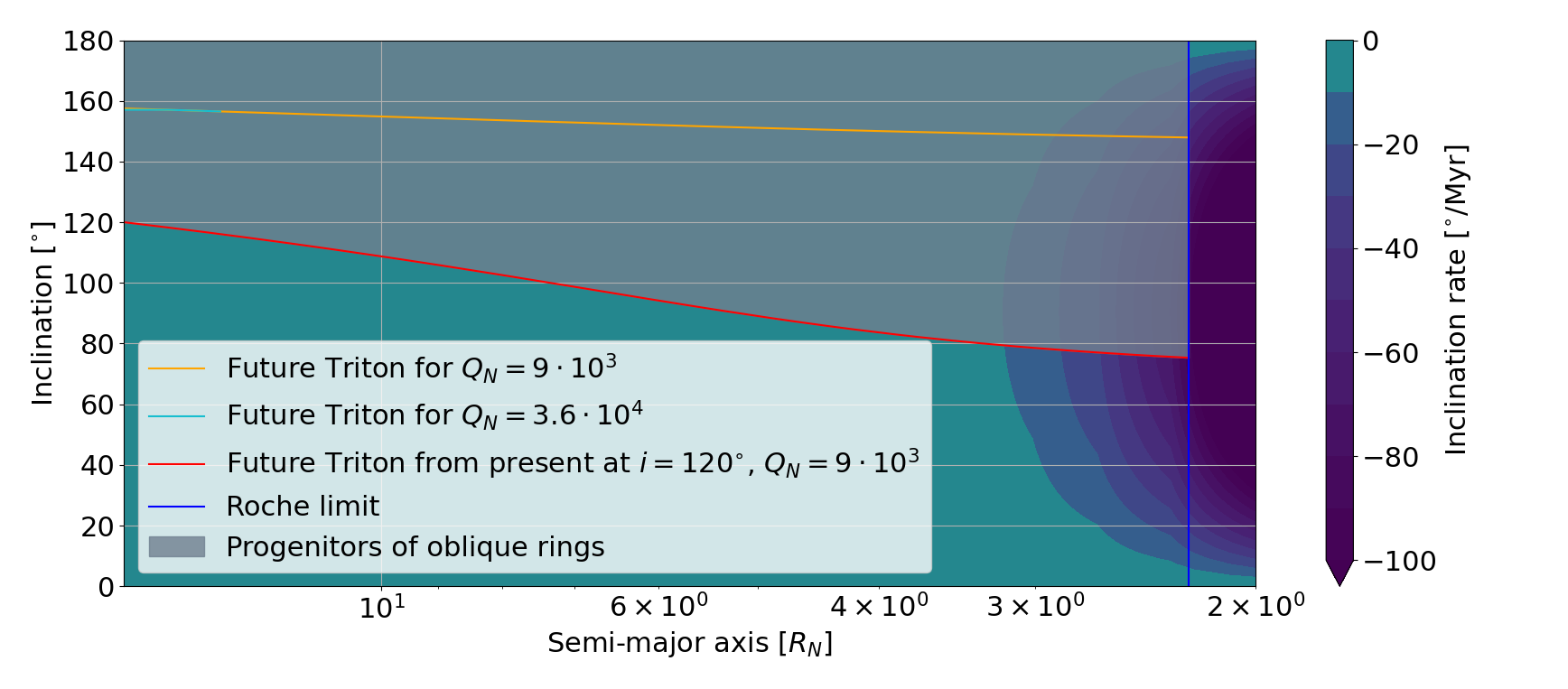}
    \caption{The inclination evolution over a representative portion of $a-i$ phase space. Triton analogues in the shaded regions will produce oblique exorings.}
    \label{fig:oblique_rings}
\end{figure}

From our results in Sec.~\ref{sec:complete_history} it becomes clear that Triton's inclination is slowly falling, a process which will only succeed in significantly tilting its inclination during the rapid evolution it undergoes while decaying into Neptune. Once it does, we should expect that Triton forms a ring on the inclination at which it enters Neptune's Roche limit. This ring will then start migrating over Neptune's poles into a prograde and eventually equatorial orbit: while the tidal bulge raised by a set of rings is not equal to that raised by a moon of equal mass, to first order we should expect that the resulting ring will follow a similar evolution to a moon on such an orbit, in which case this inclination migration would take $\sim$Myrs. Any retrograde satellite that decays into its planet will therefore create oblique rings that will live for $\sim$Myrs. To illustrate this, we have shown the rate of change of the inclination for Triton as it decays into Neptune along various semi-major axis and inclination routes in Fig.~\ref{fig:oblique_rings}. The shaded region contains all fictitious Tritons that will eventually create a set of oblique rings; for lower starting inclinations than $\sim120^{\circ}$, Triton will turn prograde before decaying into Neptune, avoiding such a fate. Hence, exo-Tritons captured at inclinations $\gtrsim120^{\circ}$ (which comprises $\sim40\%$ of them, according to \citealt{Nogueira2011ReassessingTriton}), may produce rings that migrate over the poles of their host, potentially being visible as oblique exorings over timespans of $\sim$Myrs.

Even more interestingly, we can compute what the eventual equilibrium position of such a ringed Neptune will be by assuming conservation of angular momentum between the ring and the planet (starting from the moon and the planet). For the Neptune-Triton system at present, this will result in Neptune being tilted by another $\sim25^{\circ}$; Triton seems close to a turning point in this behaviour, however, and had it been twice as massive (or Neptune half as massive), the resulting obliquity would be Uranus-like, in excess of $90^{\circ}$; this mass ratio value twice that of Triton's, coincidentally, is very similar to the mass requirement found by \citet{Saillenfest2022TiltingSatellite} for the moon they propose tilted Uranus. Interestingly, Uranus' satellite system is consistent with formation from a massive ring system \citep{Crida2012FormationSystem, Hesselbrock2019ThreeMiranda}. A second scenario for Uranus then arises: while \citet{Nogueira2011ReassessingTriton} and \citet{Vokrouhlicky2008IrregularReactions} find that Uranus also has a good chance of encountering proto-Triton-like objects, its lack of a Triton-like moon, may not, as they hypothesise, be a result of a lack of Tritons in the planetesimal disc, but we can hypothesise rather that it may have been the result of it once having such a moon, that was then destroyed upon decaying into Uranus. The resulting ring would have migrated over the poles of Uranus, but Uranus would have tilted accordingly to preserve angular momentum: eventually, this places the ring around Uranus' equator (or, more accurately, in its Laplace plane), where its present moons eventually accrete, with Uranus tilted to its oblique spin state. Regardless of the plausibility of this scenario for Uranus in particular, this gives a second mechanism by which oblique exorings can be created: capture and destruction of a sufficiently massive exo-Triton, and the consequent equilibriation of the ringed planet's spin state. The resulting oblique exorings will live for as long as the rings themselves remain, and so the prospect for observing such a system may be greater still than that we sketched before for destruction of a retrograde satellite with a Triton-like mass ratio.

\chapter{Conclusions \& recommendations}
\label{ch:conclusions_recommendations}
We now finally have the required results to start answering the questions we set out to answer. Introducing the required formalism in Ch.~\ref{ch:kaula_theory}, we have shown how the Darwin-Kaula expansion for the tidal potential can be implemented to work even for highly eccentric systems in Ch.~\ref{ch:validation}. In Ch.~\ref{ch:spin-orbit_chains} we then explored the spin-orbit behaviour of Triton depending on the interior and rheological model we use, and in Ch.~\ref{ch:initial_conditions} we put all of this together to get some idea about Triton's spin-orbit evolution over its lifetime. In Ch.~\ref{ch:discussion}, finally, we put the obtained results in the context of other scientific work.

\section{Conclusions}
With that, we are now equipped to answer the questions we posed in Ch.~\ref{ch:scientific_background}. Before answering the primary question, we will first answer each of the sub-questions we posed. The aggregate of the answers to each of those should then adequately prepare us to answer the primary question.

\subsection{How can we effectively and accurately model the high-eccentricity orbital evolution of Triton over astronomical timescales?}
In Ch.~\ref{ch:kaula_theory}, we explained how Darwin-Kaula theory is a sound choice of formalism with which to describe Triton's evolution, as it can handle the large timescales involved. This leads to an averaged version of the Lagrange planetary equations, as given by \citet{Boue2019TidalElements}. Evaluating these equations at high eccentricities can be done using a power series expression given by \citet{Proulx1988SeriesCoefficients}, and with the newly derived truncation prescriptions for the infinite sums in the Darwin-Kaula expansion of the tidal potential given in Ch.~\ref{ch:validation} it is then possible to propagate the Lagrange planetary equations forward in time in a computationally feasible manner.

\subsection{How do we appropriately model the spin-orbit evolution of Triton?}
Appropriately modelling the spin-orbit evolution of Triton requires properly accounting for its rheological and interior behaviour: previous orbital evolution models for Triton required the use of a particular rheological model, the constant time lag model, on the basis of the mathematical simplifications allowable for high eccentricities in that case. We have shown in Ch.~\ref{ch:spin-orbit_chains} that this simplified approach (as well as the constant phase lag model, another simplified model often found in literature) is not sufficient to fully capture the intricacies of spin-orbit evolution at non-zero eccentricities that are predicted by more advanced rheological models such as the Maxwell model, and that failing to heed this advice can lead one to predict qualitatively different behaviour than is realistic.

Additionally, we find that the behaviour that is missed out on, namely the progression through half-integer spin-orbit resonances, can at times lead to significant underestimation of the true tidal heating rates of Triton-like bodies, even though the associated dissipated spin energy is negligible compared to the dissipated orbital energy. For thermal modelling of such bodies, it is thus important that the evolution of the spin rate be taken into account.

\subsection{What bounds can we put on the Neptune-Triton system's past and future dynamical evolution?}
With these model prescriptions in place, in Ch.~\ref{ch:initial_conditions} we then computed a history of Triton that now incorporates, for the first time, the high-eccentricity evolution of Triton for more advanced rheology. Based on this, we can make an attempt at a qualitative dynamical history of Triton that is consistent with the predictions of more realistic rheological models, proposing the following new timeline:
\begin{itemize}
    \item Triton was likely captured initially onto a high-eccentricity orbit with an inclination close to its present-day value $\sim4.5$ Gyr ago;
    \item Over a period of $\gtrsim10$ Myr, it then underwent tidal circularisation, potentially passing through spin-orbit resonances as it does;
    \item Triton is deposited, synchronously rotating, onto a circular orbit close to its present-day semi-major axis and inclination, and undergoes slow damping of its semi-major axis and inclination to present-day values forced by the tides it raises in Neptune;
    \item Depending on the interior properties of Neptune, Triton may decay into its host $\gtrsim3$ Gyr from now.
\end{itemize}
We note that the timescale of circularisation is likely to be extended by the partial or full melting of Triton under the force of its tidal heating, and so this should strictly be taken as a lower bound. Additionally, we have concluded that Neptune's obliquity is likely primordial, and not the result of Triton's inward migration.
\newline
\newline
We are then equipped to answer our overall, primary question:
\begin{quote}
    \textbf{What constraints can be put on the tidally forced trans- and post-capture dynamical and thermal evolution of Triton?}
\end{quote}
by combining the products of these three sub-questions; the first two questions give us a functioning simulation tool, and the final question gives us, using this simulation tool, a dynamical history from which we can derive the thermal history of Triton. We then conclude as follows: Triton can have been captured by tidal action of Neptune without requiring auxiliary mechanisms such as disc drag or collisions. Once it did, it is plausible that tidal heating immediately reached Io-like levels, and it seems likely that the ice shell on Triton will have receded to be metres to kilometres thick at most in the process, if indeed it remained at all, with catastrophic (though spectacular) planetological consequences, such as the potential raising of a Titan-like atmosphere. The circularisation process will have taken at least several million years, depending on the evolution of Triton's interior state, and deposited sufficient energy into Triton to melt its mantle $\sim80$ times over. During this process, Triton may have experienced rapid transitions between spin-orbit resonances, with associated tidal heating spikes. After Triton reaches an orbit that is close to its present-day circular orbit and synchronises its rotation rate to its orbital rate, it will slowly migrate to its present-day position, its inclination likely still close to that with which it was captured. The thermal and geological evolution is, over this period, likely dominated by obliquity tides in Triton's ocean, as it still is today. This slow evolution will continue into the future and, depending on Neptune's interior, may result in Triton decaying into Neptune's Roche limit several Gyr from now.

\section{Recommendations for follow-up work}
In our conclusion, we do not put exact numbers to the constraints on the dynamical and thermal evolution of Triton: this is deliberate, as we do not expect that the model we have is sufficiently high-fidelity that such quantitative results tell the full story, and so we prefer to stay with qualitative results (which we do deem trustworthy). To obtain a model that we expect to give quantitatively useful results, there are several additions that must be made, which we intend to explore in future work.

\subsection{Higher-fidelity, evolving interiors}
From our results it is clear that we expect Triton's structure to be altered on a planetary scale by the tidal heating it experiences during the circularisation phase. Hence, it is clear that the assumption of a fixed, homogeneous interior is certainly not consistent. Similarly, we expect that Neptune is better represented with a relatively simple time-varying interior (see Sec.~\ref{subsec:evolving_interior}). With such interior models available, we can employ a matrix propagator code to produce self-consistent values for the tidal quality function, and give reasonable estimates for the ocean tides (or at least reasonably estimate whether they are relevant).

\subsection{Dynamical additions}
Previous work on Triton by \citet{Correia2009SecularTriton} and \citet{Nogueira2011ReassessingTriton} has shown that the asphericity of Triton and Neptune and the Kozai mechanism result in dynamical behaviour that is either significant for the orbital evolution of Triton (in the case of the Kozai mechanism) or that underlays the obliquity that powers geological activity on modern-day Triton (in the case of the oblateness that drives Triton into a Cassini-state with nonzero obliquity). Adding these dynamical elements to the model will therefore give a more reliable estimate of the dynamical evolution of Triton, or increase the predictive power in the context of modern-day observations of the moon. To do so in full requires that the anomalistic mean motion as well as the nodal and apsidal rates be computed and included, as it is in those terms that the asphericity-related terms enter the Lagrange planetary equations (see e.g. \citealt{Luna2020TheSystem}). This additionally entails doing away with the approximation given in Eq.~\ref{eq:omega_lmpq_def_approx}, which neglects the contribution of these terms to the tidal Fourier mode.

An additional addition to the dynamical model that we recommend be explored in future work is the addition of resonance spikes in Neptune's tidal quality function. While there is currently seemingly no consensus in literature on whether such spikes are likely to arise in the quality function of ice giants, including a toy version of them is relatively straightforward, and initial exploratory results seem to indicate that they are not important for inward-migrating moons. A relatively trivial check may thus be sufficient to put this matter to rest for retrograde tidally evolving objects.

\subsection{Planetological phenomena}
Finally, there are several planetological phenomena that may be worth exploring in future work. Linking a simple atmospheric model like that used by \citet{Lunine1992ATriton} to the dynamical-interior model is likely to lead to valuable new results, and introduces new avenues by which a feedback mechanism between the dynamical and interior evolution might arise. The associated estimates on the escaped mass of volatile species is a promising avenue by which we might be able to put hard observational constraints on Triton's past despite the fact that its geological activity has wiped away all but the last 50 Myr worth of evidence. For similar reasons, and by the fact that such impurities can significantly affect the melting point of water ice, it may be worth tracking the concentrations of ammonia in Triton's oceans (cf. \citealt{Bagheri2022TheSystem}).


\appendix 

\chapter{Mathematical addenda}
\label{app:mathematical_addenda}

\section{Integral expression for the Kaula eccentricity functions}
\label{sec:eccentricity_function_integral}
While the required computational efficiency required in simulation does not allow for numerical integration as a feasible method by which to calculate the Kaula eccentricity functions $G_{lpq}(e)$, such slow computation is permissible for validation purposes. We will therefore derive and discuss an integral expression that can be used to compute the Kaula eccentricity functions up to arbitrary $l$, $p$ and $q$.

\subsection{Kaula's eccentricity functions as a subset of the Hansen coefficients}
\label{subsec:hansen_coeff_intro}
Let us first recall how the Kaula eccentricity functions $G_{lpq}$ arise in Darwin-Kaula theory: they permit the expression of particular forms of functions of the true anomaly $v$, radius $r$ and semi-major axis $a$ in terms of the mean anomaly $M$ and eccentricity $e$ (e.g. \citealt{Kaula1961AnalysisSatellites}):
\begin{equation}
    \left(\frac{r}{a}\right)^{-(l+1)}\exp{\left(i(l-2p)v\right)} = \sum_{q=-\infty}^{+\infty}G_{lpq}(e)\exp{\left(i(l-2p+q)M\right)}.
\end{equation}
Hence, it is abundantly clear that the $G_{lpq}$ are a particular subset of the Hansen coefficients $X^{n,k}_{s}$ (e.g. \citealt{Tisserand1889TraiteCeleste, Plummer1918AnAstronomy, Proulx1988SeriesCoefficients, Gooding1989ExplicitOrbits}):
\begin{equation}
    \left(\frac{r}{a}\right)^{n}\exp{(ikv)} = \sum_{s=-\infty}^{+\infty}X_{s}^{n,k}\exp{(is M)}
\end{equation}
where $n=-(l+1)$, $k=l-2p$ and $s=l-2p+q$, whence one obtains
\begin{equation}
\label{eq:fourier_hansen}
    X^{n,k}_{s}=\frac{1}{2\pi}\int_{-\pi}^{\pi}\left(\frac{r}{a}\right)^n\exp{i(kv-sM)}\dd{M}
\end{equation}
through the usual Fourier property. 

Classically, one would expand the Hansen coefficients into a double infinite series involving the generalised Laguerre polynomials (e.g. \citealt{Proulx1988SeriesCoefficients}) or a similar double infinite series involving the Bessel function (e.g. \citealt{Giacaglia1976ATheory, Renaud2021TidalTRAPPIST-1e}), but this requires an infinite double summation, evaluation to convergence of which is computationally expensive (and moreover difficult to prove). \citet{Renaud2021TidalTRAPPIST-1e} purported to have solved this by precomputing a power series expansion in the square of the eccentricity based on this expression, but this inadvertently refactored their expression into another well-known form: a power series in the square of the eccentricity (e.g. \citealt{Izsak1964ConstructionComputer}), whose coefficients can be efficiently precomputed using a recurrence relationship  in Newcomb operators derived from the Von Zeipel differential equation (e.g. \citealt{Izsak1964ConstructionComputer, Cherniack1972ComputationCoefficients, Proulx1988SeriesCoefficients, Branham1990RecursiveCoefficients}), but whose convergence for large $e$ is particularly slow \citep{Proulx1988SeriesCoefficients}. An adaptation of the Newcomb-operator approach based on work by \citet{Proulx1988SeriesCoefficients} forms the basis for our simulation routine procedure, but we wish to verify the proper convergence thereof for the high eccentricities we encounter for early Triton. A series approach is not suitable in this context, as the convergence rate for such series is difficult to constrain, and so we must look elsewhere.

\subsection{An integral expression for the Hansen coefficients}
For a similar purpose, \citet{Gooding1989ExplicitOrbits} used numerical quadrature based on the following version of Eq.~\ref{eq:fourier_hansen} (obtained by taking the real part and using the symmetry of the cosine):
\begin{equation}
\label{eq:G_lpq_integral}
    G_{lpq} = \frac{1}{\pi}\int_0^{\pi}\left(\frac{a}{r}\right)^{l+1}\cos{((l-2p)v - (l-2p+q)M)}\dd{M}
\end{equation}
which is in principle sufficient to allow for numerical evaluation of $G_{lpq}$; this was the approach taken by \citet{Gooding1989ExplicitOrbits} and \citet{Wagner1979GravitationalOrbiters.}, however \citet{Wnuk1997HIGHLYORBITS} rightfully remarks that this form does require significant computation time and might run into numerical instabilities for high $l$, $p$ and $q$ as a result of the wildly oscillatory nature of the integrand. Moreover, though none of the preceding authors appear to have taken explicit note of this (with the exception of \citet{Proulx1988SeriesCoefficients}, though in a power series rather than integral context), the factor $a/r=(1-e\cos E)^{-1}$ in the integrand introduces a pole at $e=1$, which will slow convergence for high eccentricities. Our aim will therefore be to implement a version of Eq.~\ref{eq:G_lpq_integral} that does not suffer these issues.

The basis here shall be formed by the same integral expression from \citet{Proulx1988SeriesCoefficients} that is at the basis of the power series expression used in the simulation routine:
\begin{equation}
\label{eq:integral_G_lpq}
    G_{lpq} = (1-e^2)^{1/2-l} \frac{1}{\pi}\int_0^{\pi}\left(1+e\cos{v}\right)^{l-1}\cos{((l-2p)v - (l-2p+q)M)} \dd{v}.
\end{equation}
We note that the integrand will grow arbitrarily large for increasing $l$ due to the factor of $(1+e\cos{v})^{l-1}$, and so one may be inclined to factor out a $(1+e)^{l-1}$ to damp this growth. However, an initial trial shows that this additional factor in fact helps convergence for larger values of $l$, where the integral may otherwise be too small to compute efficiently. For applications where very large values of $l$ and $e$ are expected to occur in tandem, one may wish to consider extracting this factor, but we shall not worry about it for our purposes.

\subsection{Numerical instability of the integral expression}
Now, as mentioned, \citet{Wnuk1997HIGHLYORBITS} remarked that the integral expression that \citet{Gooding1989ExplicitOrbits} used is not numerically stable. While we have removed the pole at $e=1$ from the integral expression, two other sources of numerical problems remain, depending on the values of $l$, $p$, and $q$, namely (1) the fact that the zeroes of the integrand are ordered in a nearly uniform manner, which means that catastrophic alignment of the integration grid becomes a possibility, and (2) the fact that the integrand is highly oscillatory, which leads to similarly oscillatory behaviour of the integral approximation with increasing numbers of gridpoints.

(1) is easily solved by taking a grid large enough; (2), however, is more devious. While in principle convergence could be brute-forced despite (2), the memory allocation costs are immense and at times infeasible. We note, however, that the oscillatory part of the integrand in Eq.~\ref{eq:integral_G_lpq} is of a particularly friendly form: the factor $(1+e\cos{v})^{l-1}$ out front does not oscillate, so only $\cos{((l-2p)v - (l-2p+q)M)}$ remains. If we were able to separate the domain along the zeroes of this term, that would greatly help convergence as we would be able to integrate over well-behaved regions separately. While this term does not admit analytical computation of its zeroes (and numerical computation thereof would be prohibitively expensive), we instead note that we can rewrite the integral to be more receptive to analytic consideration:
\begin{align}
\label{eq:G_lpq_separated_integral}
    G_{lpq} = (1-e^2)&^{1/2-l}\left[\frac{1}{\pi}\int_0^{\pi}\left(1+e\cos{v}\right)^{l-1}\cos{(l-2p)v}\cos{(l-2p+q)M} \dd{v} \right. \\
    & + \left.\frac{1}{\pi}\int_0^{\pi}\left(1+e\cos{v}\right)^{l-1}\sin{(l-2p)v}\sin{(l-2p+q)M} \dd{v} \right] \nonumber
\end{align}
where now all zeroes can be computed analytically, at the cost of having to compute two integrals rather than one. Additionally, the use of $v$ as the independent variable forces us to solve Kepler's equation to compute the zeroes corresponding to the analytically computed values of the roots in $M$; however, here we make use of the method by \citet{Markley1995KeplerSolver}, which is fast, always terminates, and has error no greater than machine precision, such that it is acceptable for this purpose. In fact, for values of $l\leq5$ and $q\leq20$, the integrator does not run into any issues if we use solely the zeroes of the part due to $(l-2p)v$ for the integration bounds, such that we can do away with the need for a solver for Kepler's equation entirely. As this results in a significant speedup of the computation without introducing any apparent convergence issues, we will use this variant for now.

We can thus proceed to integrate each of the integrals in Eq.~\ref{eq:G_lpq_separated_integral} piecewise between zeroes, and in this manner guarantee that we have well-behaved intervals of integration. To be able to estimate the convergence rate of the integral, we stick with Boole's rule as used by \citet{Gooding1989ExplicitOrbits}, which allows efficient reuse of previously calculated function values.

\section{Series expression for the Kaula eccentricity functions}
\label{sec:power_series_appendix}
As evaluation to convergence of the integral expression in Eq.~\ref{eq:G_lpq_integral} is not deterministic and consequently not reliable, but above all not particularly efficient, we would like a more robust alternative for in our simulation routine; the integral expression may serve as a validation method. We opt to employ a power series solution first proposed by \citet{Proulx1988SeriesCoefficients}, akin to Newcomb-Poincaré expression used by \citet{Izsak1964ConstructionComputer} and \citet{Cherniack1972ComputationCoefficients} but with the pole at $e=1$ removed. Hence, \citet{Proulx1988SeriesCoefficients} were able to demonstrate for what they dubbed the ``Y-series" far faster convergence across a range of eccentricities $e\in[0,1)$ than for the Newcomb-Poincaré series, while maintaining efficient recursive computation of the power series coefficients using recurrence relations derived from the Von Zeipel differential equation \citep{VonZeipel1912SurNewcomb}\footnote{While \citet{VonZeipel1912SurNewcomb} wrote in French, the differential equation is derived in English in a largely similar manner in appendix A of \citet{McClain1978AAveraging}, and an English-language discussion of the method as a whole can be found in \citet{Proulx1988SeriesCoefficients}.}. 

\subsection{Motivation behind the power series expression}
Contrary to what \citet{Renaud2021TidalTRAPPIST-1e} state, a Newcomb operator-style approach allows retention of arbitrary-level precision, as the power series coefficients produced in this manner are rational and can be computed exactly. The motivation for our power series expression is found in Eq.~\ref{eq:G_lpq_integral}, which we repeat here:
\begin{equation}
    G_{lpq} = (1-e^2)^{1/2-l} \frac{1}{\pi}\int_0^{\pi}\left(1+e\cos{v}\right)^{l-1}\cos{((l-2p)v - (l-2p+q)M)} \dd{v}.
\end{equation}
The important take-away is the fact that the pole at $e=1$ is now contained in the factor in front of the integral. A general form of the power series representation of this integral can be given by (e.g. \citealt{Proulx1988SeriesCoefficients}):
\begin{equation}
\label{eq:power_series_for_X}
    X^{n,k}_{s} = (1-e^2)^{n+3/2} e^{|s-k|}Y^{n, k}_s
\end{equation}
with
\begin{equation}
\label{eq:Y_series_def}
    Y^{n,k}_s = \sum_{r\geq0}Y^{n,k}_{r+a,r+b}e^{2r}
\end{equation}
where $a=\max(0,s-k)$ and $b=\max(0,k-s)$. For the particular forms $G_{lpq}$ this becomes:
\begin{equation}
\label{eq:power_series_for_G}
    G_{lpq} = (1-e^2)^{1/2-l} e^{|q|}\sum_{r\geq0}Y^{-(l+1),(l-2p)}_{r+a,r+b}e^{2r}
\end{equation}
where $a=\max{(0, q)}$ and $b=\max{(0,-q)}$. Note that the factor of $e^{|s-k|}=e^{|q|}$ simply reflects the d'Alembert characteristic (see e.g. \citealt{Sadov2008AnalyticCoefficients}), and may be included in the power series if one so desires.
A second justification of the form in Eq.~\ref{eq:power_series_for_G} can be found if one considers the form one obtains for $G_{lpq}$ if $l-2p+q=0$ \citep{Hughes1981THECOEFFICIENTS, Kozai1973AMOTIONS}:
\begin{equation}
\label{eq:l-2p+q_zero}
    G_{lpq} = (1-e^2)^{1/2-l}e^{|l-2p|} \sum_{r = 0}^{\frac{l-|l-2p|}{2}-1}2^{-2r-|l-2p|}\binom{l-1}{2r + |l-2p|} \binom{2r+|l-2p|}{r}e^{2r}
\end{equation}
which one will note is of the same form as Eq.~\ref{eq:power_series_for_G} if one sets $|l-2p|=|q|$, which follows straightforwardly from $l-2p+q=0$. This can be derived from by setting $l-2p+q=0$ in Eq.~\ref{eq:G_lpq_integral} and expanding the factor $(1+e\cos{v})^{l-1}$ binomially:
\begin{equation}
\label{eq:cos_to_the_i}
    G_{lpq} = (1-e^2)^{1/2-l} \frac{1}{\pi}\sum_{i=0}^{l-1}\binom{l-1}{i}e^i\int_0^{\pi}\cos^{i}{v}\cos{((l-2p)v)} \dd{v}
\end{equation}
where we note that \citep{Gradshteyn2007TableProducts}
\begin{equation}
\label{eq:cos_powers}
    \cos^i{v} = 
    \begin{cases}
    \frac{1}{2^{2r-1}}\sum_{k=0}^{r-1} \binom{2r}{k} \cos{((2r - 2k)v)} + \frac{1}{2}\binom{2r}{r}\textrm{ for $i=2r$, $r\in\mathbb{N}$} \\
    \frac{1}{2^{2r-2}}\sum_{k=0}^{r-1} \binom{2r-1}{k} \cos{((2r - 2k - 1)v)}\textrm{ for $i=2r-1$, $r\in\mathbb{N}$}.
    \end{cases}
\end{equation}
As $\int_0^{\pi}\cos{mv}\dd{v}=0$ for all $m\in\mathbb{N}$, upon insertion of Eq.~\ref{eq:cos_powers} we can write for Eq.~\ref{eq:cos_to_the_i} that
\begin{equation}
    G_{lpq} = (1-e^2)^{1/2-l} \frac{2}{\pi}\sum_{i=1}^{l-1}\sum_{k=0}^{\lfloor\frac{i-1}{2}\rfloor}\binom{l-1}{i}\binom{i}{k}\left(\frac{e}{2}\right)^i\int_0^{\pi}\cos{(i - 2k)v}\cos{(l-2p)v} \dd{v}.
\end{equation}
Now, $\int_0^{\pi}\cos{(i - 2k)v}\cos{(l-2p)v} \dd{v}=\pi/2$ for $i = 2k + |l - 2p|$, and $0$ otherwise: we note that this requires that $i$ has the same parity as $|l-2p|$, and that for all such $i$ the fact that $0\leq i-2k \leq i$ means that precisely one $k$, given by $k=\frac{i-|l-2p|}{2}$, then yields a non-zero value of the integral. Therefore, only the values of $k=0,1,\ldots, \lfloor\frac{l-1-|l-2p|}{2}\rfloor$ and the corresponding $i=2k + |l-2p|$ yield a non-zero value of the integral. Hence, we can see that the only non-zero values can be written in the form of Eq.~\ref{eq:l-2p+q_zero}.

Not just does this expression justify the form of the power series, but the uniqueness of power series representations then guarantees that we have
\begin{equation}
    Y^{-(l+1),|q|}_{r, r+|q|} = \begin{cases}
        2^{-2r-|q|}\binom{l-1}{2r + |q|} \binom{2r+|q|}{r} \textrm{ if $0\leq r\leq\lfloor(l-1-|q|)/2\rfloor$} \\
        0\textrm{ else}
    \end{cases}
\end{equation}
and that the series thus terminates if $G_{lpq}$ is of the form $G_{lp(2p-l)}$ (or, equivalently, of the form $G_{l\frac{1}{2}(l+q)q}$). \citet{Hughes1981THECOEFFICIENTS} rightfully remarks that this means that the Newcomb operator approach fails to reproduce exactly any eccentricity functions for which an exact expression exists if one truncates the power series: however, by factoring out the term $(1-e^2)^{1/2-l}e^{|q|}$ in front it is clear that the greatest power one finds in the exact power series expression is $e^{l-1}$. Using the Y-series factorisation, this will therefore only be a concern if one neglects powers less than $e^{l_{\max}-1}$ (where $l_{\max}$ is the maximum degree considered), which is not a particularly stringent scenario.

\subsection{Properties of the power series coefficients}
The coefficients $Y^{n,k}_{\rho,\sigma}$ possess the symmetry property $Y^{n,k}_{\rho,\sigma}=Y^{n,-k}_{\sigma,\rho}$ as can be derived from Eq.~\ref{eq:fourier_hansen}: consequently, we can write
\begin{equation}
\label{eq:Y_to_G}
    G_{lpqr} = Y^{-(l+1), (l-2p)}_{\sigma+a,\sigma+b} = \begin{cases}
        Y^{-(l+1), |l-2p|}_{r + |q|,r}\textrm{ if }\textrm{sgn}(l-2p)=\textrm{sgn}(q) \\
        Y^{-(l+1), |l-2p|}_{r, r + |q|}\textrm{ if }\textrm{sgn}(l-2p)\neq\textrm{sgn}(q)
    \end{cases}
\end{equation}
where $a=\max{(0, q)}$ and $b=\max{(0,-q)}$, $\textrm{sgn}(x)$ is the signum or sign function, which returns 1 if $x\geq0$ and -1 if $x<0$, and we have introduced the notation $G_{lpqr}$ for convenience, such that we can write
\begin{align}
\label{eq:G_lpqr_def}
    G_{lpq} = (1-e^2)^{1/2-l} e^{|q|}\sum_{r\geq0}G_{lpqr}e^{2r}.
\end{align}
This allows us a correspondence between the particular coefficients $G_{lpqr}$ that are relevant to us and the broader set of coefficients $Y^{n,k}_{\rho,\sigma}$ through a parallel relation to Eq.~\ref{eq:Y_to_G}:
\begin{align}
\label{eq:G_to_Y}
    Y^{n,k}_{\rho,\sigma} &= G_{lpqr}\textrm{ where} \\
    l &=-(n+1) \nonumber\\
    p &=-\frac{1}{2}(n+k-1) \nonumber \\
    q &= \rho-\sigma \nonumber \\
    r &= \min(\rho, \sigma) \nonumber
\end{align}
where we note that for $p$ to be an integer (such that these correspond to the coefficients for a proper eccentricity function $G_{lpq}$), we require that $n+k$ is odd. Through Eq.~\ref{eq:Y_to_G}, each of the coefficients of interest to us can thus be reduced to a form $Y^{n,k}_{\rho,\sigma}$ with:
\begin{align}
    3\leq -&n \leq l_{\textrm{max}} + 1 \label{eq:n_condition} \\
    0\leq  &k \leq l=-(n+1),\; n+k\textrm{ odd} \label{eq:k_condition} \\
    0\leq  &\rho \leq r_{\textrm{max}} \label{eq:rho_condition} \\
    0\leq  &\sigma \leq r_{\textrm{max}} \label{eq:sigma_condition}
\end{align}
where $l_{max}$ is the maximum value of the degree $l$ that we consider and $2r_{max}$ is the maximum power of $e$ that we wish to include in our expansions of $G_{lpq}^2$ (in the term $e^{2|q|}\left(\sum_{r\geq 0} G_{lpqr}e^{2r}\right)^2$, so not incuding the prefactor $(1-e^2)^{1-2l}$, which we do not expand into a power series). While the latter definition may seem somewhat convoluted, it is in fact $G_{lpq}^2$ that appears in our tidal potential formulae, and so this definition is more meaningful than if we had phrased it in terms of powers of e in $G_{lpq}$.

An interesting corollary of the derivation for Eq.~\ref{eq:l-2p+q_zero} is that the same reasoning holds when one considers the relation
\begin{equation}
    e^{|q|}\sum_{r\geq0}Y^{-(l+1),(l-2p)}_{r+a,r+b}e^{2r} = \frac{1}{\pi}\int_0^{\pi}\left(1+e\cos{v}\right)^{l-1}\cos{((l-2p)v - (l-2p+q)M)} \dd{v}
\end{equation}
which follows straightforwardly from Eq.~\ref{eq:G_lpq_integral} and Eq.~\ref{eq:power_series_for_G}. It follows from Kepler's equation and the eccentric anomaly-mean anomaly relationship that $M\to 0$ for $e\to 1$ everywhere except at $v=\pi$, which is a set of measure zero and therefore does not affect the value of the integral\footnote{Of course, normally one would use the parabolic version of Kepler's equation for $e=1$; here we are interested in the limit of the elliptic case as $e\to 1$, and so this reasoning is valid.}; consequently, one can use the same logic as used in the case $l-2p+q=0$ to derive that
\begin{equation}
    \frac{1}{\pi}\int_0^{\pi}\left(1+\cos{v}\right)^{l-1}\cos{((l-2p)v)} \dd{v} = \sum_{r = 0}^{\lfloor\frac{l-1-|l-2p|}{2}\rfloor}\frac{\binom{l-1}{2r + |l-2p|} \binom{2r+|l-2p|}{r}}{2^{2r+|l-2p|}}.
\end{equation}
While at face value this seems like an interesting but useless fact, the fact that the integral is finite implies that in fact the partial sum of the series $\left(Y^{-(l+1),(l-2p)}_{r+a,r+b}\right)_{r\geq 0}$ converges to a known value
\begin{equation}
    Y^{-(l+1),(l-2p)}_{l-2p+q}(1)=\sum_{r\geq0}Y^{-(l+1),(l-2p)}_{r+a,r+b} = \sum_{r = 0}^{\frac{l-|l-2p|}{2}-1}\frac{\binom{l-1}{2r + |l-2p|} \binom{2r+|l-2p|}{r}}{2^{2r+|l-2p|}} = Y^{-(l+1),(l-2p)}_0(1)
\end{equation}
and consequently is bounded with terms that eventually decay to zero, a promise which the Newcomb series does not offer (and in fact it cannot; it will in general not have a finite limit of the partial sum as the Newcomb series expression diverges as $e\to 1$). Additionally, this identity provides a useful verification tool.

It now remains to find a method by which the coefficients $G_{lpqr}$ (or, equivalently, the coefficients $Y^{n,k}_{\rho,\sigma}$ satisfying conditions \ref{eq:n_condition}~-~\ref{eq:sigma_condition}) can be computed.

\subsection{Von Zeipel-type recurrence relations for the power series coefficients}
From the Von Zeipel differential equation \citep{VonZeipel1912SurNewcomb} and the power series representation Eq.~\ref{eq:power_series_for_G} one can extract the following recurrence relation for the coefficients in Eq.~\ref{eq:power_series_for_G} \citep{Proulx1988SeriesCoefficients}\footnote{N.B.: \citet{Proulx1988SeriesCoefficients} forgot the factor of two in front of the last term, involving the summation. Nonetheless, their reported convergence rates match with ours, and so we presume their error was only typographical.}:
\begin{align}
\label{eq:recurrence_Y}
    4\rho Y^{n,k}_{\rho, \sigma} &= 2(2k-n)Y^{n, k+1}_{\rho-1, \sigma} + (k-n)Y^{n, k+2}_{\rho-2,\sigma} \\
    &+ (5\rho-\sigma+3n+4k+2)Y^{n,k}_{\rho-1,\sigma-1} \nonumber \\
    &- 2(k + \rho - \sigma)\sum_{\tau=2}^{\min(\rho,\sigma)}\Tilde{c}_{\tau} Y^{n,k}_{\rho-\tau,\sigma-\tau}, \nonumber
    \textrm{where } \\
    \Tilde{c}_{\tau} &= 3\cdot 2^{2-2\tau} \frac{(2\tau-4)!}{\tau!(\tau-2)!}.
\end{align}
Note that \citet{Proulx1988SeriesCoefficients} give the coefficient $\Tilde{c}_{\tau}$ in the form $\Tilde{c}_{\tau}=(-1)^{\tau}\binom{3/2}{\tau}$ instead; while this is more compact, we will prefer the expanded version as it is more receptive to direct computation and interpretation, though only valid for $\tau\geq2$. The initial values for the recurrence relation are given as:
\begin{align}
    Y^{n,k}_{0,0}&=1 \\
    Y^{n,k}_{\rho,\sigma} &= 0\textrm{ if $\rho<0$ or $\sigma<0$,}
\end{align}
whence it follows that $Y^{n,k}_{1,0}=k-n/2$ and $Y^{n,k}_{0,1}=-k-n/2$. It is worthwhile to note that the same transform that \citet{Izsak1964ConstructionComputer} and \citet{Cherniack1972ComputationCoefficients} used to reduce the problem of computing the Newcomb-Poincaré series into an integer arithmetic problem:
\begin{equation}
    J^{n,k}_{\rho, \sigma} = X^{n,k}_{\rho,\sigma}2^{\rho+\sigma}\rho!\sigma!
\end{equation}
turns Eq.~\ref{eq:recurrence_Y} into an integer arithmetic problem as well, if one instead sets
\begin{equation}
    J^{n,k}_{\rho, \sigma} = Y^{n,k}_{\rho,\sigma}2^{\rho+\sigma}\rho!\sigma!.
\end{equation}
The integer coefficients $J^{n,k}_{\rho,\sigma}$ grow large rather quick, so we will prefer to remain in the rational domain and take the additional computation time the rational problem takes for lower values as necessary, as it prevents (or rather delays the onset of) storage issues for larger indices, which necessitate use of multi-precision values the products of which are expensive to compute. For real-time computation (where it is unlikely that one will go to the large indices where $J^{n,k}_{\rho,\sigma}$ becomes too large to represent as a reasonable multi-precision integer), one might prefer the added speed and simplicity that the integer coefficients bring; for lower-accuracy application, floats will probably suffice. As we precompute the coefficients, and the algorithm is reasonably efficient still, we opt to use the multi-precision rationals allowed by \textit{gmpy2}'s \textit{mpq} class.

If one uses the symmetry $Y^{n,k}_{\rho,\sigma}=Y^{n,-k}_{\sigma,\rho}$, a second recurrence relation can be obtained which only uses coefficients with equal or smaller $k$:
\begin{align}
\label{eq:recurrence_Y_mirror}
    4\sigma Y^{n,k}_{\rho, \sigma} &= -2(2k+n)Y^{n, k-1}_{\rho, \sigma-1} - (k+n)Y^{n, k-2}_{\rho,\sigma-2} \\
    &+ (5\sigma-\rho+3n-4k+2)Y^{n,k}_{\rho-1,\sigma-1} \nonumber \\
    &+ 2(k + \rho - \sigma)\sum_{\tau=2}^{\min(\rho,\sigma)}\Tilde{c}_{\tau} Y^{n,k}_{\rho-\tau,\sigma-\tau},\nonumber
\end{align}
and combining these two yields a final relation\footnote{\citet{Proulx1988SeriesCoefficients} were lucky insofar as their error did not propagate into this final relation, which means that a naive implementation using only this relation would not result in errors in the coefficients. However, efficient computations requires targeted re-use of previously computed and cached coefficients, which becomes increasingly less efficient for higher $\rho$ and $\sigma$ using solely this expression. Perhaps this is why their expression never found use or even mention in more recent literature; the only mention in recent literature in the context of natural satellite theory is by \citet{Luna2020TheSystem}, who nonetheless refer to them only concerning the Newcomb series, not the Y-series.}:
\begin{align}
\label{eq:recurrence_Y_sum}
    4(\rho+\sigma)Y^{n,k}_{\rho,\sigma} &= 2(2k-n)Y^{n,k+1}_{\rho-1,\sigma} + (k-n)Y^{n,k+2}_{\rho-2,\sigma}-2(2k+n)Y^{n,k-1}_{\rho,\sigma-1} \\
    &- (k+n)Y^{n,k-2}_{\rho,\sigma-2} + (4(\rho+\sigma)+6n + 4)Y^{n,k}_{\rho-1,\sigma-1}. \nonumber
\end{align}
The advantage in this final relation is that it does not require a long summation, limiting the number of terms that must be known for the computation of $Y^{n,k}_{\rho,\sigma}$ to 5 terms rather than $\min(\rho,\sigma)+1$ terms; the disadvantage is that one can always use one of Eqs.~\ref{eq:recurrence_Y} or \ref{eq:recurrence_Y_mirror} to stay within the set of conditions \ref{eq:n_condition}-\ref{eq:sigma_condition} (with the partial exception of the condition that $k=l_{\max}-2m$ for some $m\in\mathbb{N}$), while this is not always possible with Eq.~\ref{eq:recurrence_Y_sum}. To guarantee that we need not compute any terms with $k\notin [0, l]$, we can therefore use Eq.~\ref{eq:recurrence_Y_sum} in all cases except those with $k=l,l-1,l-2$, in which case we can use Eq.~\ref{eq:recurrence_Y_mirror}; for the cases that yield $k<0$, we can use the symmetry property $Y^{n,-k}_{\rho,\sigma}=Y^{n,k}_{\sigma,\rho}$.

One final problem still remains, however: if $n+k$ is odd, as is required for $Y^{n,k}_{\rho,\sigma}$ to correspond to a $G_{lpqr}$, the terms $Y^{n,k+1}_{\rho-1,\sigma}$ and $Y^{n,k-1}_{\rho,\sigma-1}$ do not correspond to a $G_{lpqr}$ as $n+k\pm 1$ are ostensibly not odd. Hence, if we use relations \ref{eq:recurrence_Y}, \ref{eq:recurrence_Y_mirror} and \ref{eq:recurrence_Y_sum}, we must also calculate redundant coefficients that do not correspond to any $G_{lpqr}$. To remedy this, we could make use of some general properties of the Hansen coefficients: \citet{Vakhidov2001SOMECOEFFICIENTS} computed through computer algebra a recursion scheme that involves solely the Hansen coefficients that correspond to the eccentricity functions, but the scheme is quite involved and does not lend itself particularly well to computation of the power series coefficients we are after due to the large expressions involved. Consequently, we choose to accept this inefficiency.

\subsection{An algorithm to compute the relevant $Y^{n,k}_{\rho,\sigma}$ recursively}
We now have all we need to compute efficiently the power series coefficients necessary to compute $G_{lpq}$: in particular, we now have the required methods to efficiently compute and store the coefficients $Y^{n,k}_{\rho,\sigma}$ that satisfy criteria \ref{eq:n_condition}-\ref{eq:sigma_condition}, though as discussed we must relax the condition on $k$ to instead be $0\leq k \leq l$, without the requirement that $n+k$ be odd. Our renewed criteria are therefore:
\begin{align}
    3\leq -&n \leq l_{\textrm{max}} + 1 \label{eq:n_condition_2} \\
    0\leq  &k \leq l=-(n+1) \label{eq:k_condition_2} \\
    0\leq  &\rho \leq r_{\textrm{max}} \label{eq:rho_condition_2} \\
    0\leq  &\sigma \leq r_{\textrm{max}} \label{eq:sigma_condition_2}.
\end{align}
By using the proper symmetries and recurrence relations, we can guarantee that any terms we compute will satisfy conditions \ref{eq:n_condition_2}-\ref{eq:sigma_condition_2}, such that we can guarantee an efficient computation of the coefficients of interest: we will never have to compute and store more than $\frac{1}{2}(l_{\max}-1)^2(l_{\max}+4)(r_{\max}+1)^2$ values. Let us go through the algorithm step by step: we shall take as starting point some arbitrary $Y^{n,k}_{\rho,\sigma}$ satisfying these relaxed conditions, and discuss what steps must be undertaken to compute it.

\subsubsection{1. Check if we need symmetries because of $-2\leq k<0$}
If the previous calculation had $k\leq 1$, we may encounter a term with $-2\leq k<0$: while nominally outside our criteria \ref{eq:n_condition_2}-\ref{eq:sigma_condition_2}, if this is the case we can use the symmetry $Y^{n,-k}_{\rho,\sigma}=Y^{n,k}_{\sigma,\rho}$ to recover a term that satisfies our criteria, so as to avoid computation of unnecessary terms. As $l\geq2$, we will in this manner always recover a term that satisfies $0 \leq k\leq l$.

\subsubsection{2. Test for trivial or previously computed cases}
If the provided $Y^{n,k}_{\rho,\sigma}$ has $\rho\leq1$ and $\sigma\leq0$ or $\sigma\leq1$ and $\rho\leq0$, we can directly compute the result and return it. If we have previously computed and cached this result, we can also directly return it.

\subsubsection{3. Check if $k\leq l-2$}
If $k\leq l-2$, we can use Eq.~\ref{eq:recurrence_Y_sum}, which does not have the dynamic sum and should therefore (for large values of $\rho$ and $\sigma$) be faster than Eqs.~\ref{eq:recurrence_Y} or \ref{eq:recurrence_Y_mirror}. For convenience, we repeat it here:
\begin{align}
\label{eq:recurrence_Y_sum_repeat}
    4(\rho+\sigma)Y^{n,k}_{\rho,\sigma} &= 2(2k-n)Y^{n,k+1}_{\rho-1,\sigma} + (k-n)Y^{n,k+2}_{\rho-2,\sigma}-2(2k+n)Y^{n,k-1}_{\rho,\sigma-1} \\
    &- (k+n)Y^{n,k-2}_{\rho,\sigma-2} + (4(\rho+\sigma)+6n + 4)Y^{n,k}_{\rho-1,\sigma-1}. \nonumber
\end{align}
This therefore necessitates the computation or retrieval of $Y^{n,k+1}_{\rho-1,\sigma}$, $Y^{n,k+2}_{\rho-2,\sigma}$, $Y^{n,k-1}_{\rho,\sigma-1}$, $Y^{n,k-2}_{\rho,\sigma-2}$ and $Y^{n,k}_{\rho-1,\sigma-1}$, which means the recursion loop returns to the first step and will go a layer deeper. If $l-2<k\leq l$, we will need another relation so as to make sure that we need only compute coefficients that still satisfy criteria \ref{eq:n_condition_2}-\ref{eq:sigma_condition_2}.

\subsubsection{4. Compute the case $l-2 < k \leq l$ with $\sigma>0$}
Only one possible case for $k$ remains, which is that $l-2<k\leq l$: now, we require an expression that does not involve $Y^{n,k+1}_{\rho-1,\sigma}$ or $Y^{n,k+2}_{\rho-2,\sigma}$, as at least one and possibly both of these would not satisfy the conditions we have set on the terms we wish to calculate. Of course, just such an expression is given by Eq.~\ref{eq:recurrence_Y_mirror}, which we repeat here for reference:
\begin{align}
\label{eq:recurrence_Y_mirror_repeat}
    4\sigma Y^{n,k}_{\rho, \sigma} &= -2(2k+n)Y^{n, k-1}_{\rho, \sigma-1} - (k+n)Y^{n, k-2}_{\rho,\sigma-2} \\
    &+ (5\sigma-\rho+3n-4k+2)Y^{n,k}_{\rho-1,\sigma-1} \nonumber \\
    &+ 2(k + \rho - \sigma)\sum_{\tau=2}^{\min(\rho,\sigma)}\Tilde{c}_{\tau} Y^{n,k}_{\rho-\tau,\sigma-\tau}.\nonumber
\end{align}
Now, we must compute or retrieve $Y^{n,k-1}_{\rho-1,\sigma}$, $Y^{n,k-2}_{\rho-2,\sigma}$, $Y^{n,k}_{\rho-1,\sigma-1}$ and $Y^{n,k}_{\rho-\tau,\sigma-\tau}$ for $\tau=2,\ldots,\min(\rho,\sigma)$, and so the recursion loop starts again.

\subsubsection{5. Compute the case $l-2<k\leq l$ with $\sigma=0$}
We must make one explicit exception to the previous part of the routine, as Eq.~\ref{eq:recurrence_Y_mirror_repeat} becomes uninformative for $\sigma=0$. At this point, we must unfortunately make an excursion from our closed parameter space defined by the conditions \ref{eq:n_condition_2}-\ref{eq:sigma_condition_2}: knowing that we have a closed-form expression for any $Y^{n,k}_{1,0}$ and $Y^{n,k}_{0,0}$, we will strive to reduce any $Y^{n,k}_{\rho,0}$ to these forms. This is possible using Eq.~\ref{eq:recurrence_Y}, which reduces down to:
\begin{align}
\label{eq:recurrence_Y_sigma0}
    4\rho Y^{n,k}_{\rho, 0} &= 2(2k-n)Y^{n, k+1}_{\rho-1, 0} + (k-n)Y^{n, k+2}_{\rho-2,0}
\end{align}
such that it is possible to reduce any such form to the trivial cases $Y^{n,k}_{1,0}$ and $Y^{n,k}_{0,0}$. Effectively, we must therefore further loosen our conditions \ref{eq:n_condition_2}-\ref{eq:sigma_condition_2} to include also all coefficients of the form $Y^{n, k}_{\rho, 0}$ with $n$ and $\rho$ satisfying conditions \ref{eq:n_condition_2} and \ref{eq:rho_condition_2}, while $l < k \leq l + r_{\max}$. This adds a further $r_{\max} (r_{\max}+1) (l_{\max}-1)$ power series coefficients which must (at most) be computed, which is permissible even if undesirable, given that it is a relatively small number compared to the total number of coefficients that have to be computed already.

With this ``closing act", the algorithm can reduce all power series coefficients down to sums of the trivial cases, and so we can construct an exact expression for any desired coefficient $G_{lpqr}$ by computing the corresponding $Y^{n,k}_{\rho,\sigma}$ while remaining (for the most part) in the nearly-closed set of power series coefficients defined by conditions \ref{eq:n_condition_2}-\ref{eq:sigma_condition_2}.

\subsection{The limiting behaviour of $G_{lpq}$ for large $|q|$}
\label{app:limiting_behaviour_G_lpq_large_q}
To appropriately truncate the sum over $q$ in Eqs.~\ref{eq:partial_potential}, we must understand the behaviour of $G^2_{lpq}$ and consequently $G_{lpq}$ as $q$ grows large (be it negative or positive). \citet{Szeto1982ONORBITS} already noted that the $G_{lpq}$ appear to decay outside a given interval of values of $q$, and computed these intervals for low values of $l$ and $p$, though they were explicitly not able to find any analytical support of this idea. We note that the Riemann-Lebesgue lemma (see \citealt[p. 1067]{Gradshteyn2007TableProducts}) guarantees that this decay does indeed occur as $|q|$ becomes large, but we should wish to put a useful bound on its decay as well. Inspired by \citet{Breiter2004GENERALIZEDCOEFFICIENTS}, who showed that a great deal of properties of the Hansen coefficients carry over once one expands the Hansen coefficients $X^{n,j}_k$ to the form $X^{\gamma,j}_k$, where $\gamma$ is a real number, we examine once more the general integral expression for the eccentricity functions, Eq.~\ref{eq:G_lpq_integral}:
\begin{equation}
    G_{lpq} = \frac{1}{\pi}\int_0^{\pi}\left(\frac{a}{r}\right)^{l+1}\cos{((l-2p)v - (l-2p+q)M)}\dd{M}.
\end{equation}
We note that there is nothing in particular about this expression that mandates that the values $l$, $p$ and $q$ take integral values. In fact, it is not difficult to see that upon relaxation of these requirements (that is, allowing these quantities to take any real value) $G_{lpq}$ becomes continuous and differentiable in $l$, $p$ and $q$, and in fact the integer values of $l$, $p$ and $q$ are simply particular values of the real-valued function. It is therefore meaningful to take derivatives with respect to each of these: in particular, we are interested in the behaviour of $q$, which motivates us to examine the derivative of $G_{lpq}$ with respect to $q$:
\begin{align}
    \derivative{G_{lpq}}{q} &= \frac{1}{\pi}\int_0^{\pi}\left(\frac{a}{r}\right)^{l+1}\sin{((l-2p)v - (l-2p+q)M)}M\dd{M} \nonumber \\
    &= \frac{1}{\pi}\left[\int_0^{\pi}\left(\frac{a}{r}\right)^{l+1}\sin{((l-2p)v)}\frac{M}{l-2p+q}\dd{\sin{((l-2p+q)M)}} \right. \nonumber \\
    &+ \left. \int_0^{\pi}\left(\frac{a}{r}\right)^{l+1}\cos{((l-2p)v)}\frac{M}{l-2p+q}\dd{\cos{((l-2p+q)M)}}\right]
\end{align}
which, upon integration by parts can be rewritten to
\begin{align}
\label{eq:G_deriv_q}
    \derivative{G_{lpq}}{q} &= -\frac{G_{lpq}}{l-2p+q} + \frac{1}{l-2p+q} \left[(1+e)^{-(l+1)}\cos{q\pi} \right. \nonumber \\
    &- \left. \frac{e(l+1)}{\pi\sqrt{1-e^2}}\int_0^{\pi}\left(\frac{a}{r}\right)^{l+2}\sin{v}\cos{((l-2p)v - (l-2p+q)M)}M \dd{M}\right. \nonumber \\
    &+ \left. \frac{(l-2p)\sqrt{1-e^2}}{\pi}\int_0^{\pi}\left(\frac{a}{r}\right)^{l+3}\sin{((l-2p)v - (l-2p+q)M)}M \dd{M} \right].
\end{align}
The quantity in square brackets can be strictly bounded in absolute value by a function of the eccentricity and $l$ alone, and so for sufficiently large values of $q$ this expression will become arbitrarily close to
\begin{equation}
    \derivative{G_{lpq}}{q} = -\frac{G_{lpq}}{l-2p+q}
\end{equation}
which suggests a limiting behaviour of $G_{lpq}\propto (l-2p+q)^{-1}$ (and in fact, for the Hansen coefficients in general we therefore have that $X^{n,m}_{k}\propto k^{-1}$ for sufficiently large $k$). Indeed, the Hansen coefficients in general satisfy (e.g. \citealt{Giacaglia1976ATheory, Vakhidov2001SOMECOEFFICIENTS}):
\begin{equation}
    X^{n,m}_k = \frac{1}{k}\left[m\sqrt{1-e^2}X^{n-2,m}_k - \frac{ne}{2\sqrt{1-e^2}}\left(X^{n-1,m+1}_k - X^{n-1,m-1}_k\right)\right]
\end{equation}
and as $|X^{n,m}_k| < (1-e)^{n}$ for $n<0$, this implies that
\begin{equation}
    |X^{n,m}_k| < \frac{(1-e)^{n-3/2}}{|k|}(|m|+|n|e)
\end{equation}
such that for the particular cases $G_{lpq}$ we have
\begin{equation}
    |G_{lpq}| < \frac{|l-2p| + |l+1|e}{|l-2p+q|(1-e)^{l+5/2}}.
\end{equation}
Indeed, at the very worst $G_{lpq}$ can therefore decay as $G_{lpq}\propto (l-2p+q)^{-1}$, and the factor $(1-e)$ in the denominator suggests that decay occurs later for larger $e$.

\section{Finite expressions for the constant time-lag sum in $q$}
\label{sec:q_series_convergence}
We note that each of the sums over $l$ in Eq.~\ref{eq:partial_potential} can be bounded term-wise by a scalar sum of one or more of the three following similar but subtly different sums (note the appearance of another factor of $p$ respectively $q$ in the second and third sums):
\begin{equation}
    \left(\frac{R_j}{a}\right)^{2l+1}\sum_{m=0}^{l}  \frac{(l-m)!}{(l+m)!}(2-\delta_{0m}) \sum_{p=0}^{l} F_{lmp}^2(i_j) \sum_{q=-\infty}^{\infty}G_{lpq}^2(e) K_{l,j}(\omega_{j,lmpq}),
\end{equation}
\begin{equation}
    \left(\frac{R_j}{a}\right)^{2l+1}\sum_{m=0}^{l}  \frac{(l-m)!}{(l+m)!}(2-\delta_{0m}) \sum_{p=0}^{l} F_{lmp}^2(i_j) p \sum_{q=-\infty}^{\infty}G_{lpq}^2(e) K_{l,j}(\omega_{j,lmpq})
\end{equation}
or
\begin{equation}
    \left(\frac{R_j}{a}\right)^{2l+1}\sum_{m=0}^{l}  \frac{(l-m)!}{(l+m)!}(2-\delta_{0m}) \sum_{p=0}^{l} F_{lmp}^2(i_j) \sum_{q=-\infty}^{\infty}q G_{lpq}^2(e) K_{l,j}(\omega_{j,lmpq}).
\end{equation}
This implies that the convergence rates of these sums with $q$ will dictate that of the sums in Eq.~\ref{eq:partial_potential}. Consequently, the convergence behaviour of each of the summands for a particular value of $l$ in the sum in Eqs.~\ref{eq:partial_potential} is governed solely by the convergence behaviour of the sums
\begin{equation}
    \sum_{q=-\infty}^{\infty}G_{lpq}^2(e) K_{l,j}(\omega_{j,lmpq})
\end{equation}
and
\begin{equation}
    \sum_{q=-\infty}^{\infty}(l-2p+q)G_{lpq}^2(e) K_{l,j}(\omega_{j,lmpq}).
\end{equation}
We need to make some assumption on the behaviour of $K_{l,j}(\omega_{j,lmpq})$ with $q$ to be able to say anything about the convergence rate of these sums, but in the interest of generality we should wish to make as little constraining an assumption as possible. To make this assumption as weak as possible, we will use the qualitative form given by \citet{Efroimsky2012TidalSuper-earths}, who divides the behaviour of $K_{l,j}$ into a low-frequency region that is well-described by Maxwell rheology and a high-frequency region that behaves according to Andrade rheology. Qualitatively, this means that up to some peak frequency, $K_{l,j}\propto \omega_{j,lmpq}$; be
yond this frequency, $K_{l,j}\propto \omega_{j,lmpq}^{-1}$ up to some frequency, from which onward $K_{l,j}\propto \omega_{j,lmpq}^{-\alpha}$. As $\omega_{j,lmpq}$ is defined by Eq.~\ref{eq:omega_lmpq_def}, the relative convergence behaviour of \textit{all} of the sums appearing in Eqs.~\ref{eq:partial_potential} for Andrade or Maxwell rheology is approximately bounded by that of the following three series:
\begin{align}
\label{eq:ctl_potential_sum}
    &\sum_{q=-\infty}^{\infty}G_{lpq}^2(e) \\
    &\sum_{q=-\infty}^{\infty}(l-2p+q) G_{lpq}^2(e) \\
    &\sum_{q=-\infty}^{\infty}(l-2p+q)^2 G_{lpq}^2(e)
\end{align}
which one may recognise as the convergence behaviour of the inner sum under the constant time-lag assumption. The shortcomings of the constant time-lag model for frequencies outside the linear regime of $K_{l,j}$ (see e.g. \citealt{Efroimsky2013TidalTorque, Bagheri2022TidalOverview} for a discussion) are precisely our saving grace; the constant time-lag model significantly overestimates the contribution of frequencies outside the linear regime to the potential. Consequently, when Eq.~\ref{eq:ctl_potential_sum} achieves a given level of relative accuracy upon truncation at some maximum value of $|q|$, we can be certain that Eqs.~\ref{eq:partial_potential} are approximated to better than that relative accuracy at the same truncation level.

Fortunately, \citet{Correia2022TidalCoefficients} were able to show using some ingenious mathematics that (note the difference in our notation)
\begin{align}
    \sum_{s=-\infty}^{+\infty} X^{n,k}_sX^{n,m}_s &= X^{2n, m-k}_0 \\
    \sum_{s=-\infty}^{+\infty} sX^{n,k}_sX^{n,m}_s &= \frac{m+k}{2}\sqrt{1-e^2}X^{2n-2, m-k}_0 \\
    \sum_{s=-\infty}^{+\infty} s^2X^{n,k}_sX^{n,m}_s &= n^2\left(2X^{2n-3,k-m}_0 - X^{2n-2,k-m}_0\right) \nonumber \\
    &+\left(\frac{n(k-m)^2}{2n-2} + km - n^2\right)(1-e^2)X^{2n-4,k-m}_0
\end{align}
whence in particular, as $G_{lpq}=X^{-(l+1), l-2p}_{l-2p+q}$:
\begin{align}
    \sum_{q=-\infty}^{\infty}G_{lpq}^2 &= X^{-2(l+1),0}_0 \\
    \sum_{q=-\infty}^{\infty}(l-2p+q)G_{lpq}^2 &= (l-2p)\sqrt{1-e^2}X^{-2l-4,0}_0 \\
    \sum_{q=-\infty}^{\infty}(l-2p+q)^2G_{lpq}^2 &= (l+1)^2\left(2X^{-2l-5,0}_0 - X^{-2l-4,0}_0\right) \nonumber \\
    &+ \left[(l-2p)^2-(l+1)^2\right](1-e^2)X^{-2l-6,0}_0.
\end{align}
such that the limiting value of this series can be computed to arbitrary accuracy at relatively little computational cost. These finite expressions therefore allow us to gauge how far out the inner sum in the Kaula expansion must be accounted for at a given eccentricity; this will then provide a conservative guess which can provide a trusted starting point for a more optimistic assessment.

A useful corollary of these expressions is the fact that the equilibrium rotation rate in the constant time-lag model with $l=2$ can be expressed in an analytical form, with Hansen coefficients that can be expressed as finite power series:
\begin{equation}
\label{eq:CTL_eq_rotrate}
    \frac{\dot{\theta}_{\textrm{eq, CTL}}}{n} = \sqrt{1-e^2}\frac{X^{-8,0}_0}{X^{-6,0}_0}/
\end{equation}

\section{A power series expression for the inclination functions}
\label{app:inclination_functions}
In general, expressions for the inclination functions are given by finite polynomials of trigonometric functions, and so are not as expensive to compute up to arbitrary accuracy as the eccentricity functions. Yet, existing series expressions do not lend themselves well to being used in tandem with efficient array operations when extended to higher degrees $l$; the expression originally used by \citet{Kaula1961AnalysisSatellites} involves a triple summation, while the expression originally derived by \citet{Allan1965OnSatellites} involves sums of various products of powers of $\sin{i/2}$ and $\cos{i/2}$. \citet{Gooding2008OnDerivatives} derived an efficient recursive procedure and showed it to operate well for degrees out until at least $l=1023$; however, we do not need to go out this far, and as we will need to compute the functions for different inclinations each timestep an expression that involves minimal recomputation for different inclinations is preferred. Hence, we will rewrite the inclination functions into a form that is receptive to implementation as a power series in some term involving the inclination with precomputed power series coefficients. We start from the expression given by \citet{Renaud2021TidalTRAPPIST-1e}:
\begin{align}
    F_{lmp}(i) &= \frac{(l+m)!}{2^l p! (l-p)!}\nonumber \\
    &\quad \times \sum_{\lambda=\sigma_{lmp}}^{v_{lmp}}(-1)^{\lambda}\binom{2(l-p)}{\lambda}\binom{2p}{l-m-\lambda}\left(\cos{\frac{i}{2}}\right)^{3l-m-2p-2\lambda}\left(\sin{\frac{i}{2}}\right)^{m-l+2p+2\lambda}
\end{align}
where
\begin{align}
    v_{lmp} = \min{\left(l-m,2(l-p)\right)} \\
    \sigma_{lmp} = \max{\left(0, l-m-2p\right)}
\end{align}
though as noted by \citet{Gooding2008OnDerivatives}, the indices of summation may for the sake of simplicity be taken to be $-\infty$ and $=-\infty$ if the common definition of the binomial coefficients is adopted (which set the summand to zero outside this range). For the sake of making explicit the loops that will have to be programmed into a computer programme, however, we will keep track of these bounds of summation. We wish to work toward a finite power series in $\cos{i/2}$, as that can be expressed through elementary trigonometric relations as a simple function of $\cos{i}$, which is one of our state variables: we can achieve this form by noting that even powers of $\sin{i/2}$ can be written as a finite sum of powers of $\cos{i/2}$, and by noting that the powers of $\cos{i/2}$ and $\sin{i/2}$ in this expression can be shown to be equal to or greater than $\abs{l-2p+m}$. Hence, we can rewrite this to a form in which $\sin{i/2}$ appears only with even powers (excepting a single multiplication out front, which we deal with later):
\begin{align}
    F_{lmp}(i) &= \frac{(l+m)!\left(\sin^{2}{\frac{i}{2}}\right)^{\delta/2}}{2^l p! (l-p)!}\nonumber \\
    &\quad \times \sum_{\lambda=\sigma_{lmp}}^{v_{lmp}}(-1)^{\lambda}\binom{2(l-p)}{\lambda}\binom{2p}{l-m-\lambda}\left(\cos{\frac{i}{2}}\right)^{3l-m-2p-2\lambda}\left(\sin^2{\frac{i}{2}}\right)^{\frac{m-l+2p+2\lambda - \delta}{2}}
\end{align}
where
\begin{equation}
    \delta =
    \begin{cases}
    1 \quad \textrm{if $l-m$ is odd} \\
    0 \quad \textrm{otherwise.}
    \end{cases}
\end{equation}
This can then be re-arranged and expanded binomially:
\begin{align}
    F_{lmp}(i) &= \frac{(l+m)!\left(1-\cos^2{i}\right)^{\delta/2}\left(\cos^2{\frac{i}{2}}\right)^{\frac{\abs{l-2p+m}-\delta}{2}}}{2^{l+\delta} p! (l-p)!} \sum_{\lambda=\sigma_{lmp}}^{v_{lmp}}\sum_{r=0}^{\alpha_{lmp} +\lambda} (-1)^{\lambda+r}\binom{2(l-p)}{\lambda}\nonumber \\ &\quad \times \binom{2p}{l-m-\lambda}\binom{\alpha_{lmp}+\lambda}{r}\left(\cos{\frac{i}{2}}\right)^{3l-m-2p - \abs{l-2p+m} -2\lambda + 2r}
\end{align}
where
\begin{equation}
    \alpha_{lmp} = \frac{2p - (l-m) - \delta}{2}.
\end{equation}
We then note that
\begin{equation}
    3l-m-2p-\abs{l-2p+m} = 2 v_{lmp}
\end{equation}
which seems suggestive of the re-indexing onto a new indexing variable $h$ through the transformation $\lambda\mapsto v_{lmp}-h$. Indeed:
\begin{align}
    F_{lmp}(i) &= \frac{(l+m)!\left(1-\cos^2{i}\right)^{\delta/2}\left(\cos^2{\frac{i}{2}}\right)^{\frac{\abs{l-2p+m}-\delta}{2}}}{2^{l+\delta} p! (l-p)!} \sum_{h=0}^{v_{lmp} - \sigma_{lmp}}\sum_{r=0}^{\alpha_{lmp} + v_{lmp} - h} (-1)^{v_{lmp}+r - h}\nonumber \\ &\quad \times\binom{2(l-p)}{v_{lmp}-h} \binom{2p}{l-m-v_{lmp}+h}\binom{\alpha_{lmp}+v_{lmp}-h}{r}\left(\cos^2{\frac{i}{2}}\right)^{h+r}.
\end{align}
This form suggests that we should be able to reindex to a power series in terms of $\cos^2{i/2}$ running over the exponents $[0, \alpha_{lmp}+v_{lmp}]$. Indeed, adding up all terms contributing to the power $n$ (which we will now use as indexing variable) and doing some reindexing we can hence derive the following expression:
\begin{align}
    F_{lmp}(i) = \left(-1\right)^{v_{lmp}}\frac{\left(1-\cos^2{i}\right)^{\delta/2}\left(\cos^2{\frac{i}{2}}\right)^{\frac{\abs{l+m-2p}-\delta}{2}}}{2^{l+\delta}}\sum_{n=0}^{\alpha_{lmp}+v_{lmp}}a_{lmpn}\left(\cos^2{\frac{i}{2}}\right)^{n}
\end{align}
where
\begin{align}
    a_{lmpn} &= \nonumber \\
    &\quad (-1)^{n}(l+m)_m \binom{l}{p}\sum_{r=0}^{n}\binom{2(l-p)}{v_{lmp}-r}\binom{2p}{l-m-v_{lmp}+r}\binom{\alpha_{lmp}+v_{lmp}-r}{n-r}
\end{align}
and where $(l+m)_m$ denotes the falling factorial. As a sanity check, we note that here too the summation bounds can in principle be ignored as the binomial coefficients evaluate to zero for all summation indices outside these bounds. This expression allows for the coefficients $a_{lmpn}$ to be pre-computed (noting additionally that the $a_{lmpn}$ are integral, meaning they can be evaluated to arbitrary precision); to evaluate the inclination functions, one then only needs the value of $\cos^2{i/2}$. This can then easily be derived from the state variable $\cos{i}$:
\begin{equation}
    \cos^2{\frac{i}{2}} = \frac{1}{2}\left(1+\cos i\right).
\end{equation}
In principle, one could use this relation to find a power series in $\cos{i}$: the resulting power series will have rational, not integral coefficients, however, and contain as many terms as this series. Consequently, we deem this series preferable. Finally, as we need the square of the inclination functions, we express it as:
\begin{align}
    F_{lmp}^2(i) = 2^{-2l-2\delta}\left(1-\cos^2{i}\right)^{\delta}\left(\cos^2{\frac{i}{2}}\right)^{\abs{l+m-2p}-\delta}\left[\sum_{n=0}^{\alpha_{lmp}+v_{lmp}}a_{lmpn}\left(\cos^2{\frac{i}{2}}\right)^{n}\right]^2
\end{align}
where not expanding the square prevents the number of terms involved in the sum from growing as $l^2$.

\section{The tidal dissipation in transition between equilibrium rotation states}
\label{sec:equilibrium_dissipation_transition}
In computing the long-term evolution according to the equations of motion, we assume that once Triton has reached some rotational equilibrium, it will remain in just such a state (though we do compute to which rotational equilibrium state this corresponds at every timestep). Consequently, our long-term equations of motion assume that $\ddot{\theta}_j$ and as a result $\partial{U_{j}}/\partial{\Omega_j}$ are zero. While this greatly expedites the process of numerical integration, this also means that we neglect the short-term variations due to the jumps between rotational equilibrium states that go into the tidal heating rate. In particular, we note that we can write the following for the tidal heating rate in body $j$ (with partner body $k$), as can be derived from Eq.~\ref{eq:partial_potential}:
\begin{equation}
    \dot{E}_j = -n^2a^2\beta \frac{M_k}{M_j} \left[\pdv{U_{j}}{\varpi_j} \dot{\varpi}_j + \pdv{U_{j}}{\mathcal{M}} n + \pdv{U_{j}}{\Omega_j} \left(\dot{\Omega}_j - \dot{\theta}_j\right)\right].
\end{equation}
If we assume that the contributions due to the nodal and apsidal rates are negligible, what then remains can be written as follows:
\begin{align}
    \dot{E}_j &\approx -n^2a^2\beta \frac{M_k}{M_j} \left[\pdv{U_{j}}{\mathcal{M}} n - \pdv{U_{j}}{\Omega_j} \dot{\theta}_j\right] \nonumber \\
    & = -\frac{n^2a\beta}{2}\left<\derivative{a}{t}\right>_j - C_j\dot{\theta}_j\left<\derivative{\dot{\theta}_j}{t}\right>
\end{align}
where
\begin{equation}
    \left<\derivative{a}{t}\right>_j = 2na\frac{M_k}{M_j}\left<\pdv{U_j}{\mathcal{M}}\right>
\end{equation}
is the contribution to the semi-major axis rate due to the tidal potential of body $j$. We can then, assuming that any mass changes or rearrangements in the body occur only on sufficiently large timescales, write:
\begin{align}
    \dot{E}_j &\approx -\left<\derivative{}{t}\left(\frac{-\mathcal{G}M_jM_k}{2a}\right)\right>_j - \left<\derivative{}{t}\left(\frac{1}{2}C_j\dot{\theta}^2_j\right)\right>.
\end{align}
That is, the tidal heating rate is the sum of the rate of loss of the potential energy of body $k$ in the tidal potential of body $j$ and the rate of loss of the rotational energy of body $j$. The change of the first term occurs on large timescales, and so is retained in the long-period formulation of the equations of motion; the second term, however, is assumed zero implicitly when the rotation rate is set to an equilibrium vale. Nonetheless, in the jumps between resonance states this term is significant: we will therefore have to manually reintroduce it in the long-term equations of motion, by adding the difference in rotational energy between successive resonances to the dissipated energy. We additionally correct the tidal heating \textit{rate} by adding the resulting average dissipation over the timestep in which the transition takes place: this means that our tidal heating rates become underestimates.

\section{Verification expressions for the equations of motion}
\label{sec:verification_expressions}
To find simplified expressions that can be used to verify the proper workings of the full implementation of the equations of motion, we derive two sets of additional formulae to complement those derived by \citet{Boue2019TidalElements}: Sec.~\ref{subsec:potential_derivatives_retrograde} presents a relation between the potential derivatives for a retrograde orbit and equivalent quantities for a prograde orbit, and Sec.~\ref{subsec:approx_rotational_rate} comprises the derivation of an expression for the time evolution of the rotational rate for low eccentricities and inclinations comparable to those derived for the semi-major axis, inclination and eccentricity in \citet{Boue2019TidalElements}. Additionally, we derive some analytic solutions to the equations of motion under some simplifying assumptions.

\subsection{The potential derivatives and equations of motion for retrograde orbits}
\label{subsec:potential_derivatives_retrograde}
While in principle the full expressions presented in Sec.~\ref{sec:darwin-kaula_expansion} are already valid for retrograde orbits, simplified expressions in literature are oftentimes only given for prograde orbits. Fortunately, symmetry inherent to the formulation by \citet{Kaula1964TidalEvolution} allows us to rewrite expressions for retrograde orbits such that they can be related to those for prograde orbits. Amongst others, this allows us to use the low-eccentricity, low-inclination expressions given by \citet{Boue2019TidalElements} to derive expressions for near-equatorial, low-eccentricity retrograde orbits: these will allow us to verify the proper working of our code in a domain more representative of the Neptune-Triton system.

The starting point here is a symmetry present for the inclination functions $F_{lmp}(i)$, as given by \citet{Gooding2008OnDerivatives} (rewritten into our notation):
\begin{equation}
    F_{lmp}(i) = (-1)^{l-m}F_{lm(l-p)}(\pi-i) \implies F^2_{lmp}(i) = F^2_{lm(l-p)}(\pi-i).
\end{equation}
Introducing the ``retrograde inclination'' $i'=\pi-i$, as well as reindexing to $p'=l-p$, we can then rewrite the inclination functions for the retrograde orbit with inclination $i$ into the equivalent of those for a prograde one with inclination $i'$:
\begin{equation}
    F^2_{lmp}(i) = F^2_{lmp'}(i').
\end{equation}
Similarly, the eccentricity functions can be rewritten to match this new index $p'$: it is also convenient here to introduce the index $q'=-q$. Then:
\begin{equation}
    G_{lpq}(e)=X^{-(l+1),l-2p}_{l-2p+q}(e)=X^{-(l+1),-(l-2p')}_{-(l-2p'+q')}(e)=X^{-(l+1),l-2p'}_{l-2p'+q'}(e)=G_{lp'q'}(e).
\end{equation}
Finally, we can rewrite the Fourier modes $\omega_{j,lmpq}$:
\begin{equation}
    \omega_{j,lmpq} = -\omega_{j,l(-m)p'q'}
\end{equation}
which, we will notice, corresponds to $\omega_{j,lmp'q'}$ if we insert $-\dot{\theta}_j$ for $\dot{\theta}_j$. Consequently, we can now write Eq.~\ref{eq:partial_potential} as:
\begin{align}
\label{eq:partial_potential_retrograde}
    \begin{bmatrix}
    \partial{U_{j}}/\partial{\mathcal{M}} \\
    \partial{U_{j}}/\partial{\varpi_j} \\
    \partial{U_{j}}/\partial{\Omega_j} \\
    \dot{E}_j
    \end{bmatrix} =& -\sum_{l\geq2}\left(\frac{R_j}{a}\right)^{2l+1}\sum_{m=0}^{l}  \frac{(l-m)!}{(l+m)!}(2-\delta_{0m}) \sum_{p'=0}^{l} F_{lmp'}^2(i'_j) \nonumber \\
    &\times\; \sum_{q'=-\infty}^{\infty}G_{lp'q'}^2(e) K_{l,j}(\omega_{j,l(-m)p'q'})
    \begin{bmatrix}
        l-2p'+q' \\
        l-2p' \\
        -m \\
        -\omega_{j,l(-m)p'q'}n^2a^2\beta M_k/M_j
    \end{bmatrix}
\end{align}
where we have used the fact that $K_{j,l}(\omega)$ is an odd function. It should be noted that this expression is is identical to that for the prograde case, with the exception of the sign accompanying the terms involving $\dot{\theta}_j$ and the sign of $\pdv{U_j}{\varpi_j}$. It then follows straightforwardly that the expressions for the derivatives of the semi-major axis and eccentricity remain unchanged when expressed in terms of the retrograde inclination rather than the prograde inclination, with the exception of the sign of the $\dot{\theta_j}$-term in the quality function (where $j$ is the body whose obliquity is retrograde: confusingly, in Kaula's formulation, this is Neptune, not Triton), which is flipped. While it requires some additional algebra, it can also be shown that the time-derivative of the retrograde inclination $i'_j$ in the retrograde case follows an identical relation to the prograde inclination $i_j$ in the prograde case, but again with the sign of $\dot{\theta}_j$ flipped. The equations of motion of the retrograde case can thus in general be obtained from those for the prograde case by interchanging the retrograde inclination $i'_j=\pi-i_j$ for the proper prograde inclination $i_j$ and flipping the sign of $\dot{\theta_j}$: one can check that this corresponds to invariance of the equations with respect to a mirroring in the plane formed by the line of nodes and the spin axis of the primary, as expected. In particular, this means that the leading-order expressions derived by \citet{Boue2019TidalElements} for the evolution of the semi-major axis, the eccentricity and the inclination can be transformed straightforwardly into equivalent expressions for a near-equatorial, low-eccentricity retrograde orbit by flipping the sign of $\dot{\theta}_j$.

\subsection{An approximate expression for the derivative of the rotational rate}
\label{subsec:approx_rotational_rate}
While \citet{Boue2019TidalElements} did not derive an expression for the evolution of the rotational rate at low eccentricities and low inclinations, we will find that such an expression will prove useful in verification of our code. Therefore, we will derive such an expression in a manner analogous to the derivation of the leading-order expressions given by \citet{Boue2019TidalElements}. We start from the expression for the $\pdv{U_j}{\Omega_j}$ as given in Eq.~\ref{eq:partial_potential}, copied here for convenience:
\begin{align}
    \pdv{U_{j}}{\Omega_j} =& -\sum_{l\geq2}\left(\frac{R_j}{a}\right)^{2l+1}\sum_{m=0}^{l}  \frac{(l-m)!}{(l+m)!}(2-\delta_{0m}) \sum_{p=0}^{l} F_{lmp}^2(i_j) \nonumber \\
    &\times\; \sum_{q=-\infty}^{\infty}G_{lpq}^2(e) K_{l,j}(\omega_{j,lmpq})m
\end{align}
To $\mathcal{O}(i_j^2)$, the only non-zero squares of the inclination functions are given by $(lmp)=(201)$ and $(220)$: the factor of $m$ in the sum for $\pdv{U_j}{\Omega_j}$ eliminates the contribution of $(lmp)=(201)$, and so we are left with the term due to $(lmp)=(220)$, which is $F_{220}^2=9+\mathcal{O}(i^2)$. In general, close to $e=0$ we have that $G_{lpq}=\mathcal{O}(e^{|q|})$, and so accounting for terms up to (but not including) $\mathcal{O}(e^6)$ like in the other expressions given by \citet{Boue2019TidalElements} requires that we include $q\in\{-2, -1, 0, 1, 2\}$. The squares of these eccentricity functions are given as follows (as can be verified from \citealt{Boue2019TidalElements} and \citealt{Renaud2021TidalTRAPPIST-1e}):
\begin{align}
    G_{20(-2)}^2&=0 \nonumber \\
    G_{20(-1)}^2&=\frac{e^2}{4}-\frac{e^4}{16} + \mathcal{O}(e^6) \nonumber \\
    G_{200}^2&=1-5e^2+\frac{63}{8}e^4 + \mathcal{O}(e^6) \nonumber \\
    G_{201}^2&=\frac{49}{4}e^2-\frac{861}{16}e^4 + \mathcal{O}(e^6) \nonumber \\
    G_{202}^2&=\frac{289}{4}e^4 + \mathcal{O}(e^6).
\end{align}
Consequently, we may write for the rotational acceleration:
\begin{align}
\label{eq:rotational_acceleration_approx}
    \left<\derivative{\dot{\theta}_j}{t}\right> =& \frac{n^2R_j^5\beta M_k}{a^3C_jM_j}\left[\frac{3e^2}{8}\left(1-\frac{e^2}{4}\right)K_{2j}(n-2\dot{\theta}_j)\right.\nonumber \\
    &\;+\frac{3}{2}\left(1-5e^2+\frac{63}{8}\right)K_{2j}\left(2(n-\dot{\theta}_j)\right)\nonumber \\
    &\;+\frac{21}{8}e^2\left(7-\frac{123}{4}e^2\right)K_{2j}(3n-2\dot{\theta}_j) \nonumber \\
    &\left.\;+\frac{867}{8}e^4K_{2j}\left(2(2n-\dot{\theta}_j)\right)\right] + \mathcal{O}(e^6) + \mathcal{O}(i_j^2).
\end{align}
By the reasoning laid out in Sec.~\ref{subsec:potential_derivatives_retrograde}, we can then also derive the rotational acceleration for an object orbited (not orbiting: recall that $i_j$ is the obliquity of the object) in a retrograde fashion:
\begin{align}
    \left<\derivative{\dot{\theta}_j}{t}\right>^{\textrm{retro}} =& -\frac{n^2R_j^5\beta M_k}{a^3C_jM_j}\left[\frac{3e^2}{8}\left(1-\frac{e^2}{4}\right)K_{2j}(n+2\dot{\theta}_j)\right.\nonumber \\
    &\;+\frac{3}{2}\left(1-5e^2+\frac{63}{8}e^4 \right)K_{2j}\left(2(n+\dot{\theta}_j)\right)\nonumber \\
    &\;+\frac{21}{8}e^2\left(7-\frac{123}{4}e^2\right)K_{2j}(3n+2\dot{\theta}_j) \nonumber \\
    &\left.\;+\frac{867}{8}e^4K_{2j}\left(2(2n+\dot{\theta}_j)\right)\right] + \mathcal{O}(e^6) + \mathcal{O}(i_j^{'2}).
\end{align}
Under the assumption of constant time lag, Eq.~\ref{eq:rotational_acceleration_approx} becomes particularly simple:
\begin{align}
\label{eq:rotational_acceleration_CTL}
    \left<\derivative{\dot{\theta}_j}{t}\right>^{\textrm{CTL}} =& 3k_{2j}\Delta t_j\frac{n^2R_j^5\beta M_k}{a^3C_jM_j}\left[\left(1+81e^2+\frac{11187}{16}e^4\right)n - \left(1+\frac{45}{2}e^2+\frac{315}{4}e^4\right)\dot{\theta}_j\right]
\end{align}
which has an analytic solution if we assume that the timescale of change of $\dot{\theta}_j$ is sufficiently small that the other orbital elements can be taken constant:
\begin{equation}
    \dot{\theta}_j^{\textrm{CTL}}(t)=\dot{\theta}_{\textrm{eq}} + \left(\dot{\theta}_{\textrm{eq}}-\dot{\theta}(0)\right)\exp\left(-Bt\right)
\end{equation}
where
\begin{align}
    B &= 3k_{2j}\Delta t_j\frac{n^2R_j^5\beta M_k}{a^3C_jM_j}\left(1+\frac{45}{2}e^2+\frac{315}{4}e^4\right) \nonumber \\
    \dot{\theta}_{\textrm{eq}} &= \left(1+\frac{6+\frac{3309}{16}e^2}{1+\frac{15}{2}e^2+\frac{105}{4}e^4}e^2\right)n \sim (1+6e^2)n
\end{align}
which is in agreement with the equilibrium value for the constant time lag-model found in literature (e.g. \citealt{Makarov2013NoMoons}). To first order, we can thus derive that the characteristic timescale of change of the rotational rate at low eccentricities is given by
\begin{equation}
    \tau \approx B^{-1} \sim \frac{1}{30\pi^2}\left(\frac{a}{R_j}\right)^3\left(\frac{M_j}{M_k}\right)\left(\frac{P}{\Delta t_j}\right) P.
\end{equation}

\subsection{An analytic expression for the damping of the inclination}
For a constant time lag-rheology, homogeneous secondary at synchronous rotation for very low eccentricities, the expression given for the inclination rate by \citet{Boue2019TidalElements} can be further simplified to
\begin{equation}
    \derivative{i_j}{t} = -Ai_j
\end{equation}
where
\begin{equation}
    A=\frac{3}{2}n^2\frac{M_k}{M_j}\left(\frac{R_j}{a}\right)^5k_{2j}\Delta t_j \left(1+\frac{5\beta a^2}{2M_j R_j^2}\right)
\end{equation}
where we have used additionally that $\sin{i_j}\approx i_j$. This has an analytic solution of the form
\begin{equation}
    i_j(t) = i_j(0)\exp{-At}.
\end{equation}

\subsection{An analytic expression for the damping of the eccentricity}
For a constant time-lag rheology, homogeneous secondary at synchronous rotation and low eccentricity, and where the mass of the secondary is negligible compared to that of the primary, the eccentricity damping is given by Eq.~(159) in \citet{Boue2019TidalElements}:
\begin{equation}
    \derivative{e}{t} = -Be
\end{equation}
where
\begin{equation}
    B = \frac{21}{2}n^2\frac{M_N}{M_T}\left(\frac{R_T}{a}\right)^5k_{2T}\Delta t_T.
\end{equation}
Assuming that the semi-major axis evolution is small, $B$ remains roughly constant. In that case, an analytic expression exists for $e(t)$:
\begin{equation}
\label{eq:eccentricity_damping}
    e(t) = e_0\exp{-Bt}
\end{equation}
where $e_0=e(t=0)$.

\subsection{An analytic expression for the evolution of the semi-major axis}
Under the same assumptions as those used to derive the eccentricity damping, we can use Eq.~(151b) in \citet{Boue2019TidalElements} to describe the evolution of the semi-major axis:
\begin{equation}
    \derivative{a}{t} = -C\left[D+Ee^2\right]a
\end{equation}
where
\begin{align}
    C &= 3n\frac{M_T}{M_N}\left(\frac{R_N}{a}\right)^5 \\
    D &= 2k_{2N}\Delta t_N (n-\dot{\theta}_N) \\
    E &= 7\left(\frac{R_T}{R_N}\right)^5\left(\frac{M_N}{M_T}\right)^2 k_{2T}\Delta t_{T}n.
\end{align}
Note that we have a factor of $7$ rather than $19$ in the definition of $E$: this discrepancy with the result given by \citet{Boue2019TidalElements} is a result of the fact that pseudo-synchronous rotation in a constant time lag-body results in $\dot{\theta_j}-n=6e^2n$ rather than $\dot{\theta_j}-n=0$ in the equilibrium case. The addition of this term results in the factor of $-21$ rather than $-57$ as \citet{Boue2019TidalElements} remark is, they thought, erroneously present in various literature. With $e(t)$ as given by Eq.~\ref{eq:eccentricity_damping}, this becomes a separable differential equation with the solution as follows:
\begin{align}
    a(t) &= a(0)\exp{-CDt - \frac{CEe_0^2}{2B}\left(1-\exp{-2Bt}\right)} \\
    &= a(0)\exp{-CDt - \frac{CE}{2B}\left(e_0^2-e(t)^2\right)}.
\end{align}
It should be noted that the assumption that $n$ is constant here is inherently contradictory, meaning that this expression should only be trusted whenever the evolution of $a$ is small.
\chapter{Auxiliary results}
\label{app:auxiliary_results}
Some additional results have been generated that the author deems useful to include, but not of sufficient importance to include in the main body. These are appended here.

\section{Degree truncation for $l\leq7$}
Aside from the runs presented in Sec.~\ref{sec:truncating_degree}, another set of runs was performed up to and including $l_{\max}=7$, which seem to agree with the results presented in these sections. Additionally, these figures seem to indicate that no significant differences exist between the $l_{\max}=6$ and $l_{\max}=7$ cases, which would suggest that all higher terms can certainly be neglected. As the convergence of the eccentricity functions and the validity of the empirical upper bound on $|q_{\max}|$ presented in Sec.~\ref{sec:truncating_kaula_general} out to $e\approx0.74$ was only validated for $l_{\max}\leq 5$, we cannot describe equal weight to these results as to those presented in Ch.~\ref{ch:validation}, however.
\label{sec:appendix_truncation_l7}
\begin{figure}
    \centering
    \includegraphics[width=1\linewidth]{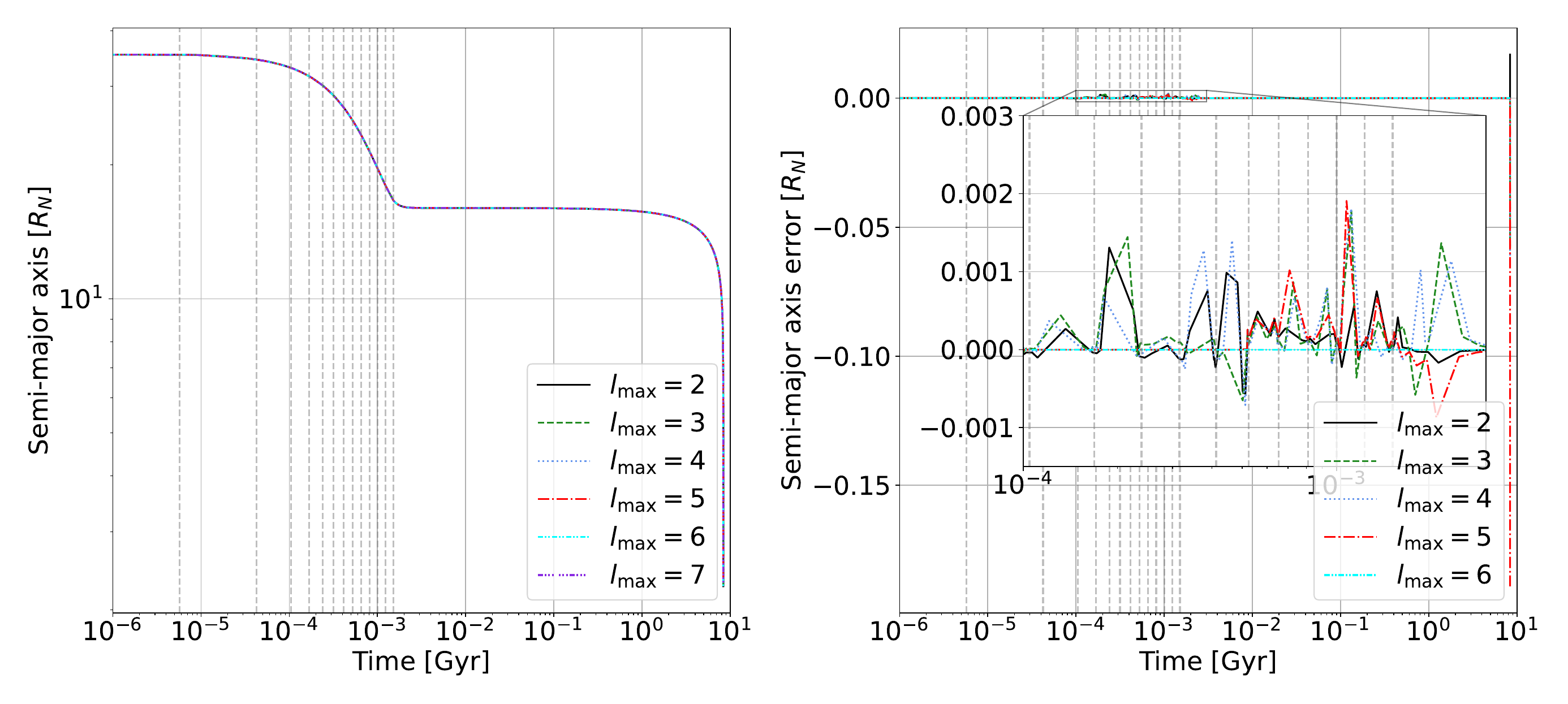}
    \caption{Evolution of the semi-major axis for the reference scenario described in Sec.~\ref{sec:truncating_degree} up to $l_{\max}=7$, as well as the error with respect to the $l\leq 7$ case. The dashed grey lines indicate the epochs at which Triton's rotational rate drops between resonances.}
    \label{fig:semi_major_axis_l7}
\end{figure}

\begin{figure}
    \centering
    \includegraphics[width=1\linewidth]{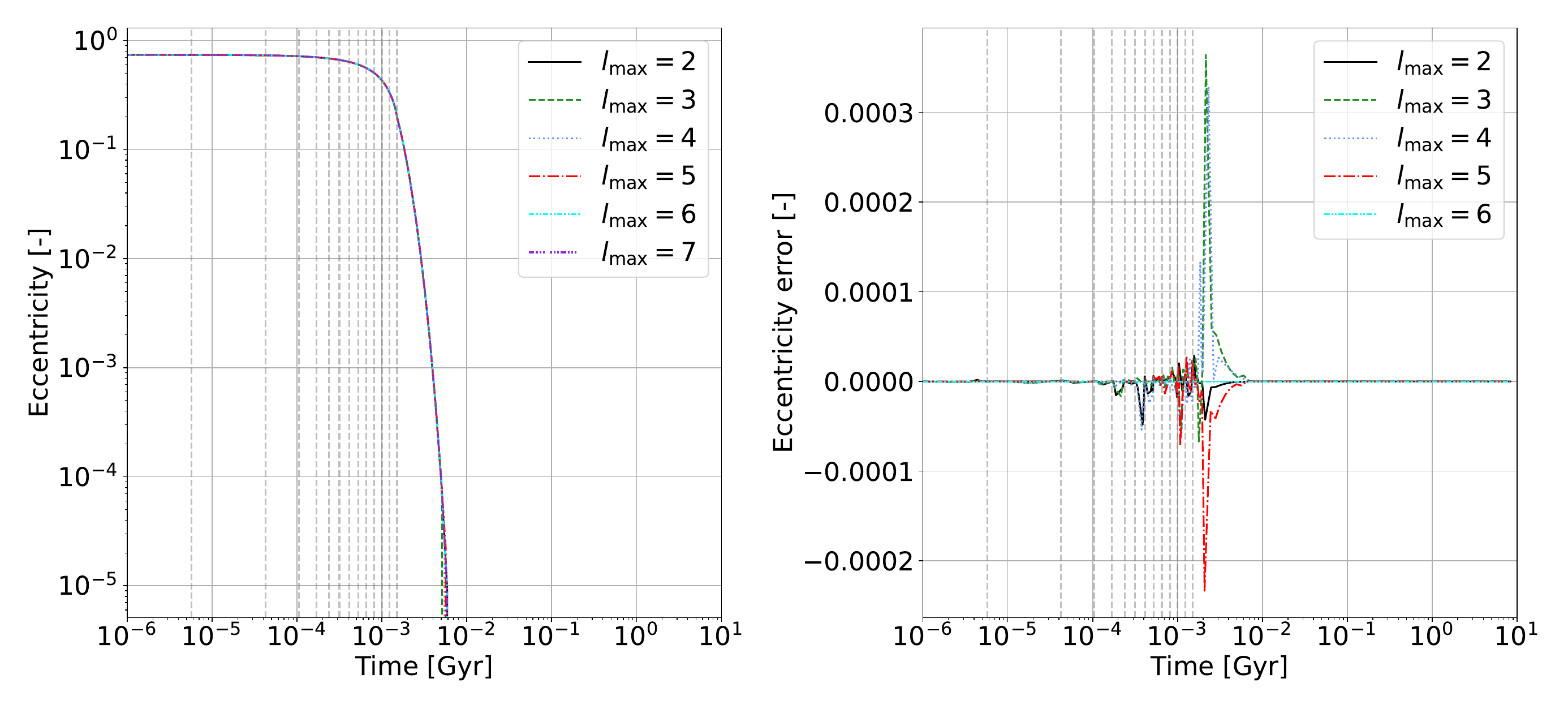}
    \caption{Evolution of the eccentricity for the reference scenario described in Sec.~\ref{sec:truncating_degree} up to $l_{\max}=7$, as well as the error with respect to the $l\leq 7$ case. The dashed grey lines indicate the epochs at which Triton's rotational rate drops between resonances.}
    \label{fig:eccentricity_l7}
\end{figure}

\begin{figure}
    \centering
    \includegraphics[width=1\linewidth]{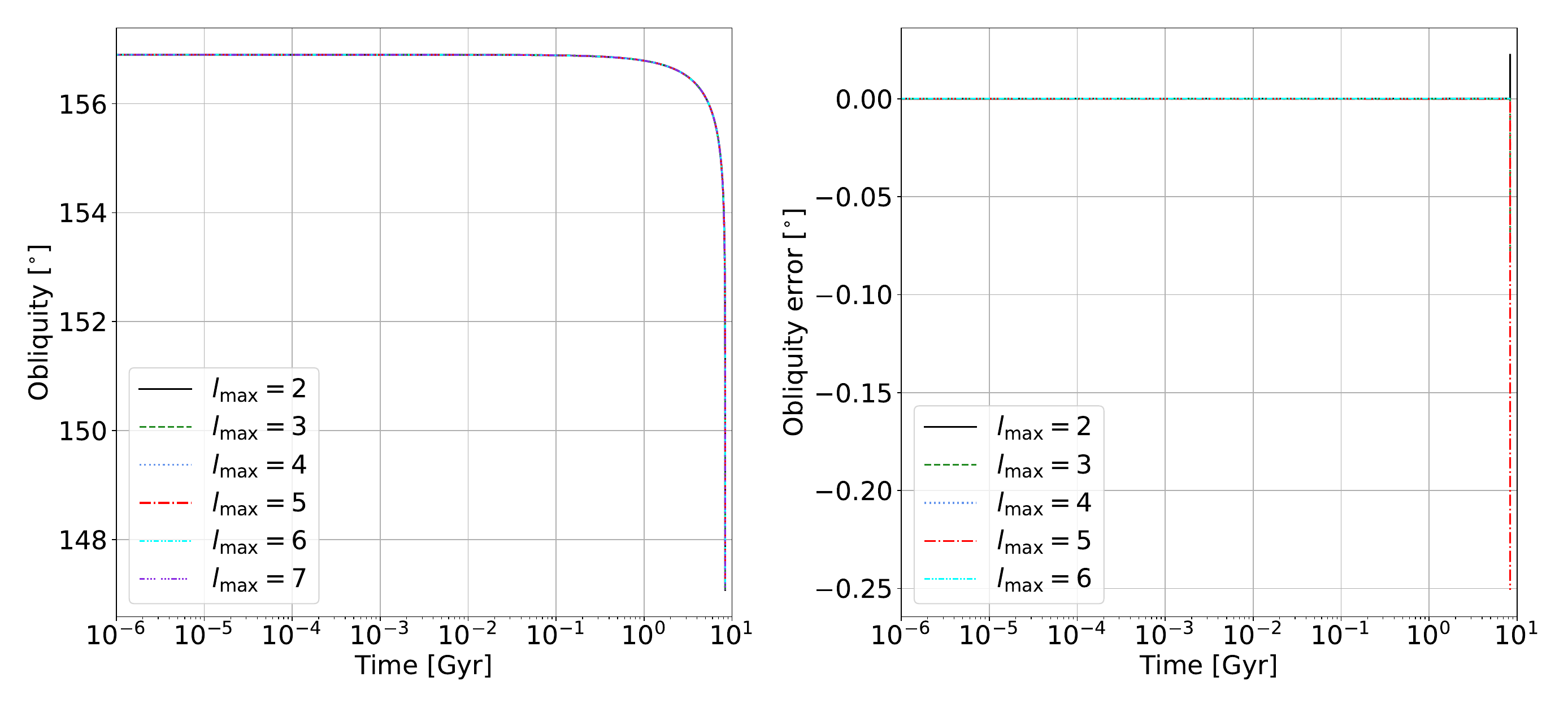}
    \caption{Evolution of the obliquity of Neptune (equivalent to the inclination of Triton's orbit) for the reference scenario described in Sec.~\ref{sec:truncating_degree} up to $l_{\max}=7$, as well as the error with respect to the $l\leq 7$ case. The dashed grey lines indicate the epochs at which Triton's rotational rate drops between resonances.}
    \label{fig:obliquity_l7}
\end{figure}

\begin{figure}
    \centering
    \includegraphics[width=1\linewidth]{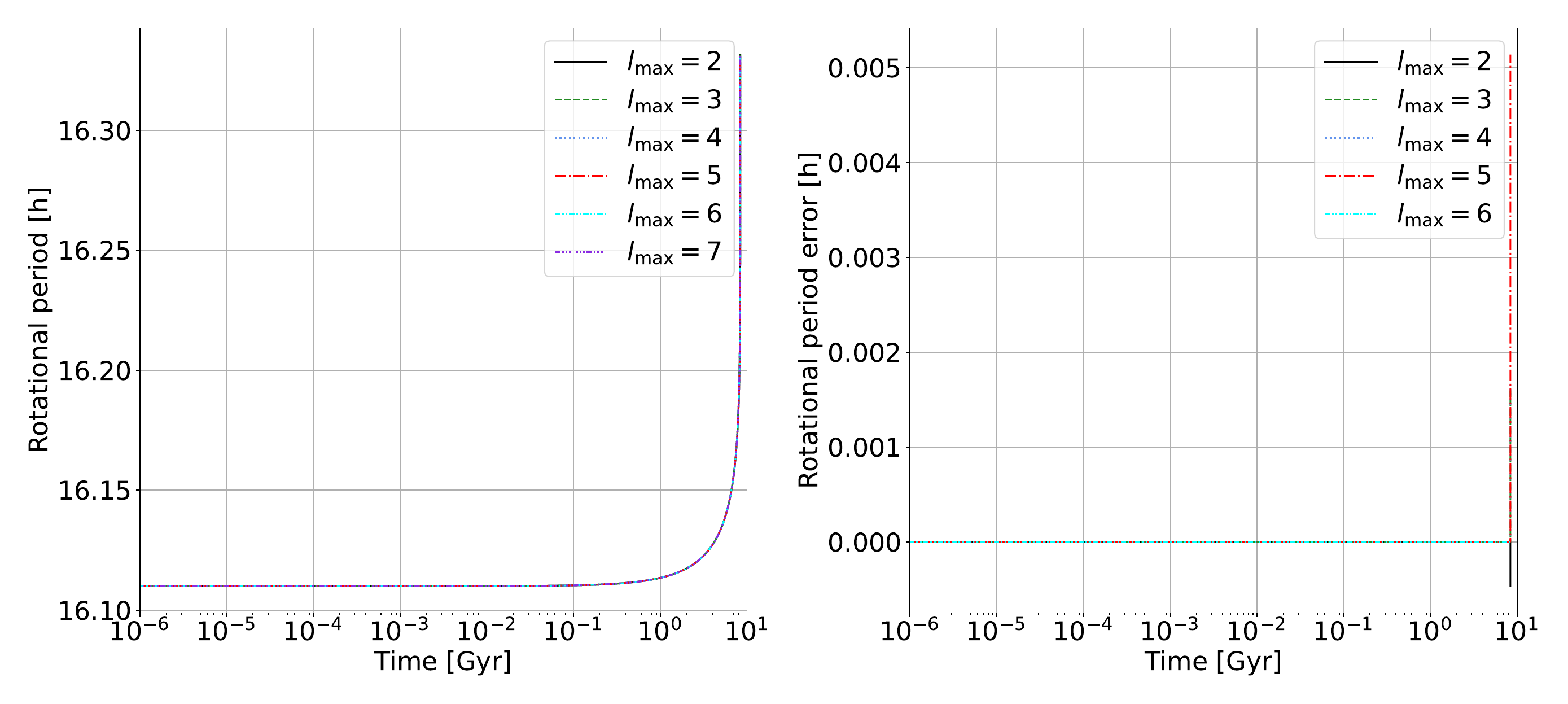}
    \caption{Evolution of the period of revolution of Neptune for the reference scenario described in Sec.~\ref{sec:truncating_degree} up to $l_{\max}=7$, as well as the error with respect to the $l\leq 7$ case. The dashed grey lines indicate the epochs at which Triton's rotational rate drops between resonances.}
    \label{fig:rot_period_l7}
\end{figure}

\begin{figure}
    \centering
    \includegraphics[width=1\linewidth]{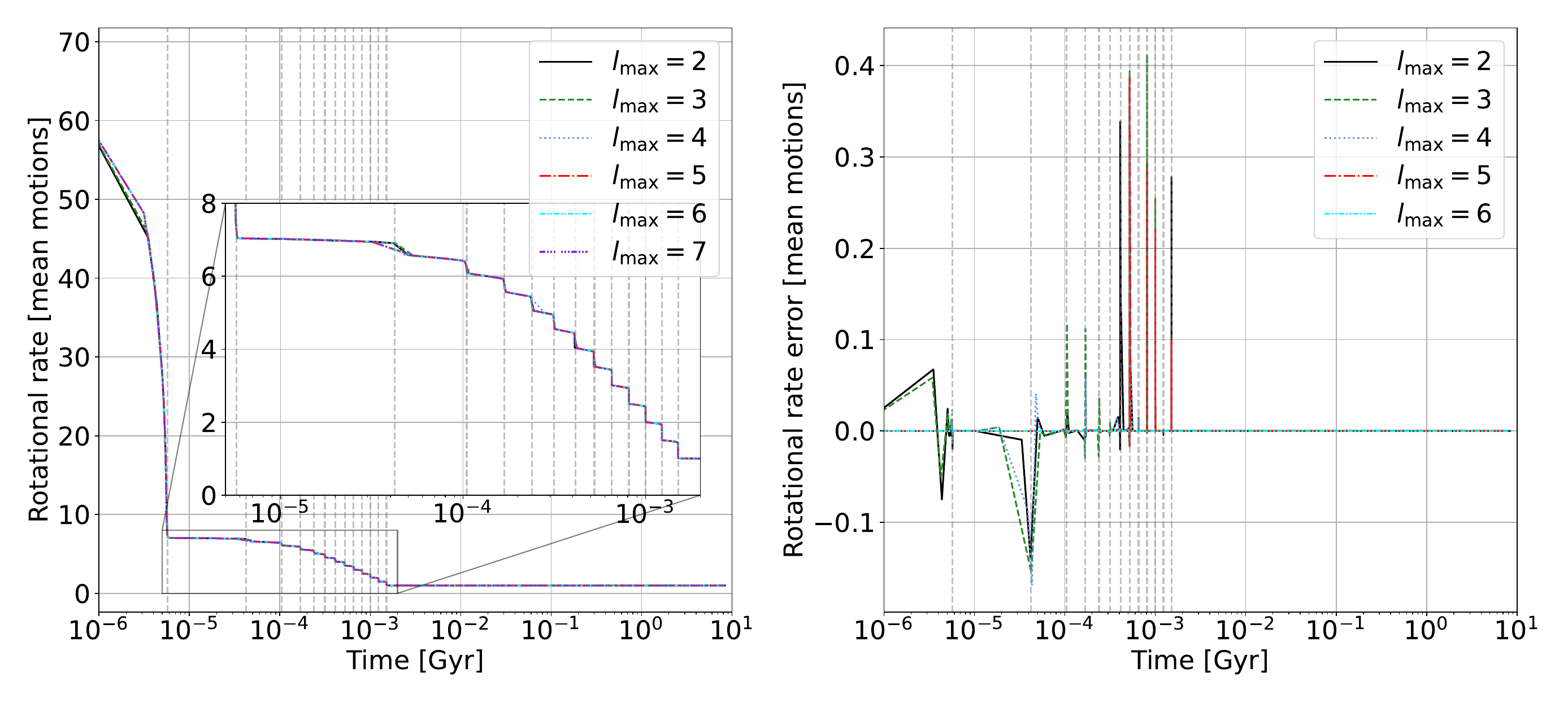}
    \caption{Evolution of the rotational rate of Triton for the reference scenario described in Sec.~\ref{sec:truncating_degree} up to $l_{\max}=7$, as well as the error with respect to the $l\leq 7$ case. The dashed grey lines mark the epochs at which Triton's rotational rate drops between resonances. Note that the discrepant feature that occurs from $\sim 10^{-6}-10^{-5}$ Gyr is a sampling artefact, and not a true discrepancy between the various degrees.}
    \label{fig:rot_rate_l7}
\end{figure}
\chapter{Verification}
\label{ch:verification}
To verify that our implementation of the equations of motion works correctly, we have tested several analytic and simplified expressions that are available in literature or newly derived. The results of this analysis are presented here.

\section{A constant time lag Io-analogue}
For a constant time lag model, analytic expressions to approximate the equations of motion can be derived for low eccentricities and inclinations. While the applicability of this model to real objects has been debated (e.g. \citealt{Makarov2013NoMoons}), this can therefore at least provide a test for the correct working of the equations of motion in our code. To showcase that our code works as intended, we will therefore compare the results it produces for a Triton-Neptune-like system on an Io-analogue orbit (where tidal evolution is expected to be rapid and severe, even for low eccentricities) to the results produced by such analytic expressions and the simplified equations of motion for low eccentricities (up to and including $\mathcal{O}(e^4)$) and inclinations (to first order) given by \citet{Boue2019TidalElements}. To complement their expressions, we also derive such a simplified expression for the rotational rate. The relevant simplified and analytic expressions, as well as how they are derived and transformed to the retrograde case, are presented in Sec.~\ref{sec:verification_expressions}.

\subsection{Damping of the rotation rate and inclination}
In the constant time lag-model, non-equilibrium rotation rates and non-zero obliquities (for sufficiently low obliquities) of the secondary will dampen out exponentially and decoupled from the other equations of motion. We therefore perform a run for a Triton-like moon on an Io-like equatorial, prograde orbit at 6 Neptune radii from a Neptune-like planet with an initial eccentricity of $e=0.05$ and propagate it over a period of 100 years with (1) an initial rotation rate of 5 times its mean motion, at zero obliquity and (2) an initial obliquity of $10^{\circ}$ and initially synchronous rotation. The values of the Love number $k_2$ and the time lag $\Delta t$ are those used by \citet{Nogueira2011ReassessingTriton}: $k_{2T}=0.1$ and $\Delta t_T=808$ s for Triton and $k_{2N}=0.407$ and $\Delta t_{N}=1.02$ s for Neptune. The object masses and radii are those of Neptune and Triton, and for simplicity the moments of inertia have been set to that of a homogeneous sphere.

\begin{figure}
    \centering
    \includegraphics[width=1\linewidth]{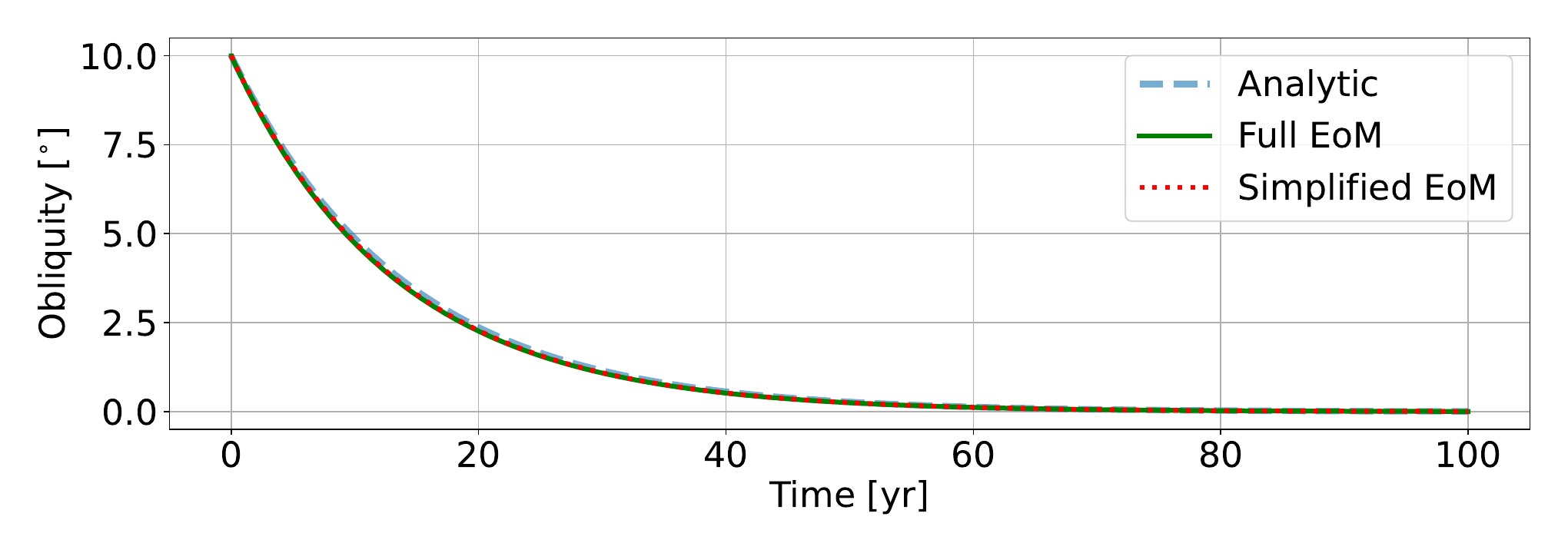}
    \caption{Damping of the obliquity for an Io-analogue using our implementation of the full equations of motion, the simplified equations of motion given by \citet{Boue2019TidalElements} and the analytic solution. Initial conditions and simulation values are given in the text.}
    \label{fig:obliquity_damping}
\end{figure}

\begin{figure}
    \centering
    \includegraphics[width=1\linewidth]{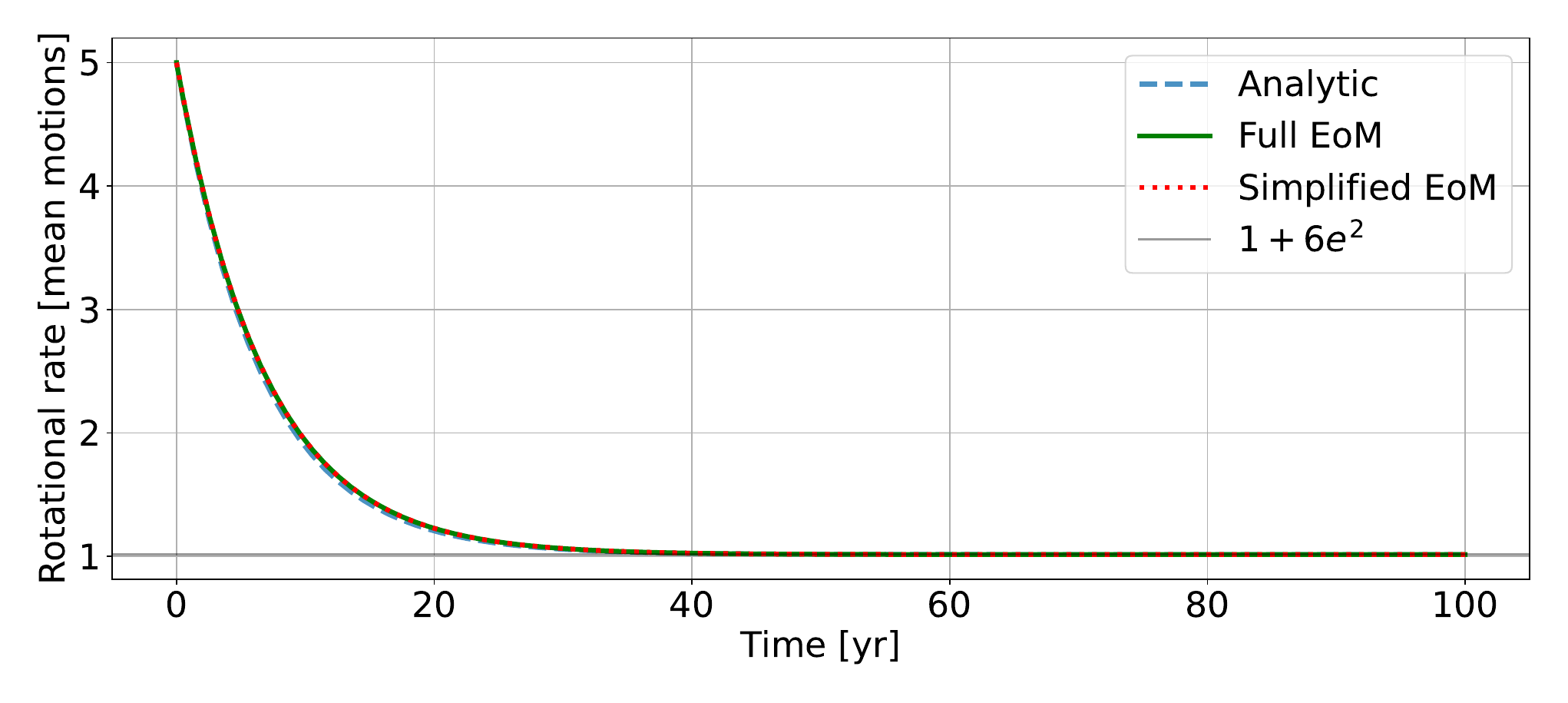}
    \caption{Damping of the rotation rate for an Io-analogue using our implementation of the full equations of motion, the simplified equations of motion given by \citet{Boue2019TidalElements} and the analytic solution. Also shown is the equilibrium value of the rotation rate in the constant time lag model (i.e. pseudosynchronous rotation). Initial conditions and simulation values are given in the text.}
    \label{fig:rotation_damping}
\end{figure}

The results for the two runs are shown in Figs.~\ref{fig:obliquity_damping} and \ref{fig:rotation_damping}; the orbital elements aside from the damped ones are very nearly constant, and therefore omitted. As can be seen, the correspondence between the three implementations is excellent, and the obliquity and rotation rate damp to their expected equilibria.

\subsection{Coupled evolution of the semi-major axis and eccentricity}
While the inclination and rotational rate in the low-eccentricity, low-inclination case evolve sufficiently quickly (and independently from one another) that their evolution can be decoupled from the other orbital elements, this is not the case for the eccentricity and semi-major axis. These quantities evolve on similar timescales and are coupled: their implementation can therefore not be checked separately. Fortunately, at sufficiently low eccentricities and over short timescales (such that the semi-major axis evolves little), their coupled evolution is described by an integrable differential equation (see Sec.~\ref{sec:verification_expressions} for details).

\begin{figure}
    \centering
    \includegraphics[width=1\linewidth]{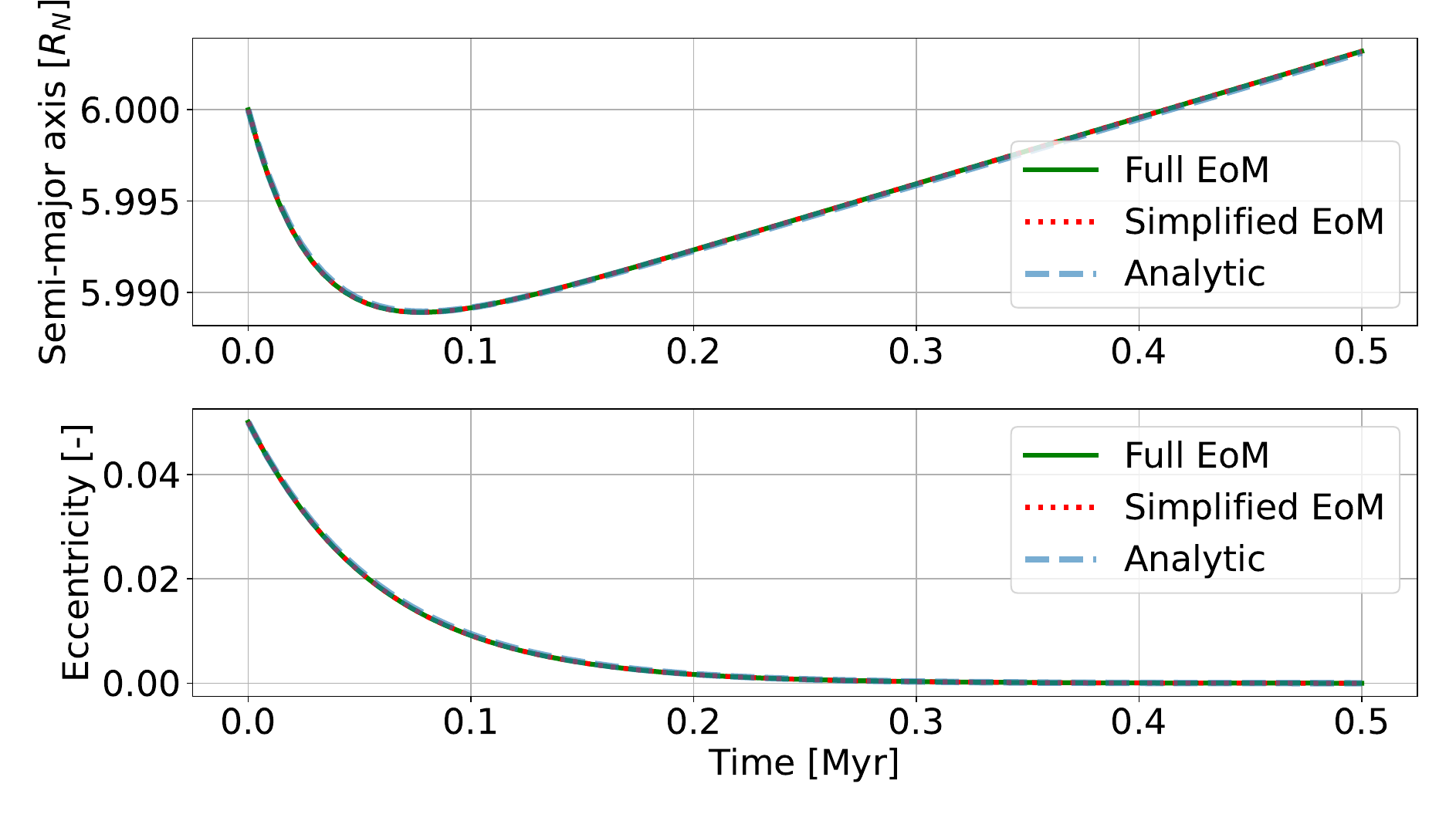}
    \caption{Coupled evolution of the semi-major axis and eccentricity for an Io-analogue on a prograde orbit using our implementation of the full equations of motion, the simplified equations of motion given by \citet{Boue2019TidalElements} and the analytic solution. Initial conditions and simulation values are given in the text.}
    \label{fig:a_e_prograde}
\end{figure}

\begin{figure}
    \centering
    \includegraphics[width=1\linewidth]{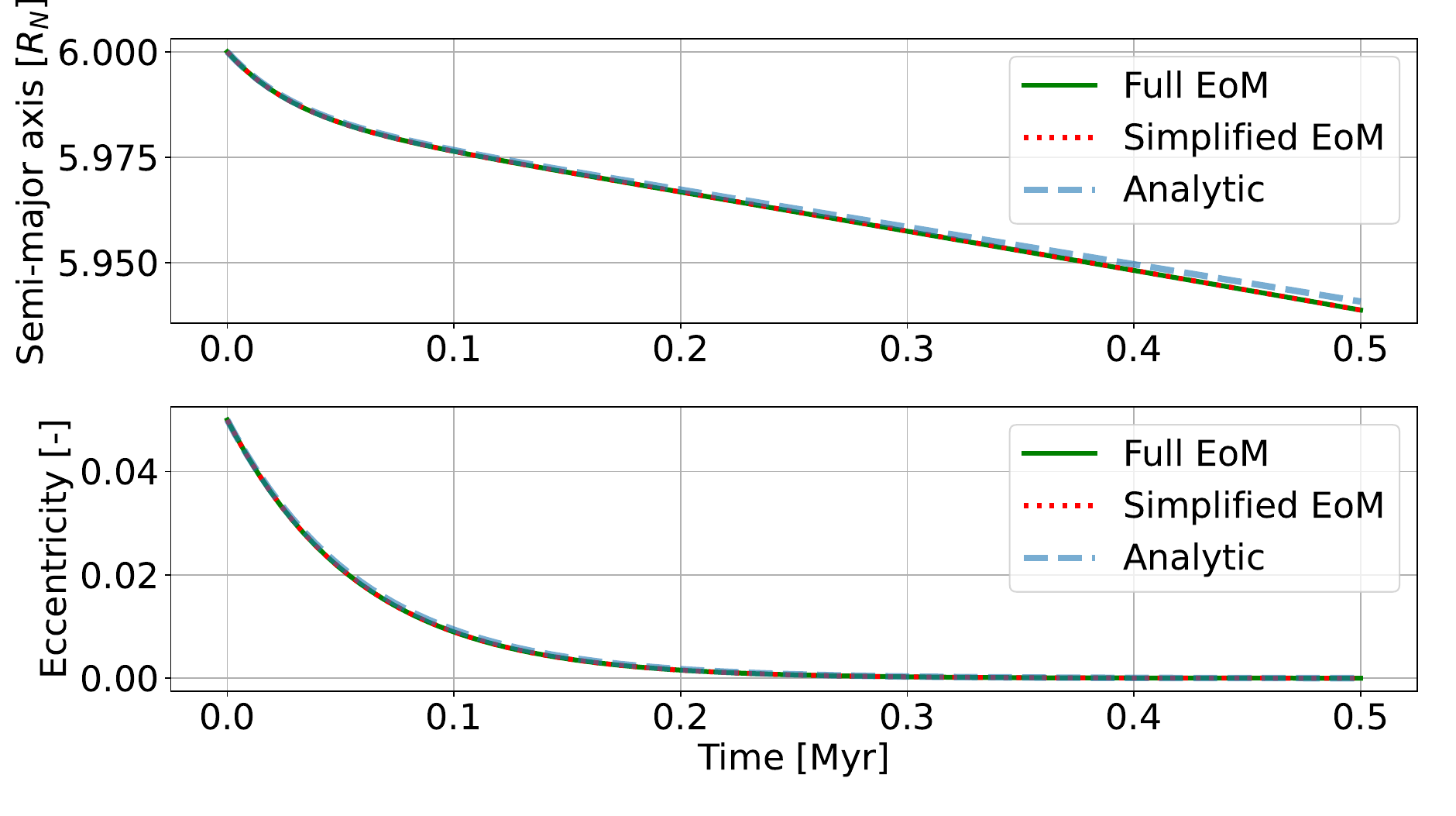}
    \caption{Coupled evolution of the semi-major axis and eccentricity for an Io-analogue on a retrograde orbit using our implementation of the full equations of motion, the simplified equations of motion given by \citet{Boue2019TidalElements} and the analytic solution. Initial conditions and simulation values are given in the text.}
    \label{fig:a_e_retrograde}
\end{figure}

Again, we propagate the full, simplified and analytic expressions and compare them: this time, the satellite starts on an equatorial orbit at $6$ Neptune radii with zero obliquity, an eccentricity of $e=0.05$ and rotating synchronously. As the semi-major axis and eccentricity evolve more slowly than the rotational rate and obliquity, even for such a close-in satellite, we propagate these equations over a duration of $0.5$ Myr: once for a prograde orbit, shown in Fig.~\ref{fig:a_e_prograde}, and once for a retrograde orbit, shown in Fig.~\ref{fig:a_e_retrograde}. The physical properties of the planet and moon are the same as those used for the inclination and obliquity damping.

Here too the agreement between the full, simplified and analytic expressions is excellent. It should be noted, however, that the analytic expression for the semi-major axis deviates significantly if one uses Eq.~(151) in \citet{Boue2019TidalElements} directly: in the constant time lag-case, their expression neglects the term owing to pseudo-synchronous rotation, and so a correction for this fact must be applied. The final, corrected expression is given in Sec.~\ref{sec:verification_expressions}.

\section{A constant time lag Triton-analogue}
While it will then no longer be possible to analytically verify the correct working of our code, we should also wish to check that the code works correctly in more involved scenarios involving the coupled evolution of all quantities, as it is likely that Triton will in the past have been in a state that does not satisfy the strict necessary assumptions for the analytical expressions to hold. Most importantly, its orbit about Neptune is non-equatorial: while it is likely that misalignment of its spin and orbit-normal as well as a non-equilibrium rotation rate may plausibly have damped out quickly after capture, it cannot be assumed that this is the case for the obliquity of Neptune with respect to the orbit.

To be able to verify the correct working of the code in such scenarios nonetheless, we will compare its performance against an implementation using the expressions given by \citet{Boue2019TidalElements} expanded out to $\mathcal{O}(e^6)$ and $\mathcal{O}(i^2)$ by hand, which can therefore be assumed to hold roughly out to $i\sim 10^{\circ}$ and $e\sim 0.1$. To verify that the code works in the retrograde case, too, equivalent expressions are derived for near-equatorial retrograde orbits, as presented in Sec.~\ref{subsec:potential_derivatives_retrograde}. As our code in the constant time lag case has been shown to work correctly in the preceding section, we will also assume a constant time lag-rheology here, with physical parameters identical to those used there.

\subsection{The short-period motion in obliquity and rotational rate}
\begin{figure}
    \centering
    \includegraphics[width=1\linewidth]{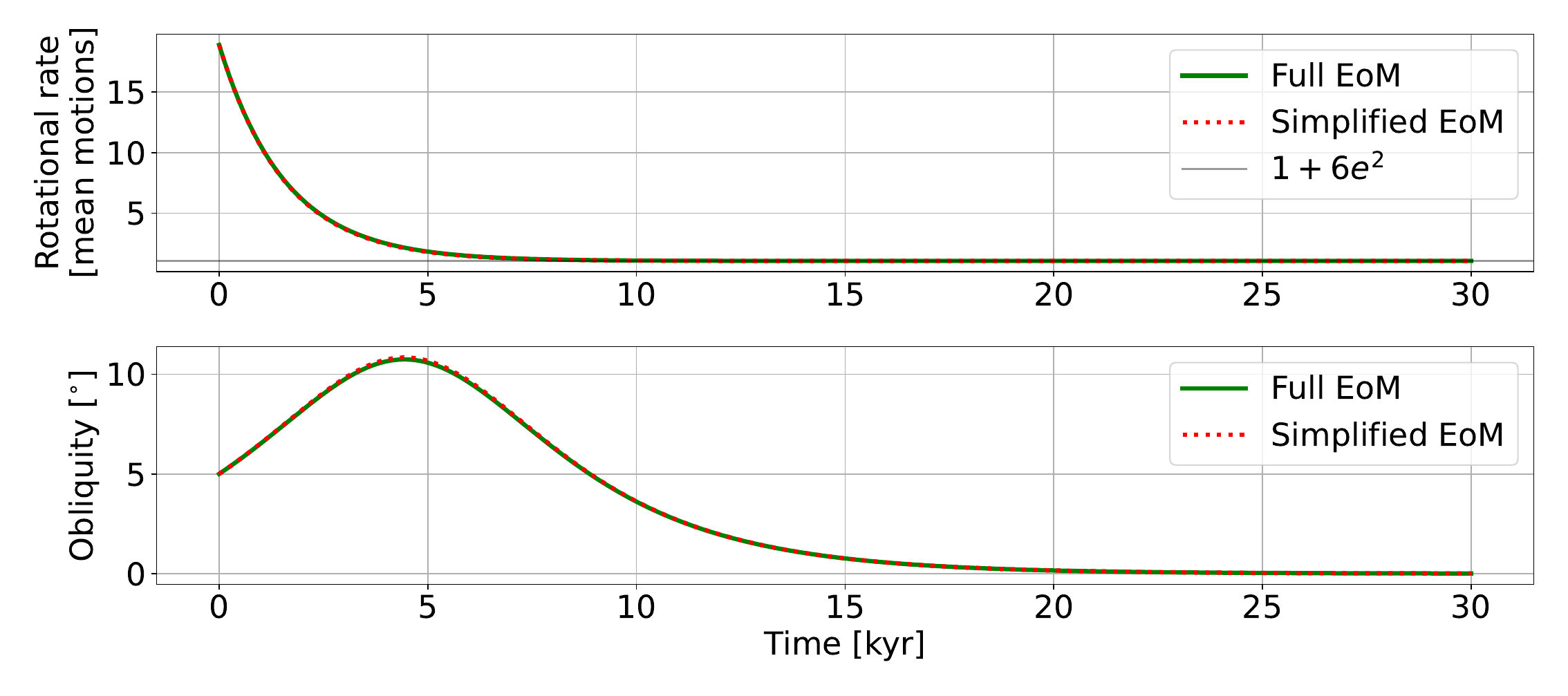}
    \caption{Short-period evolution of the obliquity and rotation rate for a Triton-analogue as propagated using the fully coupled system: both the results using the full equations of motion as well as those using the simplified equations of motion given by \citet{Boue2019TidalElements} are presented. An equivalent retrograde scenario was also propagated, but its short-period evolution does not differ significantly and it is therefore not presented.}
    \label{fig:short_period_angles}
\end{figure}

\begin{figure}
    \centering
    \includegraphics[width=1\linewidth]{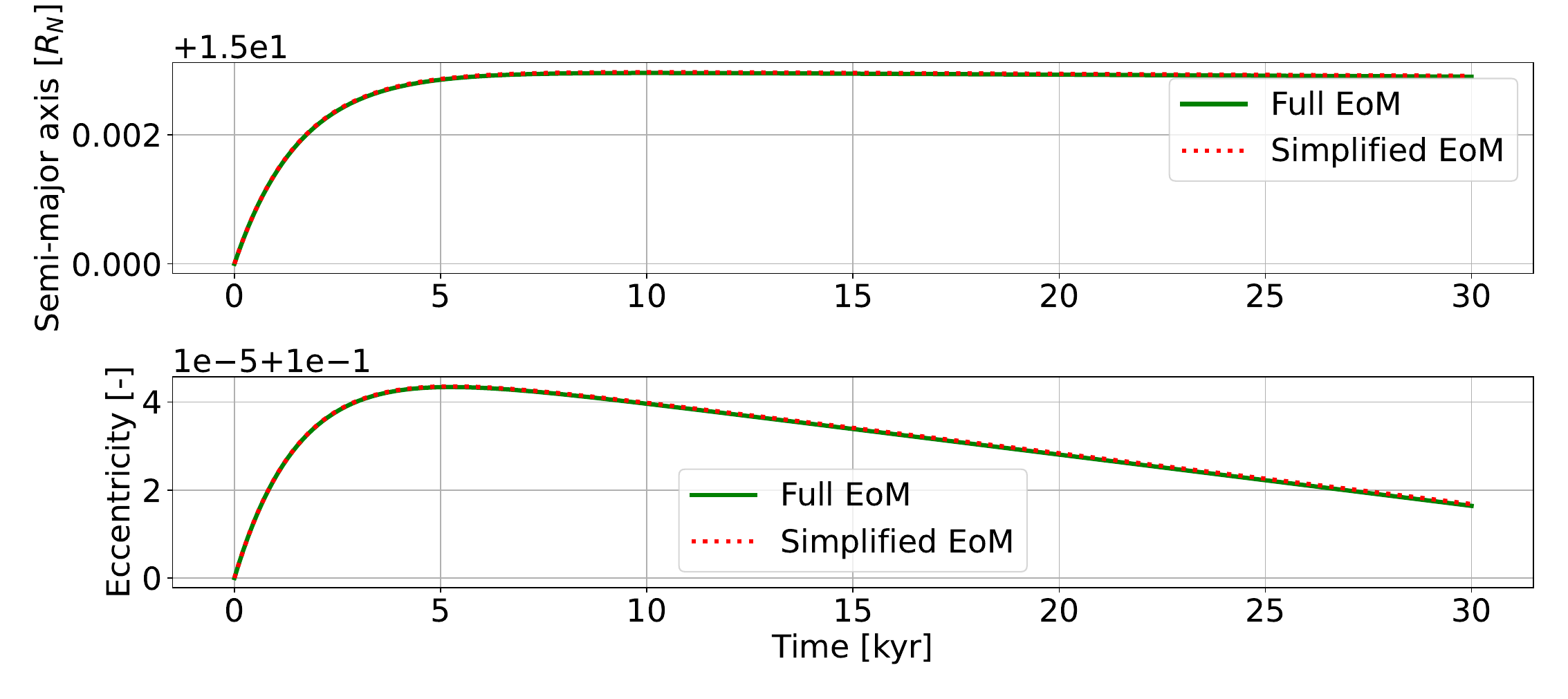}
    \caption{Short-period evolution of the semi-major axis and eccentricity for a Triton-analogue as propagated in the fully coupled system: both the results using the full equations of motion as well as those using the simplified equations of motion given by \citet{Boue2019TidalElements} are presented.}
    \label{fig:short_period_a_e}
\end{figure}

As the simplified equation for the obliquity is ill-conditioned for numerical integration with a variable timestep integrator when it starts in a spin state that is not orbit-normal, we check separately that the short-period coupled motion in the inclination and obliquity and the long-period coupled motion in the semi-major axis and eccentricity are computed correctly: we first check the short-period coupled motion when starting in a perturbed state, and then separately perform a long-period run starting in equilibrium positions. To obtain results representative for Triton, we now position the satellite in an orbit at 15 Neptune radii with an initial eccentricity of $e=0.1$ and inclination about Neptune of $i_N=5^{\circ}$, and give it an initial spin period of $8$ h (representative of that typically found for trans-Neptunian objects: see e.g. \citealt{Perna2009RotationsObjects, Thirouin2014RotationalBelt}) and obliquity of $5^{\circ}$: an additional retrograde run is performed using an obliquity of $175^{\circ}$ to show that the equations of motion are implemented as intended in the retrograde case, too. The resulting evolution for the prograde case is illustrated in Figs.~\ref{fig:short_period_angles} and \ref{fig:short_period_a_e}; the retrograde case is nearly fully identical, and so is not illustrated. In either scenario, the agreement between the full and simplified equations of motion is excellent and well within the error bounds provided to the variable timestep integrator. While the evolution in eccentricity and semi-major axis is negligible, it is nonetheless shown to illustrate that the agreement even for such small changes is satisfactory. The obliquity and rotation rate of Neptune do not evolve appreciably in this time-frame, and so their evolution is not shown.

\subsection{The long-period evolution of the semi-major axis and eccentricity}
\begin{figure}
    \centering
    \includegraphics[width=1\linewidth]{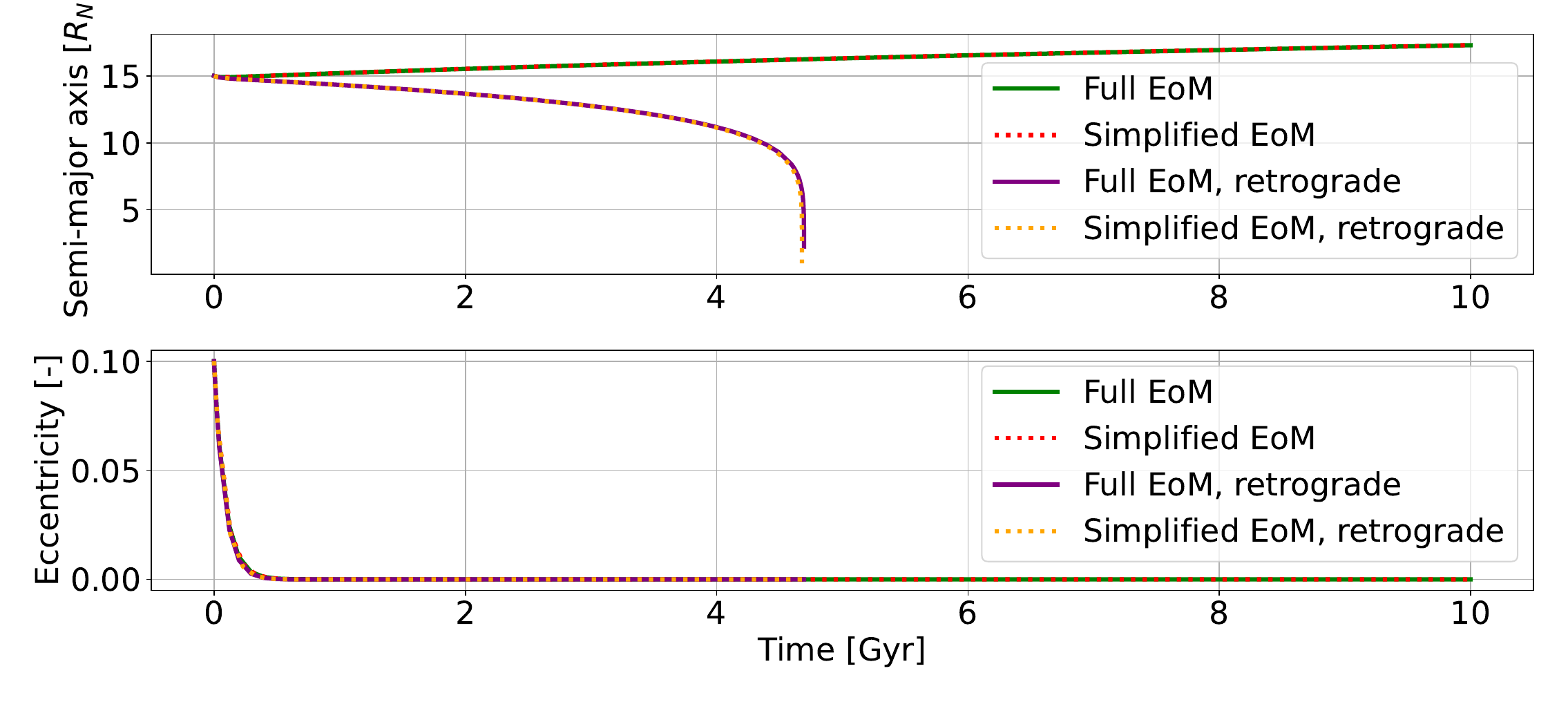}
    \caption{Long-period evolution of the semi-major axis and eccentricity for a Triton-analogue as propagated in the fully coupled system: both the results using the full equations of motion as well as those using the simplified equations of motion given by \citet{Boue2019TidalElements} are presented.}
    \label{fig:long_period_a_e}
\end{figure}

\begin{figure}
    \centering
    \includegraphics[width=1\linewidth]{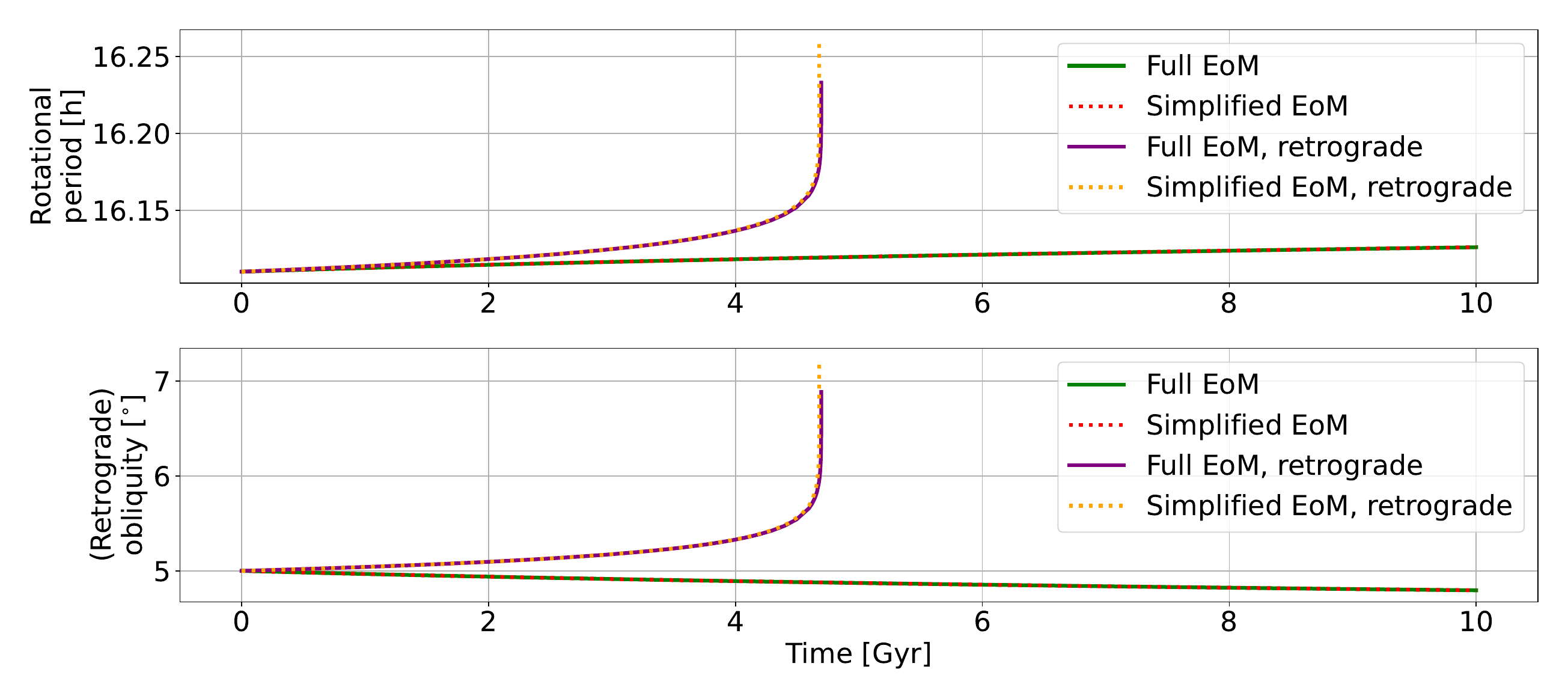}
    \caption{Long-period evolution of the rotation rate and obliquity of the primary of the Triton-analogue, as propagated in the fully coupled system: both the results using the full equations of motion as well as those using the simplified equations of motion given by \citet{Boue2019TidalElements} are presented. Note that the obliquity for the primary in the retrograde case is presented as the retrograde obliquity $i'=180^{\circ}-i$.}
    \label{fig:long_period_obliquity_rotation_primary}
\end{figure}

As shown, the short-period evolution towards equilibrium values of the obliquity and rotational rate of the secondary is properly represented by these equations of motion. After despinning, the integration module therefore instead computes the rotational rate of the secondary as its equilibrium value, comparable to the method employed by \citet{Walterova2020ThermalExoplanets}. From Fig.~\ref{fig:long_period_a_e} it is clear that our full coupled equations of motion properly represent the semi-major axis and eccentricity-evolution over long timescales, both in the prograde and retrograde case. Over these long timescales, the rotation rate and obliquity of the primary also start to evolve significantly: their evolution is shown in Fig.~\ref{fig:long_period_obliquity_rotation_primary}. It is clear that this long-period evolution for a Triton-like object is thus also properly represented by our implementation of the equations of motion.

\bibliographystyle{layout/aasjournal}
\bibliography{mendeley}
\end{document}